\documentclass[aps,prd,showpacs,superscriptaddress,nofootinbib,preprintnumbers,notitlepage]{revtex4-1} 

\makeatletter
\def\p@subsection{}
\makeatother

\usepackage{color,graphicx,amsmath,amssymb,lipsum}

\definecolor{jade}{rgb}{0.0, 0.66, 0.42}
\definecolor{coralred}{rgb}{1.0, 0.25, 0.25}

\usepackage{enumitem}
\usepackage[colorlinks=true,citecolor=jade,linkcolor=jade,urlcolor=jade, backref=false,pdfborder={0 0 0}]{hyperref}

\usepackage[normalem]{ulem}
\usepackage[T1]{fontenc}
\usepackage[utf8]{inputenc} 
\usepackage{mathtools}
\usepackage{wrapfig}
\usepackage{siunitx}
\usepackage{placeins}

\usepackage{booktabs}
\renewcommand{\arraystretch}{1.5}
\definecolor{vlightgray}{gray}{0.9}
\usepackage{float}
\usepackage{ulem}
\usepackage{diagbox}
\usepackage{bm}
\usepackage{bbm}
\renewcommand\thesection{\arabic{section}}

\newcommand{\be}{\begin{equation}}
\newcommand{\ee}{\end{equation}}
\newcommand{\beqa}{\begin{eqnarray}}
\newcommand{\eeqa}{\end{eqnarray}}

\renewcommand\k{ {\bf k}}

\newcommand{\fnl}{f_{\rm NL}}

\newcommand{\hMpc}{h \text{Mpc}^{-1}}
\newcommand{\q}{{\bf q}}

\newcommand{\bseq}{\begin{subequations}}
\newcommand{\eseq}{\end{subequations}}

\renewcommand{\ln}{\mathop{\rm ln}\nolimits}

\newcommand{\Tr}{{\rm Tr}}

\def\gsim{\raise0.3ex\hbox{$\;>$\kern-0.75em\raise-1.1ex\hbox{$\sim\;$}}}
\def\lsim{\raise0.3ex\hbox{$\;<$\kern-0.75em\raise-1.1ex\hbox{$\sim\;$}}}

\def\beqn#1{\begin{equation}\label{#1}}
\def\eeqn{\end{equation}}

\def\beqa#1{\begin{eqnarray}\label{#1}}
\def\eeqa{\end{eqnarray}}

\newcommand{\collA}{{\dot{\pi}^2\sigma}}
\newcommand{\collB}{{({\bm\nabla}\pi)^2\sigma}}
\newcommand{\equil}{{\rm equil}}
\newcommand{\ortho}{{\rm ortho}}

\makeatletter
\newlength{\apb@width}
\newcommand{\autoparbox}[2][c]{\settowidth{\apb@width}{#2}\parbox[#1]{\apb@width}{#2}}
\newcommand{\includegraphicsbox}[2][]{\autoparbox{\includegraphics[#1]{#2}}}
\makeatother

\def\Z2{$\mathcal{Z_2}$}

\newcommand {\ignore}[1]{}

\renewcommand{\arraystretch}{1.3}

\begin{document}

\preprint{RBI-ThPhys-2024-21}
\preprint{MIT-CTP/5698}

\title{\texorpdfstring{BOSS Constraints on Massive Particles during Inflation: \\
The Cosmological Collider in Action}{BOSS Constraints on Massive Particles during Inflation: 
The Cosmological Collider in Action}}

\author{Giovanni Cabass}
\email{gcabass@irb.hr}
\affiliation{Division of Theoretical Physics, Ru{\dj}er Bo{\v s}kovi{\'c} Institute, Zagreb HR-10000, Croatia}

\author{Oliver H.\,E. Philcox}
\email{ohep2@cantab.ac.uk}
\affiliation{Center for Theoretical Physics, Columbia University, New York, NY 10027, USA}
\affiliation{Simons Society of Fellows, Simons Foundation, New York, NY 10010, USA}

\author{Mikhail M. Ivanov}
\email{ivanov99@mit.edu}
\affiliation{Center for Theoretical Physics, Massachusetts Institute of Technology, Cambridge, MA 02139, USA}

\author{Kazuyuki Akitsu}
\affiliation{Theory Center, Institute of Particle and Nuclear Studies, High Energy Accelerator Research Organization (KEK), Tsukuba, Ibaraki 305-0801, Japan}

\author{Shi-Fan Chen}
\affiliation{School of Natural Sciences, Institute for Advanced Study, 1 Einstein Drive, Princeton, NJ 08540, USA}

\author{Marko Simonovi\'c}
\affiliation{Dipartimento di Fisica e Astronomia, Universit{\`a} di Firenze; Via G. Sansone 1; I-50019 Sesto Fiorentino, Italy}
\affiliation{INFN, Sezione di Firenze; Via G. Sansone 1; I-50019 Sesto Fiorentino, Italy}

\author{Matias Zaldarriaga}
\affiliation{School of Natural Sciences, Institute for Advanced Study, 1 Einstein Drive, Princeton, NJ 08540, USA}

\begin{abstract} 
\noindent 
Massive particles leave imprints on primordial non-Gaussianity via couplings to the inflaton, even despite their exponential dilution during inflation: practically, the Universe acts as a Cosmological Collider. We present the first dedicated search for spin-zero particles using BOSS redshift-space galaxy power spectrum and bispectrum multipoles, as well as \emph{Planck} CMB non-Gaussianity data. We demonstrate that some Cosmological Collider models are well approximated by the standard equilateral and orthogonal parametrization; assuming negligible inflaton self-interactions, this facilitates us translating \textit{Planck} non-Gaussianity constraints into bounds on Collider models. Many models have signatures that are not degenerate with equilateral and orthogonal non-Gaussianity and thus require dedicated searches. Here, we constrain such models using BOSS three-dimensional redshift-space galaxy clustering data, focusing on spin-zero particles in the principal series (\emph{i.e.}~with mass $m\geq 3H/2$) and constraining their couplings to the inflaton at varying speed and mass, marginalizing over the unknown inflaton self-interactions. This is made possible through an improvement in Cosmological Bootstrap techniques and the combination of perturbation theory and halo occupation distribution models for galaxy clustering. Our work sets the standard for inflationary spectroscopy with cosmological observations, providing the ultimate link between physics on the largest and smallest scales. 
\end{abstract}

\maketitle

\section{Introduction} 
\label{sec:intro}

\noindent With a Hubble scale as high as $10^{14}\,{\rm GeV}$, inflation may be the highest-energy process observable in the Universe. As such, the primordial signatures imprinted on the large-scale distribution of matter makes performing cosmological observations effectively equivalent to running a particle accelerator at energies many orders of magnitude above those of present-day experiments. 

Searches for hitherto unknown massive particles are perhaps \textit{la ragione d'essere} of particle accelerators. In this instance, one proceeds by carefully measuring the cross section for the scattering of light degrees of freedom. As the center-of-mass-energy approaches the mass of the new particle, a resonance appears in the cross section: a shape that is not degenerate with local operators in the low-energy theory. In cosmology, the situation is morally similar: integrating out particles with masses greater than the Hubble scale ($m>H$) yields self-interactions in the (light) inflaton field, and correspondingly primordial non-Gaussianity (PNG) well described by the single-field Effective Field Theory (EFT) of Inflation \cite{Creminelli:2006xe,Cheung:2007st,Cabass:2022avo}. However, such particles can also be spontaneously produced in the time-dependent inflationary background: this leads to features in the squeezed limit of the inflaton bispectrum that cannot be mimicked by spatially-local inflaton interactions. Though the massive particles decay exponentially fast during inflation, they can still, at least in principle, be observed via this smoking gun in the non-Gaussianity of the initial conditions of the primordial curvature perturbation $\zeta$ (which is set by the inflaton fluctuations). A rich field of study has been developed for characterizing this kind of PNG and searching for it in cosmological data, generally known as ``Cosmological Collider Physics'' \cite{Ghosh:2014kba,Chen:2009we,Chen:2009zp,Chen:2010xka,Byrnes:2010xd,Achucarro:2010da,Baumann:2011nk,Chen:2012ge,Noumi:2012vr,Assassi:2012zq,Arkani-Hamed:2015bza,Lee:2016vti,Kehagias:2017cym,An:2017hlx,Kumar:2017ecc,Bordin:2018pca,Kumar:2018jxz,Goon:2018fyu,Goon:2018fyu,Hook:2019zxa,Kumar:2019ebj,Liu:2019fag,Wang:2019gbi,Bodas:2020yho,Aoki:2020zbj,Lu:2021wxu,Pinol:2021aun,Reece:2022soh,MoradinezhadDizgah:2017szk,MoradinezhadDizgah:2018ssw,MoradinezhadDizgah:2019xun}. 

On the theory side, much effort has been devoted to developing tools to compute these PNG shapes from first principles. This goal (amongst loftier brethren such as an understanding of Quantum Gravity in de Sitter spacetime) is the goal of the ``Cosmological Bootstrap'' program~\cite{Arkani-Hamed:2018kmz,Goodhew:2020hob,Pajer:2020wxk,Cabass:2021fnw,Melville:2021lst,Jazayeri:2021fvk,Baumann:2021fxj,Bonifacio:2021azc,Hogervorst:2021uvp,DiPietro:2021sjt,Baumann:2022jpr,Cabass:2022rhr}. In this framework, the PNG shapes arising from massive particles are obtained as solution to boundary differential equations that are consequence of local time evolution in the bulk of de Sitter spacetime, and whose singularity structure is fixed by unitarity and the choice of vacuum. A great deal of progress has been recently made in this direction \citep[e.g.,][]{Baumann:2019oyu,Baumann:2020dch,Pimentel:2022fsc,Jazayeri:2022kjy,Wang:2022eop}. 

Where should we search for such massive particle signatures? The two most popular options are the large-scale structure (LSS) of the Universe (seen through the lens of galaxy clustering) and the anisotropies of the Cosmic Microwave Background (CMB). A three-point function of $\zeta$ leaves a direct imprint in the three-point functions of CMB temperature and polarization; to date, these observables have been the main source of constraints on PNG \cite{Planck:2019kim}. However, the (primary) CMB is limited by the fact that the number of accessible Fourier modes is quickly saturating: large scales are dominated by cosmic variance, and, for short scales, diffusion damping washes away anisotropies and secondary effects such as Sunyaev-Zel'dovich distortions dominate \citep{Kalaja:2020mkq}. Upcoming galaxy surveys will contain orders of magnitude more cosmological information than present datasets \citep[e.g.,][]{Chudaykin:2019ock,Sailer:2021yzm,MoradinezhadDizgah:2020whw}, and will soon surpass that of the primary CMB anisotropies, counting the number of accessible Fourier modes (recalling that for $3$D surveys the number of modes scales with the cube of the shortest scale that can be accessed in the data). As such, LSS promises to become the leading source of constraining power on PNG \cite{Alvarez:2014vva,Meerburg:2019qqi,Biagetti:2019bnp,Achucarro:2022qrl,Green:2023uyz,Chen:2024bdg}. In galaxy clustering, the principal effect from massive particles is to imprint a specific shape dependence in the galaxy bispectrum, which also modulates the power spectrum through loop corrections. These effects can be constrained by a systematic analysis based on the consistently-modeled power spectrum and bispectrum data using the effective field theory approach to perturbation theory (PT, see~e.g.~\cite{
Baumann:2010tm,
Carrasco:2012cv,
Porto:2013qua,
Blas:2013bpa,
Carrasco:2013sva,
Pajer:2013jj,
Mercolli:2013bsa,
Carrasco:2013mua,
Carroll:2013oxa,
Baldauf:2014qfa,
Angulo:2014tfa,
Senatore:2014vja,
Lewandowski:2014rca,
Senatore:2014via,
Mirbabayi:2014zca,
Senatore:2014eva,
Assassi:2014fva,
Angulo:2015eqa,
Abolhasani:2015mra,
Assassi:2015jqa,
Foreman:2015uva,
Assassi:2015fma,
Blas:2015qsi,
Baldauf:2015xfa,
Lewandowski:2015ziq,
Foreman:2015lca,
Vlah:2015zda,
Vlah:2015sea,
Baldauf:2015aha,
Bertolini:2015fya,
Lazeyras:2015lgp,
Baldauf:2015tla,
Baldauf:2015zga,
Desjacques:2016bnm,
Fasiello:2016qpn,
Blas:2016sfa,
Bertolini:2016bmt,
Vlah:2016bcl,
Nadler:2017qto,
Lewandowski:2017kes,
Senatore:2017pbn,
Lazeyras:2017hxw,
Senatore:2017hyk,
Ivanov:2018gjr,
Lewandowski:2018ywf,
Vlah:2018ygt,
MoradinezhadDizgah:2018ssw,
Eggemeier:2018qae,
Konstandin:2019bay,
Lazeyras:2019dcx,
Abidi:2018eyd,
Cabass:2018hum,
Schmittfull:2018yuk,
Schmidt:2018bkr,
deBelsunce:2018xtd,
Desjacques:2018pfv,
Vlah:2019byq,
Elsner:2019rql,
Beutler:2019ojk,
Ivanov:2019pdj,
DAmico:2019fhj,
Cabass:2019lqx,
Vasudevan:2019ewf,
Schmidt:2020ovm,
Cabass:2020nwf,
Schmidt:2020viy,
Schmidt:2020tao,
Nguyen:2020hxe,
Cabass:2020jqo,
Fujita:2020xtd,
Nishimichi:2020tvu,
MoradinezhadDizgah:2020whw,
Chudaykin:2020hbf,
Chen:2020zjt,
Chen:2020fxs,
Donath:2020abv,
Nishimichi:2020tvu,
Braganca:2020nhv,
Steele:2020tak,
Schmittfull:2020trd,
Chen:2020ckc,
DAmico:2021rdb,
Steele:2021lnz,
Ivanov:2021kcd,
Baldauf:2021zlt,
Cabass:2022avo,
Ivanov:2022mrd} and references therein). 

In this work, we carry out the first search for inflationary massive \emph{scalars} (with mass in the de Sitter principal series $m\geq 3H/2$, $H$ being the Hubble rate during inflation) by searching for the non-Gaussianities generated by their coupling to the inflaton, as a function of their mass and speed of propagation. Such an approach can be used to answer multiple different questions, with the first motivated by the following discussion. As above, the ``ideal'' parallel with Earth-based collider physics would be that the ``EFT low-energy background'' has already been detected and we are now hunting for resonances on top of this background, keeping in mind that large inflaton self-interactions naturally accompany sizable Cosmological Collider non-Gaussianities \cite{Lee:2016vti,MoradinezhadDizgah:2017szk}. Unfortunately we are not yet in this scenario, since there has not been a detection of primordial non-Gaussianity described by the equilateral and orthogonal templates $\smash{\fnl^{\rm equil}}$ and $\smash{\fnl^{\rm ortho}}$ \cite{Senatore:2009gt}. Given that such templates have already been constrained (most notably by \emph{Planck} \cite{Planck:2019kim}), one can indirectly assess the coupling of the inflaton to massive particles through a translation of the public equilateral and orthogonal bounds (making the important assumption that \textit{bona-fide} inflaton self-interactions are a subleading source of primordial non-Gaussianity). 

The second approach is to carry out a full combined analysis of $\smash{\fnl^{\rm equil}}$, $\smash{\fnl^{\rm ortho}}$ and the Cosmological Collider PNG, which has not yet been attempted in cosmological data analysis. From a particle physics point of view this allows us to assess how degenerate the signature from massive particle exchange is with equilateral and orthogonal non-Gaussianities in current data: we are searching for particles by looking at resonances on top of the low-energy EFT of Inflation background. We will carry out this study using the publicly available state-of-the-art redshift clustering data from the Baryon Oscillation Spectroscopic Survey (BOSS), marking an important step towards performing inflationary spectroscopy with large-scale structure observations.\footnote{Constraints on parity-violating interactions with massive spinning particles from BOSS data \cite{Cahn:2021ltp,Hou:2022wfj,Philcox:2022hkh} were derived in \cite{Cabass:2022oap} (see also \citep{Philcox:2023ffy} for CMB bounds), whilst constraints on single-field inflation and multi-field inflation (with negligible mass) were derived in \cite{Cabass:2022wjy,DAmico:2022gki,Cabass:2022ymb}.} 

An important part of this work is the development of improved Cosmological Bootstrap techniques that allow for accurate prediction of the primordial bispectrum over the full range of scales covered by BOSS and \emph{Planck}. Important work in the context of the Cosmological Collider and Cosmological Bootstrap has been carried out in \cite{Xianyu:2023ytd,Qin:2023ejc,Pinol:2023oux,Werth:2023pfl,Pinol:2023oux,Werth:2023pfl,Jazayeri:2023xcj,Chakraborty:2023eoq,Werth:2024aui}. 
To connect these to observations, we require both a theoretical model for the galaxy bispectrum~\cite{Ivanov:2021kcd,Philcox:2021kcw,Ivanov:2023qzb}~(see also~\cite{Scoccimarro:2000sn,Sefusatti:2006pa,Baldauf:2014qfa,Angulo:2014tfa,Eggemeier:2018qae,Ivanov:2019pdj,DAmico:2019fhj,Oddo:2019run,Eggemeier:2021cam,Alkhanishvili:2021pvy,Oddo:2021iwq,Chen:2021wdi,Baldauf:2021zlt,Xu:2021rwg,MoradinezhadDizgah:2017szk,MoradinezhadDizgah:2018ssw}), as well as an estimator for the spectra \cite{Philcox:2020vbm,Philcox:2021ukg} that robustly accounts for the effects of the survey window function. These ingredients have facilitated previous $\Lambda$CDM constraints to be wrought with the redshift-space bispectrum multipoles \cite{Ivanov:2023qzb}, and have yielded the first constraints on the Effective Field Theory of Inflation and massless fields from galaxy clustering \cite{Cabass:2022wjy,DAmico:2022gki,Cabass:2022ymb}. Finally, constraints on primordial non-Gaussianity from large-scale structure are affected by the marginalization over the PT bias coefficients, encoding the uncertainty over the short-scale dynamics. In this work we use newly-derived priors on these coefficients obtained by matching to halo-occuptation distribution (HOD) models of galaxy formation to mitigate these uncertainties (as described in \cite{future_kaz}, see also \citep{Ivanov:2024hgq,Ivanov:2024xgb}). 
Importantly, we do not include the loop corrections to the galaxy bispectrum in this work. As shown in \cite{Philcox:2022frc}, the gain on constraining power on PNG is negligible (assuming broad 
conservative 
priors), and it comes at the cost of many more free PT coefficients (necessary to absorb the UV dependence of the bispectrum loop integrals) that make the analysis prior-dependent. We find it more robust to use the tree-level bispectrum in combination with priors on bias coefficients motivated by galaxy formation simulations. 

\vskip 4 pt

The remainder of this paper is structured as follows. \S\ref{sec:cosmo_collider_observables} discusses the observables and ``microphysical'' parameters that we wish to constrain (or hopefully one day detect) in the context of the Cosmological Collider. In \S\ref{sec:planck_equivalent} we translate \emph{Planck} bounds on $\smash{\fnl^{\rm equil}}$ and $\smash{\fnl^{\rm ortho}}$ to constraints on the couplings of the massive particle $\sigma$ to the inflaton marginalizing over the mass $m$ and speed of propagation $c_\sigma$, under the assumption of zero inflaton self-interactions. In \S\ref{sec:boss_constraints} we carry out a full BOSS analysis using the one-loop power spectrum and tree-level bispectrum multipoles in conjunction with galaxy-formation priors on bias parameters. We discuss the impact of marginalizing over $\smash{\fnl^{\rm equil}}$ and $\smash{\fnl^{\rm ortho}}$, practically assessing the price paid if inflaton self-interactions are not known. We also discuss the case of the ``DBI Cosmological Collider'', where inflaton interactions and their couplings to $\sigma$ are constrained by a higher-dimensional boost symmetry (see e.g.~\cite{Bonifacio:2019rpv}). We conclude in \S\ref{sec:conclusions}. Appendix \ref{app:collider_solution} contains a description of the improved solution of Cosmological Bootstrap equations used to derive the PNG shapes from massive particle exchange, whilst Appendix \ref{app:HOD} contains a brief description of the connection between galaxy-formation modeling and PT bias parameters (which is expanded upon in \cite{future_kaz}), and Appendix \ref{app:SVD} contains a Principal Component Analysis study, based on Ref.~\cite{Philcox:2020zyp}, assessing the level of degeneracy between the Cosmological Collider bispectra and the equilateral and orthogonal templates. 

{\emph{Note added.} Shortly after publication of this paper on arXiv, the work \cite{Sohn:2024xzd} appeared, which carries out an analysis of the Cosmological Collider templates (both for scalar particles in the principal series, but also beyond this case), using 
the \emph{Planck} CMB data. This is a very important work, which together with the present one represents the first constraints on the Cosmological Collider scenario using cosmological data.}

\section{Observables in the Cosmological Collider} 
\label{sec:cosmo_collider_observables}

\subsection{The Inflationary Action}

\noindent At leading order in the Effective Field Theory expansion, the action for a massive scalar $\sigma$ and the inflaton perturbation $\pi$ takes the form \cite{Lee:2016vti,Pimentel:2022fsc,Jazayeri:2022kjy} 
\begin{equation}
\label{full_action}
    \begin{split}
        S &= \frac{f^4_\pi}{2}\int{\rm d}^4x\,a^3\bigg(\dot{\pi}^2 - c^2_{\rm s}\frac{({\bm\nabla}\pi)^2}{a^2}\bigg) + \frac{f^4_\pi}{2}(c^2_{\rm s}-1)\int{\rm d}^4x\,a^3\bigg[\frac{\dot{\pi}({\bm\nabla}\pi)^2}{a^2} - \bigg(1+\frac{2\tilde{c}_3}{3c^2_{\rm s}}\bigg)\dot{\pi}^3\bigg] \\
        &\;\;\;\; + \frac{1}{2}\int{\rm d}^4x\,a^3\bigg(\dot{\sigma}^2 - \frac{({\bm\nabla}\sigma)^2}{a^2} - m^2\sigma^2\bigg) + \int{\rm d}^4x\,a^3\,\bigg[{-2\omega_0^3}\dot{\pi}+(4\tilde{\omega}_0^3-{\omega}_0^3)\dot{\pi}^2+\omega_0^3\,\frac{({\bm\nabla}\pi)^2}{a^2}\bigg]\sigma \\
        &\;\;\;\; + \frac{1-c^2_\sigma}{2c^2_\sigma}\int{\rm d}^4x\,a^3\,\dot{\sigma}^2\,\,, 
    \end{split}
\end{equation} 
where $a$ is the scale factor. In addition, we have also the action for the graviton, with propagation speed $c_{\rm \gamma}=1$ \cite{Creminelli:2014wna,Bordin:2017hal,Bordin:2020eui}. Hence, in the following we consider the propagation speeds for the inflaton and massive scalar $c_{\rm s},c_\sigma \leq 1$. In Eq.~\eqref{full_action}, the parameter $f_\pi$, along with $c_{\rm s}$, sets the normalization of the power spectrum for the metric curvature perturbation $\zeta$, related to the inflaton perturbation $\pi$ via $\zeta={-H}\pi$. More precisely, we have 
\be
\label{PS}
\Delta^2_\zeta = \frac{H^4}{2f^4_\pi c^3_{\rm s}}\,\,, 
\ee
where the amplitude $\Delta^2_\zeta$ of the primordial power spectrum is defined by $k^3P_\zeta(k)=\Delta^2_\zeta (k/k_\ast)^{n_{\rm s}-1}$. From \emph{Planck} data we have $\Delta^2_\zeta\approx 4.1\times 10^{-8}$, $n_{\rm s}\approx 0.96$~\cite{Aghanim:2018eyx} for a pivot scale $k_\ast=0.05\,{\rm Mpc}^{-1}$. The remaining terms in the action of Eq.~\eqref{full_action} are, respectively, the cubic self-interactions of the inflaton, the quadratic action for $\sigma$, the linear mixing and leading interactions between $\pi$ and $\sigma$.\footnote{Notice that the relation between the $\pi\sigma$ mixing and the $\pi\pi\sigma$ interaction is dictated by the nonlinear realization of Lorentz boosts.} The last term gives $\sigma$ a propagation speed different from $1$. It arises from the unitary-gauge operator $({-\alpha^2}/2)\,n^\mu n^\nu\nabla_\mu\sigma\nabla_\nu\sigma$, where $\alpha^2 = (c^2_\sigma-1)/c^2_\sigma$ and $n^\nu$ is the unit normal vector to the constant-inflaton hypersurfaces \cite{Cabass:2022avo}.

In order to connect with the existing literature, we rescale spatial coordinates ${\bf x}\to c_{\sigma}{\bf x}$ to measure all velocities in terms of $c_\sigma$. 
After canonically-normalizing $\smash{\sigma\to\sigma/\sqrt{c_\sigma}}$ we find the following action\footnote{The different powers of $c_\sigma$ appearing in the $\pi\sigma$ mixing and $\pi\pi\sigma$ interactions are due to the fact that one of the three operators carries spatial derivatives whilst the other two do not.} 
\begin{equation}
\label{rescaled}
    \begin{split}
        S &= \frac{f^4_\pi c^3_\sigma}{2}\int{\rm d}^4x\,a^3\bigg(\dot{\pi}^2 - \frac{c^2_{\rm s}}{c^2_\sigma}\frac{({\bm\nabla}\pi)^2}{a^2}\bigg) + \frac{f^4_\pi c^3_\sigma}{2}(c^2_{\rm s}-1)\int{\rm d}^4x\,a^3\bigg[\frac{\dot{\pi}({\bm\nabla}\pi)^2}{c^2_\sigma a^2} - \bigg(1+\frac{2\tilde{c}_3}{3c^2_{\rm s}}\bigg)\dot{\pi}^3\bigg] \\
        &\;\;\;\; + \frac{1}{2}\int{\rm d}^4x\,a^3\bigg(\dot{\sigma}^2 - \frac{({\bm\nabla}\sigma)^2}{a^2} - c^2_\sigma m^2\sigma^2\bigg) + \int{\rm d}^4x\,a^3\,\bigg[{-2c^{\frac52}_\sigma\omega_0^3}\dot{\pi}+(4c^{\frac52}_\sigma\tilde{\omega}_0^3-c^{\frac52}_\sigma{\omega}_0^3)\dot{\pi}^2+c_\sigma^{\frac12}\omega_0^3\,\frac{({\bm\nabla}\pi)^2}{a^2}\bigg]\sigma\,\,. 
    \end{split}
\end{equation}

\subsection{Correlation Functions}

\noindent The goal of cosmological observations is to measure the parameters in the action \eqref{rescaled} (here, $f_\pi$, $c_{\rm s}$, $\tilde{c}_3$, $m$, $c_{\sigma}$, $\omega_0$ and $\tilde{\omega}_0$) by measuring correlation functions of $\zeta$. Firstly, it is customary to parameterize the bispectrum of $\zeta$ as 
\begin{equation} 
\label{eq:B_primordial_master_definition}
B_\zeta(k_1,k_2,k_3) = \frac{18}{5}f_{\rm NL}\Delta^4_\zeta\frac{{\cal S}(k_1,k_2,k_3)}{k_1^2k_2^2k_3^2}\,\,, 
\end{equation}
where $k_1,k_2,k_3$ are the moduli of the three momenta $\k_1,\k_2,\k_3$ and $\cal S$ is a dimensionless shape function normalized to $1$ in the equilateral configuration $k_1=k_2=k_3$. 
As is well known, the $\pi\pi\pi$ self-interaction terms in the Lagrangian give rise to a superposition of smooth bispectra that are well captured by the equilateral and orthogonal templates \cite{Babich:2004gb,Senatore:2009gt}, with amplitudes $f_{\rm NL}$ that depend on $c_{\rm s}$ and $\tilde{c}_3$. The mass and speed of $\sigma$ can be measured via the bispectrum coming from the diagram 
\be
\label{diagram}
\raisebox{-0.0cm}{\includegraphicsbox[scale=0.5]{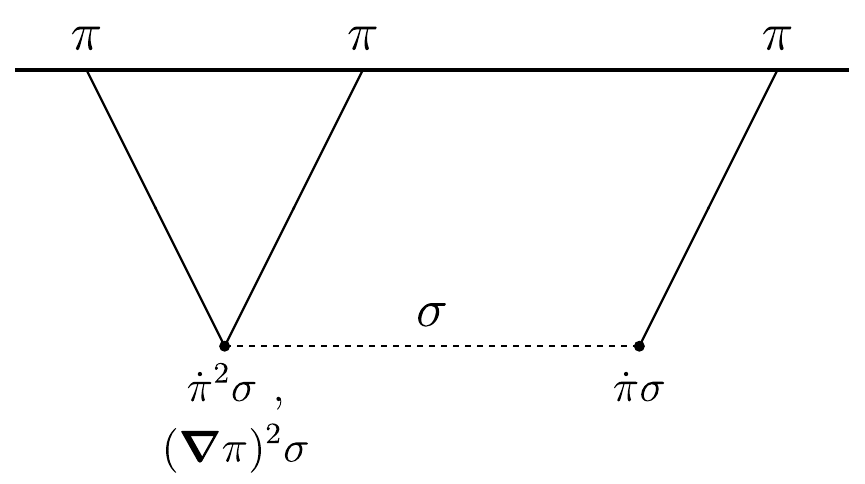}}
\ee
generated by the quadratic mixing between $\pi$ and $\sigma$ and the two $\pi\pi\sigma$ cubic vertices in \eqref{rescaled}. More precisely, the main feature is that the shape resulting from this diagram is composed of two distinct contributions: a piece non-analytic in momenta (the signature of spontaneous particle production) and one analytic in momenta (representing the part of the diagram that, in the limit of a very heavy $\sigma$, is degenerate with the self-interactions of $\pi$). From the peculiar non-analytic dependence on momenta one can, {in principle}, measure $c_\sigma m$ and $c_{\rm s}/c_\sigma$, whilst the amplitude of the signal depends on $\omega_0$ and $\tilde{\omega}_0$. Notice that with only this diagram we are blind to the speed $c_\sigma$ itself: this parameter can in principle be measured via the interactions that accompany the term $\smash{\frac{1-c^2_\sigma}{2c^2_\sigma}\int{\rm d}^4x\,a^3\,\dot{\sigma}^2}$ in \eqref{full_action}. These give rise to the diagram 
\be
\label{diagram_c_sigma}
\raisebox{-0.0cm}{\includegraphicsbox[scale=0.5]{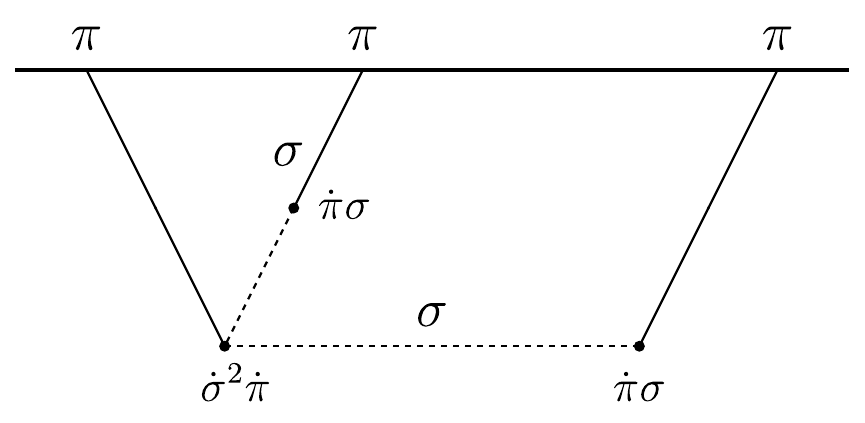}}\,\,, 
\ee
which does not come with new EFT coefficients due to the nonlinear realization of Lorentz boosts. 

Given the different goals of this paper, discussed in \S\ref{sec:intro}, it is worth to discuss the different contributions to ``$\smash{\fnl}$'' in more detail. First, let us discuss the bispectra arising from the diagram \eqref{diagram}. These take the form 
\begin{equation} 
\label{eq:B_primordial_master_definition-collider}
B^{\dot{\pi}^2\sigma}_\zeta(k_1,k_2,k_3) = {\cal B}^{\dot{\pi}^2\sigma}\,\frac{{\cal S}^{\collA}(k_1,k_2,k_3)}{(k_1k_2k_3)^2}\qquad\text{and}\qquad B^{({\bm\nabla}\pi)^2\sigma}_\zeta(k_1,k_2,k_3) = {\cal B}^{({\bm\nabla}\pi)^2\sigma}\,\frac{{\cal S}^{\collB}(k_1,k_2,k_3)}{(k_1k_2k_3)^2}\,\,. 
\end{equation} 
Here we notice that, whilst we use the same symbol $\cal S$ as in \eqref{eq:B_primordial_master_definition} to denote the shapes, $\smash{{\cal S}^{\collA}}$ and $\smash{{\cal S}^{\collB}}$ 
are not normalized to $1$ in the equilateral configuration. These functions 
are computed by acting with appropriate weight-shifting operators on the seed three-point function $\smash{{\cal I}(u)}$, which obeys the differential equation \cite{Pimentel:2022fsc,Jazayeri:2022kjy}
\be
\label{seed_ODE}
u^2(1-u^2){\cal I}'' - 2u^3{\cal I}' + \bigg(\mu^2+\frac{1}{4}\bigg){\cal I} = \frac{c^2_{\rm s}u/c^2_\sigma}{1+c_{\rm s}u/c_\sigma}\,\,, 
\ee
where 
\be
\mu\equiv\sqrt{\frac{c^2_\sigma m^2}{H^2} - \frac{9}{4}}\,\,. 
\ee
In order to have the shape function for all physical triangles the solution must be known for $u$ between $0$ and $c_\sigma/c_{\rm s}$, and the boundary conditions are fixed at $u=0$ \cite{Pimentel:2022fsc,Jazayeri:2022kjy}.\footnote{The variable $u$ can be related to the ``energy'' of the exchanged particle in the trispectrum, $\smash{s = |{\bf k}_1+{\bf k}_2| = |{\bf k}_3+{\bf k}_4|}$ in the soft limit $\smash{k_4\to 0}$.} 
From ${\cal I}(u)$ we obtain the shapes as 
\be
\label{shape_from_seed}
{\cal S}(k_1,k_2,k_3) = {\cal W}_{12}{\cal I}(u_{12}) + \text{$2$ perms.}\,\,, 
\ee
where $u_{12} = c_\sigma k_3/(c_{\rm s}k_{12})$, $k_{12}=k_1+k_2$, and
\begin{align}
{\cal W}_{12}^{\dot{\pi}^2\sigma} = \frac{c^2_\sigma k_1k_2}{c^2_{\rm s}}\frac{\partial^2}{\partial k_{12}^2}\,\,, \qquad {\cal W}_{12}^{({\bm\nabla}\pi)^2\sigma} = \frac{k_3^2-k_1^2-k_2^2}{k_1k_2}\bigg(1-k_1\frac{\partial}{\partial k_1}\bigg)\bigg(1-k_2\frac{\partial}{\partial k_2}\bigg)\,\,. 
\end{align}
In appendix \ref{app:collider_solution} we summarize the method used to solve \eqref{seed_ODE}. 
In this work we will scan over different choices of $c_{\rm s}/c_\sigma$ and $c_\sigma m$, and constrain the overall amplitudes of the PNG signals. We will focus on the range $\mu\in[0,5]$, $c_{\rm s}/c_\sigma\in\{0.1,1,10\}$, motivated by the following behavior of ${\cal S}$, which is present for both diagrams of Eq.~\eqref{diagram} \cite{Pimentel:2022fsc,Jazayeri:2022kjy}: 
\be
\label{squeezed_behavior_shape}
\text{${\cal S}(k_1,k_2,k_3)\sim
\bigg(\dfrac{k_1}{k_3}\bigg)^{\frac12}\cos\bigg(\mu\ln\dfrac{c_\sigma k_1}{c_{\rm s}k_3}\bigg)$
  for $k_1\ll \frac{c_{\rm s}}{c_\sigma}k_3$}\,\,. 
\ee 
For $c_{\rm s}\gg c_\sigma$ we are in the regime of the ``equilateral collider'' \cite{Pimentel:2022fsc}, where the non-analytic behavior of the shape due to spontaneous production of $\sigma$ can be seen also close to equilateral triangles $k_1\sim k_2\sim k_3$ and not only in the squeezed limit $k_1\ll k_2\sim k_3$. On the other hand, for supersonic $\sigma$, we are in the ``slow-speed collider'' regime \cite{Jazayeri:2022kjy}: for $c_{\rm s}\ll c_\sigma$ the oscillatory behavior of the shape is encountered only for extremely squeezed triangles, and the main signature is instead given by features around $k_1\sim c_{\rm s}k_3/c_\sigma$.\footnote{Whilst not a sign of spontaneous particle production in de Sitter spacetime, these features still cannot be captured by a finite number of local operators in the local Lagrangian of $\pi$ \cite{Gwyn:2012mw,Jazayeri:2022kjy}.} 

We are now in position to go back to the overall size of the PNG signal. Isolating the ratio of $\pi$ and $\sigma$ speeds where possible, we have \cite{Jazayeri:2022kjy}
\be
{\cal B}^{\dot{\pi}^2\sigma} = {-4 c^5_\sigma}\,\frac{c^4_{\rm s}}{c^4_\sigma}\,\frac{\Delta^6_\zeta\omega^3_0(\omega^3_0-4\tilde{\omega}^3_0)}{H^6}\qquad\text{and}\qquad{\cal B}^{({\bm\nabla}\pi)^2\sigma} = 2c^3_\sigma\frac{\Delta^6_\zeta\omega^6_0}{H^6}\,\,. 
\ee
It is useful to recast these amplitudes in terms of two parameters $\smash{\beta^{\dot{\pi}^2\sigma}}$ and $\smash{\beta^{({\bm\nabla}\pi)^2\sigma}}$. We define them as 
\be
\label{conversion}
\beta^{\dot{\pi}^2\sigma}\equiv c^5_\sigma\frac{c^3_{\rm s}}{c^3_\sigma}\frac{\Delta^2_\zeta \omega_0^3(\omega_0^3 - 4\tilde{\omega}_0^3)}{H^6} = {-\frac{9}{10}}\bigg(\frac{c_{\rm s}}{c_\sigma}\bigg)^{-1}\frac{\fnl^{\dot{\pi}^2\sigma}}{{\cal S}^{\dot{\pi}^2\sigma}(1,1,1)}\qquad\text{and}\qquad\beta^{({\bm\nabla}\pi)^2\sigma}\equiv c^3_\sigma\frac{c_{\rm s}}{c_\sigma}\frac{\Delta^2_\zeta\omega^6_0}{H^6} = \frac{9}{5}\frac{c_{\rm s}}{c_\sigma}\frac{\fnl^{({\bm\nabla}\pi)^2\sigma}}{{\cal S}^{({\bm\nabla}\pi)^2\sigma}(1,1,1)}\,\,,
\ee
such that the bispectrum can be written
\be
\label{equilateral_normalization}
    B_\zeta(k_1,k_2,k_3)\supset \frac{18}{5}\frac{f^{\dot{\pi}^2\sigma}_{\rm NL}\Delta^4_\zeta}{(k_1k_2k_3)^2}\frac{{\cal S}^{\dot{\pi}^2\sigma}(k_1,k_2,k_3)}{{\cal S}^{\dot{\pi}^2\sigma}(1,1,1)} + \frac{18}{5}\frac{f^{({\bm\nabla}\pi)^2\sigma}_{\rm NL}\Delta^4_\zeta}{(k_1k_2k_3)^2}\frac{{\cal S}^{({\bm\nabla}\pi)^2\sigma}(k_1,k_2,k_3)}{{\cal S}^{({\bm\nabla}\pi)^2\sigma}(1,1,1)}\,\,,
\ee
where the definition of $\smash{\fnl^\collA}$ and $\smash{\fnl^\collB}$ is manifest from comparing to Eq.~\eqref{eq:B_primordial_master_definition}. 
One can gain intuition on the meaning of these parameters by looking at the limit of very heavy $\sigma$. In this limit we effectively have $\smash{\sigma\sim {-2}\sqrt{c_\sigma}\,\omega_0^3\dot{\pi}/m^2}$ and the two $\pi\pi\sigma$ couplings reduce to 
\be
\int{\rm d}^4x\,a^3\,\frac{\omega_0^3}{H^2}\bigg[({-8}c^{5}_\sigma\tilde{\omega}_0^3+2c^{5}_\sigma{\omega}_0^3)\dot{\pi}^2-2c^3_\sigma\omega_0^3\,\frac{({\bm\nabla}\pi)^2}{a^2}\bigg]\dot{\pi}\,\,, 
\ee
where we have replaced $c_\sigma m$ with $H$ (we will come back to this choice in a moment). 
It is straightforward to check that the non-Gaussianities from these two interactions have typical $\smash{\fnl}$ (evaluated by taking the ratio of the cubic and quadratic Lagrangian at horizon crossing) scaling as $\smash{\beta^\collA}$ and $\smash{\beta^\collB}$ respectively. Hence the parameters $\smash{\beta^\collA}$ and $\smash{\beta^\collB}$ are defined as the effective $\smash{\fnl}$ that would result from integrating out a particle with $c_\sigma m\sim H$ (noting that $\smash{\beta^\collB}$ is forced to be positive). 
This tells us that an order-of-magnitude bound that we can put on $\smash{\beta^\collA}$ and $\smash{\beta^\collB}$ is
\be\label{eq: collider-expectation}
\beta^\collA\lesssim10^5\,\,,\qquad\beta^\collB\lesssim10^5\,\,.
\ee
The reader can compare the constraints that we will obtain in the next sections to these bounds (keeping in mind that the above are only a very rough estimate of the size of Cosmological Collider PNG: we refer to \cite{Jazayeri:2023xcj} for more accurate theory bounds). 

Let us now turn to inflaton self-interactions. As mentioned in \S\ref{sec:intro}, whilst a dedicated search for both Cosmological Collider PNG and single-field inflation PNG has not yet been performed, we can translate constraints on equilateral and orthogonal non-Gaussianity to bounds on linear combinations of the parameters controlling the size of the primordial bispectra, namely $\smash{(1-1/c^2_{\rm s})(\tilde{c}_3 + 3c^2_{\rm s}/2)}$, $\smash{1-1/c^2_{\rm s}}$ for the $\smash{\dot{\pi}^3}$ and $\smash{\dot{\pi}({\bm\nabla}\pi)^2}$ interactions in \eqref{rescaled}, and $\smash{\beta^\collA}$, $\smash{\beta^\collB}$ for the diagram \eqref{diagram}. However, given that these linear combinations are model-dependent (since the equilateral and orthogonal templates are just phenomenological templates) we prefer to proceed as follows. We will remain agnostic about the inflaton self-interactions and the bispectra of the curvature perturbation $\zeta$ they generate, and we will parametrize such effects as a superposition of the equilateral and orthogonal templates with sizes $\smash{f_{{\rm NL},\pi\pi\pi}^{\rm equil}}$ and $\smash{f_{{\rm NL},\pi\pi\pi}^{\rm ortho}}$ that are unrelated to the same microphysical parameters entering in the $\pi$ Lagrangian. Then, we have a contribution to $B_\zeta$ of the form 
\be
B_\zeta(k_1,k_2,k_3)\supset \frac{18}{5}f_{{\rm NL}}^{{\rm equil}}\Delta^4_\zeta\frac{{\cal S}^{\rm equil}(k_1,k_2,k_3)}{(k_1k_2k_3)^2} + \frac{18}{5}f_{{\rm NL}}^{{\rm ortho}}\Delta^4_\zeta\frac{{\cal S}^{\rm ortho}(k_1,k_2,k_3)}{(k_1k_2k_3)^2}\,\,, 
\ee
where we recall that \cite{Babich:2004gb,Senatore:2009gt}\footnote{Note that we use the orthogonal template from Appendix B of \cite{Senatore:2009gt}, 
which has a physically 
correct suppression 
in the squeezed limit $k_1\ll k_2\approx k_3$, where
it goes as $\mathcal{S}_{\rm ortho}\propto k_1/k_3$. 
It can be
contrasted with the commonly used approximate template that 
features an unphysical enhancement in that limit~\cite{Planck:2019kim}.
} 
\begin{align}
       {\cal S}^{\rm equil}(k_1,k_2,k_3) &= \bigg(\frac{k_1}{k_2} + \text{$5$ perms.}\bigg)- \bigg(\frac{k_1^2}{k_2k_3} + \text{$2$ perms.}\bigg) - 2\,\,, \\
       {\cal S}^{\rm ortho}(k_1,k_2,k_3) &= (1+p)\,\frac{\Delta}{e_3} - p\,\frac{\Gamma^3}{e_3^2}\,\,,
\end{align}
where $\frac{27}{p} = \frac{743}{7(20\pi^2-193)} - 21$, $\Delta = (k_T-2k_1)(k_T-2k_2)(k_T-2k_3)$, and 
\be 
\begin{split}
    k_T = k_1+k_2+k_3\,\,,\quad e_2=k_1k_2+k_2 k_3 + k_1 k_3\,\,,
    \quad e_3 = k_1 k_2 k_3\,\,,\quad 
    \Gamma = \frac{2}{3}e_2 - \frac{1}{3}(k_1^2+k_2^2+k_3^2)\,\,.
\end{split}
\ee 

In \S\ref{sec:planck_equivalent} we will fix both $\smash{f_{{\rm NL}}^{\rm equil}}$ and $\smash{f_{{\rm NL}}^{\rm ortho}}$ to zero and focus only on $\smash{\beta^\collA}$, $\smash{\beta^\collB}$. 
In \S\ref{sec:boss_constraints} we instead ask how well current data can constrain the Cosmological Collider PNG amplitudes, given that the self-interactions of the inflaton itself have not yet been pinned down. In this case, we will take both $\smash{f_{{\rm NL}}^{\rm equil}}$ and $\smash{f_{{\rm NL}}^{\rm ortho}}$ within the perturbative regime $f_{\rm NL}\Delta_\zeta\lesssim 1$, as in the analysis of \cite{Cabass:2022wjy}, and marginalize over them.

\subsection{The DBI Cosmological Collider}
\label{sec:DBI_theory}

\noindent As discussed in the introduction, in our BOSS analysis we will consider the case where the interactions of $\pi$ and its couplings to $\sigma$ are protected by a higher-dimensional boost symmetry \cite{Alishahiha:2004eh}. In the $\pi$ sector, this symmetry forces \cite{Alishahiha:2004eh,Cheung:2007st,Senatore:2009gt}
\be
\label{eq:DBI_cs_c3}
\tilde{c}_3 = \frac{3(1-c^2_{\rm s})}{2}\,\,.
\ee

We find it worth to discuss in detail what this symmetry implies for the couplings of $\sigma$ to the foliation, expanding the analysis of \cite{Bonifacio:2019rpv} to the context of the Cosmological Collider. Working in flat spacetime (which is sufficient for this analysis), the action of $\sigma$ reads
\be
S_\sigma = \frac{1}{2}\int{\rm d}^4x\,\sqrt{-\tilde{g}}\,\big({-\tilde{g}^{\mu\nu}}\partial_\mu\sigma\partial_\nu\sigma - m^2\sigma^2\big) + \Omega^3_0\int{\rm d}^4x\,\sqrt{-\tilde{g}}\,\sigma\,\,,
\ee
where
\be
\tilde{g}_{\mu\nu} = \eta_{\mu\nu} + \partial_\mu\phi\partial_\nu\phi\,\,,\qquad\sqrt{-\tilde{g}} = \sqrt{1+\eta^{\mu\nu}\partial_\mu\phi\partial_\nu\phi}\,\,.
\ee
This action is invariant under the transformation 
\be
\delta\phi = b^\mu\phi\partial_\mu\phi + b_\mu x^\mu
\ee
for small $b^\mu$ if\footnote{The invariance of the kinetic term has been checked in \cite{Bonifacio:2019rpv}. It is straightforward to check the invariance of the linear mixing term, given that $\delta\sigma$ is the Lie derivative $\smash{{\cal L}_v\sigma}$ along $\smash{v^\mu = b^\mu \phi}$. Using $\smash{\delta\tilde{g}_{\mu\nu} = {\cal L}_v \tilde{g}_{\mu\nu}}$, and $\smash{\delta(\sqrt{-\tilde{g}}) = \frac{1}{2}\sqrt{-\tilde{g}}\,\tilde{g}^{\mu\nu}\delta\tilde{g}_{\mu\nu}}$, we find that 
\be
\delta(\sqrt{-\tilde{g}}\,\sigma) = \partial_\mu\big(v^\mu\sqrt{-\tilde{g}}\big)\sigma + \sqrt{-\tilde{g}}\,v^\mu\partial_\mu\sigma = \partial_\mu\big(v^\mu\sqrt{-\tilde{g}}\,\sigma\big)\,\,,
\ee
which is a total derivative.} 
\be
\delta\sigma = \phi b^\mu\partial_\mu\sigma\,\,. 
\ee
Expanding around the Lorentz-breaking profile of $\phi$, \emph{i.e.}~$\phi = \kappa (t + \pi)$ for some dimensionless $\kappa$, we find that the following terms in the action: 
\begin{itemize}[leftmargin=*]
    \item From the kinetic term of $\sigma$, using $\smash{\sqrt{-\tilde{g}} = \sqrt{1-\kappa^2}}$ at leading order in $\pi$, and 
    \be
    \tilde{g}^{\mu\nu} = \eta^{\mu\nu} - \frac{1}{1+\eta^{\rho\sigma}\partial_\rho\phi\partial_\sigma\phi}\partial^\mu\phi\partial^\nu\phi
    \ee
    so that
    \be
    {-\tilde{g}^{\mu\nu}} = {-\eta^{\mu\nu}} + \frac{\kappa^2\delta^\mu_0\delta^\nu_0}{1-\kappa^2}
    \ee
    at leading order, we get
    \be
    S_\sigma\supset \frac{\sqrt{1-\kappa^2}}{2}\int{\rm d}^4x\,\big(\dot{\sigma}^2 - {({\bm\nabla}\sigma)^2} - m^2\sigma^2\big) + \frac{\kappa^2}{2\sqrt{1-\kappa^2}}\int{\rm d}^4x\,\dot{\sigma}^2\,\,.
    \ee
    \item From the $\smash{\sqrt{-\tilde{g}}\,\sigma}$ term we get
    \be
    \label{eq:DBI_mixing}
    S_\sigma\supset\int{\rm d}^4x\,\Omega^3_0\bigg[{-\frac{\kappa^2\dot{\pi}}{\sqrt{1-\kappa^2}}} - \frac{\kappa^2\dot{\pi}^2}{2(1-\kappa^2)^{\frac{3}{2}}}+\frac{\kappa^2({\bm\nabla}\pi)^2}{2\sqrt{1-\kappa^2}}\bigg]\sigma\,\,,
    \ee
    where we have used 
    \be
    \sqrt{-\tilde{g}} = \sqrt{1-\kappa^2 -2\kappa^2\dot{\pi} + \kappa^2\eta^{\mu\nu}\partial_\mu\pi\partial_\nu\pi}\,\,.
    \ee
    Notice that in Eq.~\eqref{eq:DBI_mixing} we neglect the tadpole term for $\sigma$: this is not justified in general, and in a cosmological setting one must check whether it leads to $\sigma$ developing a v.e.v.~and making inflation not single-clock. Discussing this is far beyond the purposes of this paper and we leave it to future investigation (indeed our goal here is mainly to show that it is possible to analyse a model where symmetries restrict the number of free parameters).
\end{itemize}
Rescaling $\smash{\sigma\to\sigma/(1-\kappa^2)^{1/4}}$, we then find the terms of the $\smash{S_\sigma}$ action that are relevant for our analysis to be
\be
\begin{split}
    S_\sigma &\supset \frac{1}{2}\int{\rm d}^4x\,\big(\dot{\sigma}^2 - {({\bm\nabla}\sigma)^2} - m^2\sigma^2\big) + \frac{\kappa^2}{2(1-\kappa^2)}\int{\rm d}^4x\,\dot{\sigma}^2 \\
    &\;\;\;\; 
    + \int{\rm d}^4x\,\Omega^3_0\bigg[{-\frac{\kappa^2\dot{\pi}}{(1-\kappa^2)^{\frac{3}{4}}}} - \frac{\kappa^2\dot{\pi}^2}{2(1-\kappa^2)^{\frac{7}{4}}}+\frac{\kappa^2({\bm\nabla}\pi)^2}{2(1-\kappa^2)^{\frac{3}{4}}}\bigg]\sigma\,\,.
\end{split}
\ee
Comparing with the second line of Eq.~\eqref{full_action} we find 
\be
\beta^\collB = \bigg(\frac{c^2_{\rm s}}{c^2_\sigma}\bigg)^{-1}\beta^\collA\qquad\text{and}\qquad\beta^\collA>0\,\,.
\ee
Using the relation 
\be 
\label{eq:decomposition} 
\begin{split}
\begin{pmatrix} f_{\rm NL}^{\rm equil} \\
f_{\rm NL}^{\rm ortho}
\end{pmatrix}
 & = 
\begin{pmatrix}
1.04021 & 1.21041 \\
-0.0395140 & -0.175685
\end{pmatrix}
\begin{pmatrix}
f_{\rm NL}^{\dot{\pi}({\bm\nabla}\pi)^2} \\
f_{\rm NL}^{\dot{\pi}^3}
\end{pmatrix}\,\,, 
\\
f_{\rm NL}^{\dot{\pi}({\bm\nabla}\pi)^2} & =\frac{85}{324}(1-c_{\rm s}^{-2})\,\,,\\
f_{\rm NL}^{\dot{\pi}^3}
&=\frac{10}{243}(1-c_{\rm s}^{-2})\left(\tilde{c}_3+\frac{3}{2}c_{\rm s}^2\right) 
\end{split}
\ee
between $\smash{f_{\rm NL}^{\rm equil},f_{\rm NL}^{\rm ortho}}$ and $\smash{c_{\rm s},\tilde{c}_3}$ \cite{Babich:2004gb,Senatore:2009gt}, together with Eq.~\eqref{eq:DBI_cs_c3}, we then find
\be
f_{\rm NL}^{\rm equil} = {0.347611}\times\frac{c^2_{\rm s}-1}{c^2_{\rm s}}\qquad\text{and}\qquad f_{\rm NL}^{\rm ortho} = {-0.0212111}\times\frac{c^2_{\rm s}-1}{c^2_{\rm s}}\,\,.
\ee
Hence in our DBI analysis we vary only $\smash{\beta^\collA}$, $\smash{\log_{10}{c_{\rm s}\in[{-3},0]}}$, $\mu$, and $\log_{10}$ of the speed ratio\footnote{Recall that the speed ratio enters also in the Cosmological Collider shapes: this is where the degeneracy between $\smash{\beta^\collA}$ and $\smash{c^2_{\rm s}/c^2_\sigma}$ is broken.} such that $\smash{c_{\rm s}/c_\sigma}$ remains between $0.1$ and $10$. Notice that we follow the \emph{Planck} PNG analysis when deciding to vary the logarithm of speeds.

\section{\texorpdfstring{$\text{\emph{Planck}}$ constraints from $\smash{\fnl^{\rm equil}}$ and $\smash{\fnl^{\rm ortho}}$}{\textit{Planck} constraints from Equilateral and Orthgonal Non-Gaussianity}}
\label{sec:planck_equivalent}

\noindent Whilst the main goals of this paper are to constrain the Cosmological Collider using galaxy surveys, it is interesting to first consider the extent to which such models are already constrained by CMB observations. In particular, we here investigate how one can indirectly constrain massive particle interactions by translating the \textit{Planck} bounds on the conventional equilateral and orthogonal bispectra to limits on the collider amplitudes discussed above. 

\subsection{Relating Collider Amplitudes to Template Coefficients}
\noindent Let us posit that inflaton self-interactions are a subleading source of PNG, \emph{i.e.}~$f_{\rm NL}^{\dot{\pi}(\nabla\pi)^2}\approx f_{\rm NL}^{\dot{\pi}^3}\approx 0$ {(Ref.~\cite{Sohn:2024xzd} does not do this assumption. The authors there, instead, carry out a new analysis, similar to the \emph{Planck} one on equilateral and orthogonal non-Gaussianity, looking for the Cosmological Collider templates instead.)}
In this scenario, the non-Gaussianity is generated only by the massive particle interactions, with shapes $\smash{\mathcal{S}^{\dot{\pi}^2\sigma}}$ and $\smash{\mathcal{S}^{({\bm\nabla}\pi)^2\sigma}}$. Since these bispectra have significant overlap with the canonical equilateral and orthogonal templates (despite the different physical sources), one can recast the measured bounds on the empirical $f_{\rm NL}^{\rm equil}$ and $f_{\rm NL}^{\rm ortho}$ coefficients as constraints on $\smash{\beta^\collA}$ and $\smash{\beta^\collB}$. Importantly, this translation requires us to ignore self-interactions, since these also source the equilateral and orthogonal templates, thus their inclusion would yield a near-perfect degeneracy. (Simply put, one cannot hope to measure four physical amplitudes from two $f_{\rm NL}$ observations). Since a direct search for the collider parameters $\smash{\beta^\collA}$ and $\smash{\beta^\collB}$ in CMB data has yet to be performed, this is the only way to proceed, at least for now.

A crucial assumption in the above discussion is that the Cosmological Collider shapes, $\smash{\mathcal{S}^{\dot{\pi}^2\sigma}}$ and $\smash{\mathcal{S}^{({\bm\nabla}\pi)^2\sigma}}$, take a similar physical form to the standard forms, ${\cal S}^{\rm equil}$ and ${\cal S}^{\rm ortho}$. The degree of similarity can be assessed via the inner product between scale-invariant primordial shapes introduced in \cite{Babich:2004gb,Senatore:2009gt} and defined as
\be
\label{eq:CV_3D_cosine}
\langle{\cal S}^A | {\cal S}^B\rangle = \int{\rm d}x_1{\rm d}x_2\,{\cal S}^A(x_1,x_2,1){\cal S}^B(x_1,x_2,1)\,\,, 
\ee 
where the integral is carried out over $x_1,x_2$ in a simplex. This further motivates the definition of a cosine
\be\label{eq:cosine}
    \cos(a,b) \equiv \frac{\langle{\cal S}^a | {\cal S}^b\rangle}{\sqrt{\langle{\cal S}^a | {\cal S}^a\rangle\langle{\cal S}^b | {\cal S}^b\rangle}}\,\,,
\ee
where $\cos(a,b) = 1$ $(0)$ indicates that shapes $a$ and $b$ are completely correlated (uncorrelated). If the cosines of the Cosmological Collider shapes with the equilateral and orthogonal templates are large, one can efficiently reconstruct $\smash{\beta^\collA}$ and $\smash{\beta^\collB}$ from the \textit{Planck} $f_{\rm NL}^{\rm equil}$ and $f_{\rm NL}^{\rm ortho}$ bounds. In contrast, if the cosines are small, this projection is highly inefficient, and the model would greatly benefit from a dedicated collider analysis (as we do for the BOSS survey in the remainder of this work).

Using the inner product of \eqref{eq:CV_3D_cosine}, the Cosmological Collider shapes can be written in terms of the equilateral and orthogonal templates as 
\begin{align}
\label{proj-1}
    {\cal S}^{\collA} &= \frac{\langle{\cal S}^\collA | {\cal S}^{\rm equil}\rangle}{\langle{\cal S}^{\rm equil} | {\cal S}^{\rm equil}\rangle}{\cal S}^{\rm equil} + \frac{\langle{\cal S}^\collA | {\cal S}^{\rm ortho}\rangle}{\langle{\cal S}^{\rm ortho} | {\cal S}^{\rm ortho}\rangle}{\cal S}^{\rm ortho} + \Delta {\cal S}^{\collA}
    \,\,, \\
\label{proj-2}
    {\cal S}^{\collB} &= \frac{\langle{\cal S}^\collB | {\cal S}^{\rm equil}\rangle}{\langle{\cal S}^{\rm equil} | {\cal S}^{\rm equil}\rangle}{\cal S}^{\rm equil} + \frac{\langle{\cal S}^\collB | {\cal S}^{\rm ortho}\rangle}{\langle{\cal S}^{\rm ortho} | {\cal S}^{\rm ortho}\rangle}{\cal S}^{\rm ortho} + \Delta {\cal S}^{\collB}
    \,\,, 
\end{align}
where $\smash{\Delta\cal S}$ is a residual shape, which is orthogonal to both equilateral and orthogonal PNG (such that $\langle{\Delta{\cal S}|{\cal S}^{\rm equil, ortho}\rangle} = 0$). Inserting into the full bispectrum form via \eqref{equilateral_normalization} and projecting onto the equilateral and orthogonal templates via the inner product, we can compute the effective $\fnl^{\rm equil}$ and $\fnl^{\rm ortho}$ parameters corresponding to the collider shapes:\footnote{Here, we note a technical point: whilst the official \emph{Planck} analyses constrained the orthogonal template defined by Eq.~(53) of Ref.~\cite{Senatore:2009gt}, in this work, we use the template defined in Appendix~B of the same reference (which is that utilized in the BOSS analyses of \cite{Cabass:2022wjy,DAmico:2022gki}). We make this choice since calculation of the dot product between the Cosmological Collider shapes and the orthogonal shape would otherwise depend strongly on the lower limit of $x_1 = k_1/k_3$ in \eqref{eq:CV_3D_cosine}, given that the \textit{Planck} orthogonal shape diverges in the squeezed limit. Since the overlap between the two shapes (which is insenstive to the minimum value of $x_1$, see \S3 of Ref.~\cite{Senatore:2009gt}) is $0.24$, our constraints are suboptimal only by the square root of this factor.} 

\begin{align}
    \fnl^{\rm equil}(\fnl^{\collA},\fnl^{\collB}) &= \frac{1}{{\cal S}^\collA(1,1,1)}\frac{\langle{\cal S}^\collA | {\cal S}^{\rm equil}\rangle}{\langle{\cal S}^{\rm equil} | {\cal S}^{\rm equil}\rangle}\fnl^{\collA}+\frac{1}{{\cal S}^\collB(1,1,1)}\frac{\langle{\cal S}^\collB | {\cal S}^{\rm equil}\rangle}{\langle{\cal S}^{\rm equil} | {\cal S}^{\rm equil}\rangle}\fnl^{\collB}\\
    \fnl^{\rm ortho}(\fnl^{\collA},\fnl^{\collB}) &= \frac{1}{{\cal S}^\collA(1,1,1)}\frac{\langle{\cal S}^\collA | {\cal S}^{\rm ortho}\rangle}{\langle{\cal S}^{\rm ortho} | {\cal S}^{\rm ortho}\rangle}\fnl^{\collA}+\frac{1}{{\cal S}^\collB(1,1,1)}\frac{\langle{\cal S}^\collB | {\cal S}^{\rm ortho}\rangle}{\langle{\cal S}^{\rm ortho} | {\cal S}^{\rm ortho}\rangle}\fnl^{\collB}\,\,.
\end{align}
Here, we have noted that $\fnl^{X} = {\cal S}^X(1,1,1)\langle{B_{\zeta}|{\cal S}^X\rangle}/\langle{{\cal S}^X|{\cal S}^X\rangle}$ (for $X=\mathrm{equil/ortho}$) from \eqref{equilateral_normalization}, and additionally assumed zero self-interactions (such that $B_\zeta$ receives contributions only from the collider shapes). Finally, we can write the amplitudes in terms of $\smash{\beta^\collA}$ and $\smash{\beta^\collB}$ using \eqref{conversion}, which yields
\begin{align}
\label{transformation-1}
    \fnl^{\rm equil}\big(\beta^{\collA},\beta^{\collB}\big) &= \bigg({-\frac{10}{9}\frac{c_{\rm s}}{c_\sigma}}\beta^\collA\bigg) \frac{\langle{\cal S}^\collA | {\cal S}^{\rm equil}\rangle}{\langle{\cal S}^{\rm equil} | {\cal S}^{\rm equil}\rangle} + \bigg({\frac{5}{9}\frac{c_\sigma}{c_{\rm s}}}\beta^\collB\bigg) \frac{\langle{\cal S}^\collB | {\cal S}^{\rm equil}\rangle}{\langle{\cal S}^{\rm equil} | {\cal S}^{\rm equil}\rangle}\,\,, \\ 
\label{transformation-2}
    \fnl^{\rm ortho}\big(\beta^{\collA},\beta^{\collB}\big) &= \bigg({-\frac{10}{9}\frac{c_{\rm s}}{c_\sigma}}\beta^\collA\bigg) \frac{\langle{\cal S}^\collA | {\cal S}^{\rm ortho}\rangle}{\langle{\cal S}^{\rm ortho} | {\cal S}^{\rm ortho}\rangle} + \bigg({\frac{5}{9}\frac{c_\sigma}{c_{\rm s}}}\beta^\collB\bigg) \frac{\langle{\cal S}^\collB | {\cal S}^{\rm ortho}\rangle}{\langle{\cal S}^{\rm ortho} | {\cal S}^{\rm ortho}\rangle}\,\,.
\end{align}
This expression will be used to obtain the bounds on $\smash{\beta^{\collA}}$ and $\smash{\beta^{\collB}}$ below.

\begin{figure}[t]
    \centering
    \includegraphics[width=\textwidth]{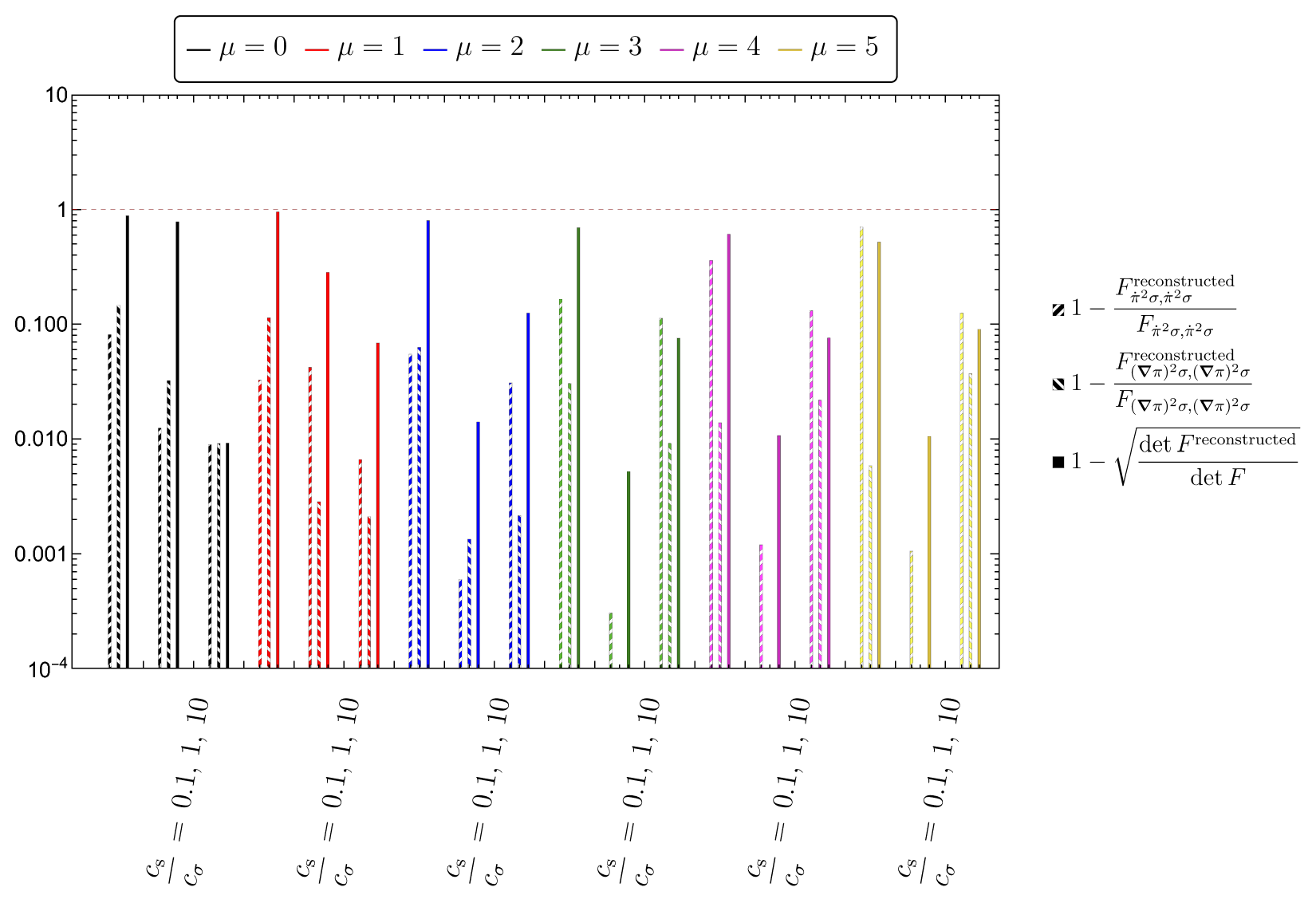}
    \caption{Overlap of spin-zero massive particle non-Gaussianity with the classical equilateral and orthogonal templates. We show three sets of results, each measuring the (idealized) fractional information loss induced by indirectly inferring a different set of collider amplitudes from the equilateral and orthogonal bounds, instead of constraining them directly from the data. These comprise: (1, north-east hatching) analyzing only the first collider amplitude ($\beta^{\collA}$); (2, south-east hatching) analyzing only the second collider amplitude ($\beta^{\collB}$); (3, full bars); jointly analyzing both collider amplitudes ($\beta^{\collA}$ and $\beta^{\collB}$). Values far below unity indicate that the relevant templates are well described by the equilateral and orthogonal shapes {(notice that $\mu = 3$ is the template best described by $\smash{{\cal S}^{\rm equil}}$ and $\smash{{\cal S}^{\rm ortho}}$)}, whilst those approaching unity indicate differences in the templates. Results are shown for five choices of mass parameter $\mu$ (indicated by color), and three choices of sound-speed ratio, $c_{\rm s}/c_{\sigma}$ (left to right within each block). As $\mu$ increases, the collider templates show considerable overlap with the canonical $\smash{{\cal S}^\equil}$ and $\smash{{\cal S}^\ortho}$ shapes (particularly joint analyses at $c_{\rm s}=c_\sigma$, corresponding to the central solid lines in each block), since the heavy particle can be efficiently integrated out. We additionally find that, whilst a single shape can usually be well described by the standard templates, the combination of two cannot (except at high mass).}    \label{fig:planck_cosine_plot}
\end{figure} 

\subsection{Fisher Forecasts}
\noindent To understand the efficacy of this approach, it is instructive to perform a simplified Fisher matrix analysis, comparing the optimal bounds on $\smash{\beta^{\collA}}$ and $\smash{\beta^{\collB}}$ to those obtained from the above projection scheme. In the ideal limit of a three-dimensional cosmic-variance-limited survey, which directly probes the $\zeta$ bispectrum (\emph{i.e.}~assuming a linear transfer function), the inverse of the Cram\'{e}r-Rao covariance (the Fisher matrix) scales as
\be
    F_{AB} \propto \smash{\langle{\cal S}^A | {\cal S}^B\rangle},
\ee
where $A,B$ index the templates of interest (expressed in terms of $f_{\rm NL}$) and we assume scale invariance in the region of interest.\footnote{The proportionality factor includes the number of pixels of the survey, various collider factors and $\smash{\Delta^2_\zeta}$: these will be irrelevant for our discussion}. As such, the optimal error-bar on a parameter $A$ from a dedicated analysis scales as $F^{-1/2}_{AA}$. If instead one computes the  parameter constraints by first analyzing the equilateral and orthogonal templates, then translating via \eqref{transformation-1}\,\&\,\eqref{transformation-2}, the Fisher matrix is given by
\begin{align}
F^{\rm reconstructed}_{AB} &\propto \sum_{X,Y\in\{\rm equil,ortho\}}\frac{\partial f_{\rm NL}^A}{\partial f_{\rm NL}^X}F_{\rm XY}\frac{\partial f_{\rm NL}^B}{\partial f_{\rm NL}^Y}\propto \frac{\langle{\cal S}^A|{\cal S}^{\equil}\rangle\langle{\cal S}^B|{\cal S}^{\equil}\rangle}{\langle{\cal S}^{\equil}|{\cal S}^{\equil}\rangle}+\frac{\langle{\cal S}^A|{\cal S}^{\ortho}\rangle\langle{\cal S}^B|{\cal S}^{\ortho}\rangle}{\langle{\cal S}^{\ortho}|{\cal S}^{\ortho}\rangle}\,\,,
\end{align}
simply propagating errors, with $\mathrm{cov}(f_{\rm NL}^{X},f_{\rm NL}^Y) = F_{XY}^{-1}$. The ``information loss'' associated with this indirect analysis is thus given by the ratio
\begin{align}
\frac{F^{\rm reconstructed}_{AB}}{F_{AB}} &= \frac{\langle{\cal S}^A|{\cal S}^{\equil}\rangle\langle{\cal S}^B|{\cal S}^{\equil}\rangle}{\langle{\cal S}^A|{\cal S}^B\rangle\langle{\cal S}^{\equil}|{\cal S}^{\equil}\rangle}+\frac{\langle{\cal S}^A|{\cal S}^{\ortho}\rangle\langle{\cal S}^B|{\cal S}^{\ortho}\rangle}{\langle{\cal S}^A|{\cal S}^B\rangle\langle{\cal S}^{\ortho}|{\cal S}^{\ortho}\rangle}\,\,,
\end{align}
for $\smash{(A,B) = (\collA,\collB)}$, as before. 

Given the above discussion, one can understand the projection efficiency using the following quantities: 
\begin{itemize}[leftmargin=*]
    \item $\smash{F^{\rm reconstructed}_{AA}/F_{AA}}$ and $\smash{F^{\rm reconstructed}_{BB}/F_{BB}}$. This indicates how well one can recover a single amplitude ($\smash{\beta^\collA}$ \emph{or} $\smash{\beta^\collB}$) from the equilateral and orthogonal observations. In terms of the cosines \eqref{eq:cosine}, this is simply the sum of squares:
    \be
    \label{AA}
    \frac{F^{\rm reconstructed}_{AA}}{F_{AA}} = \cos^2(A,\equil) + \cos^2(A,\ortho)\,\,.
    \ee
    \item $\smash{\sqrt{\det F^{\rm reconstructed}/\det F}}$. This assesses how well one can jointly recover both amplitudes ($\smash{\beta^\collA}$ \emph{and} $\smash{\beta^\collB}$) from the equilateral and orthogonal observations. Physically, this gives the (dimensionally-adjusted) of the ratio of the posterior volumes.
\end{itemize} 
Of course, these do not fully represent the information loss in any real experiment, since we have neglected binning, covariances, transfer functions, projection onto the sky, and beyond. However, as shown in \cite{Babich:2004gb}, the above is a good approximation for the large-scale \textit{Planck} behavior.

\def\prec{1}
\setlength{\tabcolsep}{4pt}
\renewcommand{\arraystretch}{1.3}
\begin{table*}[ht!]
\centering
\begin{tabular}{c|ccc|c}
 & \multicolumn{3}{c|}{${\beta^\collA}$} & \multicolumn{1}{c}{${\beta^\collB}$}\\ \hline
 \textbf{Model} & mean & \quad 95\% lower \quad & 95\% upper & 95\% upper \\ 
 \hline

$\mu=0,c_{\rm s}/c_\sigma = 0.1$ & $\num[round-mode=places,round-precision=\prec]{2.6923972720803105e+0005}$ & $\num[round-mode=places,round-precision=\prec]{1.1195377596118808e+0004}$ & $\num[round-mode=places,round-precision=\prec]{5.204820104620948e+0005}$ & $\num[round-mode=places,round-precision=\prec]{8.1152e+0004}$ \\

$\mu=0,c_{\rm s}/c_\sigma = 1.0$ & $\num[round-mode=places,round-precision=\prec]{2.492929876800598e+03}$ & $\num[round-mode=places,round-precision=\prec]{-4.959647511853991e+02}$ & $\num[round-mode=places,round-precision=\prec]{7.092388743470132e+03}$ & $\num[round-mode=places,round-precision=\prec]{3.4052215e+02}$ \\

$\mu=0,c_{\rm s}/c_\sigma = 10$ & $\num[round-mode=places,round-precision=\prec]{-3.3176630582707463e+2}$ & $\num[round-mode=places,round-precision=\prec]{-6.599983298803484e+2}$ & $\num[round-mode=places,round-precision=\prec]{-2.0433231055797023e+1}$ & $\num[round-mode=places,round-precision=\prec]{6.44714e+1}$ \\ \hline

$\mu=1,c_{\rm s}/c_\sigma = 0.1$ & $\num[round-mode=places,round-precision=\prec]{1.2632304081559654e+005}$ & $\num[round-mode=places,round-precision=\prec]{-5.364115202517423e+003}$ & $\num[round-mode=places,round-precision=\prec]{3.532172283278743e+005}$ & $\num[round-mode=places,round-precision=\prec]{5.46795e+0004}$ \\

$\mu=1,c_{\rm s}/c_\sigma = 1.0$ & $\num[round-mode=places,round-precision=\prec]{-4.777310610227196e+03}$ & $\num[round-mode=places,round-precision=\prec]{-9.613502566677413e+03}$ & $\num[round-mode=places,round-precision=\prec]{-2.5099509673616012e+02}$ & $\num[round-mode=places,round-precision=\prec]{7.840705e+02}$ \\

$\mu=1,c_{\rm s}/c_\sigma = 10$ & $\num[round-mode=places,round-precision=\prec]{-1.4732824879324723e+3}$ & $\num[round-mode=places,round-precision=\prec]{-2.9358522114315474e+3}$ & $\num[round-mode=places,round-precision=\prec]{-5.3577435518368475e+1}$ & $\num[round-mode=places,round-precision=\prec]{9.64186e+02}$ \\ \hline

$\mu=2,c_{\rm s}/c_\sigma = 0.1$ & $\num[round-mode=places,round-precision=\prec]{2.8356358415436003e+004}$ & $\num[round-mode=places,round-precision=\prec]{-7.909165966618188e+003}$ & $\num[round-mode=places,round-precision=\prec]{8.174641250464946e+004}$ & $\num[round-mode=places,round-precision=\prec]{1.0534255e+0004}$ \\

$\mu=2,c_{\rm s}/c_\sigma = 1.0$ & $\num[round-mode=places,round-precision=\prec]{-3.587221551570343e+3}$ & $\num[round-mode=places,round-precision=\prec]{-7.22614642603421e+3}$ & $\num[round-mode=places,round-precision=\prec]{-9.78655933668615e+1}$ & $\num[round-mode=places,round-precision=\prec]{2.205823e+003}$ \\

$\mu=2,c_{\rm s}/c_\sigma = 10$ & $\num[round-mode=places,round-precision=\prec]{-3.5122357901925807e+02}$ & $\num[round-mode=places,round-precision=\prec]{-2.2819754216041592e+03}$ & $\num[round-mode=places,round-precision=\prec]{1.4041415171019325e+03}$ & $\num[round-mode=places,round-precision=\prec]{5.503304999999999e+02}$ \\ \hline

$\mu=3,c_{\rm s}/c_\sigma = 0.1$ & $\num[round-mode=places,round-precision=\prec]{2.1885556999335007e+0004}$ & $\num[round-mode=places,round-precision=\prec]{-1.6138897819308608e+0004}$ & $\num[round-mode=places,round-precision=\prec]{7.054755061976006e+0004}$ & $\num[round-mode=places,round-precision=\prec]{6.449645e+003}$ \\

$\mu=3,c_{\rm s}/c_\sigma = 1.0$ & $\num[round-mode=places,round-precision=\prec]{-7.652038194515584e+03}$ & $\num[round-mode=places,round-precision=\prec]{-1.5438668561776698e+04}$ & $\num[round-mode=places,round-precision=\prec]{-1.477863455414954e+02}$ & $\num[round-mode=places,round-precision=\prec]{5.191830000000001e+003}$ \\

$\mu=3,c_{\rm s}/c_\sigma = 10$ & $\num[round-mode=places,round-precision=\prec]{1.6919980452851569e+003}$ & $\num[round-mode=places,round-precision=\prec]{-1.077814e+003}$ & $\num[round-mode=places,round-precision=\prec]{4.4592455e+003}$ & $\num[round-mode=places,round-precision=\prec]{5.51558e+02}$ \\ \hline

$\mu=4,c_{\rm s}/c_\sigma = 0.1$ & $\num[round-mode=places,round-precision=\prec]{1.6158617996922094e+0004}$ & $\num[round-mode=places,round-precision=\prec]{-4.07730167603858e+0004}$ & $\num[round-mode=places,round-precision=\prec]{7.895452557458665e+0004}$ & $\num[round-mode=places,round-precision=\prec]{4.2978085e+003}$ \\

$\mu=4,c_{\rm s}/c_\sigma = 1.0$ & $\num[round-mode=places,round-precision=\prec]{-1.6131964285711956e+04}$ & $\num[round-mode=places,round-precision=\prec]{-3.2427758549120437e+04}$ & $\num[round-mode=places,round-precision=\prec]{-3.675958191978789e+02}$ & $\num[round-mode=places,round-precision=\prec]{9.95208e+003}$ \\

$\mu=4,c_{\rm s}/c_\sigma = 10$ & $\num[round-mode=places,round-precision=\prec]{2.9137788925683044e+03}$ & $\num[round-mode=places,round-precision=\prec]{-1.843195e+02}$ & $\num[round-mode=places,round-precision=\prec]{6.005620000000001e+03}$ & $\num[round-mode=places,round-precision=\prec]{1.1094305e+003}$ \\ \hline

$\mu=5,c_{\rm s}/c_\sigma = 0.1$ & $\num[round-mode=places,round-precision=\prec]{-6.3884352043525854e+0004}$ & $\num[round-mode=places,round-precision=\prec]{-1.9655455000000002e+0005}$ & $\num[round-mode=places,round-precision=\prec]{7.11866e+0004}$ & $\num[round-mode=places,round-precision=\prec]{2.52929e+003}$ \\

$\mu=5,c_{\rm s}/c_\sigma = 1.0$ & $\num[round-mode=places,round-precision=\prec]{-2.795367529638354e+04}$ & $\num[round-mode=places,round-precision=\prec]{-5.605831209010871e+04}$ & $\num[round-mode=places,round-precision=\prec]{-7.033162202728563e+02}$ & $\num[round-mode=places,round-precision=\prec]{1.6484455e+0004}$ \\

$\mu=5,c_{\rm s}/c_\sigma = 10$ & $\num[round-mode=places,round-precision=\prec]{3.5154641225976034e+03}$ & $\num[round-mode=places,round-precision=\prec]{1.378211e+02}$ & $\num[round-mode=places,round-precision=\prec]{6.975255e+03}$ & $\num[round-mode=places,round-precision=\prec]{2.0325894999999998e+003}$ \\

 \end{tabular}
 \caption{\textit{Planck} constraints on spin-zero massive particles, obtained from the published $\smash{\fnl^{\rm equil}}$ and $\smash{\fnl^{\rm ortho}}$ bounds, assuming negligible inflaton self-interactions. We display results for the microphysical collider amplitudes $\smash{\beta^\collA}$ and $\smash{\beta^\collB}$, for various choices of the mass $\mu$ and relative sound-speed $c_{\rm s}/c_\sigma$. We show the mean and $95\%$ CL bound for the first amplitude and the $95\%\,{\rm CL}$ upper limit for the second, noting that $\smash{\beta^\collB}\geq 0$. These constraints are visualized in Figs.~\ref{fig:planck_scan-dotpi2_nablapi2}~\&~\ref{fig:planck}.} 
 \label{tab:planck}
\end{table*}

In Fig.~\ref{fig:planck_cosine_plot}, we plot the quantities given in the above bullets. Our first conclusion is that, if one wishes to measure a single shape (\emph{i.e.}~$\smash{\beta^\collA}$ \textit{or} $\smash{\beta^\collB}$) from $f_{\rm NL}^{\rm equil}$ and $f_{\rm NL}^{\rm orthog}$ measurements, the projection is very efficient, such that $F_{AA}^{\rm reconstructed}/F_{AA}$ close to unity. In other words, a single collider shape can be well described by a linear combination of $\smash{{\cal S}^\equil}$ and $\smash{{\cal S}^\ortho}$. In contrast, the joint analysis of $\smash{\beta^\collA}$ \textit{and} $\smash{\beta^\collB}$ usually suffers from significant information loss, since both Cosmological Collider shapes project similarly onto the equilateral and orthogonal basis, thus cannot be easily distinguished. To understand this, we consider the ratio of posterior volumes, which can be written
\be
\label{det_ratio}
\frac{\det F^{\rm reconstructed}}{\det F} = \frac{\big[\cos(A,\equil)\cos(B,\ortho) - \cos(A,\ortho)\cos(B,\equil)\big]^2}{1-\cos^2(A,B)}
\ee
(using the relation $\smash{\det(M_1 + M_2) = \det(M_1)+\det(M_2)+\det(M_1)\Tr(M_1^{-1}\cdot M_2)}$ for $2\times 2$ matrices $M_{1,2}$).
Unlike the scenario in which we reconstruct a single shape from measurements of equilateral and orthogonal PNG \eqref{AA}, our ability to reconstruct both shapes depends not only on the overlap of the shapes with the equilateral and orthogonal templates, but also on how close $\smash{\cos(B,\equil)}$ and $\smash{\cos(B,\ortho)}$ are to $\smash{\cos(A,\equil)}$ and $\smash{\cos(A,\ortho)}$. If the two Cosmological Collider shapes project similarly onto the equilateral and orthogonal templates, right-hand side of Eq.~\eqref{det_ratio} becomes small, and the projection is inefficient, sourcing the close-to-unity values seen in the $1-\sqrt{\det F^{\rm reconstructed}/\det F}$ plot of Fig.~\ref{fig:planck_cosine_plot} (and also the strong degeneracies observed in the two-dimensional contours of Fig.~\ref{fig:planck-mu_marginalized}, especially for small $c_{\rm s}/c_\sigma$).

Secondly, we can assess how the template overlap fares for different values of the mass $\mu$ and relative sound-speed $c_{\rm s}/c_\sigma$. For both single and dual parameter analyses with $c_{\rm s}=c_\sigma$, the projection becomes highly efficient at large-$\mu$ (recovering $>99\%$ of the posterior volume by $\mu=3$). This matches expectations, since, as the mass increases, the bispectrum arising from the diagram of \eqref{diagram} can be better and better described by integrating out the massive particle $\sigma$. This is particularly true for $\smash{{\cal S}^\collB}$: integrating out the massive particle results in a $({\bm\nabla}\pi)^2\dot{\pi}$ inflaton self-interaction, itself notoriously well described by a superposition of equilateral and orthogonal templates \cite{Senatore:2009gt}. For the ``equilateral collider'' regime with $c_{\rm s}\gg c_\sigma$, the projections are somewhat efficient, though significant differences (and oscillatory signatures) appear at large $\mu$. For the ``low-speed'' Cosmological Collider (with $c_{\rm s}\ll c_\sigma$), the projection is inefficient for joint analyses with all $\mu$, with the standard templates missing at least half of the posterior volume \citep[cf.,][]{Jazayeri:2023xcj}. This motivates the need for a dedicated low-speed collider-analysis of CMB data.

\subsection{\textit{Planck} Analyses}

\begin{figure}[t]
\centering
\includegraphics[width = 0.49\textwidth]{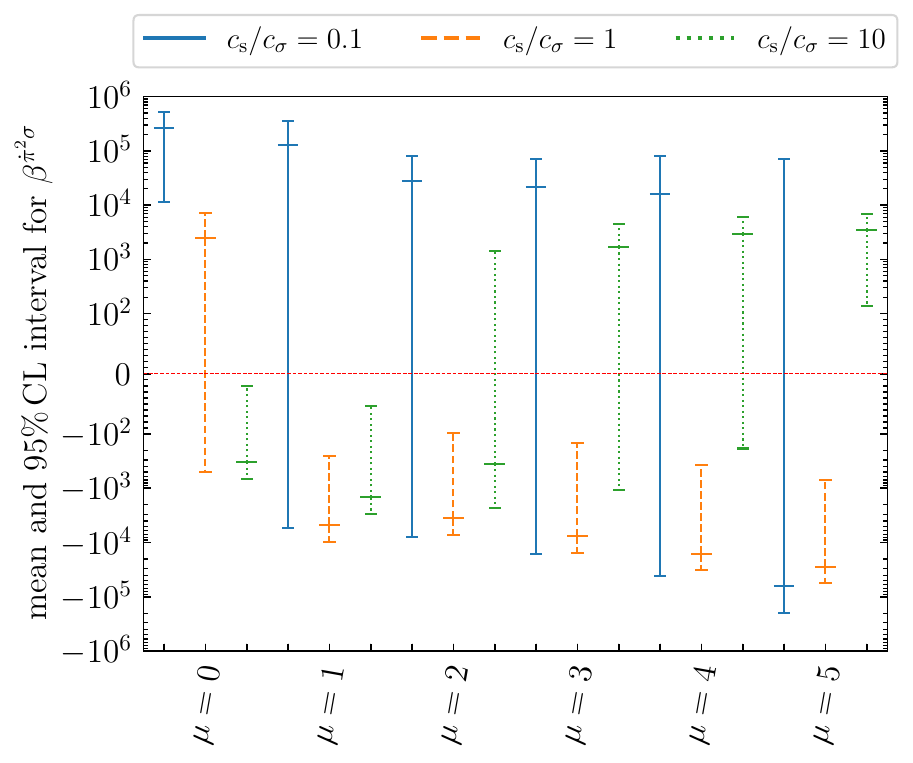} 
\includegraphics[width = 0.49\textwidth]{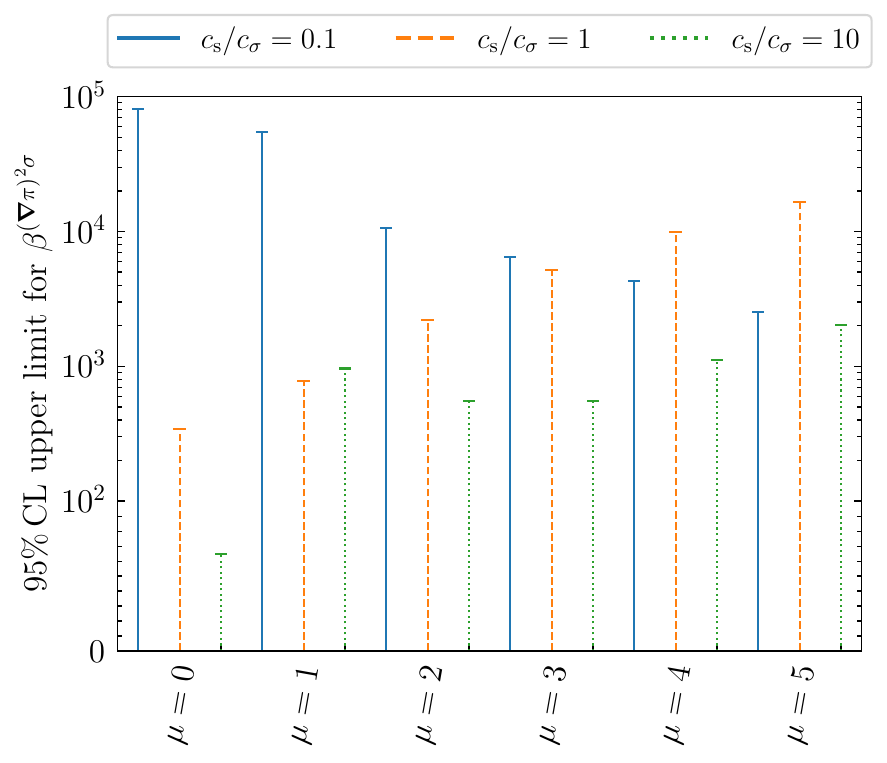} 
\caption{Mean and 95\% CL limits on the microphysical collider amplitude $\smash{\beta^\collA}$ (left plot) and $95\%\,{\rm CL}$ upper limit for the $\smash{\beta^\collB}$ amplitude (right plot) from \textit{Planck} data, assuming negligible inflaton self-interactions. We perform $18$ analyses, varying both the mass $\mu$ and relative sound-speed $c_{\rm s}/c_\sigma$, as indicated by the captions. The numerical values are given in Tab.~\ref{tab:planck}. Although we find some slight preference for $\beta^\collA<0$ for some choices of mass $\mu$ (at $c_{\rm s}=c_\sigma$) this is largely driven by projection effects (due to the degenerate amplitudes), and the slight preference for $\fnl^{\rm ortho}\neq 0$ in \textit{Planck}. We do not find evidence of non-zero $\smash{\beta^\collB}$.} 
\label{fig:planck_scan-dotpi2_nablapi2}
\end{figure}

\setlength{\tabcolsep}{4pt}
\renewcommand{\arraystretch}{1.3}
\begin{table*}[t]
\centering
\begin{tabular}{ccc|c|c}
 \multicolumn{3}{c|}{${\beta^\collA}$} & \multicolumn{1}{c|}{${\beta^\collB}$} & \multicolumn{1}{c}{${\mu}$}\\ \hline
 mean & \quad 95\% lower \quad & 95\% upper & 95\% upper & 95\% lower \\ 
 \hline 
 
$\num[round-mode=places,round-precision=\prec]{-1.9336767701577468e+04}$ & $\num[round-mode=places,round-precision=\prec]{-4.404711890147392e+04}$ & $\num[round-mode=places,round-precision=\prec]{6.71379010587727e+02}$ & $\num[round-mode=places,round-precision=\prec]{1.3295195e+0004}$ & $\num[round-mode=places,round-precision=\prec]{2.842476}$ \\

 \end{tabular}
 \caption{\textit{Planck} constraints on the microphysical collider amplitudes $\smash{\beta^\collA}$ and $\smash{\beta^\collB}$, fixing $\smash{c_{\rm s} = c_\sigma}$ and marginalizing over the mass parameter $\mu\in[0,5]$. As before, we give the mean and $95\%$ CL bound for $\smash{\beta^\collA}$ and the 95\% upper bound on the positive parameter $\smash{\beta^\collB}$, and assume negligible inflaton self-interactions in all cases.} 
 \label{tab:planck-mu_marginalized}
\end{table*}

\begin{figure}
  \begin{minipage}{0.45\textwidth}
    \centering
    \includegraphics[height=0.8\textwidth]{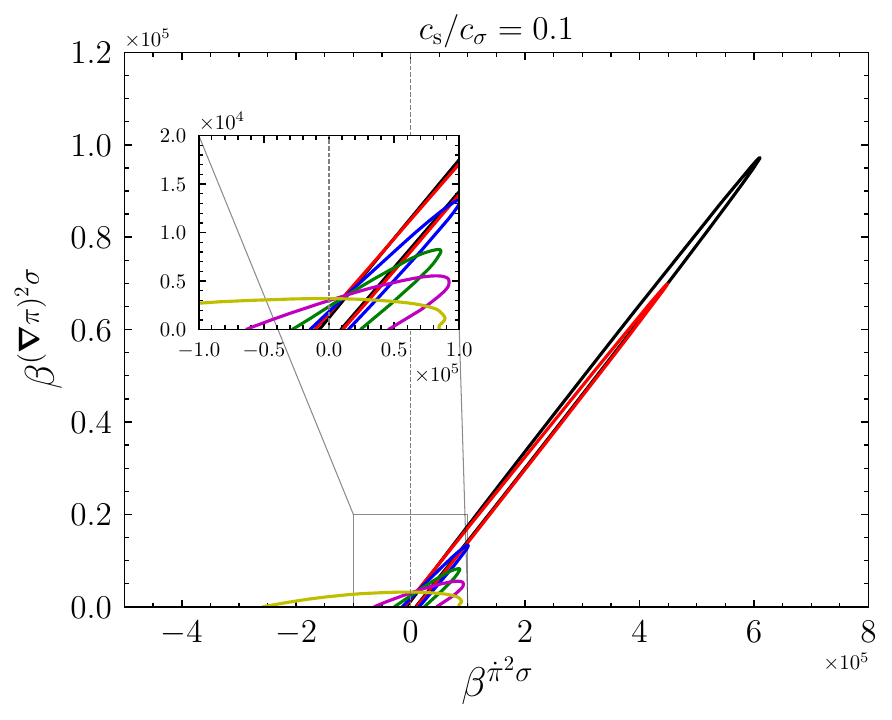}
  \end{minipage}
  \hfill
  \begin{minipage}{0.45\textwidth}
    \centering
    \includegraphics[height=0.8\textwidth]{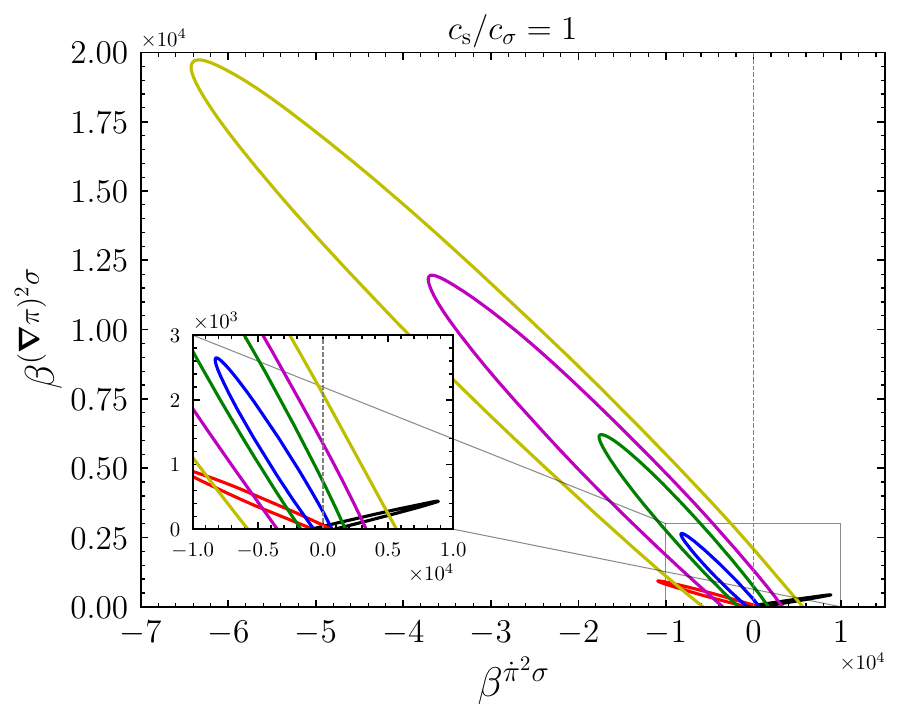}
  \end{minipage}
  \hfill
  \begin{minipage}{0.45\textwidth}
    \centering
    \includegraphics[height=0.8\textwidth]{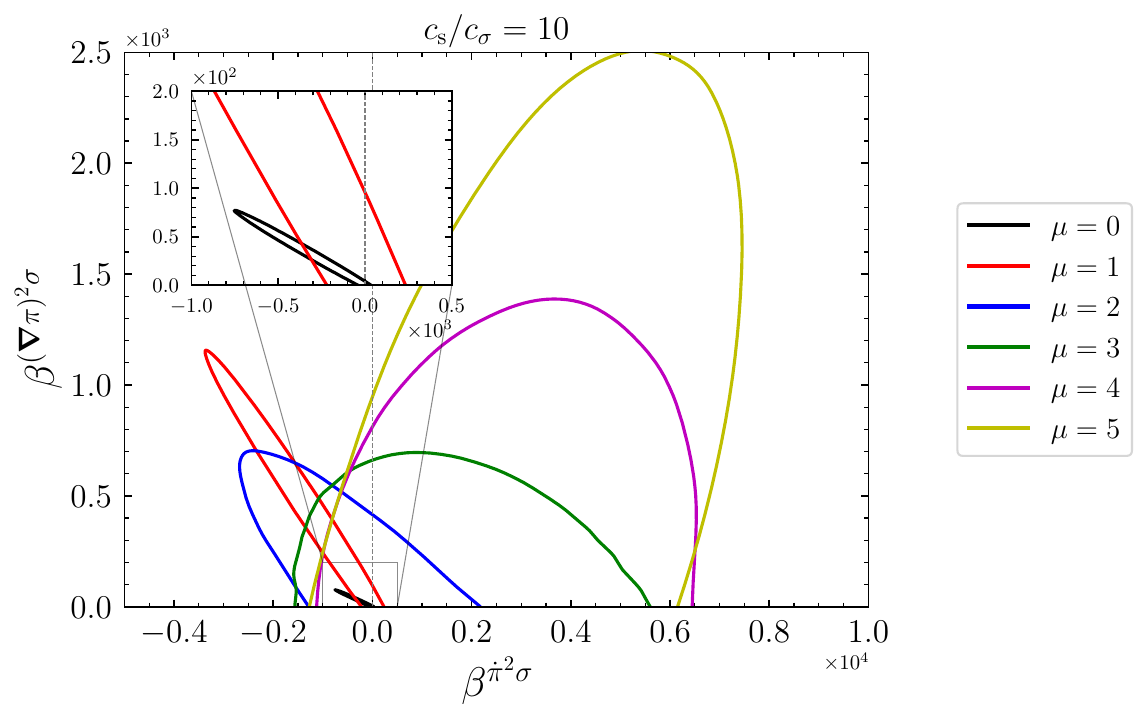}
  \end{minipage}
  \caption{Two-dimensional \textit{Planck} constraints on the spin-zero Cosmological Collider, displayed in terms of the microphysical amplitudes $\smash{\beta^\collA}$ and $\smash{\beta^\collB}$. We show $95\%$ CLs, obtained by translating the published \emph{Planck} constraints on $\smash{\fnl^{\rm equil}}$ and $\smash{\fnl^{\rm ortho}}$ using \eqref{transformation-1}\,\&\,\eqref{transformation-2}, assuming negligible inflaton self-interactions. Numerical constraints are given in Tab.~\ref{tab:planck}, and we show results marginalizing over $\mu$ (for $\smash{c_{\rm s} = c_\sigma}$) in Fig.~\ref{fig:planck-mu_marginalized}.}
  \label{fig:planck}
\end{figure}

\begin{figure}[t!]
    \centering    \includegraphics[width=0.8\textwidth]{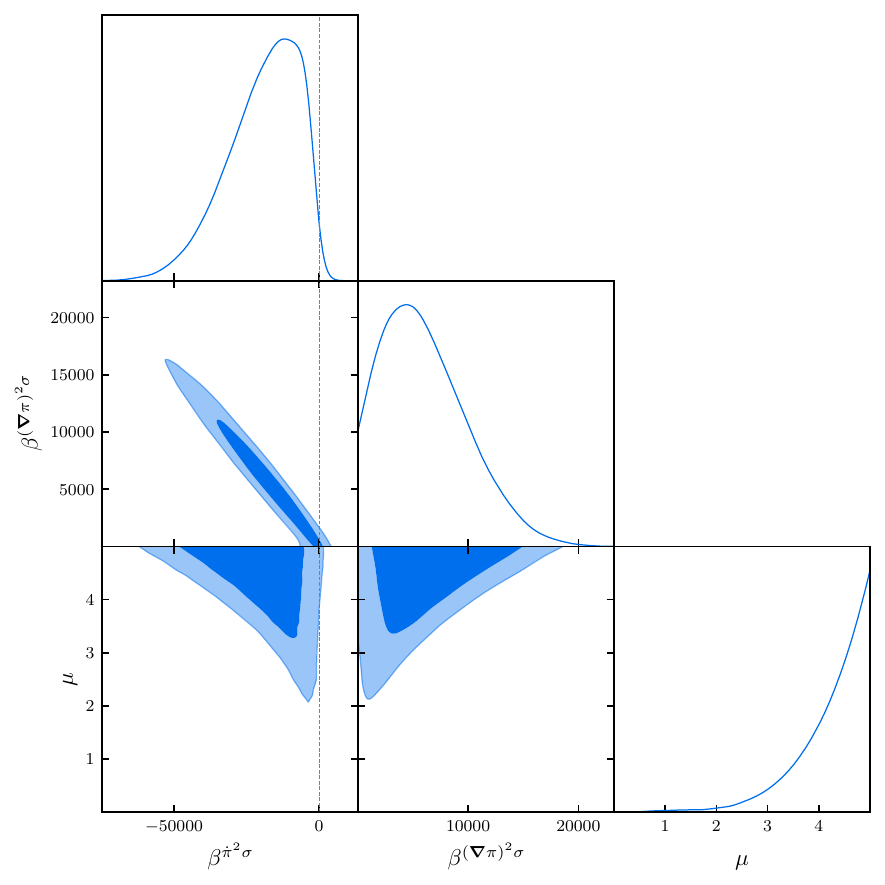}
    \caption{$68\%\,{\rm CL}$ and $95\%\,{\rm CL}$ contours and marginalized $1{\rm D}$ posteriors for $\smash{\beta^\collA,\beta^\collB,\mu}$ obtained by translating \emph{Planck} constraints on $\smash{\fnl^{\rm equil}}$ and $\smash{\fnl^{\rm ortho}}$, assuming negligible inflaton self-interactions and fixing $\smash{c_{\rm s} = c_\sigma}$. We note considerable degeneracy between $\smash{\beta^\collA}$ and $\smash{\beta^\collB}$, with constraints rapidly degrading as $\mu$ increases.}
    \label{fig:planck-mu_marginalized}
\end{figure}

\noindent We now proceed to compute the bounds on $\beta^\collA$ and $\beta^\collB$ from the \textit{Planck} $\smash{\fnl^{\rm equil}}$ and $\smash{\fnl^{\rm ortho}}$ limits from Ref.~\cite{Planck:2019kim}. The \textit{Planck} PNG posterior is well approximated by the following (Gaussian) form
\be
\label{planck_likelihood}
{\cal L}({{\bm\theta}})\propto\exp\bigg[{-\frac{({\bm\theta}-{\bm\theta}_0)\cdot(R^T\cdot D\cdot R)\cdot({\bm\theta}-{\bm\theta}_0)}{2}}\bigg]\,\,, 
\ee
where $\smash{{\bm\theta}\equiv(\fnl^{\rm equil},\fnl^{\rm ortho})}$, and the mean and covariance are specified by 
\be
\sqrt{D}=\begin{pmatrix}
    \dfrac{1}{47.1} & 0 \\
    0   & \dfrac{1}{23.9}
\end{pmatrix}\,\,,\qquad{\bm\theta}_0 = ({-26},{-38})
\qquad\text{and}\qquad 
R=\begin{pmatrix}
    \cos\alpha & {-\sin\alpha} \\
    \sin\alpha   & \cos\alpha
\end{pmatrix}
\qquad\text{with}\qquad\pi-\alpha=0.06\,\,.
\ee
from \cite{Planck:2019kim}, where we note that the empirical correlation between $\smash{\fnl^{\rm equil}}$ and $\smash{\fnl^{\rm ortho}}$ (encoded by $\sin\alpha$) is very small. By sampling this likelihood in conjunction with the Collider-to-template relations of \eqref{transformation-1}\,\&\,\eqref{transformation-2} we can obtain constraints on the collider parameters $\beta^\collA$ and $\beta^\collB$. We consider two main analyses: 
\begin{enumerate}[leftmargin=*]
    \item Sampling the amplitude parameters $\{\beta^\collA, \beta^\collB\}$ at fixed $\mu,c_{\rm s}/c_\sigma$, considering $6$ values of $\mu$ between $0$ and $5$ and $\smash{c_{\rm s}/c_\sigma\in\{0.1,1,10\}}$. Corresponding results are shown in Tab.~\ref{tab:planck} and Figs.~\ref{fig:planck}~\&~\ref{fig:planck_scan-dotpi2_nablapi2}.
    \item Jointly sampling the amplitude and mass parameters $\{\beta^\collA, \beta^\collB, \mu\}$, with $\smash{\mu\in[0,5]}$, at fixed $c_{\rm s}=c_\sigma$. Corresponding results are shown in Tab.~\ref{tab:planck-mu_marginalized} and Fig.~\ref{fig:planck-mu_marginalized}.
\end{enumerate}

First, we discuss the analysis at fixed $\mu,c_{\rm s}/c_\sigma$. From the constraints given in Tab.~\ref{tab:planck} (visualized in Fig.~\ref{fig:planck_scan-dotpi2_nablapi2}), we find that the CMB bounds on most parameters are relevant, \emph{i.e.}~they are similar to or stronger than the order-of-magnitude expectation $\beta^{\collA}, \beta^{\collB}\lesssim 10^5$ \eqref{eq: collider-expectation}. At fixed $\mu$, the bounds on both parameters strengthen significantly as $c_{\rm s}/c_\sigma$ increases (and we reach the ``equilateral collider'' limit). In contrast, the constraints somewhat weaken as $\mu$ increases (as is clear for $\smash{c_{\rm s}/c_\sigma = 1,10}$), though the trends are less obvious for $\smash{c_{\rm s}/c_\sigma = 0.1}$. Notably, these scalings encode two phenomena: (1) the varying size of inflationary bispectra produced by unit $\beta^{\collA}$ and $\beta^{\collB}$; (2) the information loss induced by our indirect measurement of the collider amplitudes (as discussed above). In this vein, we note that the weakest constraints (from small $\mu$ and low $c_{\rm s}/c_\sigma$) correspond also to the regimes where the template projection is least efficient.

Fig.~\ref{fig:planck} shows the two-dimensional posteriors corresponding to the above \textit{Planck} collider analyses. Notably, the correlation strength and direction between parameters varies strongly as a function of $c_{\rm s}/c_\sigma$ and $\mu$. For small $c_{\rm s}/c_\sigma$ and low $\mu$, the correlation becomes exceptionally tight, implying an almost-perfect degeneracy between the parameters. This matches the idealized conclusions from Fig.~\ref{fig:planck_cosine_plot}; low-mass and low-speed regimes are not well described by a combination of the equilateral and orthogonal templates. At higher mass, and for $c_{\rm s}\gg c_\sigma$, the correlations greatly reduce, whereupon the projected analyses become close to optimal. These figures can also be used to understand a curious feature of the Tab.~\ref{tab:planck} results: the $\beta^{\collA}$ posteriors are non-zero at $2\sigma-3\sigma$ for several parameter values. This occurs since the \textit{Planck} $f_{\rm NL}^{\rm ortho}$ constraint is already $1.6\sigma$ away from unity, and the joint posterior clearly suffers from significant projection effects, particularly given the $\beta^{\collB}>0$ restriction.

In Tab.~\ref{tab:planck-mu_marginalized} and Fig.\,\ref{fig:planck-mu_marginalized} we show results from the joint analysis of collider amplitudes and $\mu$ at fixed $c_{\rm s}=c_\sigma$. Here, we find no evidence for any detection, and note that the parameter space is highly non-Gaussian, due to the complex dependence of the bispectra on $\mu$. The marginalized constraint is fully consistent with zero, with constraints dominated by large masses (matching expectations, given that we previously found stronger constraints at larger, but fixed, $\mu$). We caution that this dependence may simply be a consequence of the incomplete projection of the collider shapes onto the conventional templates at low-$\mu$; we will later discuss the behavior in the full parameter space obtained with dedicated galaxy survey analysis.

The conclusion from this section is the following. In certain mass and speed regimes, the spin-zero Cosmological Collider is already tightly constrained by the \textit{Planck} bounds on equilateral and orthogonal templates. However, at small mass parameter $\mu$, and for low relative sound-speeds $c_{\rm s}\lesssim c_\sigma$, the collider shapes cannot be well-represented by the classical templates, in which case the \textit{Planck} bounds above are strongly pessimistic. In such cases, one could obtain significantly stronger constraints on the collider formalism via a dedicated analysis, such as via more optimal templates \citep[cf.][]{Jazayeri:2023xcj}, or with Large-Scale Structure data (as we perform below). The indirect analysis has a second drawback: if, in the future, one obtained a detection of $f_{\rm NL}^{\rm equil}$ or $f_{\rm NL}^{\rm ortho}$, it would not be possible to say whether this arises from massive particle interactions or \textit{bona fide} self-interactions (\emph{i.e.}~from the $f_{\rm NL}^{\dot{\pi}(\nabla\pi)^2}$ or $f_{\rm NL}^{\dot{\pi}^3}$ terms neglected above). To find distinctive signatures of massive particle non-Gaussianity, one must instead jointly analyze both collider and self-interaction signatures, allowing marginalization over the latter. We demonstrate this approach below.

\section{BOSS Full-Shape Constraints} 
\label{sec:boss_constraints}

\noindent Having whet our appetite with the CMB analyses, we now turn to the main topic of interest: constraining the spin-zero Cosmological Collider with the BOSS galaxy dataset. Since we here perform dedicated analyses (rather than reinterpretations of old data, as in \S\ref{sec:planck_equivalent}), we can simultaneously vary both $\smash{\fnl^{\rm equil}}$, $\smash{\fnl^{\rm ortho}}$ and Cosmological Collider parameters, and thus assess the possibility of massive-particle-specific signatures in the galaxy dataset. This is additionally motivated by noting that BOSS constraints on $f_{\rm NL}^{\rm equil}$ and $f_{\rm NL}^{\rm ortho}$ have already been reported in \citep{Cabass:2022epm,Cabass:2022wjy}, and exhibit much broader constraints (and thus weaker bounds on derived Cosmological Collider parameters) than those of \textit{Planck}. To assess the degeneracy between the two sets of parameters (describing inflaton self-interactions and the Cosmological Collider PNG) in BOSS data we have also performed a Principal Component Analysis based on Ref.~\cite{Philcox:2020zyp}: its results are collected in Appendix~\ref{app:SVD} and summarized below. 

To analyze the galaxy dataset, we require a physical model incorporating both gravitational and primordial effects. Here, we describe structure formation in the framework of PT, following the methodology of~\cite{Ivanov:2019pdj,DAmico:2019fhj,Philcox:2020vvt,Philcox:2021kcw,Ivanov:2023qzb}, adapted for our particular inflationary models. For Gaussian initial conditions, our perturbative expansion is identical to \cite{Cabass:2022wjy,Cabass:2022ymb}, which was extensively validated using large-volume numerical simulations in \cite{Nishimichi:2020tvu} and succeeding works \cite{Philcox:2022frc,Ivanov:2023qzb,Ivanov:2021kcd,Ivanov:2021haa,Chudaykin:2020hbf}. 

PNG affects large-scale structure via three channels: initial conditions, loop corrections, and scale-dependent galaxy bias. The implementation of these three contributions in our analysis pipeline and the estimate of their relative size compared to the Gaussian ones follows closely that of \cite{Cabass:2022wjy}. 
The theory model for the redshift-space galaxy power spectrum and bispectrum is implemented in an extension of the \texttt{CLASS-PT} code~\cite{Chudaykin:2020aoj},\footnote{The code is available at \href{https://github.com/Michalychforever/CLASS-PT}{{github.com/Michalychforever/CLASS-PT}} and custom MontePython likelihoods can be found at \href{https://github.com/oliverphilcox/full_shape_likelihoods}{github.com/oliverphilcox/full\_shape\_likelihoods}.}, which includes the non-Gaussian corrections arising from $\smash{\fnl^{\rm equil}}$ and $\smash{\fnl^{\rm ortho}}$ self-interactions, computed using the FFTlog approach~\cite{Simonovic:2017mhp}. Given the complex structure of the primordial massive particle templates, we take a different route to computing the Cosmological Collider corrections; we simply interpolate a precomputed table of power spectra and bispectra as a function of momentum, $\mu$ and $c_{\rm s}/c_\sigma$. Inclusion of this model in \texttt{CLASS-PT} is left to future work. Following \cite{Cabass:2022ymb}, we fix cosmology to the \emph{Planck} best-fit. Up to the different PNG contributions to power spectrum and bispectrum (detailed in \S\ref{sec:theory_LSS}), and the inclusion of the bispectrum multipoles, our analysis and priors on nuisance parameters are identical to~\cite{Philcox:2021kcw,Cabass:2022wjy,Cabass:2022ymb}. We describe the modeling and implementation in more detail in \S\ref{sec:theory_LSS}. 

In this analysis we vary the following PNG amplitudes: $\{\smash{\beta^\collA}, \smash{\beta^\collB}, \smash{\fnl^\equil}, \smash{\fnl^\ortho}\}$, considering the same choices of $\mu$ and $c_{\rm s}/c_\sigma$ as in \S\ref{sec:planck_equivalent}. As before, we first carry out an analysis where we scan over Collider parameters (\S\ref{sec:boss_mu_scan}), and then one in which we fix $\smash{c_{\rm s}/c_\sigma = 1}$ and marginalize over $\mu$ (\S\ref{sec:boss_mu_marginalize}). Finally, \S\ref{sec:boss_dbi} considers the DBI Cosmological Collider discussed in \S\ref{sec:DBI_theory}, for which $\smash{\smash{\fnl^\ortho} = \smash{\fnl^\ortho}(\fnl^\equil)}$ and $\smash{\beta^\collB = \beta^\collB(\beta^\collA,c_{\rm s}/c_\sigma)}$, which significantly reduces parameter degeneracies. 

Amongst the various microphysical parameters in the EFT, the quadratic galaxy biases $\smash{b_2}$, $\smash{b_{{\cal G}_2}}$ (as defined in \S\ref{sec:theory_LSS}) play the most important role, since they are the most degenerate with inflaton self-interactions \citep[e.g.,][]{Baumann:2021ykm}. As discussed in \S\ref{sec:cosmo_collider_observables}, Cosmological Collider signatures are degenerate with $\smash{\fnl^\equil}$ and $\smash{\fnl^\ortho}$ in certain limiting regimes; as such, marginalization over $\smash{b_2}$, $\smash{b_{{\cal G}_2}}$ will likely affect the constraints on $\smash{\beta^\collA}$ and $\smash{\beta^\collB}$. To study this, in \S\ref{sec:boss_mu_marginalize} and \S\ref{sec:boss_dbi} we also implement new physically-motivated priors on $b_2$, $b_{{\cal G}_2}$, derived from mapping the PT parameters to state-of-the-art halo occupation distribution modeling of galaxy formation and evolution, taking into account theoretical uncertainties. Both the priors and details on their derivation are contained in Appendix \ref{app:HOD}, and further details are presented in the companion work \cite{future_kaz}.  
In addition, see the recent works \cite{Barreira:2021ukk,Sullivan:2021sof,Lazeyras:2021dar,Ivanov:2024hgq,Ivanov:2024xgb} for calibration of perturbation theory parameters with simulations.

In terms of observations, we consider the twelfth data release (DR12) of the BOSS survey, part of SDSS-III~\cite{Alam:2016hwk,SDSS:2011jap}. The data is split in two redshift bins with effective redshifts $z=0.38,0.61$, in each of the Northern and Southern galactic caps (NGC and SGC). This results in four independent data chunks, with the union comprising $\sim 1.2\times 10^6$ galaxies across a volume of $6\,(h^{-1}\text{Gpc})^3$. From each data chunk we use the power spectrum multipoles ($\ell=0,2,4$) for $k\in [0.01,0.17)\,\hMpc$, the real-space power spectrum $Q_0$ for $k\in [0.17,0.4)\,\hMpc$ \cite{Ivanov:2021haa}, and the BAO parameters extracted from the post-reconstructed power spectrum data~\cite{Philcox:2020vvt}, as in~\cite{Philcox:2021kcw}. Moreover we include the bispectrum monopole, quadrupole and hexadecapole \cite{Ivanov:2023qzb} for each data chunk, for triangle configurations within the range of $k_i\in [0.01,0.08) \,\hMpc$, for a total of $62\times 3$ triangles. As in \cite{Cabass:2022wjy}, power spectra and bispectra are measured with the window-free estimators~\cite{Philcox:2020vbm,Philcox:2021ukg}. We measure sample covariance matrices from the $2048$ MultiDark-Patchy mocks~\cite{Kitaura:2015uqa}.

\subsection{Details of the perturbation theory model}
\label{sec:theory_LSS}

\noindent In this section, we derive the corrections to the tree-level bispectrum and one-loop power spectrum of galaxies arising from initial conditions including massive particle interactions. For this purpose, we will require the linear and quadratic redshift-space kernels relating the galaxy density field to the linear matter overdensity field $\delta^{(1)}$. Following the notation of \cite{Chudaykin:2020aoj,Cabass:2022wjy}, we have 
\be
Z_1(\k)=b_1+f (\hat{\k}\cdot \hat{\mathbf{z}})^2 
\ee
and 
\begin{equation}
\begin{split}
    Z_2(\k_1,\k_2)  &=\frac{b_2}{2}+b_{\mathcal{G}_2}\bigg(\frac{(\k_1\cdot \k_2)^2}{k_1^2k_2^2}-1\bigg)
+b_1 F_2(\k_1,\k_2)+f\mu^2 G_2(\k_1,\k_2) \\
&\;\;\;\;+\frac{f\mu k}{2}\bigg(\frac{\mu_1}{k_1}(b_1+f\mu_2^2)+
\frac{\mu_2}{k_2}(b_1+f\mu_1^2)
\bigg)\,\,,
\end{split}
\end{equation}
where $F_2$ and $G_2$ are the standard matter and velocity kernels \cite{Bernardeau:2001qr} 
\begin{equation}
\begin{split}
F_2(\k_1,\k_2) &= \frac{5}{7} + \frac{2}{7}\frac{(\k_1\cdot \k_2)^2}{k_1^2k_2^2} + \frac{1}{2}\frac{\k_1\cdot \k_2}{k_1k_2}\bigg(\frac{k_1}{k_2} + \frac{k_2}{k_1}\bigg)\,\,, \\
G_2(\k_1,\k_2) &= \frac{3}{7} + \frac{4}{7}\frac{(\k_1\cdot \k_2)^2}{k_1^2k_2^2} + \frac{1}{2}\frac{\k_1\cdot \k_2}{k_1k_2}\bigg(\frac{k_1}{k_2} + \frac{k_2}{k_1}\bigg) \,\,.
\end{split}
\end{equation}
In these equations $f$ is the logarithmic growth factor, $\hat{\mathbf{z}}$ is the line-of-sight direction unit vector, $\hat{\k}\equiv \k/k$, and $\mu_i = \hat{\mathbf{k}}_i\cdot\hat{\mathbf{z}}$. The correction to the bispectrum follows immediately:
\be 
\label{eq:galb}
\begin{split}
    B_{g,111}(\mathbf{k}_1,\mathbf{k}_2,\mathbf{k}_3) & = Z_1(\mathbf{k}_1)Z_1(\mathbf{k}_2)Z_1(\mathbf{k}_3)B_{111}(k_1,k_2,k_3)
\end{split}
\ee
where $\smash{B_{111}}$ denotes the PNG bispectrum signal~\cite{Sefusatti:2007ih,Sefusatti:2009qh}, \emph{i.e.} 
\be 
\label{eq:inib}
\begin{split}
\langle \delta^{(1)}\delta^{(1)}\delta^{(1)}\rangle & \equiv \fnl B_{111}(k_1,k_2,k_3) (2\pi)^3\delta^{(3)}_D(\k_{123})\,\,,\\
\fnl B_{111}(k_1,k_2,k_3)& =\mathcal{T}(k_1)\mathcal{T}(k_2)\mathcal{T}(k_3)B_\zeta(k_1,k_2,k_3)\,\,.
\end{split}
\ee
Here $\k_{123}\equiv\k_1+\k_2+\k_3$, and the transfer function $\smash{\cal T}$ is defined as $\smash{\mathcal{T}(k)\equiv \delta^{(1)}(\k)/\zeta(\k)=(P_{11}(k)/P_\zeta(k))^{1/2}}$, where $\smash{\langle\delta^{(1)}(\k)\delta^{(1)}(\k')\rangle=P_{11}(k)(2\pi)^3\delta^{(3)}_{\rm D}(\k+\k')}$ and we have suppressed the time dependence for clarity. In terms of the $Z_2$ kernel, the one-loop correction to the galaxy power spectrum from PNG, $\smash{P_{12}(\mathbf{k})}$, is given by \cite{Taruya:2008pg,Assassi:2015jqa,MoradinezhadDizgah:2020whw,MoradinezhadDizgah:2017szk}
\be 
\label{eq:P12_expression}
\begin{split}
\fnl P_{g,12}(\textbf{k}) = 2\fnl Z_1(\k)
\int\frac{{\rm d}^3q}{(2\pi)^3} Z_2(\q,\k-\q) B_{111}(k,q,|\k-\q|)\,\,,
\end{split}
\ee 
These formulas hold for the Cosmological Collider contributions from both the $\smash{\dot{\pi}^2\sigma}$ and the $\smash{({\bm\nabla}\pi)^2\sigma}$ vertices: in the final galaxy power spectrum and bispectrum they must be summed together, each multiplied by their respective $f_{\rm NL}$ (we refer to Eq.~\eqref{equilateral_normalization} for a definition of $\smash{f^\collA_{\rm NL}}$ and $\smash{f^\collB_{\rm NL}}$ in terms of $\smash{\beta^\collA_{\rm NL}}$ and $\smash{\beta^\collB_{\rm NL}}$, and a definition of their respective primordial bispectra $B_\zeta$). From \eqref{eq:P12_expression} we see that the PNG contribution to the power spectrum depends on the quadratic bias $b_2$ and the tidal bias $b_{\mathcal{G}_2}$ (in addition to linear bias, $b_1$), which we marginalize over in our MCMC analyses.

\setlength{\tabcolsep}{4pt}
\renewcommand{\arraystretch}{1.3}
\begin{table*}[t!]
\centering
\begin{tabular}{c|ccc|c}
 & \multicolumn{3}{c|}{${\beta^\collA}$} & \multicolumn{1}{c}{${\beta^\collB}$}\\ \hline
 \textbf{Model} & mean & \quad 95\% lower \quad & 95\% upper & 95\% upper \\ 
 \hline

$\mu=0,c_{\rm s}/c_\sigma = 0.1$ & $\num[round-mode=places,round-precision=\prec]{2.0963328347075072e+0004}$ & $\num[round-mode=places,round-precision=\prec]{-1.4548062446088338e+0005}$ & $\num[round-mode=places,round-precision=\prec]{1.9982979736274714e+0005}$ & $\num[round-mode=places,round-precision=\prec]{2.36877e+0004}$ \\

$\mu=0,c_{\rm s}/c_\sigma = 1$ & $\num[round-mode=places,round-precision=\prec]{7.92053773542419e+003}$ & $\num[round-mode=places,round-precision=\prec]{-8.802214796648754e+003}$ & $\num[round-mode=places,round-precision=\prec]{2.9119663006104558e+004}$ & $\num[round-mode=places,round-precision=\prec]{1.1657720000000002e+003}$ \\

$\mu=0,c_{\rm s}/c_\sigma = 10$ & $\num[round-mode=places,round-precision=\prec]{-4.706736506938791e+03}$ & $\num[round-mode=places,round-precision=\prec]{-1.3076936874319086e+04}$ & $\num[round-mode=places,round-precision=\prec]{8.437092038377814e+02}$ & $\num[round-mode=places,round-precision=\prec]{1.164901e+003}$ \\ \hline

$\mu=1,c_{\rm s}/c_\sigma = 0.1$ & $\num[round-mode=places,round-precision=\prec]{2.2666170874861058e+0004}$ & $\num[round-mode=places,round-precision=\prec]{-1.0864781844364892e+0005}$ & $\num[round-mode=places,round-precision=\prec]{1.9295836206162052e+0005}$ & $\num[round-mode=places,round-precision=\prec]{3.0860170000000002e+0004}$ \\

$\mu=1,c_{\rm s}/c_\sigma = 1$ & $\num[round-mode=places,round-precision=\prec]{-6.210096300100323e+003}$ & $\num[round-mode=places,round-precision=\prec]{-2.5484917552375482e+004}$ & $\num[round-mode=places,round-precision=\prec]{9.375740999903428e+003}$ & $\num[round-mode=places,round-precision=\prec]{1.8954879999999998e+003}$ \\

$\mu=1,c_{\rm s}/c_\sigma = 10$ & $\num[round-mode=places,round-precision=\prec]{-7.081472186185528e+003}$ & $\num[round-mode=places,round-precision=\prec]{-2.006251751858183e+004}$ & $\num[round-mode=places,round-precision=\prec]{2.4635356806983546e+003}$ & $\num[round-mode=places,round-precision=\prec]{5.479189e+003}$ \\ \hline

$\mu=2,c_{\rm s}/c_\sigma = 0.1$ & $\num[round-mode=places,round-precision=\prec]{8.026420507767801e+0004}$ & $\num[round-mode=places,round-precision=\prec]{-2.7801980492925097e+0004}$ & $\num[round-mode=places,round-precision=\prec]{2.2594481574684966e+0005}$ & $\num[round-mode=places,round-precision=\prec]{3.478652e+0004}$ \\

$\mu=2,c_{\rm s}/c_\sigma = 1$ & $\num[round-mode=places,round-precision=\prec]{-3.4776224862976895e+004}$ & $\num[round-mode=places,round-precision=\prec]{-9.62301172211811e+004}$ & $\num[round-mode=places,round-precision=\prec]{2.385846008188848e+003}$ & $\num[round-mode=places,round-precision=\prec]{2.4632109999999997e+0004}$ \\

$\mu=2,c_{\rm s}/c_\sigma = 10$ & $\num[round-mode=places,round-precision=\prec]{-1.8189346403356714e+004}$ & $\num[round-mode=places,round-precision=\prec]{-4.53523203703303e+004}$ & $\num[round-mode=places,round-precision=\prec]{2.220393488798756e+003}$ & $\num[round-mode=places,round-precision=\prec]{1.106223e+0004}$ \\ \hline

$\mu=3,c_{\rm s}/c_\sigma = 0.1$ & $\num[round-mode=places,round-precision=\prec]{1.1176555901932406e+005}$ & $\num[round-mode=places,round-precision=\prec]{-7.166029005143777e+003}$ & $\num[round-mode=places,round-precision=\prec]{2.6372968805494596e+005}$ & $\num[round-mode=places,round-precision=\prec]{3.320432e+0004}$ \\

$\mu=3,c_{\rm s}/c_\sigma = 1$ & $\num[round-mode=places,round-precision=\prec]{-1.2167750348403562e+005}$ & $\num[round-mode=places,round-precision=\prec]{-3.210408483211994e+005}$ & $\num[round-mode=places,round-precision=\prec]{2.416865838597063e+003}$ & $\num[round-mode=places,round-precision=\prec]{9.00535e+0004}$ \\

$\mu=3,c_{\rm s}/c_\sigma = 10$ & $\num[round-mode=places,round-precision=\prec]{-2.0678427353561096e+004}$ & $\num[round-mode=places,round-precision=\prec]{-5.109968593992335e+004}$ & $\num[round-mode=places,round-precision=\prec]{1.5016226649980526e+003}$ & $\num[round-mode=places,round-precision=\prec]{1.194472e+0004}$ \\ \hline

$\mu=4,c_{\rm s}/c_\sigma = 0.1$ & $\num[round-mode=places,round-precision=\prec]{1.57315454331324e+005}$ & $\num[round-mode=places,round-precision=\prec]{-5.358506504853402e+003}$ & $\num[round-mode=places,round-precision=\prec]{3.5880527843426226e+005}$ & $\num[round-mode=places,round-precision=\prec]{3.317346e+0004}$ \\

$\mu=4,c_{\rm s}/c_\sigma = 1$ & $\num[round-mode=places,round-precision=\prec]{-1.9131168126820147e+005}$ & $\num[round-mode=places,round-precision=\prec]{-5.0048238680046936e+005}$ & $\num[round-mode=places,round-precision=\prec]{3.043940287598176e+003}$ & $\num[round-mode=places,round-precision=\prec]{1.2740059999999999e+00005}$ \\

$\mu=4,c_{\rm s}/c_\sigma = 10$ & $\num[round-mode=places,round-precision=\prec]{-2.2666386159092974e+004}$ & $\num[round-mode=places,round-precision=\prec]{-5.6405082554297056e+004}$ & $\num[round-mode=places,round-precision=\prec]{1.4742874913798296e+003}$ & $\num[round-mode=places,round-precision=\prec]{1.3128310000000001e+0004}$ \\ \hline

$\mu=5,c_{\rm s}/c_\sigma = 0.1$ & $\num[round-mode=places,round-precision=\prec]{2.4094109699244105e+005}$ & $\num[round-mode=places,round-precision=\prec]{-8.19581536086317e+003}$ & $\num[round-mode=places,round-precision=\prec]{5.400784546784197e+005}$ & $\num[round-mode=places,round-precision=\prec]{2.809533e+0004}$ \\

$\mu=5,c_{\rm s}/c_\sigma = 1$ & $\num[round-mode=places,round-precision=\prec]{-3.4272970668060525e+005}$ & $\num[round-mode=places,round-precision=\prec]{-8.9286316888657e+005}$ & $\num[round-mode=places,round-precision=\prec]{4.869582520024385e+003}$ & $\num[round-mode=places,round-precision=\prec]{2.1872340000000002e+00005}$ \\

$\mu=5,c_{\rm s}/c_\sigma = 10$ & $\num[round-mode=places,round-precision=\prec]{-2.5099837212009603e+004}$ & $\num[round-mode=places,round-precision=\prec]{-6.160124281889525e+004}$ & $\num[round-mode=places,round-precision=\prec]{1.6373456344425795e+003}$ & $\num[round-mode=places,round-precision=\prec]{1.43277e+0004}$ \\

 \end{tabular}
 \caption{BOSS constraints on spin-zero massive particles for various values of mass parameter $\mu$ and ratio of sound-speeds $\smash{c_{\rm s}/c_\sigma}$. In all cases, we marginalize over inflaton self-interactions (unlike in the CMB analyses). As in Tab.~\ref{tab:planck} we display results for the microphysical collider amplitudes $\smash{\beta^\collA}$ and $\smash{\beta^\collB}$, displaying the mean and $95\%$ CL bound for the first amplitude and the $95\%\,{\rm CL}$ upper limit for the second (recalling that $\smash{\beta^\collB}\geq 0$). These constraints are visualized in Figs.~\ref{fig:boss_scan-dotpi2_nablapi2}~\&~\ref{fig:boss_scan}.}
  \label{tab:boss_scan}
\end{table*}

The final ingredient is the scale-dependent bias. Let us focus on the PNG contribution to the power spectrum \eqref{eq:P12_expression} proportional to $b_1 b_2$. Up to irrelevant numerical factors it takes the form \cite{Cabass:2022ymb}
\be
f_{\rm NL}P^{b_1b_2}_{g,12}({\bf{k}}) = \frac{f_{\rm NL}\Delta^4_\zeta{\cal T}(k)}{k^2}\int_{\bf q}\frac{{\cal S}(k,q,|{\bf k}-{\bf q}|){\cal T}(q){\cal T}(|{\bf k}-{\bf q}|)}{q^2|{\bf k}-{\bf q}|^2}\,\,. 
\ee
In the low-$k$ limit, using \eqref{squeezed_behavior_shape} and multiplying and dividing by a reference momentum $k'_\ast$, we can write this as 
\be
\frac{f_{\rm NL}P_{11}(k)}{{\cal T}(k)}\,k^{\frac{3}{2}}\Bigg[\cos\bigg(\mu\ln\frac{k}{k'_\ast}\bigg)\int_{\bf q}\frac{P_{11}(q)}{q^{{3}/{2}}}\cos\bigg(\mu\ln\frac{qc_{\rm s}}{k'_\ast}\bigg) + \sin\bigg(\mu\ln\frac{k}{k'_\ast}\bigg)\int_{\bf q}\frac{P_{11}(q)}{q^{{3}/{2}}}\sin\bigg(\mu\ln\frac{qc_{\rm s}}{k'_\ast}\bigg)\Bigg]\,\,. 
\ee 
This form implies the counterterm (\emph{i.e.}~a scale-dependent bias)
\be
\label{counterterm}
\frac{f_{\rm NL}P_{11}(k)}{{\cal T}(k)}\,(L_1k)^{\frac{3}{2}}\cos\bigg(\mu\ln\frac{k}{k'_\ast}\bigg) + \frac{f_{\rm NL}P_{11}(k)}{{\cal T}(k)}\,(L_2k)^{\frac{3}{2}}\sin\bigg(\mu\ln\frac{k}{k'_\ast}\bigg) = \frac{f_{\rm NL}P_{11}(k)}{{\cal T}(k)}\,(Lk)^{\frac{3}{2}}\cos\bigg(\mu\ln\frac{k}{k_\ast}\bigg)\,\,, 
\ee 
for $\smash{k_\ast = k'_\ast\exp(\varphi/\mu)}$, $\smash{\varphi = \arg(L_1+{\rm i}L_2)}$, and $\smash{L=\sqrt{L_1^2+L^2_2}}$. Both $L$ and $k_\ast$ are PT parameters that absorb the UV dependence of the loop integral (note also that the non-Gaussianities coming from the $\smash{\dot{\pi}^2\sigma}$ and $\smash{({\bm\nabla}\pi)^2\sigma}$ couplings and orthogonal PNG produce identical scale-dependent biases that can be captured by a single counterterm, as in \cite{Cabass:2022wjy}).\footnote{Notice that Eq.~\eqref{counterterm} is the leading counterterm in a derivative expansion. We refer to \cite{Schmidt:2013nsa,Cabass:2018roz} for a discussion on how to go beyond leading order (in non-Cosmological-Collider PNG scenarios).} To avoid having non-linearities in the sampling, our analysis expands Eq.~\eqref{counterterm} as 
\be
\begin{split}
    \frac{f_{\rm NL}P_{11}(k)}{{\cal T}(k)}\,(Lk)^{\frac{3}{2}}\cos\bigg(\mu\ln\frac{k}{k_\ast}\bigg) &= b^{\rm coll}_{A}\frac{f_{\rm NL}P_{11}(k)}{{\cal T}(k)}\,\bigg(\frac{k}{0.45 \,{\rm Mpc}^{-1}}\bigg)^{\frac{3}{2}}\cos\bigg(\mu\ln\frac{k}{0.45 \,{\rm Mpc}^{-1}}\bigg) \\
    &\;\;\;\;+ b^{\rm coll}_{B}\frac{f_{\rm NL}P_{11}(k)}{{\cal T}(k)}\,\bigg(\frac{k}{0.45 \,{\rm Mpc}^{-1}}\bigg)^{\frac{3}{2}}\sin\bigg(\mu\ln\frac{k}{0.45 \,{\rm Mpc}^{-1}}\bigg)\,\,,
\end{split}
\ee
and we sample $\smash{b^{\rm coll}_{A} = 1.686 \times  \frac{18}{5} (b_1-1) \tilde{b}^{\rm coll}_{A}}$, $\smash{b^{\rm coll}_{B} = 1.686 \times  \frac{18}{5} (b_1-1) \tilde{b}^{\rm coll}_{B}}$ within a Gaussian prior $\smash{ \tilde{b}^{\rm coll}_{A}\sim \mathcal{N}(1,5)}$, $\smash{\tilde{b}^{\rm coll}_{B}\sim \mathcal{N}(1,5)}$. This is motivated by the peak-background split model~\cite{Schmidt:2010gw,Cabass:2022wjy}. Note that in the analysis we consistently include also the $P_{g,12}$ contributions from equilateral and orthogonal PNG, together with the respective counterterms. Their implementation is the same as in Ref.~\cite{Cabass:2022wjy}. 

In summary, the Cosmological Collider corrections to the galaxy power spectrum and bispectrum in redshift space are 
\be
\begin{split}
P_g(\k) & \supset (f^\collA_{\rm NL}+f^\collB_{\rm NL})\Bigg[P_{g,12}(\k) + 2 b^{\rm coll}_A Z_1(\k)\bigg(\frac{k}{0.45 \,{\rm Mpc}^{-1}}\bigg)^{\frac{3}{2}}\cos\bigg(\mu\ln\frac{k}{0.45 \,{\rm Mpc}^{-1}}\bigg)\frac{P_{11}(k)}{\mathcal{T}(k)} \\
&\;\;\;\,\,\, \hphantom{(f^\collA_{\rm NL}+f^\collB_{\rm NL})\Bigg[P_{12}(\k)} + 2 b^{\rm coll}_B Z_1(\k)\bigg(\frac{k}{0.45 \,{\rm Mpc}^{-1}}\bigg)^{\frac{3}{2}}\sin\bigg(\mu\ln\frac{k}{0.45 \,{\rm Mpc}^{-1}}\bigg)\frac{P_{11}(k)}{\mathcal{T}(k)}\Bigg]\,\,, \\
B_g(\k_1,\k_2,\k_3) & \supset f^\collA_{\rm NL}Z_1(\k_1)Z_1(\k_2)Z_1(\k_3)B^\collA_{111}(k_1,k_2,k_3) + f^\collB_{\rm NL}Z_1(\k_1)Z_1(\k_2)Z_1(\k_3)B^\collB_{111}(k_1,k_2,k_3)
\,\,,
\end{split}
\ee
where we again refer to Eq.~\eqref{equilateral_normalization} for the definition of $\smash{f^\collA_{\rm NL}}$ and $\smash{f^\collB_{\rm NL}}$. 
This adds to the Gaussian power spectrum and bispectrum models that include the effects of late-time nonlinear mode coupling at one-loop and tree level respectively~\cite{Ivanov:2019hqk,Ivanov:2021kcd}, \emph{and} the contributions from equilateral and orthogonal non-Gaussianity~\cite{Cabass:2022wjy}. 
In practice, we compute the Legendre redshift-space multipoles $P_{\ell}$ ($\ell=0,2,4$) of the galaxy power spectrum and use the bispectrum monopole, quadrupole and hexadecapole as in \cite{Ivanov:2023qzb}. We also implement IR resummation in redshift space~\cite{Senatore:2014via,Baldauf:2015xfa,Blas:2015qsi,Blas:2016sfa,Ivanov:2018gjr,Vasudevan:2019ewf} (to account for long-wavelength displacement effects) and the Alcock-Paczy{\'n}ski effect~\cite{Alcock:1979mp} (to account for coordinate conversions~\cite{Ivanov:2019pdj}). 

As in \cite{Cabass:2022wjy,Cabass:2022ymb}, we estimate the relative size of various perturbative corrections to the galaxy power spectrum using the scaling universe approximation~\cite{Pajer:2013jj,Assassi:2015jqa}. Using the scale invariance of the PNG shapes $\cal S$, and for a power-law linear power spectrum $P_{11}(k)\sim (k/k_{\rm NL})^{n}k_{\rm NL}^{-3}$ with $n\approx -1.7$ for quasi-linear wavenumbers $k\simeq 0.15\,\hMpc$, the total dimensionless galaxy power spectrum $\Delta^2(k)\equiv k^3 P(k)/(2\pi^2)$ can be estimated as 
\be 
\label{scalings_scaling_universe}
\begin{split}
\Delta^2(k)& =
{\underbrace{\left(\frac{k}{k_{\rm NL}}\right)^{1.3}}_{\rm tree}} +
{\underbrace{\left(\frac{k}{k_{\rm NL}}\right)^{2.6}}_{\rm 1-loop}} 
+ {\underbrace{\left(\frac{k}{k_{\rm NL}}\right)^{3.3}}_{\rm ctr}} 
+ {\underbrace{\left(\frac{k}{k_{\rm NL}}\right)^{3}}_{\rm stoch}} 
+ {\underbrace{\fnl \Delta_\zeta\left(\frac{k}{k_{\rm NL}}\right)^{1.95}}_{\text{NG matter loops}}} 
+ {\underbrace{\fnl b_{\zeta}\Delta_\zeta\left(\frac{k}{k_{\rm NL}}\right)^{2.15}}_{\text{linear PNG bias}}}\,\,.
\end{split} 
\ee 
The only difference with \cite{Cabass:2022wjy} is in the linear PNG bias, which scales as $\smash{(k/k_{\rm NL})^{2.15}}$ instead of $\smash{(k/k_{\rm NL})^{2.65}}$. As in \cite{Cabass:2022wjy}, all higher-order corrections (including higher-derivative counterterms in the linear PNG bias) are subleading for $\fnl \Delta_\zeta \lesssim 0.1$ and $k\lesssim 0.17~\hMpc$ typical for our analysis, thus they can be neglected. Given that these scalings are very similar to the case of single-field inflation non-Gaussianity, we expect that the application of our pipeline on mock catalogs generated with Gaussian initial conditions would correctly recover zero PNG, as was shown in \cite{Cabass:2022wjy}, through application to the Nseries mocks~\cite{Alam:2016hwk} considering equilateral and orthogonal PNG.

\subsection{\texorpdfstring{Constraints at fixed $\mu,c_{\rm s}/c_\sigma$}{Constraints at fixed mass and speed}}
\label{sec:boss_mu_scan}

\noindent In this section we start collecting the results of our MCMC analyses. We first present BOSS constraints on the Cosmological Collider PNG amplitudes obtained from $18$ analyses, each with a different value of $\mu$ and $\smash{c_{\rm s}/c_\sigma}$. This mirrors the analysis of \S\ref{sec:planck_equivalent}, except that we now marginalize over self-interactions (to search for massive-particle specific shapes, given previous equilateral and orthogonal bounds).

\begin{figure}[t]
\centering
\includegraphics[width = 0.49\textwidth]{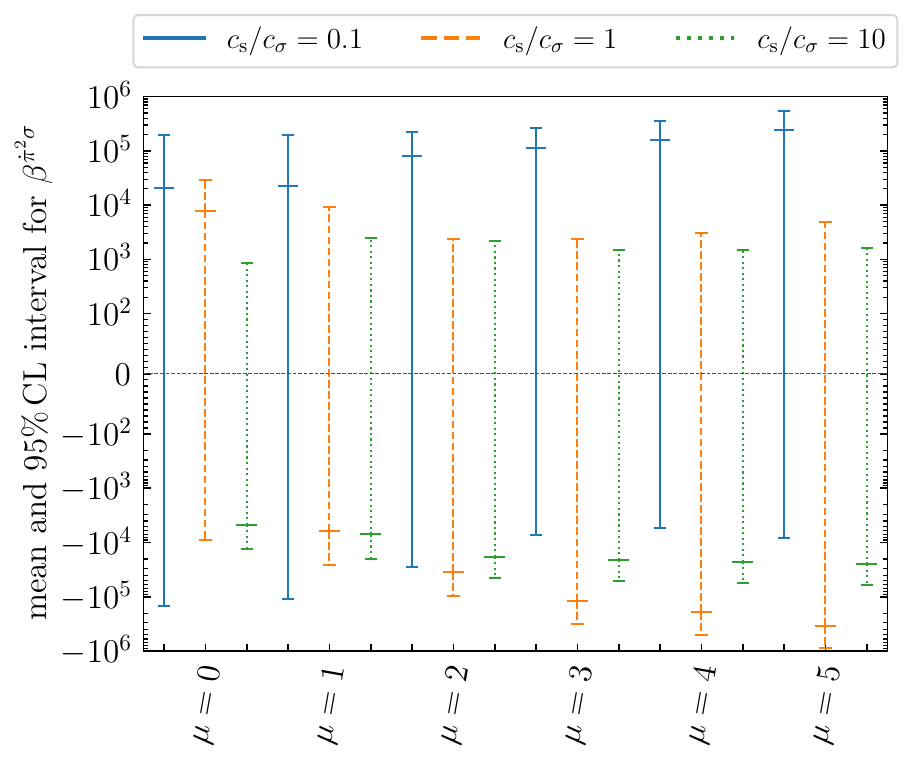} 
\includegraphics[width = 0.49\textwidth]{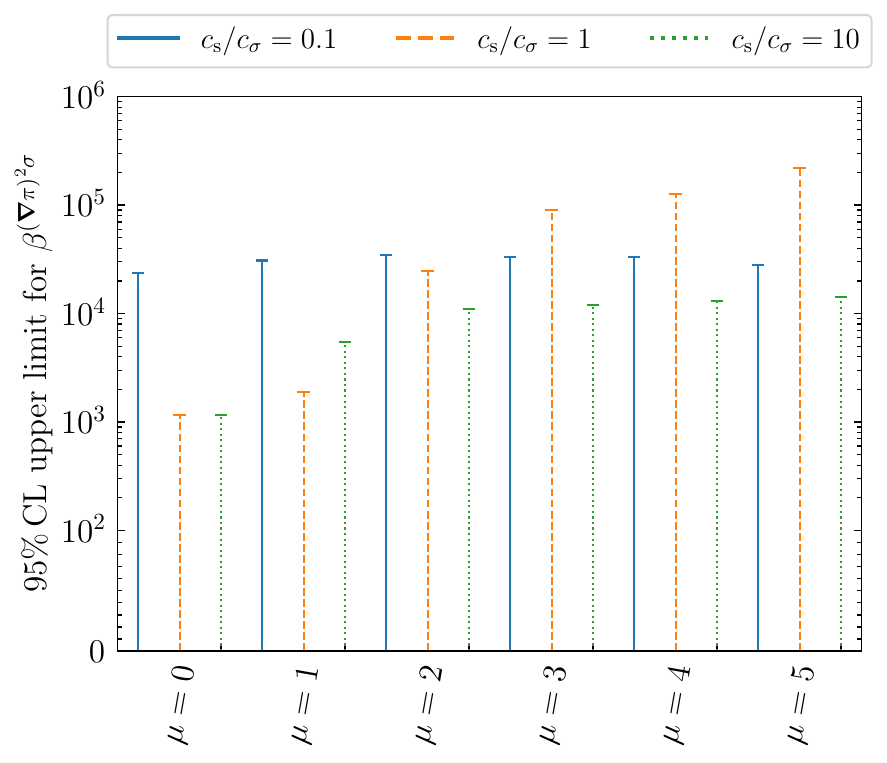} 
\caption{Mean and 95\% CL limits on $\smash{\beta^\collA}$ (left) and $95\%\,{\rm CL}$ upper limit for $\smash{\beta^\collB}$ (right) from BOSS data after marginalization over inflaton self-interactions. We plot the same choices of mass $\mu$ and relative sound-speed $c_{\rm s}/c_\sigma$ as in \S\ref{sec:planck_equivalent}. The numerical values are given in Tab.~\ref{tab:boss_scan}. We find no detection of either collider amplitude in any model.} 
\label{fig:boss_scan-dotpi2_nablapi2}
\end{figure}

\begin{figure}
  \begin{minipage}{0.45\textwidth}
    \centering
    \includegraphics[height=0.8\textwidth]{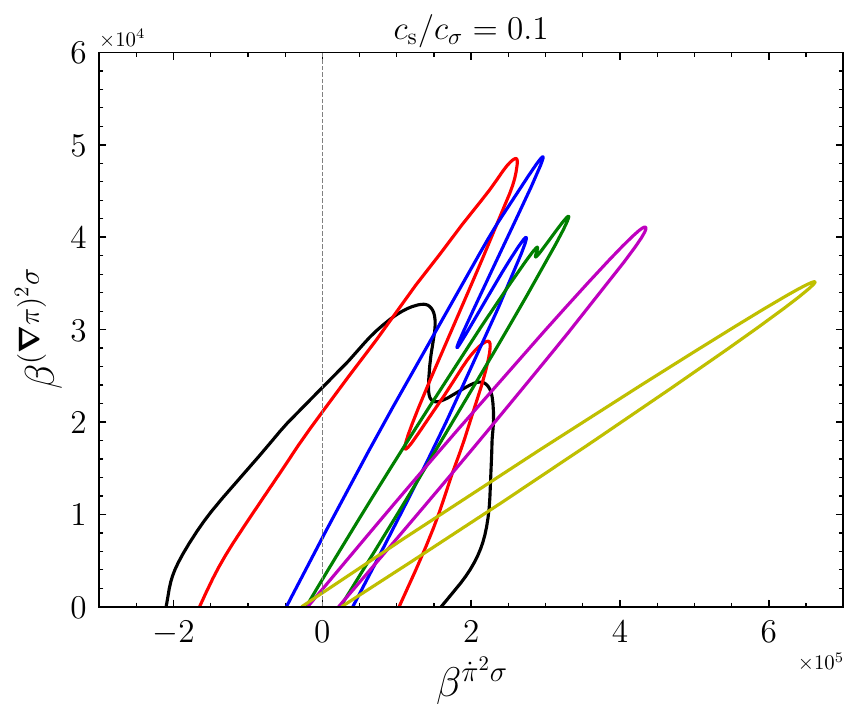}
  \end{minipage}
  \hfill
  \begin{minipage}{0.45\textwidth}
    \centering
    \includegraphics[height=0.8\textwidth]{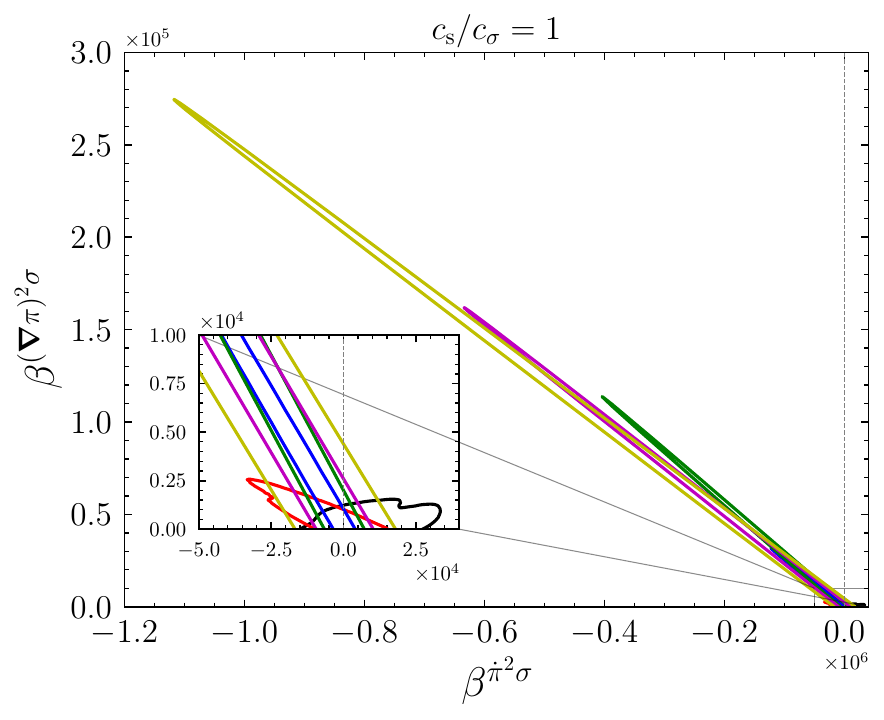}
  \end{minipage}
  \hfill
  \begin{minipage}{0.45\textwidth}
    \centering
    \includegraphics[height=0.8\textwidth]{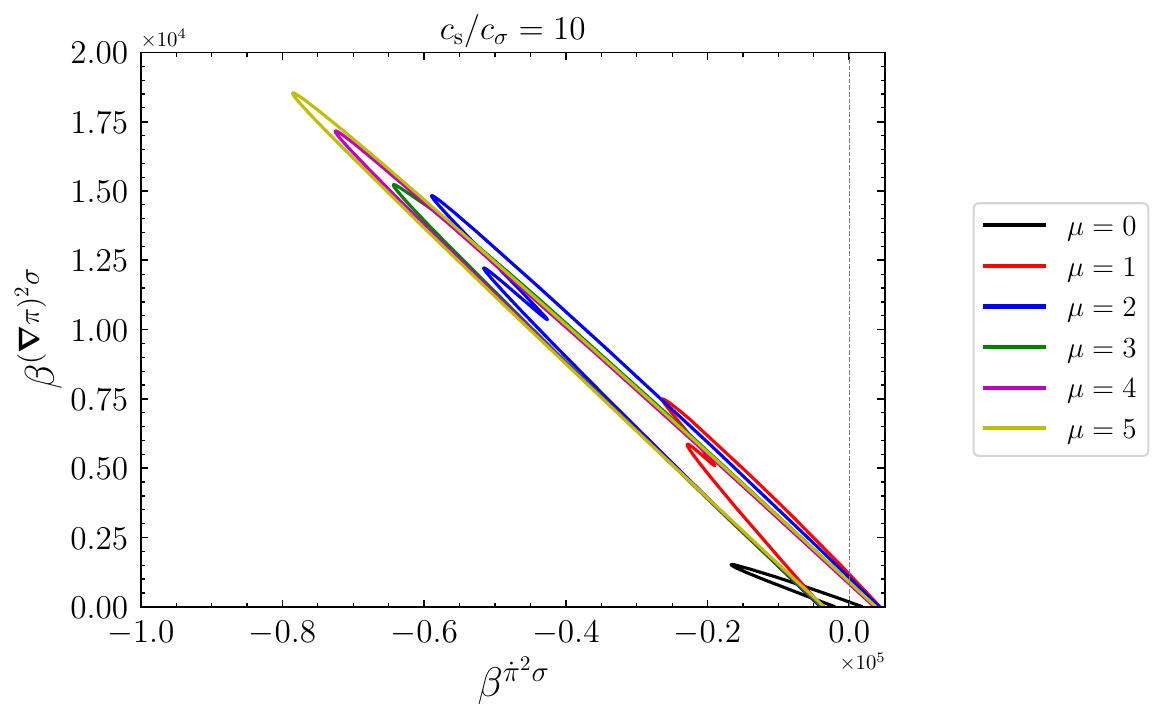}
  \end{minipage}
  \caption{Two-dimensional BOSS constraints on the spin-zero Cosmological Collider for various $\mu$ and $c_{\rm s}/c_\sigma$, displayed in terms of the microphysical amplitudes $\smash{\beta^\collA}$ and $\smash{\beta^\collB}$. In all cases, we marginalize over $\smash{\fnl^{\rm equil}}$ and $\smash{\fnl^{\rm ortho}}$, which parametrize inflaton self-interactions. We show $95\%$ CLs in all cases. Numerical constraints are given in Tab.~\ref{tab:boss_scan}.}
  \label{fig:boss_scan}
\end{figure}

\begin{figure}[t]
\centering
\includegraphics[width = 0.49\textwidth]{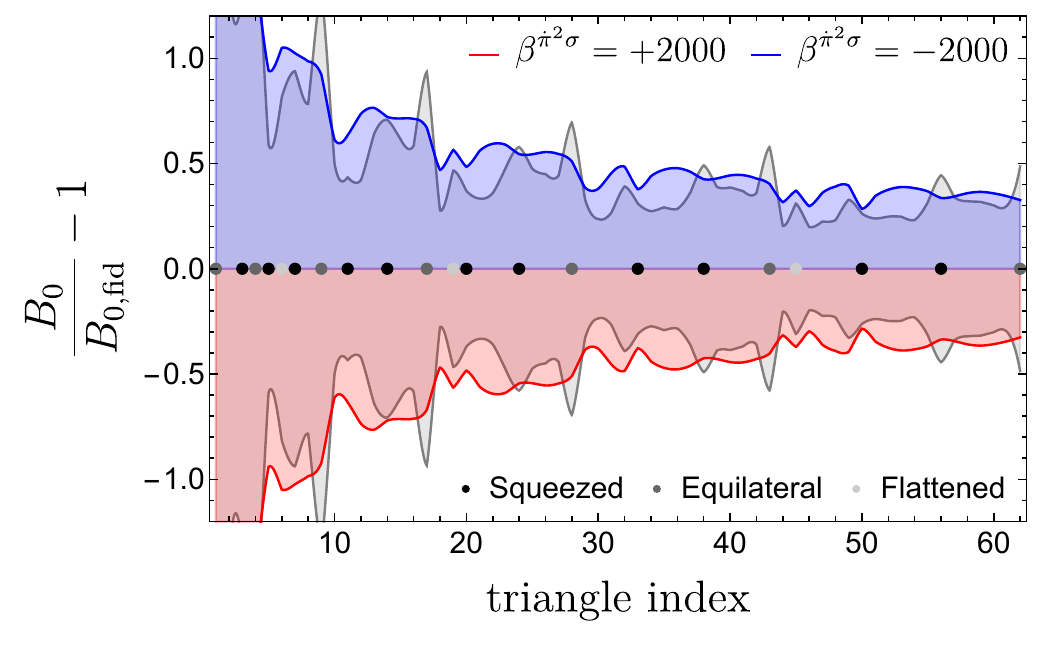} 
\includegraphics[width = 0.49\textwidth]{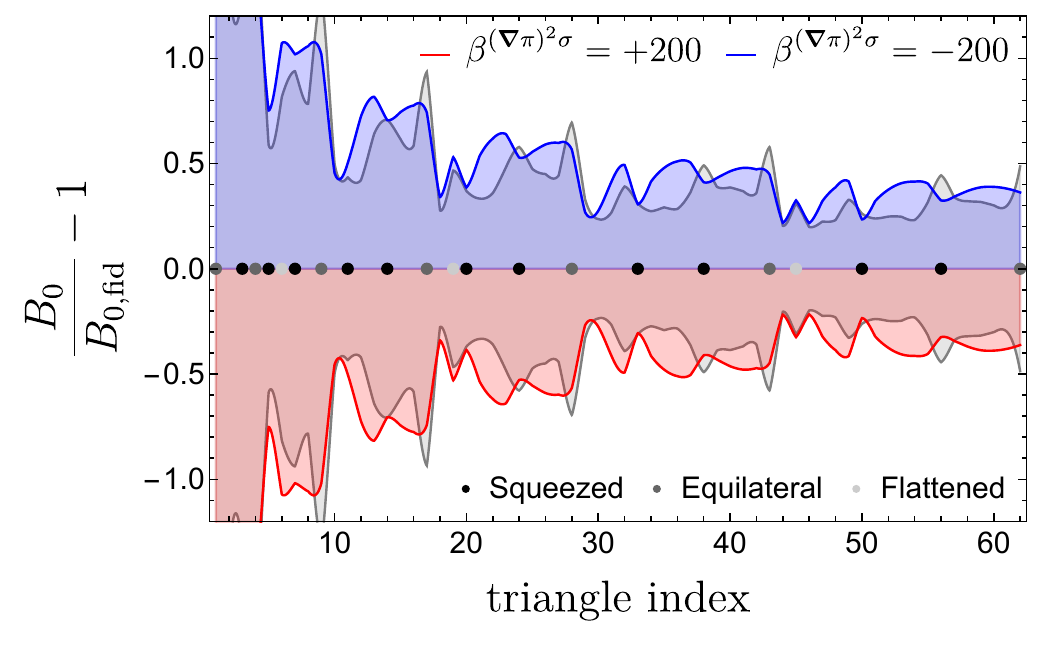} 
\caption{Residual variations of the
galaxy bispectrum monopole w.r.t. variations of $\smash{\beta^\collA}$ (left) and $\smash{\beta^\collB}$ (right). Black, dark-grey and light-gray dots denote the squeezed ($k_3 = 0.015\,\hMpc; k_1,k_2 > k_3$), equilateral ($k_1 = k_2 = k_3$), and flattened ($k_2 = k_3, 2k_2 = k_1 + 0.015\,\hMpc$) triangle configurations, respectively, as in \cite{Cabass:2022ymb}.} 
\label{fig:residuals}
\end{figure}

Results are given in Tab.~\ref{tab:boss_scan} and visualized in Figs.~\ref{fig:boss_scan-dotpi2_nablapi2}~\&~\ref{fig:boss_scan}.
In all cases, we find no detection, \textit{i.e.}~no evidence for massive particles in inflation. Our constraints are generally weaker than those from \S\ref{sec:planck_equivalent}; this is as expected since (a) the BOSS data contains far fewer primordial modes than the CMB, and (b) we additionally marginalize over self-interactions. However, given the theoretical bounds discussed in \S\ref{sec:cosmo_collider_observables} -- of $\mathcal{O}(10^5)$ -- they are still parametrically relevant. From Fig.~\ref{fig:boss_scan-dotpi2_nablapi2} we see that constraints become tighter with increasing $c_{\rm s}/c_\sigma$ (as we approach the ``equilateral collider'' regime), though the scaling is somewhat more muted than the \textit{Planck} case, and we observe some variation with mass, especially for $\smash{\beta^\collB}$. We see that the strongest constraints are for $\mu=0$ and $\smash{c_{\rm s}/c_\sigma = 10}$: this can be seen also by the two-dimensional contours plotted in Fig.~\ref{fig:boss_scan}. We present the residuals of the bispectrum monopole with respect to the best-fit of a zero-PNG analysis as a function
of the triangle index, similarly to the plots of Refs.~\cite{Cabass:2022wjy,Cabass:2022ymb}, in Fig.~\ref{fig:residuals}. Perhaps the most interesting outcome of this analysis is that, for $c_{\rm s} = c_\sigma$ and $\mu>1$, we find a strong degeneracy between $\smash{\beta^\collA}$ and $\smash{\beta^\collB}$, with degeneracy direction $\smash{\beta^\collB \approx {-\beta^\collA}/4}$. This is physically sourced by the degeneracies between the collider templates, self-interactions, and hydrodynamic physics (via $b_2$ and $b_{\mathcal{G}_2}$), which are discussed in more detail below.

Whilst this fixed-$\mu,c_{\rm s}/c_\sigma$ analysis certainly provides an intriguing manner through which to assess the overall constraining power of BOSS, it is undoubtedly more interesting to carry out a ``proper'' particle physics search, in which we marginalize over $\mu$ and $\smash{c_{\rm s}/c_\sigma}$. This allows us to avoid ``look-elsewhere'' effects that can otherwise lead to false detections. 
We find that exploring the entire parameter space using MCMC is very difficult if both $\mu$ and $\smash{c_{\rm s}/c_\sigma}$ are varied simultaneously: for this reason we fix $c_{\rm s} = c_\sigma$ in \S\ref{sec:boss_mu_marginalize}, and vary $\mu$ together with the two Cosmological Collider PNG amplitudes and $\smash{\fnl^{\rm equil},\fnl^{\rm ortho}}$. We consider a varying $\smash{c_{\rm s}/c_\sigma}$ in the DBI Cosmological Collider constraints of \S\ref{sec:boss_dbi}. In these sections we will also discuss degeneracies between PNG and quadratic bias parameters $b_2$ and $\smash{b_{{\cal G}_2}}$, and discuss the impact of priors on them.

\begin{figure}[t]
\centering
\includegraphics[width = 0.49\textwidth]{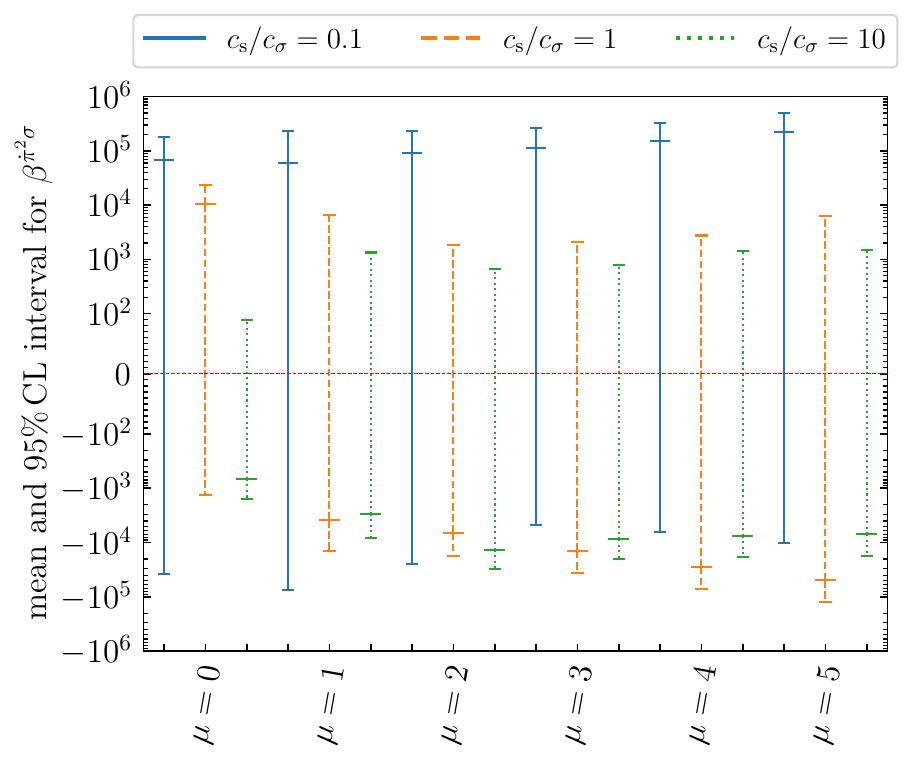} 
\includegraphics[width = 0.49\textwidth]{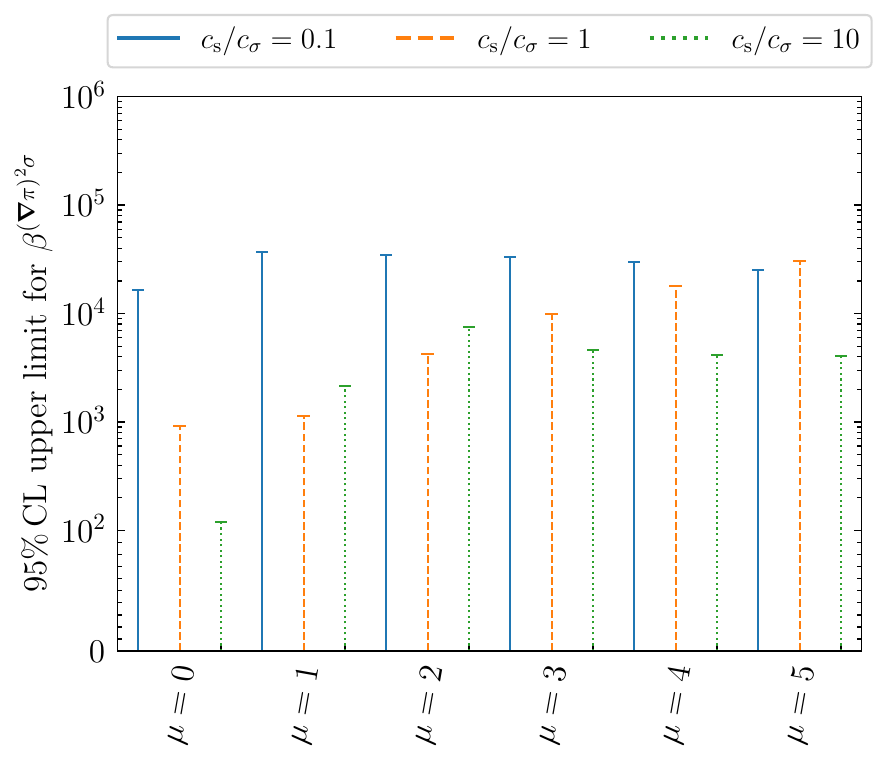} 
\caption{As Fig.~\ref{fig:boss_scan-dotpi2_nablapi2}, but including a strong Gaussian prior on $\smash{\fnl^{\rm equil}}$ and $\smash{\fnl^{\rm ortho}}$, centered in $(0,0)$ with a width $10$ times tighter than the BOSS marginalized $1\sigma$ error. This emulates the \textit{Planck} analyses where the self-interaction shapes were set to zero. We observe a tightening of the constraints across all choices of $\mu$ and $\smash{c_{\rm s}/c_\sigma}$. The most improvement is seen for $\smash{c_{\rm s}/c_\sigma = 1,10}$.} 
\label{fig:boss_scan-dotpi2_nablapi2-fix_fNLs}
\end{figure}

\begin{figure}
  \begin{minipage}{0.45\textwidth}
    \centering
    \includegraphics[height=0.8\textwidth]{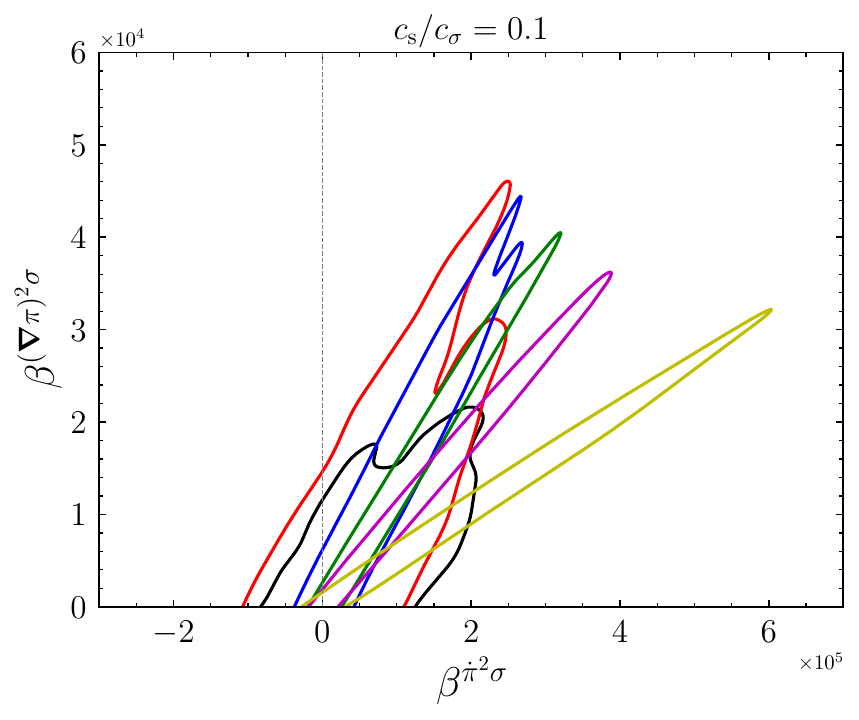}
  \end{minipage}
  \hfill
  \begin{minipage}{0.45\textwidth}
    \centering
    \includegraphics[height=0.8\textwidth]{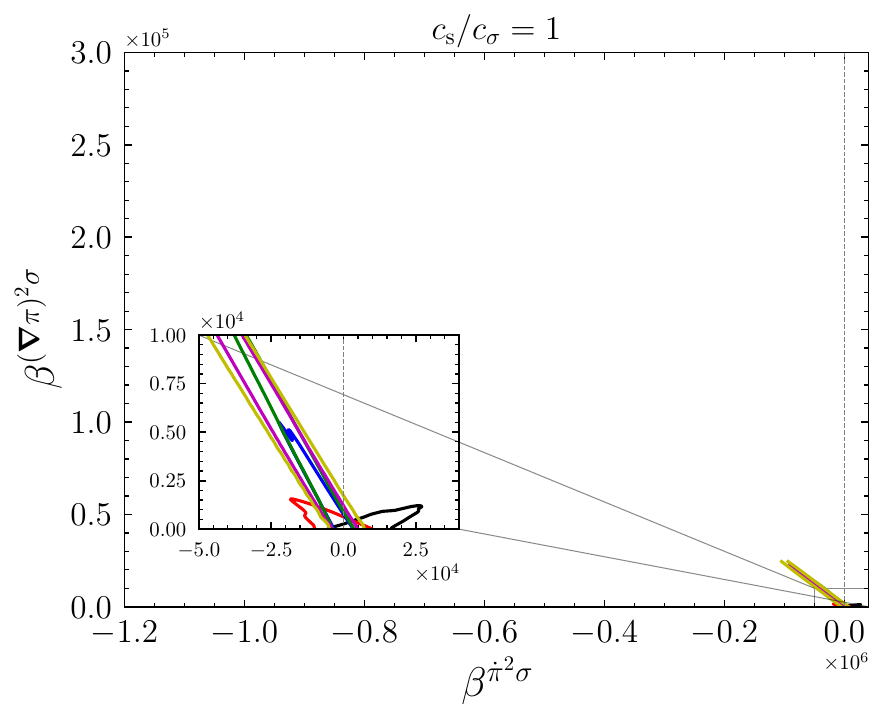}
  \end{minipage}
  \hfill
  \begin{minipage}{0.45\textwidth}
    \centering
    \includegraphics[height=0.8\textwidth]{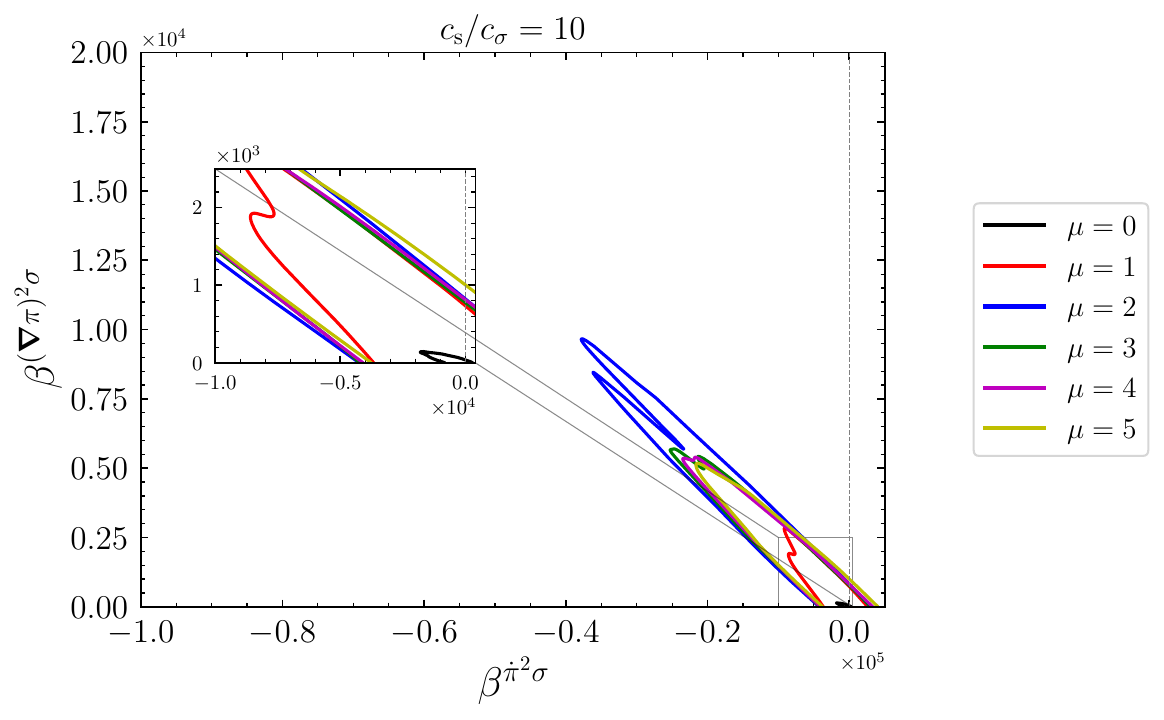}
  \end{minipage}
  \caption{As Fig.~\ref{fig:boss_scan}, but including strong priors on inflaton self-interactions, centered on zero. We use the same plot ranges as in Fig.~\ref{fig:boss_scan} to facilitate comparison.} 
  \label{fig:boss_scan-fix_fNLs}
\end{figure}

Finally, it is interesting to fix the inflaton self-interactions to zero, and thereby compare the resulting BOSS constraints to those obtained from \emph{Planck} in \ref{sec:planck_equivalent}. This is performed mainly for illustrative purposes. Instead of running a new MCMC analysis, we obtain the relevant chains by importance sampling the Markov chains for the $18$ models analyzed in this section using a prior 
\be
{-2\log\chi^2} = \frac{\big(\fnl^{\rm equil}\big)^2}{\sigma_{\fnl^{\rm equil}}^2} + \frac{\big(\fnl^{\rm ortho}\big)^2}{\sigma_{\fnl^{\rm ortho}}^2}\,\,,
\ee
where $\smash{\sigma_{\fnl}}$ is taken to be $10$ 
times smaller than the minimum between the ``upper-$\sigma$'' 
and ``lower-$\sigma$'' for the respective $\smash{\fnl}$ in the previous analysis, where the upper-$\sigma$ (lower-$\sigma$) 
is defined by the absolute value of the $68\%$ CL upper (lower) limit minus the mean, marginalized over all remaining parameters. This width has been chosen such that the prior is considerably tighter than the previously-obtained posteriors on $\smash{\fnl^{\rm equil}}$ and $\smash{\fnl^{\rm ortho}}$ but still facilitates accurate importance sampling. The corresponding results are shown in Figs.~\ref{fig:boss_scan-dotpi2_nablapi2-fix_fNLs}
and ~\ref{fig:boss_scan-fix_fNLs}. 
We find a reduction in the error-bars as expected, since we have eliminated a major source of degeneracy. {In particular, we find a 
$\sim\!10\%$ 
improvement of the constraints on both $\smash{\beta^\collA}$ and $\smash{\beta^\collB}$ 
for $\smash{c_{\rm s}/c_\sigma} = 0.1$ for all values of $\mu$, while for 
$\smash{c_{\rm s}/c_\sigma = 10}$ and $\mu = 0$ there is an order of magnitude improvement in the constraints on $\smash{\beta^\collA}$. The improvement of the constraints on $\smash{\beta^\collB}$ for $\smash{c_{\rm s}/c_\sigma = 1}$ is of a factor of a few for $\mu\leq 1$, while it becomes of an order of magnitude at large $\mu$. For $\smash{c_{\rm s}/c_\sigma = 10}$, instead, the largest improvement (of around an order of magnitude) is seen for $\mu=0$. This matches the discussion of primordial degeneracies in \ref{sec:planck_equivalent}.}

\subsection{\texorpdfstring{Constraints marginalized over $\mu$}{Constraints marginalized over mass}}
\label{sec:boss_mu_marginalize}

\noindent Let us begin this section by reviewing the results of the singular value decomposition (SVD) analysis whose details are contained in Appendix \ref{app:SVD}. Following~\cite{Philcox:2020zyp} we explore the posterior on $\smash{\beta^\collA}$, $\smash{\beta^\collB}$ and $\mu$ by generating a template bank of bispectrum multipoles (the observable that dominates the constraints on non-local-type PNG) over the ranges
\be
\{\beta^\collA,\beta^\collB,\mu\}\in[{-10^6},10^6]\times[0,10^6]\times[0,5]\,\,,
\ee 
fixing $\smash{c_{\rm s}/c_\sigma = 0.1,1,10}$ and the remaining bias parameters to the BOSS (non-PNG) best-fit values. We also fix $\smash{f_{\rm NL}^{\rm equil}=f_{\rm NL}^{\rm ortho}}=0$. By performing an SVD of such template bank, we can identify which are the directions in observable-space have largest contributions to the log-likelihood. Moreover, we can check whether these have significant overlap with $\smash{f_{\rm NL}^{\rm equil}}$ and $\smash{f_{\rm NL}^{\rm ortho}}$ by computing their cosine (defined with respect to the BOSS covariance) with bispectrum templates that have zero Cosmological Collider PNG and unit $\smash{f_{\rm NL}^{\rm equil}}$ or $\smash{f_{\rm NL}^{\rm ortho}}$ (noting that the specific value of $\smash{f_{\rm NL}^{\rm equil}}$ and $\smash{f_{\rm NL}^{\rm ortho}}$ is irrelevant in this context).

\setlength{\tabcolsep}{4pt}
\renewcommand{\arraystretch}{1.3}
\begin{table*}[ht!]
\centering
\begin{tabular}{c|ccc|ccc}
 \multicolumn{1}{c|}{} & \multicolumn{3}{c|}{without galaxy-formation priors} & \multicolumn{3}{c}{including galaxy-formation priors}\\ \hline
 \textbf{Parameter} & mean & \quad 95\% lower \quad & 95\% upper & mean & \quad 95\% lower \quad & 95\% upper \\ 
 \hline

$f_{\rm NL}^{\rm equil}$ & $\num[round-mode=places,round-precision=\prec]{1.8746229408125114e+002}$ & $\num[round-mode=places,round-precision=\prec]{-1.1573016485723092e+003}$ & $\num[round-mode=places,round-precision=\prec]{1.4314995924748819e+003}$ 

& $\num[round-mode=places,round-precision=\prec]{2.4736341470153036e+001}$ & $\num[round-mode=places,round-precision=\prec]{-1.4067406376833264e+003}$ & $\num[round-mode=places,round-precision=\prec]{1.2647859139981456e+003}$ \\

$f_{\rm NL}^{\rm ortho}$ & $\num[round-mode=places,round-precision=\prec]{5.40404609399981e+02}$ & $\num[round-mode=places,round-precision=\prec]{-1.156104141198208e+02}$ & $\num[round-mode=places,round-precision=\prec]{1.4704738783654093e+03}$ 

& $\num[round-mode=places,round-precision=\prec]{7.246203863770736e+02}$ & $\num[round-mode=places,round-precision=\prec]{-5.616695673035247e+01}$ & $\num[round-mode=places,round-precision=\prec]{1.8458112312222145e+03}$ \\

$\beta^\collA$ 
& $\num[round-mode=places,round-precision=\prec]{-1.5542520461068145e+0005}$ 
& $\num[round-mode=places,round-precision=\prec]{-5.410762868807428e+0005}$ 
& $\num[round-mode=places,round-precision=\prec]{3.7072404968490824e+0004}$ 
& $\num[round-mode=places,round-precision=\prec]{-1.9138896404528312e+0005}$ 
& $\num[round-mode=places,round-precision=\prec]{-6.320357577729938e+0005}$ 
& $\num[round-mode=places,round-precision=\prec]{4.69107555485107e+0004}$ \\

$\beta^\collB$ 
& $-$ 
& $-$ 
& $<\num[round-mode=places,round-precision=\prec]{1.3223969999999998e+00005}$
& $-$  
&  $<\num[round-mode=places,round-precision=\prec]{1.543238e+00005}$ \\

$\mu$ & $-$ & $>\num[round-mode=places,round-precision=\prec]{1.356835}$ & $-$ 

& $-$ & $>\num[round-mode=places,round-precision=\prec]{1.640695}$ & $-$ \\

 \end{tabular}
 \caption{BOSS constraints on the microphysical collider amplitudes $\smash{\beta^\collA}$ and $\smash{\beta^\collB}$ and the self-interaction templates $\smash{\fnl^{\rm equil}}$ and $\smash{\fnl^{\rm ortho}}$, fixing $\smash{c_{\rm s} = c_\sigma}$ and marginalizing over the mass parameter $\mu\in[0,5]$. As in the previous sections we give the mean and $95\%$ CL bound for $\smash{\beta^\collA}$ and the 95\% upper bound on the positive parameter $\smash{\beta^\collB}$. The first (second) three columns show constraints without (with) the HOD-inspired priors on quadratic bias parameters, which are summarized in Appendix \ref{app:HOD}.} 
 \label{tab:boss-mu_marginalized}
\end{table*}

\begin{figure}[t]
\centering
\includegraphics[width = 0.8\textwidth]{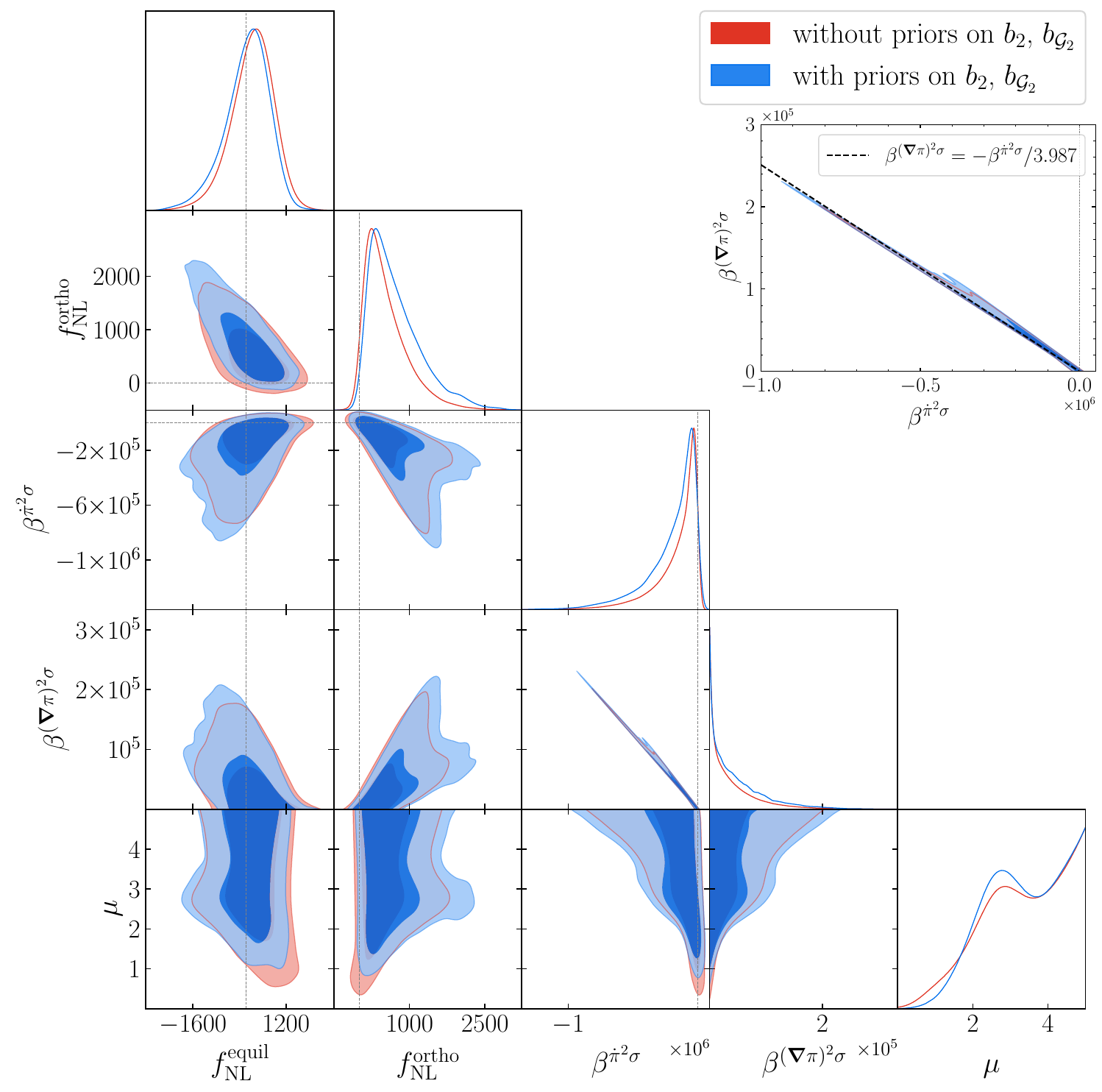} 
\caption{Marginalized constraints on parameters describing inflaton self-interactions and Cosmological Collider PNG for $\smash{c_{\rm s}/c_\sigma}=1$, varying the mass $\mu$. The inset panel shows the two-dimensional contour in the $\smash{\beta^\collA-\beta^\collB}$ plane, which exhibits a strong degeneracy. In red (blue) we show results without (with) constraining priors on $b_2$ and $\smash{b_{{\cal G}_2}}$) obtained from HOD analyses. The marginalized constraints are collected in Tab.~\ref{tab:boss-mu_marginalized}.} 
\label{fig:boss-mu_marginalized}
\end{figure}

Interestingly, we find that for all three choices of $\smash{c_{\rm s}/c_\sigma}$ there is a best-constrained direction that is, for all practical purposes, parallel to the best-constrained linear combination of equilateral and orthogonal PNG. The second best-constrained direction also exhibits very significant overlap with these self-interaction templates, whilst the third is almost entirely orthogonal. In all cases, we find a significant drop in constraining power between the first and third best-constrained directions: the relative drop in $\chi^2$ is between ${\cal O}(10^7)$ and ${\cal O}(10^5)$ for the three sound-speed ratios considered. Due to (a) the strong degeneracy between $\smash{\beta^\collA}$ and $\smash{\beta^\collB}$ present for $\smash{c_{\rm s}/c_\sigma = 1,10}$, (b) marginalization over $\mu$, (c) exclusion of the power spectrum multipoles, and (d) fixing $b_2$ and $\smash{b_{{\cal G}_2}}$, this drop in $\chi^2$ is not straightforwardly translatable in the improvement on marginalized constraints on these parameters observed in \S\ref{sec:boss_mu_scan} once we fix $\smash{\fnl^{\rm equil}}$ and $\smash{\fnl^{\rm ortho}}$ and scan over mass. Nevertheless, this analysis confirms our conclusions from the end of \S\ref{sec:boss_mu_scan}: we pay a significant price due to the fact that we can hunt for massive particles only through the signatures that are not degenerate with the EFT of Inflation background. 

We now proceed to discuss the constraints on $\smash{\beta^\collA}$ and $\smash{\beta^\collB}$ marginalized over $\mu$ (but at fixed $\smash{c_{\rm s}/c_\sigma = 1}$, as above); these are collected in the leftmost three columns of Tab.~\ref{tab:boss-mu_marginalized}. We find that marginalizing over $\mu$ degrades the constraints on both Cosmological Collider amplitudes compared to the above results. Whilst in \S\ref{sec:boss_mu_scan} we had seen that some choices of $\mu$ led to constraints tighter than the ``perturbativity bounds'' $\smash{|\beta^\collA|\lesssim 10^5}$, $\smash{\beta^\collB\lesssim 10^5}$, Tab.~\ref{tab:boss-mu_marginalized} shows that after $\mu$ marginalization this does not happen anymore, and the $95\%$ CL intervals for both $\smash{\beta^\collA}$ and $\smash{\beta^\collB}$ are of order $10^5$. Tab.~\ref{tab:boss-mu_marginalized} also shows that the constraints on equilateral and orthogonal non-Gaussianities are weaker than those obtained by Ref.~\cite{Cabass:2022wjy} for the single-field inflation case, due to the Collider degeneracy. The triangle plot for primordial parameters is shown in Fig.~\ref{fig:boss-mu_marginalized} (red contours). The most interesting features of this plot are the confirmation of a very strong degeneracy between $\smash{\beta^\collA}$ and $\smash{\beta^\collB}$. There is also a degeneracy between $\smash{\fnl}$ and $\beta$, but this is not as pronounced. {From the two-dimensional plots, the strongest correlation seems to be in the $\smash{\fnl^{\rm equil}-\beta^\collA}$ and $\smash{\fnl^{\rm ortho}-\beta^\collA}$ planes, and between $\smash{\beta^\collB}$ and $\smash{\fnl^{\rm ortho}}$}.

\begin{figure}[t]
\centering
\includegraphics[width = 0.49\textwidth]{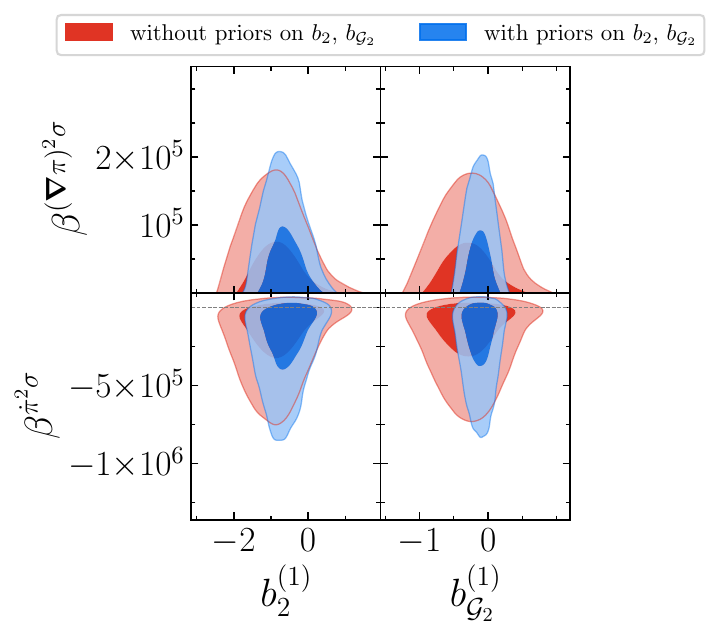} 
\includegraphics[width = 0.49\textwidth]{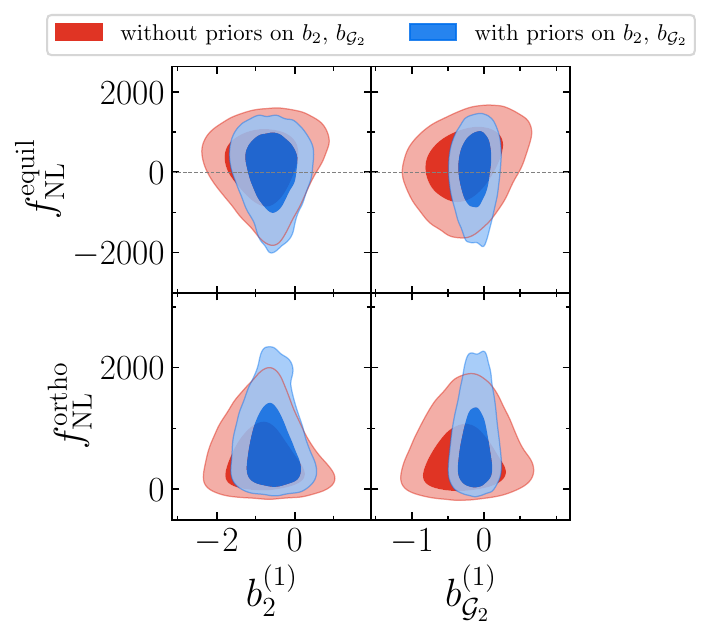} 
\caption{Two-dimensional contour plots showing the covariance between quadratic biases and Collider amplitudes (left) or self-interaction amplitudes (right). Results without (with) priors on $b_2$ and $\smash{b_{{\cal G}_2}}$ are shown in red (blue) contours. We show only the plots for the biases describing the North Galactic Cap, $z=0.61$ data chunk for simplicity: the biases of the other data chunks show very similar correlations with $\beta$ and $\smash{\fnl}$ parameters.} 
\label{fig:bias_degeneracy}
\end{figure}

In Fig.~\ref{fig:bias_degeneracy} we show the two-dimensional contours describing the correlations of PNG amplitudes with quadratic galaxy biases.\footnote{We restrict to the BOSS North Galactic Cap $z=0.61$ data chunk for clarity -- contours for other data chunks are similar.} Comparing the red contours in the rightmost plot of Fig.~\ref{fig:bias_degeneracy} to those of Fig.~S3 of Ref.~\cite{Cabass:2022wjy} (which considered only self-interactions), we find that the degeneracy between $\smash{\fnl^{\rm equil}}$ and $\smash{b_{{\cal G}_2}}$ is still present, and that the contours involving both quadratic bias and $\smash{\fnl}$ are more non-Gaussian than before. In the red contours of the leftmost plot we observe that the contours in the $\smash{\beta^\collA - b_2}$ and $\smash{\beta^\collA - b_{{\cal G}_2}}$ planes exhibit strong non-Gaussianity since the BOSS data is much more constraining in the $\smash{\beta^\collA>0}$ half of the parameter space. Moreover, we find that there is not a strong degeneracy between the Collider $\beta$ parameters and quadratic biases. Even so, the combination of the degeneracy between $\smash{\fnl^{\rm equil}}$ and $\smash{b_{{\cal G}_2}}$ and that of $\smash{\fnl^{\rm equil}}$ with $\smash{\beta^\collA}$ implies that marginalization over galaxy biases will affect the $\smash{\beta^\collA}$ constraints. 

This is addressed by the blue contours in Figs.~\ref{fig:boss-mu_marginalized} and \ref{fig:bias_degeneracy}. These are obtained by imposing HOD-motivated priors on $b_2$ and $\smash{b_{{\cal G}_2}}$ obtained by fitting an EFTofLSS model to HOD mock catalogs (see Appendix \ref{app:HOD} for details). Interestingly, we see that the effect of the priors is to tighten the constraints on $\smash{\fnl^{\rm equil}}$ but somewhat weaken those on $\smash{\fnl^{\rm ortho}}$. Even more interestingly for this work, we see that the complicated (and often, as discussed, non-Gaussian) correlations between these parameters lead to a very slight worsening of the constraints on the Cosmological Collider PNG amplitudes.

\subsection{Constraints on the DBI Cosmological Collider}
\label{sec:boss_dbi}

\noindent Finally, we study the DBI Cosmological Collider. Due to the additional boost symmetries, the number of free inflationary parameters drops from $6$ to $4$ ($\log_{10}c_{\rm s}$, $\smash{\beta^\collA}$, $\mu$, $\smash{\log_{10}c_{\rm s}/c_\sigma}$); moreover, $\smash{\beta^\collA}$ is restricted to be positive. This makes sampling the full parameter space much easier. We again restrict to $\mu\in[0,5]$, but now sample $\log_{10}c_{\rm s}\in[-3,0]$ and $\smash{\log_{10}c_{\rm s}/c_\sigma\in[-1,1]}$. We also impose a theoretical prior $\smash{\log_{10}c_\sigma\leq 0}$, which does not significantly impact the posterior, given that the bulk of the likelihood is already within the $\smash{c_{\rm s}>c_\sigma}$ half of parameter space.

\setlength{\tabcolsep}{4pt}
\renewcommand{\arraystretch}{1.3}
\begin{table*}[ht!]
\centering
\begin{tabular}{c|ccc|ccc}
 \multicolumn{1}{c|}{} & \multicolumn{3}{c|}{without galaxy-formation priors} & \multicolumn{3}{c}{including galaxy-formation priors}\\ \hline
 \textbf{Parameter} & mean & \quad 95\% lower \quad & 95\% upper & mean & \quad 95\% lower \quad & 95\% upper \\ 
 \hline

$\log_{10}c_{\rm s}$ & $-$ & $>\num[round-mode=places,round-precision=\prec]{-1.393542}$ & $-$ 

& $-$ & $>\num[round-mode=places,round-precision=\prec]{-1.334552}$ & $-$ \\

$\beta^\collA$ & $-$ & $-$ & $<\num[round-mode=places,round-precision=\prec]{2.251078e+003}$ 

& $-$ & $-$ & $<\num[round-mode=places,round-precision=\prec]{1.990399e+003}$ \\

$\mu$ & $-$ & $>\num[round-mode=places,round-precision=\prec]{1.013528}$ & $-$ 

& $-$ & $>\num[round-mode=places,round-precision=\prec]{1.067614}$ & $-$ \\

$\log_{10}c_{\rm s}/c_\sigma$ & $-$ & $>\num[round-mode=places,round-precision=\prec]{-0.08571271}$ & $-$ 

& $-$ & $>\num[round-mode=places,round-precision=\prec]{-0.1285367}$ & $-$ \\

 \end{tabular}
 \caption{As Tab.~\ref{tab:boss-mu_marginalized}, but for the DBI Cosmological Collider. In this case, given the restricted parameter space, we are able to marginalize also over the sound-speed ratio in addition to the mass $\mu$. We do not find any detection of either inflaton self-interactions or Cosmological Collider PNG, though find much tighter constraints than before.} 
 \label{tab:boss-dbi}
\end{table*}

\begin{figure}[t]
\centering
\includegraphics[width = 0.8\textwidth]{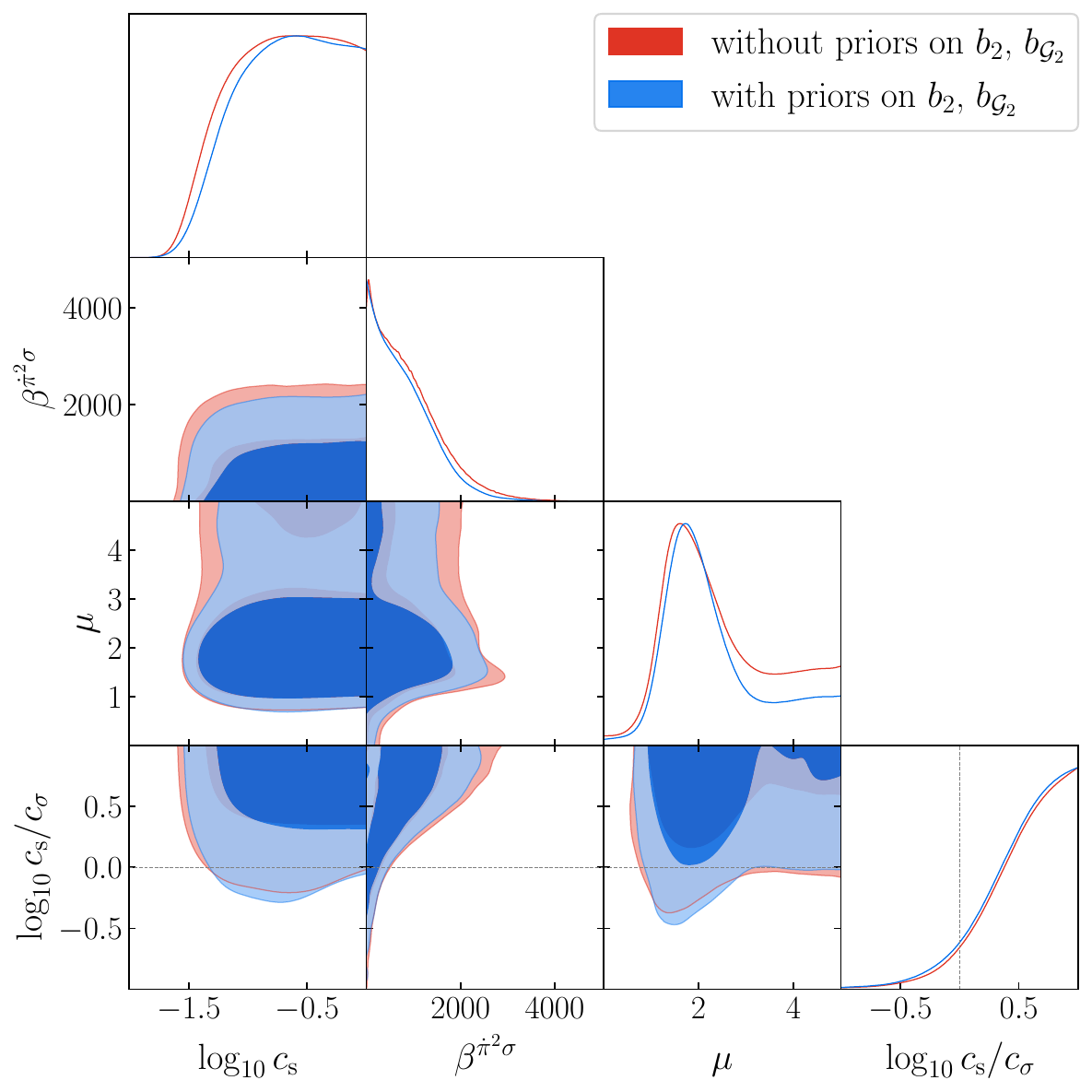} 
\caption{As Fig.~\ref{fig:boss-mu_marginalized}, but for the DBI Cosmological Collider. We show constraints on the relevant microphysical parameters both without (red) and with (blue) HOD-priors on the quadratic biases $b_2$ and $\smash{b_{{\cal G}_2}}$. Marginalized constraints are collected in Tab.~\ref{tab:boss-dbi}. We find \textit{far} tighter bounds on $\smash{\beta^\collA}$ than in \S\ref{sec:boss_mu_marginalize}, due to the additional symmetry assumptions.} 
\label{fig:boss-dbi}
\end{figure}

The corresponding constraints are collected in Tab.~\ref{tab:boss-dbi} and Fig.~\ref{fig:boss-dbi}. The most striking feature is the upper limit on $\smash{\beta^\collA}$, which is almost two orders of magnitude stronger than those given in \S\ref{sec:boss_mu_scan}~\&~\ref{sec:boss_mu_marginalize}. This is expected due to the considerable shrinking of the available parameter space, which breaks a number of degeneracies. From the blue contours in this figure we see that the impact of priors on quadratic galaxy biases, in this case, is to slightly tighten the bounds on all Cosmological Collider microphysical parameters.

\section{Conclusions}
\label{sec:conclusions}

\noindent This work has presented the first systematic cosmological data analysis targeting new massive particles during inflation. More precisely, we have placed constraints on ``Cosmological Collider'' non-Gaussianity, by probing the couplings between the inflaton and a massive spin-zero particle $\sigma$. More precisely, we have pursued two lines of attack: 
\begin{enumerate}[leftmargin=*]
    \item Translating previously-derived \emph{Planck} constraints on equilateral and orthogonal non-Gaussianity into constraints on the Cosmological Collider primordial non-Gaussianity amplitudes, $\smash{\beta^{\dot{\pi}^2\sigma},\beta^{({\bm\nabla}\pi)^2\sigma}}$, assuming negligible inflaton self-interactions.
    \item Assuming that inflaton self-interactions are present but unknown, and marginalizing over them in a search for massive particles using BOSS redshift-space galaxy clustering data. 
\end{enumerate}

Via approach (1), we can study whether there are models (described by different choices of particle mass and its speed of propagation relative to the inflaton) that are not well described by the standard equilateral and orthogonal PNG templates. We find that such models exist (mainly at low masses and speeds), and would benefit from a  future dedicated search using CMB data. {This research has been carried out in the recent work \cite{Sohn:2024xzd}.} That said, on theoretical grounds it is not easy to imagine models in which large Cosmological Collider signatures are not accompanied by large inflaton self-interactions. As such, any proper search for new particles should proceed as (2), \emph{i.e.}~marginalizing over the EFT of Inflation PNG background (described by $\smash{\fnl^{\rm equil}}$ and $\smash{\fnl^{\rm ortho}}$), and hunting for ``resonances'' (the peculiar non-analytic signatures in the primordial bispectrum coming from spontaneous particle production in a de Sitter background). 

Our search in (2) with BOSS data finds no evidence for novel massive particles during inflation: for many choices of particle mass and speed the corresponding bounds on the coupling amplitudes are comparable or (slightly) better than those from the requirement of perturbativity. The principal reasons for this are that (a) the genuine Cosmological Collider signature is a small perturbation on top of the part of the primordial bispectrum degenerate with $\smash{\fnl^{\rm equil}}$ and $\smash{\fnl^{\rm ortho}}$, and (b) these parameters ($\smash{\fnl^{\rm equil}}$ especially) are degenerate with the PT bias coefficients that parameterize our uncertainity in the physics of galaxy formation. To ameliorate these issues, we have performed analyses in which we fix $\smash{\fnl^{\rm equil}}$ and $\smash{\fnl^{\rm ortho}}$ (essentially the approach taken in (1)), or we add priors on galaxy biases arising from HOD modeling \cite{future_kaz} (see also \citep{Ivanov:2024hgq,Ivanov:2024xgb}). We find that fixing $\smash{\fnl^{\rm equil}}$ and $\smash{\fnl^{\rm ortho}}$ can indeed improve bounds on $\smash{\beta^{\dot{\pi}^2\sigma},\beta^{({\bm\nabla}\pi)^2\sigma}}$, especially in the ``equilateral collider'' regime of the massive particle being slower than the inflaton. Finally, we see that due to the high level of (often non-Gaussian) degeneracy between parameters describing primordial physics, adding priors on galaxy biases only slightly affects constraints on $\smash{\beta^{\dot{\pi}^2\sigma},\beta^{({\bm\nabla}\pi)^2\sigma}}$. 

An alternative way to treat these degeneracies is by invoking additional symmetry assumptions in the inflationary theory. This is evident if one considers the ``DBI Cosmological Collider'', in which the interactions between the massive particle and the inflaton are protected by a higher-dimensional boost symmetry. This greatly shrinks the available parameter space -- fixing $\smash{\beta^{({\bm\nabla}\pi)^2\sigma}}$, for example -- leading to very strong constraints on $\smash{\beta^\collA}$ that are comparable with those obtained at fixed $\smash{\fnl^{\rm equil}}$ and $\smash{\fnl^{\rm ortho}}$ even if both the mass and the speed of the massive particle are allowed to vary. In this case, we see that adding priors on quadratic galaxy biases slightly improves the constraining power on Cosmological Collider PNG, but by a very small margin with respect to the improvement due to the reduction of the available parameter space. 

{Whilst in this paper we do not find evidence of massive particles, it is important to emphasize that this is just the first step towards inflationary spectroscopy from cosmological observations. Our constraints will certainly improve with data from future surveys such as \textit{Euclid}~\cite{Laureijs:2011gra}, DESI~\cite{Aghamousa:2016zmz} or 
Spec-5/Megamapper. 
Based on their volume with respect to BOSS, 
we expect a reduction of error bars by a factor of 
$\mathcal{O}(3)$ with DESI and a factor $\mathcal{O}(10)$ with MegaMapper~\cite{Schlegel:2019eqc}, even without additional modeling developments~\cite{Cabass:2022epm}.}\footnote{Ref.~\cite{Cabass:2022epm} showed that the mode counting estimates
are quite accurate for the equilateral primordial non-Gaussianity. Since its shape is similar
to the cosmological collider templates, we expect the mode counting estimates to be accurate
in the collider case as well.} 
Futuristic $21$cm/intensity mapping observations will map our Universe at higher redshifts, reaping the benefits of a weaker non-Gaussianity background from nonlinear gravitational evolution. As such, they also promise to significantly improve the constraining power \cite{Karkare:2022bai,Floss:2022wkq}. Importantly, there has also been movement on the CMB side, with the recent constraints on the Cosmological Collider obtained by Ref.~\cite{Sohn:2024xzd} that, together with this work, represents the state of the art of constraints on the Cosmological Collider.

On the theory side, stronger constraints may possibly be wrought by including triangles with larger wavenumbers: the modeling of these requires one-loop bispectrum corrections, recently derived in \cite{Philcox:2022frc,DAmico:2022osl,DAmico:2022ukl}, though this comes with the addition of many more nuisance parameters. In regards to the Cosmological Collider program, it will be interesting to further develop the program of spectroscopy of the inflationary particle content, including looking for the tower of Kaluza-Klein higher-spin states that might offer us a glimpse of string theory \cite{Arkani-Hamed:2015bza}. For massive \textit{spinning} particles, it will be interesting to see if the peculiar dependence on the angle between wavevectors of large and small wavenumber in squeezed triangles can help in disentangling the Cosmological Collider non-Gaussianity from the other backgrounds (both from inflaton self-interactions and late-time evolution). It will also be interesting to include information from the recently-measured BOSS galaxy trispectrum~\cite{Philcox:2021hbm} (though this would require a dedicated PT theory extension): it is only at the level of the trispectrum that the presence of a massive spinning particle can be truly disentangled from derivative couplings of the inflaton to massive scalars. {Finally, we emphasize that while a CMB analysis of the PNG templates discussed in this work has now been performed \cite{Sohn:2023fte,Sohn:2024xzd}, it would be extremely interesting to apply the recently-developed techniques of \cite{Philcox:2023uwe,Philcox:2023psd} to this problem as well.}

\vskip 4pt

\noindent\textit{Acknowledgments ---} {\small We are grateful to Sadra Jazayeri, Enrico Pajer, Gui Pimentel, Fabian Schmidt, David Stefanyszyn, and Dong-Gang Wang for useful discussions. GC thanks the Institute for Advanced Study where a large part of this work was carried out, and the Cacao and Cookie Factory pastry shops for their excellent cake selection. OHEP thanks the Simons Foundation for support and La Ruta del Cares for their impeccable signposting. KA acknowledges support from Fostering Joint International Research (B) under Contract No.~21KK0050. SC acknowledges the support of the National Science Foundation at the Institute for Advanced Study. MZ is supported by NSF 2209991 and NSF-BSF 2207583.  Parameter estimates presented in this paper have been obtained with the \texttt{CLASS-PT} Boltzmann code \cite{Chudaykin:2020aoj} (see also \cite{Blas:2011rf}) interfaced with the \texttt{Montepython} MCMC sampler \cite{Audren:2012wb,Brinckmann:2018cvx}. The triangle plots are generated with the \texttt{getdist} package\footnote{\href{https://getdist.readthedocs.io/en/latest/}{https://getdist.readthedocs.io/en/latest/}.}~\cite{Lewis:2019xzd}.}

\appendix

\section{Solving for the Cosmological Collider Templates}
\label{app:collider_solution}

\noindent To obtain the Cosmological Collider shapes, we must solve the boundary ODE \eqref{seed_ODE} 
\be
\label{seed_ODE_recall}
u^2(1-u^2){\cal I}'' - 2u^3{\cal I}' + \bigg(\mu^2+\frac{1}{4}\bigg){\cal I} = \frac{c^2_{\rm s}u}{1+c_{\rm s}u} 
\ee
for $u\in[0,1/c_{\rm s}]$ (here we will fix $c_\sigma = 1$ to avoid cluttering the notation). The boundary conditions are fixed in the ``squeezed limit'' $u=0$, in which $\cal I$ is given by the sum of the solution to the homogeneous equation and a peculiar solution, obtained as a power series in $u$ \cite{Arkani-Hamed:2018kmz,Pimentel:2022fsc,Jazayeri:2022kjy} 
\be
\label{u_solution} 
{\cal I}(u) = {-\frac{c^2_{\rm s}}{2}}\sum_{\pm} \mathcal{C}_\pm \left(\frac{{\rm i} u}{2 \mu }\right)^{\alpha_\pm} \, _2F_1\left(a_\pm,b_\pm;c_\pm;u^2\right) + \sum_{n=0}^{+\infty}c_n u^{n+1}\,\,, 
\ee
where 
\begin{align}
\label{eq:known_stuff}
&\alpha_\pm = \frac{1}{2}\pm{\rm i} \mu\,\,, \qquad a_\pm = \frac{1}{4}\pm\frac{{\rm i} \mu}{2}\,\,, \qquad b_\pm = \frac{3}{4}\pm\frac{{\rm i} \mu}{2}\,\,, \qquad c_\pm = a_\pm+b_\pm\,\,, \\
&\Xi_0 = \Gamma(\alpha_-) \Gamma(\alpha_+) \, {_2F_1}\left(\alpha_-,\alpha_+;1;\frac{1-c_{\rm s}}{2}\right)\,\,, \\
&{\cal B}_{\pm} = \frac{\sqrt{\pi }\,\Xi_0\Gamma (\alpha_\pm) \left(1\mp\frac{{\rm i}}{\sinh \pi\mu}\right)}{\Gamma (c_\pm)}\,\,, \qquad {\cal C}_{\pm} = ({-{\rm i}} \mu )^{\alpha_\pm} \mathcal{B}_\pm\,\,, \\
\end{align}
and
\be
c_n = c^2_{\rm s}\sum _{l=0}^{\lfloor m \rfloor} \frac{m! ({-c_{\rm s}})^{m-2 l}}{(m-2 l)! \prod _{i=0}^l \left(\left(-2 i+m+\frac{1}{2}\right)^2+\mu ^2\right)}\,\,. 
\ee
A drawback of this expression for $\cal I$ is that both the homogeneous and peculiar solutions diverge as $\ln(1-u)$ in the ``folded limit'' $u\to 1$, in such a way that the full solution is regular there (as it physically should be, due to the assumption of Bunch-Davies initial conditions). To avoid having to deal with non-exact cancellation of divergences, which turn from logarithmic to power-law once we apply weight-shifting operators to obtain the PNG shapes $\cal S$ from $\cal I$ as in \eqref{shape_from_seed}, we solve \eqref{seed_ODE_recall} as a power series in $x=1-u$ around $x=0$. It is straightforward to see that the general form of the solution that is free of logarithmic divergences for $x\to 0$ is 
\be
\label{x_solution} 
\text{${\cal I}(x) = \sum_{n=0}^{+\infty}d_n x^n$ \quad with $x=1-u$}\,\,, 
\ee 
where 
\begin{subequations}
\label{eq:coefficients_x_solution}
\begin{align}
d_1&=\frac{c_{\rm s}^2}{2 c_{\rm s}+2}-\frac{1}{8} d_0 \left(4 \mu ^2+1\right)\,\,, \label{eq:coefficients_x_solution-1} \\
d_2&=\frac{1}{256} \left(d_0 \left(4 \mu ^2-23\right) \left(4 \mu ^2+1\right)+\frac{4 c_{\rm s}^2 \left(-4 (c_{\rm s}+1) \mu ^2+23 c_{\rm s}+15\right)}{(c_{\rm s}+1)^2}\right)\,\,, \label{eq:coefficients_x_solution-2} \\
d_n&=\frac{-\left(\frac{c_{\rm s}}{c_{\rm s}+1}\right)^n+d_{n-1} \left(-\mu ^2+n (5 n-9)+\frac{15}{4}\right)+(n-3) (n-2) d_{n-3}-2 (n-2) (2 n-3) d_{n-2}}{2 n^2}\,\,. \label{eq:coefficients_x_solution-4}
\end{align}
\end{subequations}
In order to determine the coefficient $d_0$, we match to \eqref{u_solution} at $u=1/2$. For $c_{\rm s}<1$ we still need the solution at $u>1$. We follow a different approach than \cite{Jazayeri:2022kjy}, that allows us to maintain regularity at $u=1$. We change variables to $z=1/u$, and solve the resulting ODE for $z\in(0,1]$ with boundary conditions at $z=1$. It is straightforward to see that the solution of the homogeneous equation is 
\be
\text{${\cal I}^{\rm homogeneous}(z) = c_{\cal P}{\cal P}_{{-\frac{1}{2}}+{\rm i}\mu}(z) +c_{\cal Q}{\cal Q}_{{-\frac{1}{2}}+{\rm i}\mu}(z)$ \quad with $z=\frac{1}{u}$}\,\,, 
\ee 
where ${\cal P}_n$ and ${\cal Q}_n$ are the Legendre functions of the first and second kind, respectively. The peculiar solution can be then written as 
\be
\label{peculiar_wronskian_integral}
{\cal I}^{\rm peculiar}(z) = {\cal Q}_{{-\frac{1}{2}}+{\rm i}\mu}(z)\int_1^z{\rm d}z'\,\frac{{\cal P}_{{-\frac{1}{2}}+{\rm i}\mu}(z')}{w(z')}\frac{c^2_{\rm s}}{(z'+c_{\rm s})(z'^2-1)} 
- {\cal P}_{{-\frac{1}{2}}+{\rm i}\mu}(z)\int_1^z{\rm d}z'\,\frac{{\cal Q}_{{-\frac{1}{2}}+{\rm i}\mu}(z')}{w(z')}\frac{c^2_{\rm s}}{(z'+c_{\rm s})(z'^2-1)} \,\,, 
\ee
where $w$ is the Wronskian. It is also straightforward to see that the peculiar solution is analytic in $z=1$: hence, given that ${\cal Q}_n(z)$ has a logarithmic divergence at $z=1$, we can automatically take $c_{\cal Q}=0$. $c_{\cal P}$ is instead fixed by matching to \eqref{x_solution} at $u=1$. Crucially, even for very high $\mu$, the integrands in \eqref{peculiar_wronskian_integral} are very smooth functions of $z'$ for $z'\in(0,1]$: this makes this method for solving the boundary ODE very simple, fast and reliable also in the case of very large masses of the exchanged particle, where instead the homogeneous solution oscillates very fast around $u=0$.

\section{HOD-derived priors on quadratic galaxy biases}
\label{app:HOD}

\noindent In this appendix we provide a brief summary on the mapping between the HOD parameters and galaxy bias parameters. Further details are presented in \cite{future_kaz}.

The halo occupation distribution (HOD) is a widely-used empirical model to populate galaxies given a set of dark matter halos. It gives a probabilistic prescription for the halo-galaxy connection, instead of the \textit{ab initio} modeling of galaxy formation. 
The standard HOD model divides galaxies into central and satellite galaxies and specifies the mean number of each type of galaxies as a function of \textit{only} host halo mass $M$:
\begin{align}
    \langle N_c \rangle (M) & = \frac12 \left[ 1 + {\rm erf}\left(\frac{\log_{10}M - \log_{10}M_{\rm min}}{\sigma_{\log_{10}M}} \right) \right]\,\,,
    \label{eq:Nc}
    \\
    \langle N_s \rangle (M) & = \langle N_c \rangle (M) \left(\frac{M - \kappa M_{\rm min}}{M_1}\right)^\alpha\,\,,
    \label{eq:Ns}
\end{align}
where ${\rm erf}(x)$ is the error function, $M_{\rm min}$ and $\sigma_{\log_{10}M}$ controls the typical minimum mass cut and the softness of this cut, and $\kappa$, $M_1$, and $\alpha$ determine the profile of the mean number of satellite galaxies. 
Given these mean numbers, galaxies are populated following a Bernoulli distribution for centrals and Poisson for satellites. We consider three different HOD models to account for model variations of HOD and the dependence of the galaxy-halo connection on other than halo mass (\emph{i.e.}~assembly bias):
\begin{itemize}[leftmargin=*]
    \item Simplified HOD: Galaxies are populated only according to Bernoulli distribution with the mean \eqref{eq:Nc} for both host and sub halos identified by \texttt{Rockstar} (see \cite{Nishimichi:2020tvu} for details). In this case we only have the two parameters; $M_{\rm min}$ and $\sigma_{\log_{10}M}$.
    \item Standard HOD: Galaxies are populated only for host halos exactly following the prescription described above.
    \item Standard HOD with concentration: This is based on the standard HOD procedure, but we add a dependence on halo concentration. Details are given in \cite{future_kaz}.
\end{itemize}

We populate galaxies to \texttt{Rockstar} halos identified in eight independent realizations of the $N$-body simulation of each $1.5~{\rm Gpc}/h$ length and with $1536^3$ particles, employed in~\cite{Akitsu:2023eqa}.
For each type of the HOD models we produce 44,550 mock catalogs of galaxies, varying the HOD parameters in the following ranges:
\begin{align}
    \log_{10}M_{\rm min} \in & [12.4, 14.2]\,\,,
    \\
    \sigma_{\log_{10}M} \in & [0.1, 1.0]\,\,,
    \\
    \kappa \in & [0.1, 1.0]\,\,,
    \\
    \log_{10} M_1 \in & [13.0, 15.0]\,\,,
    \\
    \alpha \in & [0.0, 1.6]\,\,.
\end{align}
The HOD galaxy mocks with different HOD parameter sets are generated at $z=0.38$, $0.51$ and $0.61$, corresponding to LOWZ, CMASS1 and CMASS2, respectively.

In order to precisely measure the galaxy bias parameters from the HOD mock galaxies, we take the cross-correlations between the initial conditions of the simulation and the mock galaxy field, which largely cancels the sample variance since they share the same random seed.
In particular, we employ the quadratic-field method \cite{Schmittfull:2014tca} to measure the quadratic galaxy bias parameters, $b_2$ and $b_{{\cal G}_2}$.
We also apply our analysis pipeline for the halo density field with the $b_2 - b_{K^2}$ bias basis and confirm that our measurement of $b_2$ for the halo is in great agreement with its separate-universe measurement in \cite{Lazeyras:2015lgp}.
With a large suite of HOD mock catalogs we can estimate the conditional probability distribution of the galaxy bias parameters given the HOD parameters, ${\cal P}(\boldsymbol{\theta}_{\rm bias}|\boldsymbol{\theta}_{\rm HOD})$.

\begin{figure}[ht!]
	\centering
    \includegraphics[width=0.48\textwidth]{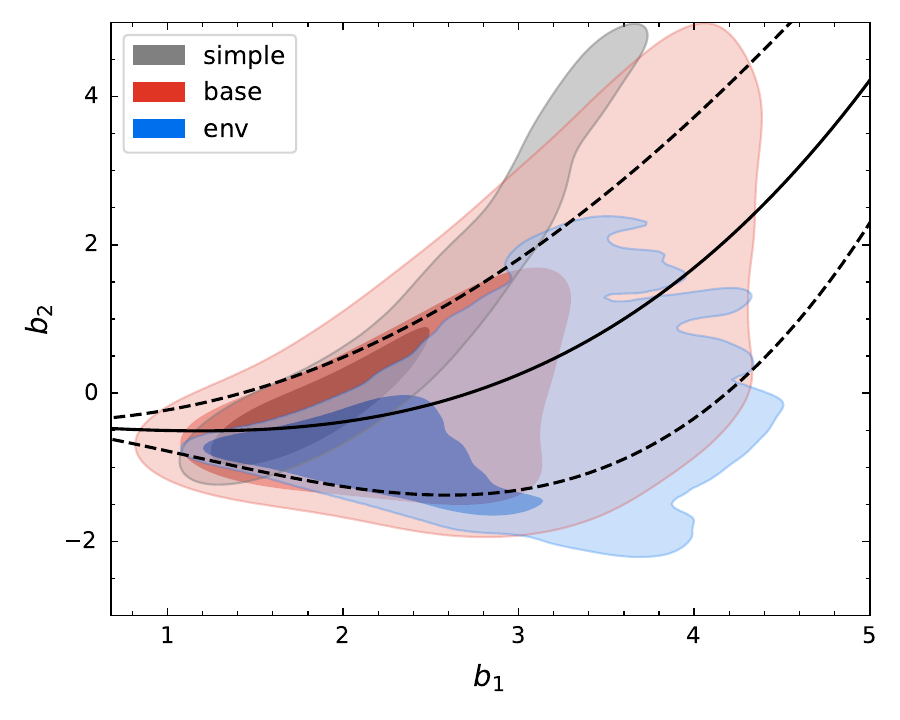}
	\includegraphics[width=0.48\textwidth]{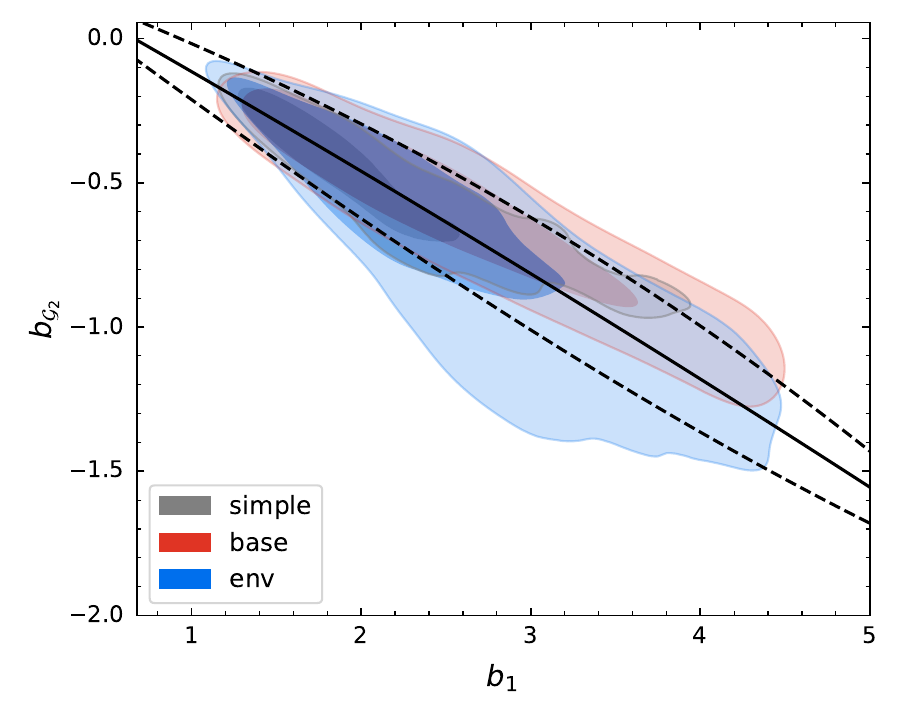}
    \caption{The probability distribution of the galaxy bias parameters, $b_1$, $b_2$ and $b_{{\cal G}_2}$, marginalized over the HOD parameters.
    The grey, red and blue contours correspond to the simplified HOD model, the standard HOD model and the standard HOD with a concentration-dependence consideration (\emph{i.e.}~the environmental dependence), respectively. Density levels correspond to $68\%$ and $95\%$ CL.
    The black solid lines show the best fit of $b_2$ and $b_{{\cal G}_2}$ as a function of $b_1$ (Eqs.~\eqref{eq:b1_b2_mean} and \eqref{eq:b1_bG2_mean}) and the black dashed lines show the similar fittings for the $1\sigma$ deviation from the mean relations (Eqs.~\eqref{eq:b1_b2_sigma} and \eqref{eq:b1_bG2_sigma}).}
	\label{fig:HOD_bias}
\end{figure}

The priors on the quadratic galaxy bias parameters are then obtained by marginalizing over the HOD parameters with uniform priors on the HOD parameters;
\begin{align}
    {\cal P}( \boldsymbol{\theta}_{\rm bias}) = 
    \int {\rm d} \boldsymbol{\theta}_{\rm HOD}~
    {\cal P}(\boldsymbol{\theta}_{\rm bias}|\boldsymbol{\theta}_{\rm HOD}) ~{\cal U}(\boldsymbol{\theta}_{\rm HOD})\,\,.
\end{align}
Note that we combine all three different redshift catalogs to get this distribution. {The scatter due to the HOD parameters is much bigger than the difference caused by the different redshifts, so combining different redshifts will have effectively no visible impact on the distribution of the biases, while allowing us to have more data points from which to fit relationships between bias parameters.} In Fig.~\ref{fig:HOD_bias}, we show the 2D projected ${\cal P}( \boldsymbol{\theta}_{\rm bias})$ onto the $b_1 - b_2$ and $b_1 - b_{{\cal G}_2}$ plane.
The different color contours correspond to the different HOD models listed above, namely the gray one is obtained from the simplified HOD, the red is from the standard HOD and the blue is from the standard HOD with the additional halo concentration dependence.
We find the restrictive prior on $b_{{\cal G}_2}$ even when the HOD parameters are varied a lot, while the prior on $b_2$ is not as restrictive as $b_{{\cal G}_2}$.
In principle this distribution of the galaxy bias parameters itself can be used as the priors on the galaxy bias parameters. 
However, it is a non-trivial task to get a functional form of this distribution in order to generate new samples under this distribution. One method to get around this problem is to use normalizing flows, see~\cite{Ivanov:2024hgq,Ivanov:2024xgb}.
Instead, in this work, we just obtain simple polynomial fitting formulas for $b_2$ and $b_{{\cal G}_2}$ as functions of $b_1$ for convenience, although this is not an optimal use of the obtained probability distribution.
Specifically, here we provide the mean relations between $b_1 - b_2$ and $b_1 - b_{{\cal G}_2}$ as well as the $1 \sigma$ deviations from these mean relations as functions of $b_1$ that approximate the distributions in the 2D planes:
\begin{align}
    b_2 &= -0.38 - 0.15~b_1 - 0.021~b_1^2 + 0.047~b_1^3\,\,,
    \label{eq:b1_b2_mean}
    \\
    \sigma(b_2) &=  0.06 ~b_1 + 0.24~b_1^2 - 0.02~b_1^3 - 0.003~b_1^4\,\,,
    \label{eq:b1_b2_sigma}
    \\
    b_{{\cal G}_2} &= 0.22 - 0.33~ b_1 - 0.005~b_1^2\,\,,
    \label{eq:b1_bG2_mean}
    \\
    \sigma(b_{{\cal G}_2}) &= 0.11~b_1 - 0.012~b_1^2 - 0.001~b_1^3\,\,.
    \label{eq:b1_bG2_sigma}
\end{align}
Note that the fittings provided here are obtained in $b_1 \in [1.0, 4.0]$, focusing on BOSS-like galaxies.

\section{BOSS SVD analysis}
\label{app:SVD}

\noindent In this section we ask the following questions: 
\begin{itemize}[leftmargin=*]
    \item What Cosmological Collider information is present in the ($\mu$-marginalized) BOSS dataset?
    \item How degenerate is it with self-interactions?
\end{itemize}
To assess this, we perform an SVD (Singular Value Decomposition) of the Cosmological Collider galaxy bispectrum multipoles, $B_\ell(k_1,k_2,k_3)$, which will allow us to extract the most well-constrained massive-particle PNG modes, which we can then check for degeneracies. We do not consider power spectra in this test, since the leading constraining power on non-local PNG arises from the bispectrum, which thus dictates the degeneracy directions. 

Following Ref.~\cite{Philcox:2020zyp}, we first generate a template bank of bispectrum multipoles, $\smash{B^{(i)}_b({\bm\theta})}$, where $b$ indexes the bins, multipoles, and the four BOSS data chunks, and $\smash{i}$ runs over the templates. We vary the following parameters:
\be
{\bm\theta} = \{\beta^\collA,\beta^\collB,\mu\} 
\ee
for three values of $\smash{c_{\rm s}/c_\sigma = 0.1,1,10}$. In total, we generate $\smash{51\times 51\times 21}$ templates with $\smash{\bm\theta}$ in the range 
\be
\{\beta^\collA,\beta^\collB,\mu\}\in[{-10^6},10^6]\times[0,10^6]\times[0,5]\,\,,
\ee 
fixing all remaining bias parameters to the BOSS (non-PNG) best-fit values, since we are here interested only in primordial-induced degeneracies. We additionally set $\smash{f_{\rm NL}^{\rm equil}=0=f_{\rm NL}^{\rm ortho}}$, and will discuss the correlations below (\textit{i.e.}~drop any inflaton self-interactions).

Next, we define the inner product between two points $\smash{{\bm\theta}^{(i)}}$ and $\smash{{\bm\theta}^{(j)}}$, via the noise-weighted distance from the mean $\smash{\bar{B}_b}$ of the template bank: 
\be
\langle\delta\!B^{(i)},\delta\!B^{(j)}\rangle_{\mathsf{C}}\equiv\sum_{b,c} \delta\!B^{(i)}_b\mathsf{C}^{-1}_{bc}\delta\!B^{(j)}_c\qquad\text{with}\qquad \delta\!B_b^{(i)} = B_b^{(i)} - \bar{B}_b\,\,,
\ee
where $\smash{\mathsf{C}^{-1}_{bc}}$ is the inverse bispectrum covariance measure from mocks. We then perform the linear transformation 
\be
\delta\!B_b({\bm\theta}) \to X_b({\bm\theta})\qquad\text{with}\qquad X_b({\bm\theta})\equiv\sum_c\mathsf{C}^{-1/2}_{bc}(B_c({\bm\theta}) - \bar{B}_c)\,\,,
\ee
where $\smash{\mathsf{C}^{1/2}_{bc}}$ is the Cholesky decomposition of the covariance; this centers and whitens the bispectra. Following this decomposition, an arbitrary rotated bispectrum can be written as linear combination of a set of $\smash{N_{\rm bin}}$ orthonormal basis vectors $\smash{V_{\alpha b}}$, \emph{i.e.}
\be
\label{svd_eq}
X_b({\bm\theta}) = \sum_\alpha c_\alpha({\bm\theta})V_{\alpha b}\qquad\text{and}\qquad c_\alpha({\bm\theta}) = \sum_bX_b({\bm\theta})V_{\alpha b}\,\,.
\ee

To compute the basis vectors we execute an SVD of our template bank. We form an $\smash{N_{\rm bank} \times N_{\rm bin}}$ matrix $\smash{X_{ib}}$, which has the SVD
\be
X_{ib} = \sum_\alpha U_{i\alpha}D_\alpha V_{\alpha b}\,\,,
\ee
where $\smash{D_\alpha}$ is a rank-$N_{\rm bin}$ diagonal matrix of the singular values (SVs) and the matrices $U$ and $V$ (of dimension $\smash{N_{\rm bank} \times N_{\rm bin}}$ and $\smash{N_{\rm bin} \times N_{\rm bin}}$) are unitary, projecting from observations to SVs, and SVs to spectra respectively. By comparing with Eq.~\eqref{svd_eq} we see that 
\be
c_\alpha^{(i)} = U_{i\alpha}D_\alpha\,\,,
\ee
and since $\smash{D_\alpha\geq |c_\alpha|}$ we can approximate every template in the bank with a finite number $\smash{N_{\rm SV} < N_{\rm bin}}$ of basis vectors. By design, these are the directions which have the largest contributions to the log-likelihood.

\begin{figure}[ht!]
    \centering
    \includegraphics[width=0.6\textwidth]{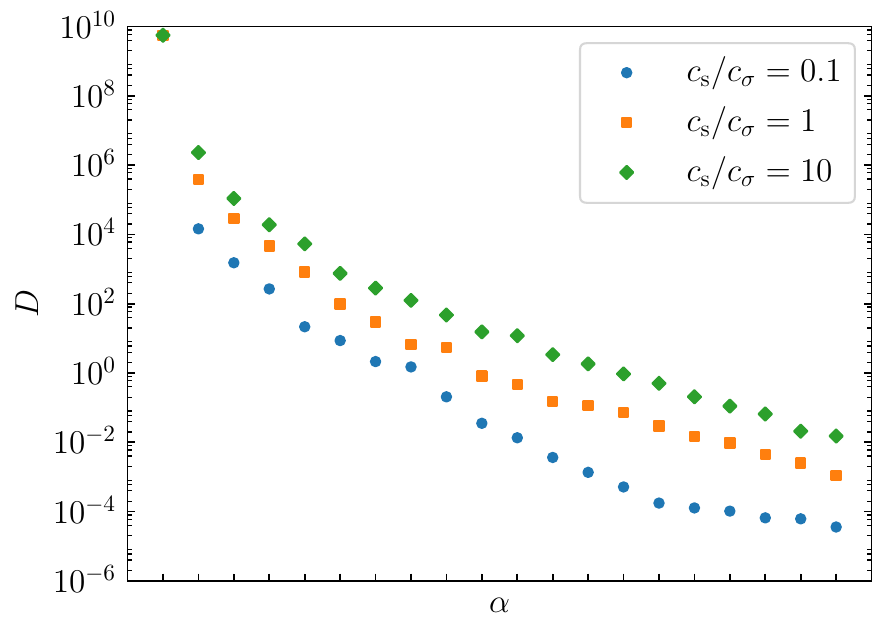}
    \caption{First $20$ entries of the diagonal matrix $D_\alpha$ for three different values of $\smash{c_{\rm s}/c_\sigma}$. We clearly observe that the data preferentially constrains a single direction (which dominates the $D$ matrix). This is the same regardless of speed: as shown in Tab.~\ref{tab:svd_cosines} it is essentially parallel to the the combination of equilateral and orthogonal non-Gaussianities that is best constrained by BOSS.} 
    \label{fig:svds_D}
\end{figure}

In Fig.~\ref{fig:svds_D} we show the contributions of the first $20$ SVs to the $D_\alpha$ matrix: we observe that the first SV strongly dominates, exhibiting the same behavior regardless of speed. Given this result, it is instructive to then study how much the first three Cosmological Collider basis vectors ($V_{1b},V_{2b},V_{3b}$) overlap with the equilateral and orthogonal templates. For this purpose, we first compute the bispectrum multipoles at unit $\smash{\fnl^{\rm equil}}$ and $\smash{\fnl^{\rm ortho}}$ and zero $\beta^{\collA}=\beta^{\collB}$, then orthogonalize with respect to the BOSS covariance using a Gram-Schmidt procedure.\footnote{Note that the measured correlations will be independent of the value of $f_{\rm NL}^{\rm equil}$ and $f_{\rm NL}^{\rm ortho}$.} The resulting self-interaction vectors, denoted $\smash{o_{1b}}$ and $\smash{o_{2b}}$, are given by 
\be
o_{1b} = X^{\fnl^{\rm equil}}_b \qquad\text{and}\qquad o_{2b} = o_{1b} - \frac{\langle X^{\fnl^{\rm ortho}}|o_{1}\rangle}{\langle o_{1}|o_{1}\rangle}\,o_{1b}\,\,,
\ee
where 
\be
X^{\fnl^{\rm equil}}_b = \sum_{c} \mathsf{C}^{-1/2}_{bc}B^{\fnl^{\rm equil}=1}_c\qquad\text{and}\qquad X^{\fnl^{\rm ortho}}_b = \sum_{c} \mathsf{C}^{-1/2}_{bc}B^{\fnl^{\rm ortho}=1}_c 
\ee
subject to the inner product 
\be
\langle u|v\rangle = \sum_b u_b v_b\,\,.
\ee

\setlength{\tabcolsep}{4pt}
\renewcommand{\arraystretch}{1.3}
\begin{table*}[ht!]
\centering
\begin{tabular}{c|ccc}
 & \multicolumn{3}{c}{$1-\cos^2(V,o_{1})-\cos^2(V,o_{2})$} \\ \hline
 \textbf{Model} & $V_{1}$ & \quad $V_{2}$ \quad & $V_{3}$ \\ 
 \hline

$c_{\rm s}/c_\sigma = 0.1$ & $\num[round-mode=places,round-precision=0]{7.101398194785306e-14}$ & $\num[round-mode=places,round-precision=3]{0.010819047478806465}$ & $\num[round-mode=places,round-precision=3]{0.9903984294306882}$ \\

$c_{\rm s}/c_\sigma = 1.0$ & $\num[round-mode=places,round-precision=0]{4.116291262538175e-12}$ & $\num[round-mode=places,round-precision=3]{0.008257321006555851}$ & $\num[round-mode=places,round-precision=3]{0.9932041716484573}$ \\

$c_{\rm s}/c_\sigma = 10$ & $\num[round-mode=places,round-precision=0]{2.1841270607712298e-10}$ & $\num[round-mode=places,round-precision=3]{0.050800654730993045}$ & $\num[round-mode=places,round-precision=3]{0.9497989031137681}$ \\

 \end{tabular}
 \caption{Overlap of the first three massive particle basis vectors with the equilateral and orthogonal templates, for various values of $\smash{c_{\rm s}/c_\sigma}$. This is defined as $\smash{\cos^2(V,o_{1})+\cos^2(V,o_{2})}$, where $V_{\{1,2,3\}b}$ is the Cosmological Collider basis vector of interest, and $\smash{o_{\{1,2\}b}}$ are an orthogonal (and suitably normalized) representation of the equilateral and orthogonal templates. Here, $\smash\cos(V,v) = \langle V|v\rangle/\sqrt{\langle V|V\rangle\langle v|v\rangle}$. Whilst the first basis vector is almost perfectly described by the self-interaction templates, the second template (and beyond) have reduced overlaps, suggesting that some additional information is present, though, from Fig.\,\ref{fig:svds_D}, it is evident that the information content is dominated by the degenerate combinations.} 
 \label{tab:svd_cosines}
\end{table*}

\begin{figure}
  \begin{minipage}{0.45\textwidth}
    \centering
    \includegraphics[height=0.7\textwidth]{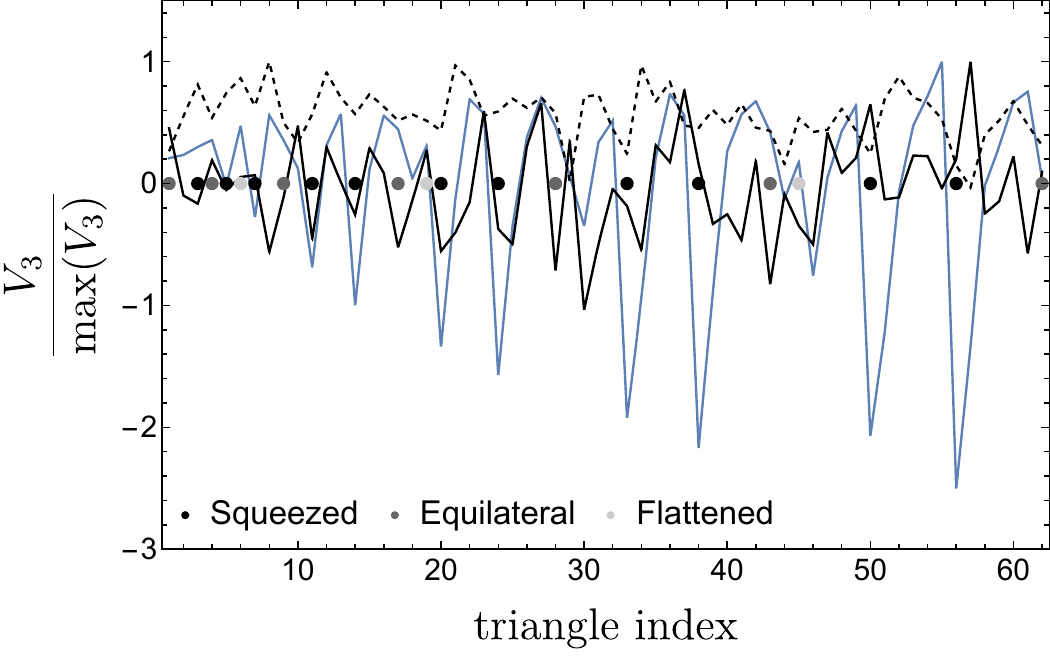}
  \end{minipage}
  \hfill
  \begin{minipage}{0.45\textwidth}
    \centering
    \includegraphics[height=0.7\textwidth]{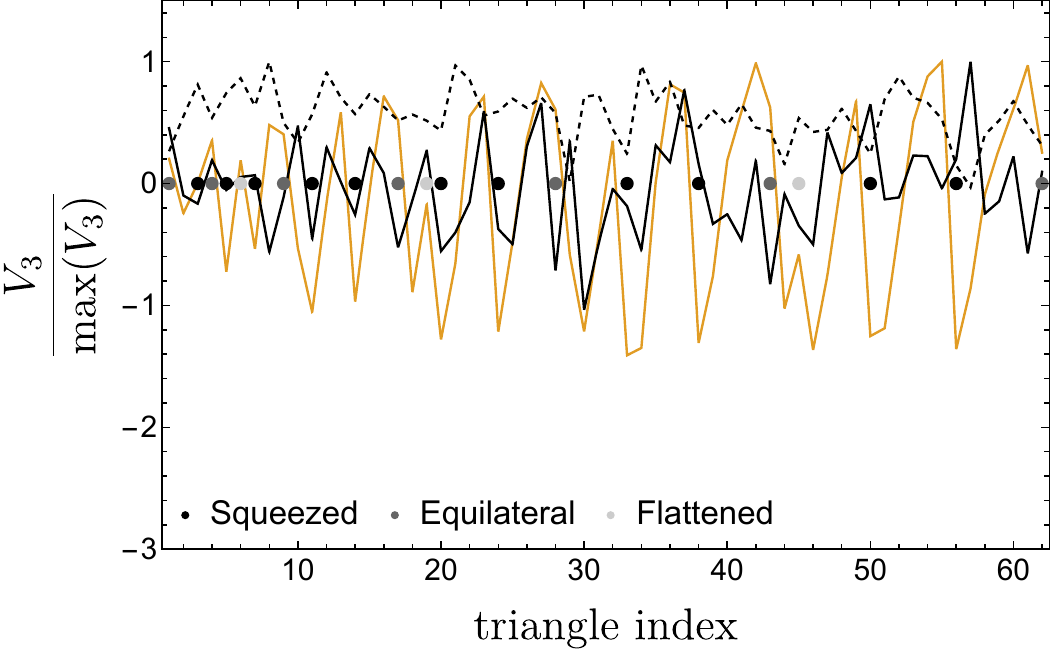}
  \end{minipage}
  \hfill
  \begin{minipage}{0.45\textwidth}
    \centering
    \includegraphics[height=0.7\textwidth]{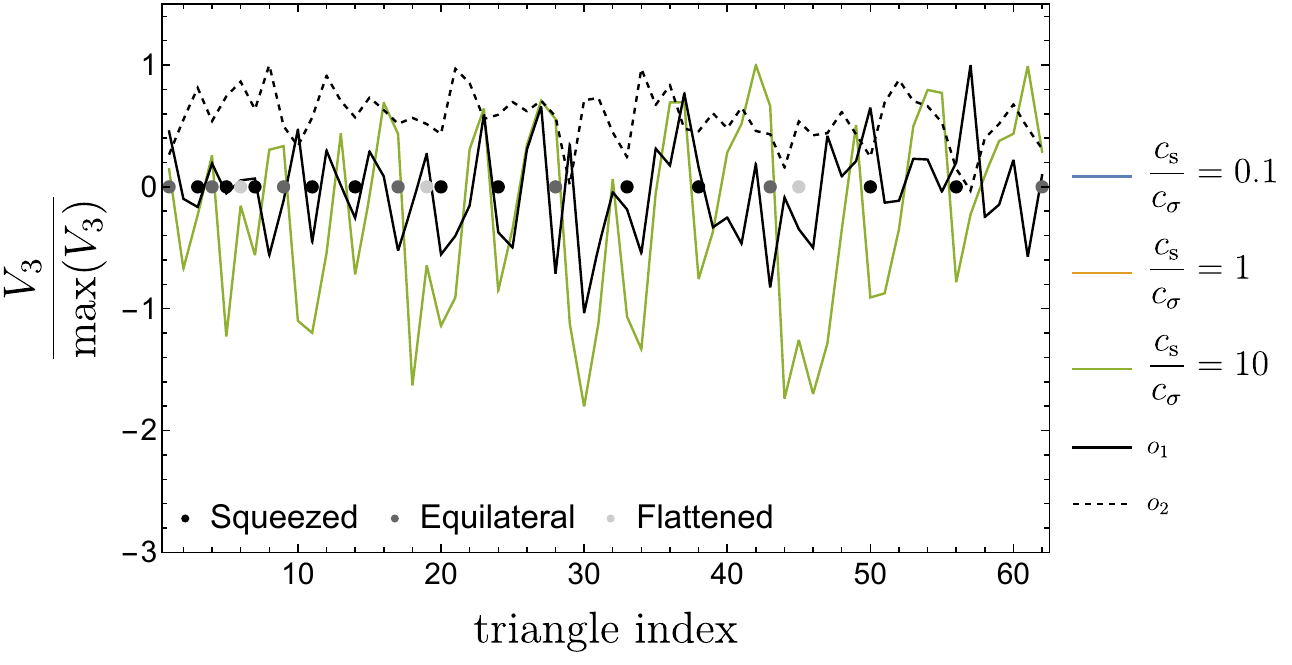}
  \end{minipage}
  \caption{Plot of the third Cosmological Collider bispectrum basis vector $V_{3b}$ (colored lines), alongside the self-interaction vectors $\smash{o_{1b}}$ (black full lines) and $\smash{o_{2b}}$ (black dashed lines), for $\smash{c_{\rm s}/c_\sigma = 0.1, 1, 10}$. We plot triangle indices up to $62$, \emph{i.e.}~only those corresponding to the bispectrum monopole. Since the purpose of this plot is only to show the triangle dependence we have normalized all shapes to have unit maximum. Triangles are ordered in increasing wavenumber, compressed into one dimension, leading to the sawtooth patterns. The magnitudes of these wavenumber satisfy $k\in [0.015, 0.08)\,\hMpc$. Black, dark-grey and light-gray dots denote the squeezed ($k_3 = 0.015\,\hMpc; k_1,k_2 > k_3$), equilateral ($k_1 = k_2 = k_3$), and flattened ($k_2 = k_3, 2k_2 = k_1 + 0.015\,\hMpc$) triangle configurations, respectively, as in \cite{Cabass:2022ymb}. The triangle-dependence of the various shapes differs considerably from {the self-interaction vectors} $\smash{o_{1b}}$ and $\smash{o_{2b}}$, as expected.}
  \label{fig:svds_V}
\end{figure}

With these definitions we can proceed to compute the cosines of the lowest-order Cosmological Collider basis vectors ($V_{1b},V_{2b},V_{3b}$) with the orthonormalized self-interaction templates $\smash{o_{1b}}$ and $\smash{o_{2b}}$. These are collected in Tab.~\ref{tab:svd_cosines}. We find that $V_{1b}$ is essentially perfectly described by a linear combination of $\smash{o_{1b}}$ and $\smash{o_{2b}}$, that $V_{2b}$ is very well described by a superposition of the two, and finally that it is only $V_{3b}$ that doesn't show overlap with either. We show the triangle dependence of $V_{3b}$ compared to $\smash{o_{1b}}$ and $\smash{o_{2b}}$ in Fig.~\ref{fig:svds_V}. From Fig.\,\ref{fig:svds_D}, it is clear that the majority of constraining power (encoded in $D_\alpha^{1/2}$) is contained within $V_{1b}$ and $V_{2b}$, and thus strongly entangled with self-interaction shapes. That said, there remains some non-trivial information in the higher-order templates, which are non-degenerate.

\bibliography{short.bib}

\begin{thebibliography}{224}%
\makeatletter
\providecommand \@ifxundefined [1]{%
 \@ifx{#1\undefined}
}%
\providecommand \@ifnum [1]{%
 \ifnum #1\expandafter \@firstoftwo
 \else \expandafter \@secondoftwo
 \fi
}%
\providecommand \@ifx [1]{%
 \ifx #1\expandafter \@firstoftwo
 \else \expandafter \@secondoftwo
 \fi
}%
\providecommand \natexlab [1]{#1}%
\providecommand \enquote  [1]{``#1''}%
\providecommand \bibnamefont  [1]{#1}%
\providecommand \bibfnamefont [1]{#1}%
\providecommand \citenamefont [1]{#1}%
\providecommand \href@noop [0]{\@secondoftwo}%
\providecommand \href [0]{\begingroup \@sanitize@url \@href}%
\providecommand \@href[1]{\@@startlink{#1}\@@href}%
\providecommand \@@href[1]{\endgroup#1\@@endlink}%
\providecommand \@sanitize@url [0]{\catcode `\\12\catcode `\$12\catcode
  `\&12\catcode `\#12\catcode `\^12\catcode `\_12\catcode `\%12\relax}%
\providecommand \@@startlink[1]{}%
\providecommand \@@endlink[0]{}%
\providecommand \url  [0]{\begingroup\@sanitize@url \@url }%
\providecommand \@url [1]{\endgroup\@href {#1}{\urlprefix }}%
\providecommand \urlprefix  [0]{URL }%
\providecommand \Eprint [0]{\href }%
\providecommand \doibase [0]{http://dx.doi.org/}%
\providecommand \selectlanguage [0]{\@gobble}%
\providecommand \bibinfo  [0]{\@secondoftwo}%
\providecommand \bibfield  [0]{\@secondoftwo}%
\providecommand \translation [1]{[#1]}%
\providecommand \BibitemOpen [0]{}%
\providecommand \bibitemStop [0]{}%
\providecommand \bibitemNoStop [0]{.\EOS\space}%
\providecommand \EOS [0]{\spacefactor3000\relax}%
\providecommand \BibitemShut  [1]{\csname bibitem#1\endcsname}%
\let\auto@bib@innerbib\@empty
\bibitem [{\citenamefont {Creminelli}\ \emph {et~al.}(2006)\citenamefont
  {Creminelli}, \citenamefont {Luty}, \citenamefont {Nicolis},\ and\
  \citenamefont {Senatore}}]{Creminelli:2006xe}%
  \BibitemOpen
  \bibfield  {author} {\bibinfo {author} {\bibfnamefont {P.}~\bibnamefont
  {Creminelli}}, \bibinfo {author} {\bibfnamefont {M.~A.}\ \bibnamefont
  {Luty}}, \bibinfo {author} {\bibfnamefont {A.}~\bibnamefont {Nicolis}}, \
  and\ \bibinfo {author} {\bibfnamefont {L.}~\bibnamefont {Senatore}},\ }\href
  {\doibase 10.1088/1126-6708/2006/12/080} {\bibfield  {journal} {\bibinfo
  {journal} {JHEP}\ }\textbf {\bibinfo {volume} {12}},\ \bibinfo {pages} {080}
  (\bibinfo {year} {2006})},\ \Eprint {http://arxiv.org/abs/hep-th/0606090}
  {arXiv:hep-th/0606090} \BibitemShut {NoStop}%
\bibitem [{\citenamefont {Cheung}\ \emph {et~al.}(2008)\citenamefont {Cheung},
  \citenamefont {Creminelli}, \citenamefont {Fitzpatrick}, \citenamefont
  {Kaplan},\ and\ \citenamefont {Senatore}}]{Cheung:2007st}%
  \BibitemOpen
  \bibfield  {author} {\bibinfo {author} {\bibfnamefont {C.}~\bibnamefont
  {Cheung}}, \bibinfo {author} {\bibfnamefont {P.}~\bibnamefont {Creminelli}},
  \bibinfo {author} {\bibfnamefont {A.~L.}\ \bibnamefont {Fitzpatrick}},
  \bibinfo {author} {\bibfnamefont {J.}~\bibnamefont {Kaplan}}, \ and\ \bibinfo
  {author} {\bibfnamefont {L.}~\bibnamefont {Senatore}},\ }\href {\doibase
  10.1088/1126-6708/2008/03/014} {\bibfield  {journal} {\bibinfo  {journal}
  {JHEP}\ }\textbf {\bibinfo {volume} {03}},\ \bibinfo {pages} {014} (\bibinfo
  {year} {2008})},\ \Eprint {http://arxiv.org/abs/0709.0293} {arXiv:0709.0293
  [hep-th]} \BibitemShut {NoStop}%
\bibitem [{\citenamefont {Cabass}\ \emph
  {et~al.}(2023{\natexlab{a}})\citenamefont {Cabass}, \citenamefont {Ivanov},
  \citenamefont {Lewandowski}, \citenamefont {Mirbabayi},\ and\ \citenamefont
  {Simonovi\'c}}]{Cabass:2022avo}%
  \BibitemOpen
  \bibfield  {author} {\bibinfo {author} {\bibfnamefont {G.}~\bibnamefont
  {Cabass}}, \bibinfo {author} {\bibfnamefont {M.~M.}\ \bibnamefont {Ivanov}},
  \bibinfo {author} {\bibfnamefont {M.}~\bibnamefont {Lewandowski}}, \bibinfo
  {author} {\bibfnamefont {M.}~\bibnamefont {Mirbabayi}}, \ and\ \bibinfo
  {author} {\bibfnamefont {M.}~\bibnamefont {Simonovi\'c}},\ }\href {\doibase
  10.1016/j.dark.2023.101193} {\bibfield  {journal} {\bibinfo  {journal} {Phys.
  Dark Univ.}\ }\textbf {\bibinfo {volume} {40}},\ \bibinfo {pages} {101193}
  (\bibinfo {year} {2023}{\natexlab{a}})},\ \Eprint
  {http://arxiv.org/abs/2203.08232} {arXiv:2203.08232 [astro-ph.CO]}
  \BibitemShut {NoStop}%
\bibitem [{\citenamefont {Ghosh}\ \emph {et~al.}(2014)\citenamefont {Ghosh},
  \citenamefont {Kundu}, \citenamefont {Raju},\ and\ \citenamefont
  {Trivedi}}]{Ghosh:2014kba}%
  \BibitemOpen
  \bibfield  {author} {\bibinfo {author} {\bibfnamefont {A.}~\bibnamefont
  {Ghosh}}, \bibinfo {author} {\bibfnamefont {N.}~\bibnamefont {Kundu}},
  \bibinfo {author} {\bibfnamefont {S.}~\bibnamefont {Raju}}, \ and\ \bibinfo
  {author} {\bibfnamefont {S.~P.}\ \bibnamefont {Trivedi}},\ }\href {\doibase
  10.1007/JHEP07(2014)011} {\bibfield  {journal} {\bibinfo  {journal} {JHEP}\
  }\textbf {\bibinfo {volume} {07}},\ \bibinfo {pages} {011} (\bibinfo {year}
  {2014})},\ \Eprint {http://arxiv.org/abs/1401.1426} {arXiv:1401.1426
  [hep-th]} \BibitemShut {NoStop}%
\bibitem [{\citenamefont {Chen}\ and\ \citenamefont
  {Wang}(2010{\natexlab{a}})}]{Chen:2009we}%
  \BibitemOpen
  \bibfield  {author} {\bibinfo {author} {\bibfnamefont {X.}~\bibnamefont
  {Chen}}\ and\ \bibinfo {author} {\bibfnamefont {Y.}~\bibnamefont {Wang}},\
  }\href {\doibase 10.1103/PhysRevD.81.063511} {\bibfield  {journal} {\bibinfo
  {journal} {Phys. Rev. D}\ }\textbf {\bibinfo {volume} {81}},\ \bibinfo
  {pages} {063511} (\bibinfo {year} {2010}{\natexlab{a}})},\ \Eprint
  {http://arxiv.org/abs/0909.0496} {arXiv:0909.0496 [astro-ph.CO]} \BibitemShut
  {NoStop}%
\bibitem [{\citenamefont {Chen}\ and\ \citenamefont
  {Wang}(2010{\natexlab{b}})}]{Chen:2009zp}%
  \BibitemOpen
  \bibfield  {author} {\bibinfo {author} {\bibfnamefont {X.}~\bibnamefont
  {Chen}}\ and\ \bibinfo {author} {\bibfnamefont {Y.}~\bibnamefont {Wang}},\
  }\href {\doibase 10.1088/1475-7516/2010/04/027} {\bibfield  {journal}
  {\bibinfo  {journal} {JCAP}\ }\textbf {\bibinfo {volume} {04}},\ \bibinfo
  {pages} {027} (\bibinfo {year} {2010}{\natexlab{b}})},\ \Eprint
  {http://arxiv.org/abs/0911.3380} {arXiv:0911.3380 [hep-th]} \BibitemShut
  {NoStop}%
\bibitem [{\citenamefont {Chen}(2010)}]{Chen:2010xka}%
  \BibitemOpen
  \bibfield  {author} {\bibinfo {author} {\bibfnamefont {X.}~\bibnamefont
  {Chen}},\ }\href {\doibase 10.1155/2010/638979} {\bibfield  {journal}
  {\bibinfo  {journal} {Adv. Astron.}\ }\textbf {\bibinfo {volume} {2010}},\
  \bibinfo {pages} {638979} (\bibinfo {year} {2010})},\ \Eprint
  {http://arxiv.org/abs/1002.1416} {arXiv:1002.1416 [astro-ph.CO]} \BibitemShut
  {NoStop}%
\bibitem [{\citenamefont {Byrnes}\ \emph {et~al.}(2010)\citenamefont {Byrnes},
  \citenamefont {Enqvist},\ and\ \citenamefont {Takahashi}}]{Byrnes:2010xd}%
  \BibitemOpen
  \bibfield  {author} {\bibinfo {author} {\bibfnamefont {C.~T.}\ \bibnamefont
  {Byrnes}}, \bibinfo {author} {\bibfnamefont {K.}~\bibnamefont {Enqvist}}, \
  and\ \bibinfo {author} {\bibfnamefont {T.}~\bibnamefont {Takahashi}},\ }\href
  {\doibase 10.1088/1475-7516/2010/09/026} {\bibfield  {journal} {\bibinfo
  {journal} {JCAP}\ }\textbf {\bibinfo {volume} {09}},\ \bibinfo {pages} {026}
  (\bibinfo {year} {2010})},\ \Eprint {http://arxiv.org/abs/1007.5148}
  {arXiv:1007.5148 [astro-ph.CO]} \BibitemShut {NoStop}%
\bibitem [{\citenamefont {Achucarro}\ \emph {et~al.}(2011)\citenamefont
  {Achucarro}, \citenamefont {Gong}, \citenamefont {Hardeman}, \citenamefont
  {Palma},\ and\ \citenamefont {Patil}}]{Achucarro:2010da}%
  \BibitemOpen
  \bibfield  {author} {\bibinfo {author} {\bibfnamefont {A.}~\bibnamefont
  {Achucarro}}, \bibinfo {author} {\bibfnamefont {J.-O.}\ \bibnamefont {Gong}},
  \bibinfo {author} {\bibfnamefont {S.}~\bibnamefont {Hardeman}}, \bibinfo
  {author} {\bibfnamefont {G.~A.}\ \bibnamefont {Palma}}, \ and\ \bibinfo
  {author} {\bibfnamefont {S.~P.}\ \bibnamefont {Patil}},\ }\href {\doibase
  10.1088/1475-7516/2011/01/030} {\bibfield  {journal} {\bibinfo  {journal}
  {JCAP}\ }\textbf {\bibinfo {volume} {01}},\ \bibinfo {pages} {030} (\bibinfo
  {year} {2011})},\ \Eprint {http://arxiv.org/abs/1010.3693} {arXiv:1010.3693
  [hep-ph]} \BibitemShut {NoStop}%
\bibitem [{\citenamefont {Baumann}\ and\ \citenamefont
  {Green}(2012)}]{Baumann:2011nk}%
  \BibitemOpen
  \bibfield  {author} {\bibinfo {author} {\bibfnamefont {D.}~\bibnamefont
  {Baumann}}\ and\ \bibinfo {author} {\bibfnamefont {D.}~\bibnamefont
  {Green}},\ }\href {\doibase 10.1103/PhysRevD.85.103520} {\bibfield  {journal}
  {\bibinfo  {journal} {Phys. Rev. D}\ }\textbf {\bibinfo {volume} {85}},\
  \bibinfo {pages} {103520} (\bibinfo {year} {2012})},\ \Eprint
  {http://arxiv.org/abs/1109.0292} {arXiv:1109.0292 [hep-th]} \BibitemShut
  {NoStop}%
\bibitem [{\citenamefont {Chen}\ and\ \citenamefont
  {Wang}(2012)}]{Chen:2012ge}%
  \BibitemOpen
  \bibfield  {author} {\bibinfo {author} {\bibfnamefont {X.}~\bibnamefont
  {Chen}}\ and\ \bibinfo {author} {\bibfnamefont {Y.}~\bibnamefont {Wang}},\
  }\href {\doibase 10.1088/1475-7516/2012/09/021} {\bibfield  {journal}
  {\bibinfo  {journal} {JCAP}\ }\textbf {\bibinfo {volume} {09}},\ \bibinfo
  {pages} {021} (\bibinfo {year} {2012})},\ \Eprint
  {http://arxiv.org/abs/1205.0160} {arXiv:1205.0160 [hep-th]} \BibitemShut
  {NoStop}%
\bibitem [{\citenamefont {Noumi}\ \emph {et~al.}(2013)\citenamefont {Noumi},
  \citenamefont {Yamaguchi},\ and\ \citenamefont {Yokoyama}}]{Noumi:2012vr}%
  \BibitemOpen
  \bibfield  {author} {\bibinfo {author} {\bibfnamefont {T.}~\bibnamefont
  {Noumi}}, \bibinfo {author} {\bibfnamefont {M.}~\bibnamefont {Yamaguchi}}, \
  and\ \bibinfo {author} {\bibfnamefont {D.}~\bibnamefont {Yokoyama}},\ }\href
  {\doibase 10.1007/JHEP06(2013)051} {\bibfield  {journal} {\bibinfo  {journal}
  {JHEP}\ }\textbf {\bibinfo {volume} {06}},\ \bibinfo {pages} {051} (\bibinfo
  {year} {2013})},\ \Eprint {http://arxiv.org/abs/1211.1624} {arXiv:1211.1624
  [hep-th]} \BibitemShut {NoStop}%
\bibitem [{\citenamefont {Assassi}\ \emph {et~al.}(2012)\citenamefont
  {Assassi}, \citenamefont {Baumann},\ and\ \citenamefont
  {Green}}]{Assassi:2012zq}%
  \BibitemOpen
  \bibfield  {author} {\bibinfo {author} {\bibfnamefont {V.}~\bibnamefont
  {Assassi}}, \bibinfo {author} {\bibfnamefont {D.}~\bibnamefont {Baumann}}, \
  and\ \bibinfo {author} {\bibfnamefont {D.}~\bibnamefont {Green}},\ }\href
  {\doibase 10.1088/1475-7516/2012/11/047} {\bibfield  {journal} {\bibinfo
  {journal} {JCAP}\ }\textbf {\bibinfo {volume} {11}},\ \bibinfo {pages} {047}
  (\bibinfo {year} {2012})},\ \Eprint {http://arxiv.org/abs/1204.4207}
  {arXiv:1204.4207 [hep-th]} \BibitemShut {NoStop}%
\bibitem [{\citenamefont {Arkani-Hamed}\ and\ \citenamefont
  {Maldacena}(2015)}]{Arkani-Hamed:2015bza}%
  \BibitemOpen
  \bibfield  {author} {\bibinfo {author} {\bibfnamefont {N.}~\bibnamefont
  {Arkani-Hamed}}\ and\ \bibinfo {author} {\bibfnamefont {J.}~\bibnamefont
  {Maldacena}},\ }\href@noop {} {\  (\bibinfo {year} {2015})},\ \Eprint
  {http://arxiv.org/abs/1503.08043} {arXiv:1503.08043 [hep-th]} \BibitemShut
  {NoStop}%
\bibitem [{\citenamefont {Lee}\ \emph {et~al.}(2016)\citenamefont {Lee},
  \citenamefont {Baumann},\ and\ \citenamefont {Pimentel}}]{Lee:2016vti}%
  \BibitemOpen
  \bibfield  {author} {\bibinfo {author} {\bibfnamefont {H.}~\bibnamefont
  {Lee}}, \bibinfo {author} {\bibfnamefont {D.}~\bibnamefont {Baumann}}, \ and\
  \bibinfo {author} {\bibfnamefont {G.~L.}\ \bibnamefont {Pimentel}},\ }\href
  {\doibase 10.1007/JHEP12(2016)040} {\bibfield  {journal} {\bibinfo  {journal}
  {JHEP}\ }\textbf {\bibinfo {volume} {12}},\ \bibinfo {pages} {040} (\bibinfo
  {year} {2016})},\ \Eprint {http://arxiv.org/abs/1607.03735} {arXiv:1607.03735
  [hep-th]} \BibitemShut {NoStop}%
\bibitem [{\citenamefont {Kehagias}\ and\ \citenamefont
  {Riotto}(2017)}]{Kehagias:2017cym}%
  \BibitemOpen
  \bibfield  {author} {\bibinfo {author} {\bibfnamefont {A.}~\bibnamefont
  {Kehagias}}\ and\ \bibinfo {author} {\bibfnamefont {A.}~\bibnamefont
  {Riotto}},\ }\href {\doibase 10.1088/1475-7516/2017/07/046} {\bibfield
  {journal} {\bibinfo  {journal} {JCAP}\ }\textbf {\bibinfo {volume} {07}},\
  \bibinfo {pages} {046} (\bibinfo {year} {2017})},\ \Eprint
  {http://arxiv.org/abs/1705.05834} {arXiv:1705.05834 [hep-th]} \BibitemShut
  {NoStop}%
\bibitem [{\citenamefont {An}\ \emph {et~al.}(2018)\citenamefont {An},
  \citenamefont {McAneny}, \citenamefont {Ridgway},\ and\ \citenamefont
  {Wise}}]{An:2017hlx}%
  \BibitemOpen
  \bibfield  {author} {\bibinfo {author} {\bibfnamefont {H.}~\bibnamefont
  {An}}, \bibinfo {author} {\bibfnamefont {M.}~\bibnamefont {McAneny}},
  \bibinfo {author} {\bibfnamefont {A.~K.}\ \bibnamefont {Ridgway}}, \ and\
  \bibinfo {author} {\bibfnamefont {M.~B.}\ \bibnamefont {Wise}},\ }\href
  {\doibase 10.1007/JHEP06(2018)105} {\bibfield  {journal} {\bibinfo  {journal}
  {JHEP}\ }\textbf {\bibinfo {volume} {06}},\ \bibinfo {pages} {105} (\bibinfo
  {year} {2018})},\ \Eprint {http://arxiv.org/abs/1706.09971} {arXiv:1706.09971
  [hep-ph]} \BibitemShut {NoStop}%
\bibitem [{\citenamefont {Kumar}\ and\ \citenamefont
  {Sundrum}(2018)}]{Kumar:2017ecc}%
  \BibitemOpen
  \bibfield  {author} {\bibinfo {author} {\bibfnamefont {S.}~\bibnamefont
  {Kumar}}\ and\ \bibinfo {author} {\bibfnamefont {R.}~\bibnamefont
  {Sundrum}},\ }\href {\doibase 10.1007/JHEP05(2018)011} {\bibfield  {journal}
  {\bibinfo  {journal} {JHEP}\ }\textbf {\bibinfo {volume} {05}},\ \bibinfo
  {pages} {011} (\bibinfo {year} {2018})},\ \Eprint
  {http://arxiv.org/abs/1711.03988} {arXiv:1711.03988 [hep-ph]} \BibitemShut
  {NoStop}%
\bibitem [{\citenamefont {Bordin}\ \emph {et~al.}(2018)\citenamefont {Bordin},
  \citenamefont {Creminelli}, \citenamefont {Khmelnitsky},\ and\ \citenamefont
  {Senatore}}]{Bordin:2018pca}%
  \BibitemOpen
  \bibfield  {author} {\bibinfo {author} {\bibfnamefont {L.}~\bibnamefont
  {Bordin}}, \bibinfo {author} {\bibfnamefont {P.}~\bibnamefont {Creminelli}},
  \bibinfo {author} {\bibfnamefont {A.}~\bibnamefont {Khmelnitsky}}, \ and\
  \bibinfo {author} {\bibfnamefont {L.}~\bibnamefont {Senatore}},\ }\href
  {\doibase 10.1088/1475-7516/2018/10/013} {\bibfield  {journal} {\bibinfo
  {journal} {JCAP}\ }\textbf {\bibinfo {volume} {10}},\ \bibinfo {pages} {013}
  (\bibinfo {year} {2018})},\ \Eprint {http://arxiv.org/abs/1806.10587}
  {arXiv:1806.10587 [hep-th]} \BibitemShut {NoStop}%
\bibitem [{\citenamefont {Kumar}\ and\ \citenamefont
  {Sundrum}(2019)}]{Kumar:2018jxz}%
  \BibitemOpen
  \bibfield  {author} {\bibinfo {author} {\bibfnamefont {S.}~\bibnamefont
  {Kumar}}\ and\ \bibinfo {author} {\bibfnamefont {R.}~\bibnamefont
  {Sundrum}},\ }\href {\doibase 10.1007/JHEP04(2019)120} {\bibfield  {journal}
  {\bibinfo  {journal} {JHEP}\ }\textbf {\bibinfo {volume} {04}},\ \bibinfo
  {pages} {120} (\bibinfo {year} {2019})},\ \Eprint
  {http://arxiv.org/abs/1811.11200} {arXiv:1811.11200 [hep-ph]} \BibitemShut
  {NoStop}%
\bibitem [{\citenamefont {Goon}\ \emph {et~al.}(2019)\citenamefont {Goon},
  \citenamefont {Hinterbichler}, \citenamefont {Joyce},\ and\ \citenamefont
  {Trodden}}]{Goon:2018fyu}%
  \BibitemOpen
  \bibfield  {author} {\bibinfo {author} {\bibfnamefont {G.}~\bibnamefont
  {Goon}}, \bibinfo {author} {\bibfnamefont {K.}~\bibnamefont {Hinterbichler}},
  \bibinfo {author} {\bibfnamefont {A.}~\bibnamefont {Joyce}}, \ and\ \bibinfo
  {author} {\bibfnamefont {M.}~\bibnamefont {Trodden}},\ }\href {\doibase
  10.1007/JHEP10(2019)182} {\bibfield  {journal} {\bibinfo  {journal} {JHEP}\
  }\textbf {\bibinfo {volume} {10}},\ \bibinfo {pages} {182} (\bibinfo {year}
  {2019})},\ \Eprint {http://arxiv.org/abs/1812.07571} {arXiv:1812.07571
  [hep-th]} \BibitemShut {NoStop}%
\bibitem [{\citenamefont {Hook}\ \emph {et~al.}(2020)\citenamefont {Hook},
  \citenamefont {Huang},\ and\ \citenamefont {Racco}}]{Hook:2019zxa}%
  \BibitemOpen
  \bibfield  {author} {\bibinfo {author} {\bibfnamefont {A.}~\bibnamefont
  {Hook}}, \bibinfo {author} {\bibfnamefont {J.}~\bibnamefont {Huang}}, \ and\
  \bibinfo {author} {\bibfnamefont {D.}~\bibnamefont {Racco}},\ }\href
  {\doibase 10.1007/JHEP01(2020)105} {\bibfield  {journal} {\bibinfo  {journal}
  {JHEP}\ }\textbf {\bibinfo {volume} {01}},\ \bibinfo {pages} {105} (\bibinfo
  {year} {2020})},\ \Eprint {http://arxiv.org/abs/1907.10624} {arXiv:1907.10624
  [hep-ph]} \BibitemShut {NoStop}%
\bibitem [{\citenamefont {Kumar}\ and\ \citenamefont
  {Sundrum}(2020)}]{Kumar:2019ebj}%
  \BibitemOpen
  \bibfield  {author} {\bibinfo {author} {\bibfnamefont {S.}~\bibnamefont
  {Kumar}}\ and\ \bibinfo {author} {\bibfnamefont {R.}~\bibnamefont
  {Sundrum}},\ }\href {\doibase 10.1007/JHEP04(2020)077} {\bibfield  {journal}
  {\bibinfo  {journal} {JHEP}\ }\textbf {\bibinfo {volume} {04}},\ \bibinfo
  {pages} {077} (\bibinfo {year} {2020})},\ \Eprint
  {http://arxiv.org/abs/1908.11378} {arXiv:1908.11378 [hep-ph]} \BibitemShut
  {NoStop}%
\bibitem [{\citenamefont {Liu}\ \emph {et~al.}(2020)\citenamefont {Liu},
  \citenamefont {Tong}, \citenamefont {Wang},\ and\ \citenamefont
  {Xianyu}}]{Liu:2019fag}%
  \BibitemOpen
  \bibfield  {author} {\bibinfo {author} {\bibfnamefont {T.}~\bibnamefont
  {Liu}}, \bibinfo {author} {\bibfnamefont {X.}~\bibnamefont {Tong}}, \bibinfo
  {author} {\bibfnamefont {Y.}~\bibnamefont {Wang}}, \ and\ \bibinfo {author}
  {\bibfnamefont {Z.-Z.}\ \bibnamefont {Xianyu}},\ }\href {\doibase
  10.1007/JHEP04(2020)189} {\bibfield  {journal} {\bibinfo  {journal} {JHEP}\
  }\textbf {\bibinfo {volume} {04}},\ \bibinfo {pages} {189} (\bibinfo {year}
  {2020})},\ \Eprint {http://arxiv.org/abs/1909.01819} {arXiv:1909.01819
  [hep-ph]} \BibitemShut {NoStop}%
\bibitem [{\citenamefont {Wang}\ and\ \citenamefont
  {Xianyu}(2020)}]{Wang:2019gbi}%
  \BibitemOpen
  \bibfield  {author} {\bibinfo {author} {\bibfnamefont {L.-T.}\ \bibnamefont
  {Wang}}\ and\ \bibinfo {author} {\bibfnamefont {Z.-Z.}\ \bibnamefont
  {Xianyu}},\ }\href {\doibase 10.1007/JHEP02(2020)044} {\bibfield  {journal}
  {\bibinfo  {journal} {JHEP}\ }\textbf {\bibinfo {volume} {02}},\ \bibinfo
  {pages} {044} (\bibinfo {year} {2020})},\ \Eprint
  {http://arxiv.org/abs/1910.12876} {arXiv:1910.12876 [hep-ph]} \BibitemShut
  {NoStop}%
\bibitem [{\citenamefont {Bodas}\ \emph {et~al.}(2021)\citenamefont {Bodas},
  \citenamefont {Kumar},\ and\ \citenamefont {Sundrum}}]{Bodas:2020yho}%
  \BibitemOpen
  \bibfield  {author} {\bibinfo {author} {\bibfnamefont {A.}~\bibnamefont
  {Bodas}}, \bibinfo {author} {\bibfnamefont {S.}~\bibnamefont {Kumar}}, \ and\
  \bibinfo {author} {\bibfnamefont {R.}~\bibnamefont {Sundrum}},\ }\href
  {\doibase 10.1007/JHEP02(2021)079} {\bibfield  {journal} {\bibinfo  {journal}
  {JHEP}\ }\textbf {\bibinfo {volume} {02}},\ \bibinfo {pages} {079} (\bibinfo
  {year} {2021})},\ \Eprint {http://arxiv.org/abs/2010.04727} {arXiv:2010.04727
  [hep-ph]} \BibitemShut {NoStop}%
\bibitem [{\citenamefont {Aoki}\ and\ \citenamefont
  {Yamaguchi}(2021)}]{Aoki:2020zbj}%
  \BibitemOpen
  \bibfield  {author} {\bibinfo {author} {\bibfnamefont {S.}~\bibnamefont
  {Aoki}}\ and\ \bibinfo {author} {\bibfnamefont {M.}~\bibnamefont
  {Yamaguchi}},\ }\href {\doibase 10.1007/JHEP04(2021)127} {\bibfield
  {journal} {\bibinfo  {journal} {JHEP}\ }\textbf {\bibinfo {volume} {04}},\
  \bibinfo {pages} {127} (\bibinfo {year} {2021})},\ \Eprint
  {http://arxiv.org/abs/2012.13667} {arXiv:2012.13667 [hep-th]} \BibitemShut
  {NoStop}%
\bibitem [{\citenamefont {Lu}\ \emph {et~al.}(2021)\citenamefont {Lu},
  \citenamefont {Reece},\ and\ \citenamefont {Xianyu}}]{Lu:2021wxu}%
  \BibitemOpen
  \bibfield  {author} {\bibinfo {author} {\bibfnamefont {Q.}~\bibnamefont
  {Lu}}, \bibinfo {author} {\bibfnamefont {M.}~\bibnamefont {Reece}}, \ and\
  \bibinfo {author} {\bibfnamefont {Z.-Z.}\ \bibnamefont {Xianyu}},\ }\href
  {\doibase 10.1007/JHEP12(2021)098} {\bibfield  {journal} {\bibinfo  {journal}
  {JHEP}\ }\textbf {\bibinfo {volume} {12}},\ \bibinfo {pages} {098} (\bibinfo
  {year} {2021})},\ \Eprint {http://arxiv.org/abs/2108.11385} {arXiv:2108.11385
  [hep-ph]} \BibitemShut {NoStop}%
\bibitem [{\citenamefont {Pinol}\ \emph
  {et~al.}(2023{\natexlab{a}})\citenamefont {Pinol}, \citenamefont {Aoki},
  \citenamefont {Renaux-Petel},\ and\ \citenamefont
  {Yamaguchi}}]{Pinol:2021aun}%
  \BibitemOpen
  \bibfield  {author} {\bibinfo {author} {\bibfnamefont {L.}~\bibnamefont
  {Pinol}}, \bibinfo {author} {\bibfnamefont {S.}~\bibnamefont {Aoki}},
  \bibinfo {author} {\bibfnamefont {S.}~\bibnamefont {Renaux-Petel}}, \ and\
  \bibinfo {author} {\bibfnamefont {M.}~\bibnamefont {Yamaguchi}},\ }\href
  {\doibase 10.1103/PhysRevD.107.L021301} {\bibfield  {journal} {\bibinfo
  {journal} {Phys. Rev. D}\ }\textbf {\bibinfo {volume} {107}},\ \bibinfo
  {pages} {L021301} (\bibinfo {year} {2023}{\natexlab{a}})},\ \Eprint
  {http://arxiv.org/abs/2112.05710} {arXiv:2112.05710 [hep-th]} \BibitemShut
  {NoStop}%
\bibitem [{\citenamefont {Reece}\ \emph {et~al.}(2022)\citenamefont {Reece},
  \citenamefont {Wang},\ and\ \citenamefont {Xianyu}}]{Reece:2022soh}%
  \BibitemOpen
  \bibfield  {author} {\bibinfo {author} {\bibfnamefont {M.}~\bibnamefont
  {Reece}}, \bibinfo {author} {\bibfnamefont {L.-T.}\ \bibnamefont {Wang}}, \
  and\ \bibinfo {author} {\bibfnamefont {Z.-Z.}\ \bibnamefont {Xianyu}},\
  }\href@noop {} {\  (\bibinfo {year} {2022})},\ \Eprint
  {http://arxiv.org/abs/2204.11869} {arXiv:2204.11869 [hep-ph]} \BibitemShut
  {NoStop}%
\bibitem [{\citenamefont {Moradinezhad~Dizgah}\ and\ \citenamefont
  {Dvorkin}(2018)}]{MoradinezhadDizgah:2017szk}%
  \BibitemOpen
  \bibfield  {author} {\bibinfo {author} {\bibfnamefont {A.}~\bibnamefont
  {Moradinezhad~Dizgah}}\ and\ \bibinfo {author} {\bibfnamefont
  {C.}~\bibnamefont {Dvorkin}},\ }\href {\doibase
  10.1088/1475-7516/2018/01/010} {\bibfield  {journal} {\bibinfo  {journal}
  {JCAP}\ }\textbf {\bibinfo {volume} {01}},\ \bibinfo {pages} {010} (\bibinfo
  {year} {2018})},\ \Eprint {http://arxiv.org/abs/1708.06473} {arXiv:1708.06473
  [astro-ph.CO]} \BibitemShut {NoStop}%
\bibitem [{\citenamefont {Moradinezhad~Dizgah}\ \emph
  {et~al.}(2018)\citenamefont {Moradinezhad~Dizgah}, \citenamefont {Lee},
  \citenamefont {Mu\~noz},\ and\ \citenamefont
  {Dvorkin}}]{MoradinezhadDizgah:2018ssw}%
  \BibitemOpen
  \bibfield  {author} {\bibinfo {author} {\bibfnamefont {A.}~\bibnamefont
  {Moradinezhad~Dizgah}}, \bibinfo {author} {\bibfnamefont {H.}~\bibnamefont
  {Lee}}, \bibinfo {author} {\bibfnamefont {J.~B.}\ \bibnamefont {Mu\~noz}}, \
  and\ \bibinfo {author} {\bibfnamefont {C.}~\bibnamefont {Dvorkin}},\ }\href
  {\doibase 10.1088/1475-7516/2018/05/013} {\bibfield  {journal} {\bibinfo
  {journal} {JCAP}\ }\textbf {\bibinfo {volume} {05}},\ \bibinfo {pages} {013}
  (\bibinfo {year} {2018})},\ \Eprint {http://arxiv.org/abs/1801.07265}
  {arXiv:1801.07265 [astro-ph.CO]} \BibitemShut {NoStop}%
\bibitem [{\citenamefont {Moradinezhad~Dizgah}\ \emph
  {et~al.}(2020)\citenamefont {Moradinezhad~Dizgah}, \citenamefont {Lee},
  \citenamefont {Schmittfull},\ and\ \citenamefont
  {Dvorkin}}]{MoradinezhadDizgah:2019xun}%
  \BibitemOpen
  \bibfield  {author} {\bibinfo {author} {\bibfnamefont {A.}~\bibnamefont
  {Moradinezhad~Dizgah}}, \bibinfo {author} {\bibfnamefont {H.}~\bibnamefont
  {Lee}}, \bibinfo {author} {\bibfnamefont {M.}~\bibnamefont {Schmittfull}}, \
  and\ \bibinfo {author} {\bibfnamefont {C.}~\bibnamefont {Dvorkin}},\ }\href
  {\doibase 10.1088/1475-7516/2020/04/011} {\bibfield  {journal} {\bibinfo
  {journal} {JCAP}\ }\textbf {\bibinfo {volume} {04}},\ \bibinfo {pages} {011}
  (\bibinfo {year} {2020})},\ \Eprint {http://arxiv.org/abs/1911.05763}
  {arXiv:1911.05763 [astro-ph.CO]} \BibitemShut {NoStop}%
\bibitem [{\citenamefont {Arkani-Hamed}\ \emph {et~al.}(2020)\citenamefont
  {Arkani-Hamed}, \citenamefont {Baumann}, \citenamefont {Lee},\ and\
  \citenamefont {Pimentel}}]{Arkani-Hamed:2018kmz}%
  \BibitemOpen
  \bibfield  {author} {\bibinfo {author} {\bibfnamefont {N.}~\bibnamefont
  {Arkani-Hamed}}, \bibinfo {author} {\bibfnamefont {D.}~\bibnamefont
  {Baumann}}, \bibinfo {author} {\bibfnamefont {H.}~\bibnamefont {Lee}}, \ and\
  \bibinfo {author} {\bibfnamefont {G.~L.}\ \bibnamefont {Pimentel}},\ }\href
  {\doibase 10.1007/JHEP04(2020)105} {\bibfield  {journal} {\bibinfo  {journal}
  {JHEP}\ }\textbf {\bibinfo {volume} {04}},\ \bibinfo {pages} {105} (\bibinfo
  {year} {2020})},\ \Eprint {http://arxiv.org/abs/1811.00024} {arXiv:1811.00024
  [hep-th]} \BibitemShut {NoStop}%
\bibitem [{\citenamefont {Goodhew}\ \emph {et~al.}(2021)\citenamefont
  {Goodhew}, \citenamefont {Jazayeri},\ and\ \citenamefont
  {Pajer}}]{Goodhew:2020hob}%
  \BibitemOpen
  \bibfield  {author} {\bibinfo {author} {\bibfnamefont {H.}~\bibnamefont
  {Goodhew}}, \bibinfo {author} {\bibfnamefont {S.}~\bibnamefont {Jazayeri}}, \
  and\ \bibinfo {author} {\bibfnamefont {E.}~\bibnamefont {Pajer}},\ }\href
  {\doibase 10.1088/1475-7516/2021/04/021} {\bibfield  {journal} {\bibinfo
  {journal} {JCAP}\ }\textbf {\bibinfo {volume} {04}},\ \bibinfo {pages} {021}
  (\bibinfo {year} {2021})},\ \Eprint {http://arxiv.org/abs/2009.02898}
  {arXiv:2009.02898 [hep-th]} \BibitemShut {NoStop}%
\bibitem [{\citenamefont {Pajer}(2021)}]{Pajer:2020wxk}%
  \BibitemOpen
  \bibfield  {author} {\bibinfo {author} {\bibfnamefont {E.}~\bibnamefont
  {Pajer}},\ }\href {\doibase 10.1088/1475-7516/2021/01/023} {\bibfield
  {journal} {\bibinfo  {journal} {JCAP}\ }\textbf {\bibinfo {volume} {01}},\
  \bibinfo {pages} {023} (\bibinfo {year} {2021})},\ \Eprint
  {http://arxiv.org/abs/2010.12818} {arXiv:2010.12818 [hep-th]} \BibitemShut
  {NoStop}%
\bibitem [{\citenamefont {Cabass}\ \emph
  {et~al.}(2022{\natexlab{a}})\citenamefont {Cabass}, \citenamefont {Pajer},
  \citenamefont {Stefanyszyn},\ and\ \citenamefont {Supel}}]{Cabass:2021fnw}%
  \BibitemOpen
  \bibfield  {author} {\bibinfo {author} {\bibfnamefont {G.}~\bibnamefont
  {Cabass}}, \bibinfo {author} {\bibfnamefont {E.}~\bibnamefont {Pajer}},
  \bibinfo {author} {\bibfnamefont {D.}~\bibnamefont {Stefanyszyn}}, \ and\
  \bibinfo {author} {\bibfnamefont {J.}~\bibnamefont {Supel}},\ }\href
  {\doibase 10.1007/JHEP05(2022)077} {\bibfield  {journal} {\bibinfo  {journal}
  {JHEP}\ }\textbf {\bibinfo {volume} {05}},\ \bibinfo {pages} {077} (\bibinfo
  {year} {2022}{\natexlab{a}})},\ \Eprint {http://arxiv.org/abs/2109.10189}
  {arXiv:2109.10189 [hep-th]} \BibitemShut {NoStop}%
\bibitem [{\citenamefont {Melville}\ and\ \citenamefont
  {Pajer}(2021)}]{Melville:2021lst}%
  \BibitemOpen
  \bibfield  {author} {\bibinfo {author} {\bibfnamefont {S.}~\bibnamefont
  {Melville}}\ and\ \bibinfo {author} {\bibfnamefont {E.}~\bibnamefont
  {Pajer}},\ }\href {\doibase 10.1007/JHEP05(2021)249} {\bibfield  {journal}
  {\bibinfo  {journal} {JHEP}\ }\textbf {\bibinfo {volume} {05}},\ \bibinfo
  {pages} {249} (\bibinfo {year} {2021})},\ \Eprint
  {http://arxiv.org/abs/2103.09832} {arXiv:2103.09832 [hep-th]} \BibitemShut
  {NoStop}%
\bibitem [{\citenamefont {Jazayeri}\ \emph {et~al.}(2021)\citenamefont
  {Jazayeri}, \citenamefont {Pajer},\ and\ \citenamefont
  {Stefanyszyn}}]{Jazayeri:2021fvk}%
  \BibitemOpen
  \bibfield  {author} {\bibinfo {author} {\bibfnamefont {S.}~\bibnamefont
  {Jazayeri}}, \bibinfo {author} {\bibfnamefont {E.}~\bibnamefont {Pajer}}, \
  and\ \bibinfo {author} {\bibfnamefont {D.}~\bibnamefont {Stefanyszyn}},\
  }\href {\doibase 10.1007/JHEP10(2021)065} {\bibfield  {journal} {\bibinfo
  {journal} {JHEP}\ }\textbf {\bibinfo {volume} {10}},\ \bibinfo {pages} {065}
  (\bibinfo {year} {2021})},\ \Eprint {http://arxiv.org/abs/2103.08649}
  {arXiv:2103.08649 [hep-th]} \BibitemShut {NoStop}%
\bibitem [{\citenamefont {Baumann}\ \emph
  {et~al.}(2022{\natexlab{a}})\citenamefont {Baumann}, \citenamefont {Chen},
  \citenamefont {Duaso~Pueyo}, \citenamefont {Joyce}, \citenamefont {Lee},\
  and\ \citenamefont {Pimentel}}]{Baumann:2021fxj}%
  \BibitemOpen
  \bibfield  {author} {\bibinfo {author} {\bibfnamefont {D.}~\bibnamefont
  {Baumann}}, \bibinfo {author} {\bibfnamefont {W.-M.}\ \bibnamefont {Chen}},
  \bibinfo {author} {\bibfnamefont {C.}~\bibnamefont {Duaso~Pueyo}}, \bibinfo
  {author} {\bibfnamefont {A.}~\bibnamefont {Joyce}}, \bibinfo {author}
  {\bibfnamefont {H.}~\bibnamefont {Lee}}, \ and\ \bibinfo {author}
  {\bibfnamefont {G.~L.}\ \bibnamefont {Pimentel}},\ }\href {\doibase
  10.1007/JHEP09(2022)010} {\bibfield  {journal} {\bibinfo  {journal} {JHEP}\
  }\textbf {\bibinfo {volume} {09}},\ \bibinfo {pages} {010} (\bibinfo {year}
  {2022}{\natexlab{a}})},\ \Eprint {http://arxiv.org/abs/2106.05294}
  {arXiv:2106.05294 [hep-th]} \BibitemShut {NoStop}%
\bibitem [{\citenamefont {Bonifacio}\ \emph {et~al.}(2021)\citenamefont
  {Bonifacio}, \citenamefont {Pajer},\ and\ \citenamefont
  {Wang}}]{Bonifacio:2021azc}%
  \BibitemOpen
  \bibfield  {author} {\bibinfo {author} {\bibfnamefont {J.}~\bibnamefont
  {Bonifacio}}, \bibinfo {author} {\bibfnamefont {E.}~\bibnamefont {Pajer}}, \
  and\ \bibinfo {author} {\bibfnamefont {D.-G.}\ \bibnamefont {Wang}},\ }\href
  {\doibase 10.1007/JHEP10(2021)001} {\bibfield  {journal} {\bibinfo  {journal}
  {JHEP}\ }\textbf {\bibinfo {volume} {10}},\ \bibinfo {pages} {001} (\bibinfo
  {year} {2021})},\ \Eprint {http://arxiv.org/abs/2106.15468} {arXiv:2106.15468
  [hep-th]} \BibitemShut {NoStop}%
\bibitem [{\citenamefont {Hogervorst}\ \emph {et~al.}(2023)\citenamefont
  {Hogervorst}, \citenamefont {Penedones},\ and\ \citenamefont
  {Vaziri}}]{Hogervorst:2021uvp}%
  \BibitemOpen
  \bibfield  {author} {\bibinfo {author} {\bibfnamefont {M.}~\bibnamefont
  {Hogervorst}}, \bibinfo {author} {\bibfnamefont {J.~a.}\ \bibnamefont
  {Penedones}}, \ and\ \bibinfo {author} {\bibfnamefont {K.~S.}\ \bibnamefont
  {Vaziri}},\ }\href {\doibase 10.1007/JHEP02(2023)162} {\bibfield  {journal}
  {\bibinfo  {journal} {JHEP}\ }\textbf {\bibinfo {volume} {02}},\ \bibinfo
  {pages} {162} (\bibinfo {year} {2023})},\ \Eprint
  {http://arxiv.org/abs/2107.13871} {arXiv:2107.13871 [hep-th]} \BibitemShut
  {NoStop}%
\bibitem [{\citenamefont {Di~Pietro}\ \emph {et~al.}(2022)\citenamefont
  {Di~Pietro}, \citenamefont {Gorbenko},\ and\ \citenamefont
  {Komatsu}}]{DiPietro:2021sjt}%
  \BibitemOpen
  \bibfield  {author} {\bibinfo {author} {\bibfnamefont {L.}~\bibnamefont
  {Di~Pietro}}, \bibinfo {author} {\bibfnamefont {V.}~\bibnamefont {Gorbenko}},
  \ and\ \bibinfo {author} {\bibfnamefont {S.}~\bibnamefont {Komatsu}},\ }\href
  {\doibase 10.1007/JHEP03(2022)023} {\bibfield  {journal} {\bibinfo  {journal}
  {JHEP}\ }\textbf {\bibinfo {volume} {03}},\ \bibinfo {pages} {023} (\bibinfo
  {year} {2022})},\ \Eprint {http://arxiv.org/abs/2108.01695} {arXiv:2108.01695
  [hep-th]} \BibitemShut {NoStop}%
\bibitem [{\citenamefont {Baumann}\ \emph
  {et~al.}(2022{\natexlab{b}})\citenamefont {Baumann}, \citenamefont {Green},
  \citenamefont {Joyce}, \citenamefont {Pajer}, \citenamefont {Pimentel},
  \citenamefont {Sleight},\ and\ \citenamefont {Taronna}}]{Baumann:2022jpr}%
  \BibitemOpen
  \bibfield  {author} {\bibinfo {author} {\bibfnamefont {D.}~\bibnamefont
  {Baumann}}, \bibinfo {author} {\bibfnamefont {D.}~\bibnamefont {Green}},
  \bibinfo {author} {\bibfnamefont {A.}~\bibnamefont {Joyce}}, \bibinfo
  {author} {\bibfnamefont {E.}~\bibnamefont {Pajer}}, \bibinfo {author}
  {\bibfnamefont {G.~L.}\ \bibnamefont {Pimentel}}, \bibinfo {author}
  {\bibfnamefont {C.}~\bibnamefont {Sleight}}, \ and\ \bibinfo {author}
  {\bibfnamefont {M.}~\bibnamefont {Taronna}},\ }in\ \href@noop {} {\emph
  {\bibinfo {booktitle} {{Snowmass 2021}}}}\ (\bibinfo {year} {2022})\ \Eprint
  {http://arxiv.org/abs/2203.08121} {arXiv:2203.08121 [hep-th]} \BibitemShut
  {NoStop}%
\bibitem [{\citenamefont {Cabass}\ \emph
  {et~al.}(2023{\natexlab{b}})\citenamefont {Cabass}, \citenamefont {Jazayeri},
  \citenamefont {Pajer},\ and\ \citenamefont {Stefanyszyn}}]{Cabass:2022rhr}%
  \BibitemOpen
  \bibfield  {author} {\bibinfo {author} {\bibfnamefont {G.}~\bibnamefont
  {Cabass}}, \bibinfo {author} {\bibfnamefont {S.}~\bibnamefont {Jazayeri}},
  \bibinfo {author} {\bibfnamefont {E.}~\bibnamefont {Pajer}}, \ and\ \bibinfo
  {author} {\bibfnamefont {D.}~\bibnamefont {Stefanyszyn}},\ }\href {\doibase
  10.1007/JHEP02(2023)021} {\bibfield  {journal} {\bibinfo  {journal} {JHEP}\
  }\textbf {\bibinfo {volume} {02}},\ \bibinfo {pages} {021} (\bibinfo {year}
  {2023}{\natexlab{b}})},\ \Eprint {http://arxiv.org/abs/2210.02907}
  {arXiv:2210.02907 [hep-th]} \BibitemShut {NoStop}%
\bibitem [{\citenamefont {Baumann}\ \emph {et~al.}(2020)\citenamefont
  {Baumann}, \citenamefont {Duaso~Pueyo}, \citenamefont {Joyce}, \citenamefont
  {Lee},\ and\ \citenamefont {Pimentel}}]{Baumann:2019oyu}%
  \BibitemOpen
  \bibfield  {author} {\bibinfo {author} {\bibfnamefont {D.}~\bibnamefont
  {Baumann}}, \bibinfo {author} {\bibfnamefont {C.}~\bibnamefont
  {Duaso~Pueyo}}, \bibinfo {author} {\bibfnamefont {A.}~\bibnamefont {Joyce}},
  \bibinfo {author} {\bibfnamefont {H.}~\bibnamefont {Lee}}, \ and\ \bibinfo
  {author} {\bibfnamefont {G.~L.}\ \bibnamefont {Pimentel}},\ }\href {\doibase
  10.1007/JHEP12(2020)204} {\bibfield  {journal} {\bibinfo  {journal} {JHEP}\
  }\textbf {\bibinfo {volume} {12}},\ \bibinfo {pages} {204} (\bibinfo {year}
  {2020})},\ \Eprint {http://arxiv.org/abs/1910.14051} {arXiv:1910.14051
  [hep-th]} \BibitemShut {NoStop}%
\bibitem [{\citenamefont {Baumann}\ \emph {et~al.}(2021)\citenamefont
  {Baumann}, \citenamefont {Duaso~Pueyo}, \citenamefont {Joyce}, \citenamefont
  {Lee},\ and\ \citenamefont {Pimentel}}]{Baumann:2020dch}%
  \BibitemOpen
  \bibfield  {author} {\bibinfo {author} {\bibfnamefont {D.}~\bibnamefont
  {Baumann}}, \bibinfo {author} {\bibfnamefont {C.}~\bibnamefont
  {Duaso~Pueyo}}, \bibinfo {author} {\bibfnamefont {A.}~\bibnamefont {Joyce}},
  \bibinfo {author} {\bibfnamefont {H.}~\bibnamefont {Lee}}, \ and\ \bibinfo
  {author} {\bibfnamefont {G.~L.}\ \bibnamefont {Pimentel}},\ }\href {\doibase
  10.21468/SciPostPhys.11.3.071} {\bibfield  {journal} {\bibinfo  {journal}
  {SciPost Phys.}\ }\textbf {\bibinfo {volume} {11}},\ \bibinfo {pages} {071}
  (\bibinfo {year} {2021})},\ \Eprint {http://arxiv.org/abs/2005.04234}
  {arXiv:2005.04234 [hep-th]} \BibitemShut {NoStop}%
\bibitem [{\citenamefont {Pimentel}\ and\ \citenamefont
  {Wang}(2022)}]{Pimentel:2022fsc}%
  \BibitemOpen
  \bibfield  {author} {\bibinfo {author} {\bibfnamefont {G.~L.}\ \bibnamefont
  {Pimentel}}\ and\ \bibinfo {author} {\bibfnamefont {D.-G.}\ \bibnamefont
  {Wang}},\ }\href {\doibase 10.1007/JHEP10(2022)177} {\bibfield  {journal}
  {\bibinfo  {journal} {JHEP}\ }\textbf {\bibinfo {volume} {10}},\ \bibinfo
  {pages} {177} (\bibinfo {year} {2022})},\ \Eprint
  {http://arxiv.org/abs/2205.00013} {arXiv:2205.00013 [hep-th]} \BibitemShut
  {NoStop}%
\bibitem [{\citenamefont {Jazayeri}\ and\ \citenamefont
  {Renaux-Petel}(2022)}]{Jazayeri:2022kjy}%
  \BibitemOpen
  \bibfield  {author} {\bibinfo {author} {\bibfnamefont {S.}~\bibnamefont
  {Jazayeri}}\ and\ \bibinfo {author} {\bibfnamefont {S.}~\bibnamefont
  {Renaux-Petel}},\ }\href {\doibase 10.1007/JHEP12(2022)137} {\bibfield
  {journal} {\bibinfo  {journal} {JHEP}\ }\textbf {\bibinfo {volume} {12}},\
  \bibinfo {pages} {137} (\bibinfo {year} {2022})},\ \Eprint
  {http://arxiv.org/abs/2205.10340} {arXiv:2205.10340 [hep-th]} \BibitemShut
  {NoStop}%
\bibitem [{\citenamefont {Wang}\ \emph {et~al.}(2022)\citenamefont {Wang},
  \citenamefont {Pimentel},\ and\ \citenamefont {Ach\'ucarro}}]{Wang:2022eop}%
  \BibitemOpen
  \bibfield  {author} {\bibinfo {author} {\bibfnamefont {D.-G.}\ \bibnamefont
  {Wang}}, \bibinfo {author} {\bibfnamefont {G.~L.}\ \bibnamefont {Pimentel}},
  \ and\ \bibinfo {author} {\bibfnamefont {A.}~\bibnamefont {Ach\'ucarro}},\
  }\href@noop {} {\  (\bibinfo {year} {2022})},\ \Eprint
  {http://arxiv.org/abs/2212.14035} {arXiv:2212.14035 [astro-ph.CO]}
  \BibitemShut {NoStop}%
\bibitem [{\citenamefont {Akrami}\ \emph {et~al.}(2020)\citenamefont {Akrami}
  \emph {et~al.}}]{Planck:2019kim}%
  \BibitemOpen
  \bibfield  {author} {\bibinfo {author} {\bibfnamefont {Y.}~\bibnamefont
  {Akrami}} \emph {et~al.} (\bibinfo {collaboration} {Planck}),\ }\href
  {\doibase 10.1051/0004-6361/201935891} {\bibfield  {journal} {\bibinfo
  {journal} {Astron. Astrophys.}\ }\textbf {\bibinfo {volume} {641}},\ \bibinfo
  {pages} {A9} (\bibinfo {year} {2020})},\ \Eprint
  {http://arxiv.org/abs/1905.05697} {arXiv:1905.05697 [astro-ph.CO]}
  \BibitemShut {NoStop}%
\bibitem [{\citenamefont {Kalaja}\ \emph {et~al.}(2021)\citenamefont {Kalaja},
  \citenamefont {Meerburg}, \citenamefont {Pimentel},\ and\ \citenamefont
  {Coulton}}]{Kalaja:2020mkq}%
  \BibitemOpen
  \bibfield  {author} {\bibinfo {author} {\bibfnamefont {A.}~\bibnamefont
  {Kalaja}}, \bibinfo {author} {\bibfnamefont {P.~D.}\ \bibnamefont
  {Meerburg}}, \bibinfo {author} {\bibfnamefont {G.~L.}\ \bibnamefont
  {Pimentel}}, \ and\ \bibinfo {author} {\bibfnamefont {W.~R.}\ \bibnamefont
  {Coulton}},\ }\href {\doibase 10.1088/1475-7516/2021/04/050} {\bibfield
  {journal} {\bibinfo  {journal} {JCAP}\ }\textbf {\bibinfo {volume} {04}},\
  \bibinfo {pages} {050} (\bibinfo {year} {2021})},\ \Eprint
  {http://arxiv.org/abs/2011.09461} {arXiv:2011.09461 [astro-ph.CO]}
  \BibitemShut {NoStop}%
\bibitem [{\citenamefont {Chudaykin}\ and\ \citenamefont
  {Ivanov}(2019)}]{Chudaykin:2019ock}%
  \BibitemOpen
  \bibfield  {author} {\bibinfo {author} {\bibfnamefont {A.}~\bibnamefont
  {Chudaykin}}\ and\ \bibinfo {author} {\bibfnamefont {M.~M.}\ \bibnamefont
  {Ivanov}},\ }\href {\doibase 10.1088/1475-7516/2019/11/034} {\bibfield
  {journal} {\bibinfo  {journal} {JCAP}\ }\textbf {\bibinfo {volume} {11}},\
  \bibinfo {pages} {034} (\bibinfo {year} {2019})},\ \Eprint
  {http://arxiv.org/abs/1907.06666} {arXiv:1907.06666 [astro-ph.CO]}
  \BibitemShut {NoStop}%
\bibitem [{\citenamefont {Sailer}\ \emph {et~al.}(2021)\citenamefont {Sailer},
  \citenamefont {Castorina}, \citenamefont {Ferraro},\ and\ \citenamefont
  {White}}]{Sailer:2021yzm}%
  \BibitemOpen
  \bibfield  {author} {\bibinfo {author} {\bibfnamefont {N.}~\bibnamefont
  {Sailer}}, \bibinfo {author} {\bibfnamefont {E.}~\bibnamefont {Castorina}},
  \bibinfo {author} {\bibfnamefont {S.}~\bibnamefont {Ferraro}}, \ and\
  \bibinfo {author} {\bibfnamefont {M.}~\bibnamefont {White}},\ }\href@noop {}
  {\  (\bibinfo {year} {2021})},\ \Eprint {http://arxiv.org/abs/2106.09713}
  {arXiv:2106.09713 [astro-ph.CO]} \BibitemShut {NoStop}%
\bibitem [{\citenamefont {Moradinezhad~Dizgah}\ \emph
  {et~al.}(2021)\citenamefont {Moradinezhad~Dizgah}, \citenamefont {Biagetti},
  \citenamefont {Sefusatti}, \citenamefont {Desjacques},\ and\ \citenamefont
  {Nore\~na}}]{MoradinezhadDizgah:2020whw}%
  \BibitemOpen
  \bibfield  {author} {\bibinfo {author} {\bibfnamefont {A.}~\bibnamefont
  {Moradinezhad~Dizgah}}, \bibinfo {author} {\bibfnamefont {M.}~\bibnamefont
  {Biagetti}}, \bibinfo {author} {\bibfnamefont {E.}~\bibnamefont {Sefusatti}},
  \bibinfo {author} {\bibfnamefont {V.}~\bibnamefont {Desjacques}}, \ and\
  \bibinfo {author} {\bibfnamefont {J.}~\bibnamefont {Nore\~na}},\ }\href
  {\doibase 10.1088/1475-7516/2021/05/015} {\bibfield  {journal} {\bibinfo
  {journal} {JCAP}\ }\textbf {\bibinfo {volume} {05}},\ \bibinfo {pages} {015}
  (\bibinfo {year} {2021})},\ \Eprint {http://arxiv.org/abs/2010.14523}
  {arXiv:2010.14523 [astro-ph.CO]} \BibitemShut {NoStop}%
\bibitem [{\citenamefont {Alvarez}\ \emph {et~al.}(2014)\citenamefont {Alvarez}
  \emph {et~al.}}]{Alvarez:2014vva}%
  \BibitemOpen
  \bibfield  {author} {\bibinfo {author} {\bibfnamefont {M.}~\bibnamefont
  {Alvarez}} \emph {et~al.},\ }\href@noop {} {\  (\bibinfo {year} {2014})},\
  \Eprint {http://arxiv.org/abs/1412.4671} {arXiv:1412.4671 [astro-ph.CO]}
  \BibitemShut {NoStop}%
\bibitem [{\citenamefont {Meerburg}\ \emph {et~al.}(2019)\citenamefont
  {Meerburg} \emph {et~al.}}]{Meerburg:2019qqi}%
  \BibitemOpen
  \bibfield  {author} {\bibinfo {author} {\bibfnamefont {P.~D.}\ \bibnamefont
  {Meerburg}} \emph {et~al.},\ }\href@noop {} {\bibfield  {journal} {\bibinfo
  {journal} {Bull. Am. Astron. Soc.}\ }\textbf {\bibinfo {volume} {51}},\
  \bibinfo {pages} {107} (\bibinfo {year} {2019})},\ \Eprint
  {http://arxiv.org/abs/1903.04409} {arXiv:1903.04409 [astro-ph.CO]}
  \BibitemShut {NoStop}%
\bibitem [{\citenamefont {Biagetti}(2019)}]{Biagetti:2019bnp}%
  \BibitemOpen
  \bibfield  {author} {\bibinfo {author} {\bibfnamefont {M.}~\bibnamefont
  {Biagetti}},\ }\href {\doibase 10.3390/galaxies7030071} {\bibfield  {journal}
  {\bibinfo  {journal} {Galaxies}\ }\textbf {\bibinfo {volume} {7}},\ \bibinfo
  {pages} {71} (\bibinfo {year} {2019})},\ \Eprint
  {http://arxiv.org/abs/1906.12244} {arXiv:1906.12244 [astro-ph.CO]}
  \BibitemShut {NoStop}%
\bibitem [{\citenamefont {Ach\'ucarro}\ \emph {et~al.}(2022)\citenamefont
  {Ach\'ucarro} \emph {et~al.}}]{Achucarro:2022qrl}%
  \BibitemOpen
  \bibfield  {author} {\bibinfo {author} {\bibfnamefont {A.}~\bibnamefont
  {Ach\'ucarro}} \emph {et~al.},\ }\href@noop {} {\  (\bibinfo {year}
  {2022})},\ \Eprint {http://arxiv.org/abs/2203.08128} {arXiv:2203.08128
  [astro-ph.CO]} \BibitemShut {NoStop}%
\bibitem [{\citenamefont {Green}\ \emph {et~al.}(2024)\citenamefont {Green},
  \citenamefont {Guo}, \citenamefont {Han},\ and\ \citenamefont
  {Wallisch}}]{Green:2023uyz}%
  \BibitemOpen
  \bibfield  {author} {\bibinfo {author} {\bibfnamefont {D.}~\bibnamefont
  {Green}}, \bibinfo {author} {\bibfnamefont {Y.}~\bibnamefont {Guo}}, \bibinfo
  {author} {\bibfnamefont {J.}~\bibnamefont {Han}}, \ and\ \bibinfo {author}
  {\bibfnamefont {B.}~\bibnamefont {Wallisch}},\ }\href {\doibase
  10.1088/1475-7516/2024/05/090} {\bibfield  {journal} {\bibinfo  {journal}
  {JCAP}\ }\textbf {\bibinfo {volume} {05}},\ \bibinfo {pages} {090} (\bibinfo
  {year} {2024})},\ \Eprint {http://arxiv.org/abs/2311.04882} {arXiv:2311.04882
  [astro-ph.CO]} \BibitemShut {NoStop}%
\bibitem [{\citenamefont {Chen}\ \emph {et~al.}(2024)\citenamefont {Chen},
  \citenamefont {Chakraborty},\ and\ \citenamefont {Dvorkin}}]{Chen:2024bdg}%
  \BibitemOpen
  \bibfield  {author} {\bibinfo {author} {\bibfnamefont {S.-F.}\ \bibnamefont
  {Chen}}, \bibinfo {author} {\bibfnamefont {P.}~\bibnamefont {Chakraborty}}, \
  and\ \bibinfo {author} {\bibfnamefont {C.}~\bibnamefont {Dvorkin}},\ }\href
  {\doibase 10.1088/1475-7516/2024/05/011} {\bibfield  {journal} {\bibinfo
  {journal} {JCAP}\ }\textbf {\bibinfo {volume} {05}},\ \bibinfo {pages} {011}
  (\bibinfo {year} {2024})},\ \Eprint {http://arxiv.org/abs/2401.13036}
  {arXiv:2401.13036 [astro-ph.CO]} \BibitemShut {NoStop}%
\bibitem [{\citenamefont {Baumann}\ \emph {et~al.}(2012)\citenamefont
  {Baumann}, \citenamefont {Nicolis}, \citenamefont {Senatore},\ and\
  \citenamefont {Zaldarriaga}}]{Baumann:2010tm}%
  \BibitemOpen
  \bibfield  {author} {\bibinfo {author} {\bibfnamefont {D.}~\bibnamefont
  {Baumann}}, \bibinfo {author} {\bibfnamefont {A.}~\bibnamefont {Nicolis}},
  \bibinfo {author} {\bibfnamefont {L.}~\bibnamefont {Senatore}}, \ and\
  \bibinfo {author} {\bibfnamefont {M.}~\bibnamefont {Zaldarriaga}},\ }\href
  {\doibase 10.1088/1475-7516/2012/07/051} {\bibfield  {journal} {\bibinfo
  {journal} {JCAP}\ }\textbf {\bibinfo {volume} {1207}},\ \bibinfo {pages}
  {051} (\bibinfo {year} {2012})},\ \Eprint {http://arxiv.org/abs/1004.2488}
  {arXiv:1004.2488 [astro-ph.CO]} \BibitemShut {NoStop}%
\bibitem [{\citenamefont {Carrasco}\ \emph {et~al.}(2012)\citenamefont
  {Carrasco}, \citenamefont {Hertzberg},\ and\ \citenamefont
  {Senatore}}]{Carrasco:2012cv}%
  \BibitemOpen
  \bibfield  {author} {\bibinfo {author} {\bibfnamefont {J.~J.~M.}\
  \bibnamefont {Carrasco}}, \bibinfo {author} {\bibfnamefont {M.~P.}\
  \bibnamefont {Hertzberg}}, \ and\ \bibinfo {author} {\bibfnamefont
  {L.}~\bibnamefont {Senatore}},\ }\href {\doibase 10.1007/JHEP09(2012)082}
  {\bibfield  {journal} {\bibinfo  {journal} {JHEP}\ }\textbf {\bibinfo
  {volume} {09}},\ \bibinfo {pages} {082} (\bibinfo {year} {2012})},\ \Eprint
  {http://arxiv.org/abs/1206.2926} {arXiv:1206.2926 [astro-ph.CO]} \BibitemShut
  {NoStop}%
\bibitem [{\citenamefont {Porto}\ \emph {et~al.}(2014)\citenamefont {Porto},
  \citenamefont {Senatore},\ and\ \citenamefont {Zaldarriaga}}]{Porto:2013qua}%
  \BibitemOpen
  \bibfield  {author} {\bibinfo {author} {\bibfnamefont {R.~A.}\ \bibnamefont
  {Porto}}, \bibinfo {author} {\bibfnamefont {L.}~\bibnamefont {Senatore}}, \
  and\ \bibinfo {author} {\bibfnamefont {M.}~\bibnamefont {Zaldarriaga}},\
  }\href {\doibase 10.1088/1475-7516/2014/05/022} {\bibfield  {journal}
  {\bibinfo  {journal} {JCAP}\ }\textbf {\bibinfo {volume} {1405}},\ \bibinfo
  {pages} {022} (\bibinfo {year} {2014})},\ \Eprint
  {http://arxiv.org/abs/1311.2168} {arXiv:1311.2168 [astro-ph.CO]} \BibitemShut
  {NoStop}%
\bibitem [{\citenamefont {Blas}\ \emph {et~al.}(2013)\citenamefont {Blas},
  \citenamefont {Garny},\ and\ \citenamefont {Konstandin}}]{Blas:2013bpa}%
  \BibitemOpen
  \bibfield  {author} {\bibinfo {author} {\bibfnamefont {D.}~\bibnamefont
  {Blas}}, \bibinfo {author} {\bibfnamefont {M.}~\bibnamefont {Garny}}, \ and\
  \bibinfo {author} {\bibfnamefont {T.}~\bibnamefont {Konstandin}},\ }\href
  {\doibase 10.1088/1475-7516/2013/09/024} {\bibfield  {journal} {\bibinfo
  {journal} {JCAP}\ }\textbf {\bibinfo {volume} {09}},\ \bibinfo {pages} {024}
  (\bibinfo {year} {2013})},\ \Eprint {http://arxiv.org/abs/1304.1546}
  {arXiv:1304.1546 [astro-ph.CO]} \BibitemShut {NoStop}%
\bibitem [{\citenamefont {Carrasco}\ \emph
  {et~al.}(2014{\natexlab{a}})\citenamefont {Carrasco}, \citenamefont
  {Foreman}, \citenamefont {Green},\ and\ \citenamefont
  {Senatore}}]{Carrasco:2013sva}%
  \BibitemOpen
  \bibfield  {author} {\bibinfo {author} {\bibfnamefont {J.~J.~M.}\
  \bibnamefont {Carrasco}}, \bibinfo {author} {\bibfnamefont {S.}~\bibnamefont
  {Foreman}}, \bibinfo {author} {\bibfnamefont {D.}~\bibnamefont {Green}}, \
  and\ \bibinfo {author} {\bibfnamefont {L.}~\bibnamefont {Senatore}},\ }\href
  {\doibase 10.1088/1475-7516/2014/07/056} {\bibfield  {journal} {\bibinfo
  {journal} {JCAP}\ }\textbf {\bibinfo {volume} {07}},\ \bibinfo {pages} {056}
  (\bibinfo {year} {2014}{\natexlab{a}})},\ \Eprint
  {http://arxiv.org/abs/1304.4946} {arXiv:1304.4946 [astro-ph.CO]} \BibitemShut
  {NoStop}%
\bibitem [{\citenamefont {Pajer}\ and\ \citenamefont
  {Zaldarriaga}(2013)}]{Pajer:2013jj}%
  \BibitemOpen
  \bibfield  {author} {\bibinfo {author} {\bibfnamefont {E.}~\bibnamefont
  {Pajer}}\ and\ \bibinfo {author} {\bibfnamefont {M.}~\bibnamefont
  {Zaldarriaga}},\ }\href {\doibase 10.1088/1475-7516/2013/08/037} {\bibfield
  {journal} {\bibinfo  {journal} {JCAP}\ }\textbf {\bibinfo {volume} {08}},\
  \bibinfo {pages} {037} (\bibinfo {year} {2013})},\ \Eprint
  {http://arxiv.org/abs/1301.7182} {arXiv:1301.7182 [astro-ph.CO]} \BibitemShut
  {NoStop}%
\bibitem [{\citenamefont {Mercolli}\ and\ \citenamefont
  {Pajer}(2014)}]{Mercolli:2013bsa}%
  \BibitemOpen
  \bibfield  {author} {\bibinfo {author} {\bibfnamefont {L.}~\bibnamefont
  {Mercolli}}\ and\ \bibinfo {author} {\bibfnamefont {E.}~\bibnamefont
  {Pajer}},\ }\href {\doibase 10.1088/1475-7516/2014/03/006} {\bibfield
  {journal} {\bibinfo  {journal} {JCAP}\ }\textbf {\bibinfo {volume} {03}},\
  \bibinfo {pages} {006} (\bibinfo {year} {2014})},\ \Eprint
  {http://arxiv.org/abs/1307.3220} {arXiv:1307.3220 [astro-ph.CO]} \BibitemShut
  {NoStop}%
\bibitem [{\citenamefont {Carrasco}\ \emph
  {et~al.}(2014{\natexlab{b}})\citenamefont {Carrasco}, \citenamefont
  {Foreman}, \citenamefont {Green},\ and\ \citenamefont
  {Senatore}}]{Carrasco:2013mua}%
  \BibitemOpen
  \bibfield  {author} {\bibinfo {author} {\bibfnamefont {J.~J.~M.}\
  \bibnamefont {Carrasco}}, \bibinfo {author} {\bibfnamefont {S.}~\bibnamefont
  {Foreman}}, \bibinfo {author} {\bibfnamefont {D.}~\bibnamefont {Green}}, \
  and\ \bibinfo {author} {\bibfnamefont {L.}~\bibnamefont {Senatore}},\ }\href
  {\doibase 10.1088/1475-7516/2014/07/057} {\bibfield  {journal} {\bibinfo
  {journal} {JCAP}\ }\textbf {\bibinfo {volume} {07}},\ \bibinfo {pages} {057}
  (\bibinfo {year} {2014}{\natexlab{b}})},\ \Eprint
  {http://arxiv.org/abs/1310.0464} {arXiv:1310.0464 [astro-ph.CO]} \BibitemShut
  {NoStop}%
\bibitem [{\citenamefont {Carroll}\ \emph {et~al.}(2014)\citenamefont
  {Carroll}, \citenamefont {Leichenauer},\ and\ \citenamefont
  {Pollack}}]{Carroll:2013oxa}%
  \BibitemOpen
  \bibfield  {author} {\bibinfo {author} {\bibfnamefont {S.~M.}\ \bibnamefont
  {Carroll}}, \bibinfo {author} {\bibfnamefont {S.}~\bibnamefont
  {Leichenauer}}, \ and\ \bibinfo {author} {\bibfnamefont {J.}~\bibnamefont
  {Pollack}},\ }\href {\doibase 10.1103/PhysRevD.90.023518} {\bibfield
  {journal} {\bibinfo  {journal} {Phys. Rev. D}\ }\textbf {\bibinfo {volume}
  {90}},\ \bibinfo {pages} {023518} (\bibinfo {year} {2014})},\ \Eprint
  {http://arxiv.org/abs/1310.2920} {arXiv:1310.2920 [hep-th]} \BibitemShut
  {NoStop}%
\bibitem [{\citenamefont {Baldauf}\ \emph
  {et~al.}(2015{\natexlab{a}})\citenamefont {Baldauf}, \citenamefont
  {Mercolli}, \citenamefont {Mirbabayi},\ and\ \citenamefont
  {Pajer}}]{Baldauf:2014qfa}%
  \BibitemOpen
  \bibfield  {author} {\bibinfo {author} {\bibfnamefont {T.}~\bibnamefont
  {Baldauf}}, \bibinfo {author} {\bibfnamefont {L.}~\bibnamefont {Mercolli}},
  \bibinfo {author} {\bibfnamefont {M.}~\bibnamefont {Mirbabayi}}, \ and\
  \bibinfo {author} {\bibfnamefont {E.}~\bibnamefont {Pajer}},\ }\href
  {\doibase 10.1088/1475-7516/2015/05/007} {\bibfield  {journal} {\bibinfo
  {journal} {JCAP}\ }\textbf {\bibinfo {volume} {1505}},\ \bibinfo {pages}
  {007} (\bibinfo {year} {2015}{\natexlab{a}})},\ \Eprint
  {http://arxiv.org/abs/1406.4135} {arXiv:1406.4135 [astro-ph.CO]} \BibitemShut
  {NoStop}%
\bibitem [{\citenamefont {Angulo}\ \emph
  {et~al.}(2015{\natexlab{a}})\citenamefont {Angulo}, \citenamefont {Foreman},
  \citenamefont {Schmittfull},\ and\ \citenamefont
  {Senatore}}]{Angulo:2014tfa}%
  \BibitemOpen
  \bibfield  {author} {\bibinfo {author} {\bibfnamefont {R.~E.}\ \bibnamefont
  {Angulo}}, \bibinfo {author} {\bibfnamefont {S.}~\bibnamefont {Foreman}},
  \bibinfo {author} {\bibfnamefont {M.}~\bibnamefont {Schmittfull}}, \ and\
  \bibinfo {author} {\bibfnamefont {L.}~\bibnamefont {Senatore}},\ }\href
  {\doibase 10.1088/1475-7516/2015/10/039} {\bibfield  {journal} {\bibinfo
  {journal} {JCAP}\ }\textbf {\bibinfo {volume} {1510}},\ \bibinfo {pages}
  {039} (\bibinfo {year} {2015}{\natexlab{a}})},\ \Eprint
  {http://arxiv.org/abs/1406.4143} {arXiv:1406.4143 [astro-ph.CO]} \BibitemShut
  {NoStop}%
\bibitem [{\citenamefont {Senatore}\ and\ \citenamefont
  {Zaldarriaga}(2014)}]{Senatore:2014vja}%
  \BibitemOpen
  \bibfield  {author} {\bibinfo {author} {\bibfnamefont {L.}~\bibnamefont
  {Senatore}}\ and\ \bibinfo {author} {\bibfnamefont {M.}~\bibnamefont
  {Zaldarriaga}},\ }\href@noop {} {\  (\bibinfo {year} {2014})},\ \Eprint
  {http://arxiv.org/abs/1409.1225} {arXiv:1409.1225 [astro-ph.CO]} \BibitemShut
  {NoStop}%
\bibitem [{\citenamefont {Lewandowski}\ \emph {et~al.}(2015)\citenamefont
  {Lewandowski}, \citenamefont {Perko},\ and\ \citenamefont
  {Senatore}}]{Lewandowski:2014rca}%
  \BibitemOpen
  \bibfield  {author} {\bibinfo {author} {\bibfnamefont {M.}~\bibnamefont
  {Lewandowski}}, \bibinfo {author} {\bibfnamefont {A.}~\bibnamefont {Perko}},
  \ and\ \bibinfo {author} {\bibfnamefont {L.}~\bibnamefont {Senatore}},\
  }\href {\doibase 10.1088/1475-7516/2015/05/019} {\bibfield  {journal}
  {\bibinfo  {journal} {JCAP}\ }\textbf {\bibinfo {volume} {1505}},\ \bibinfo
  {pages} {019} (\bibinfo {year} {2015})},\ \Eprint
  {http://arxiv.org/abs/1412.5049} {arXiv:1412.5049 [astro-ph.CO]} \BibitemShut
  {NoStop}%
\bibitem [{\citenamefont {Senatore}\ and\ \citenamefont
  {Zaldarriaga}(2015)}]{Senatore:2014via}%
  \BibitemOpen
  \bibfield  {author} {\bibinfo {author} {\bibfnamefont {L.}~\bibnamefont
  {Senatore}}\ and\ \bibinfo {author} {\bibfnamefont {M.}~\bibnamefont
  {Zaldarriaga}},\ }\href {\doibase 10.1088/1475-7516/2015/02/013} {\bibfield
  {journal} {\bibinfo  {journal} {JCAP}\ }\textbf {\bibinfo {volume} {1502}},\
  \bibinfo {pages} {013} (\bibinfo {year} {2015})},\ \Eprint
  {http://arxiv.org/abs/1404.5954} {arXiv:1404.5954 [astro-ph.CO]} \BibitemShut
  {NoStop}%
\bibitem [{\citenamefont {Mirbabayi}\ \emph {et~al.}(2015)\citenamefont
  {Mirbabayi}, \citenamefont {Schmidt},\ and\ \citenamefont
  {Zaldarriaga}}]{Mirbabayi:2014zca}%
  \BibitemOpen
  \bibfield  {author} {\bibinfo {author} {\bibfnamefont {M.}~\bibnamefont
  {Mirbabayi}}, \bibinfo {author} {\bibfnamefont {F.}~\bibnamefont {Schmidt}},
  \ and\ \bibinfo {author} {\bibfnamefont {M.}~\bibnamefont {Zaldarriaga}},\
  }\href {\doibase 10.1088/1475-7516/2015/07/030} {\bibfield  {journal}
  {\bibinfo  {journal} {JCAP}\ }\textbf {\bibinfo {volume} {1507}},\ \bibinfo
  {pages} {030} (\bibinfo {year} {2015})},\ \Eprint
  {http://arxiv.org/abs/1412.5169} {arXiv:1412.5169 [astro-ph.CO]} \BibitemShut
  {NoStop}%
\bibitem [{\citenamefont {Senatore}(2015)}]{Senatore:2014eva}%
  \BibitemOpen
  \bibfield  {author} {\bibinfo {author} {\bibfnamefont {L.}~\bibnamefont
  {Senatore}},\ }\href {\doibase 10.1088/1475-7516/2015/11/007} {\bibfield
  {journal} {\bibinfo  {journal} {JCAP}\ }\textbf {\bibinfo {volume} {1511}},\
  \bibinfo {pages} {007} (\bibinfo {year} {2015})},\ \Eprint
  {http://arxiv.org/abs/1406.7843} {arXiv:1406.7843 [astro-ph.CO]} \BibitemShut
  {NoStop}%
\bibitem [{\citenamefont {Assassi}\ \emph {et~al.}(2014)\citenamefont
  {Assassi}, \citenamefont {Baumann}, \citenamefont {Green},\ and\
  \citenamefont {Zaldarriaga}}]{Assassi:2014fva}%
  \BibitemOpen
  \bibfield  {author} {\bibinfo {author} {\bibfnamefont {V.}~\bibnamefont
  {Assassi}}, \bibinfo {author} {\bibfnamefont {D.}~\bibnamefont {Baumann}},
  \bibinfo {author} {\bibfnamefont {D.}~\bibnamefont {Green}}, \ and\ \bibinfo
  {author} {\bibfnamefont {M.}~\bibnamefont {Zaldarriaga}},\ }\href {\doibase
  10.1088/1475-7516/2014/08/056} {\bibfield  {journal} {\bibinfo  {journal}
  {JCAP}\ }\textbf {\bibinfo {volume} {1408}},\ \bibinfo {pages} {056}
  (\bibinfo {year} {2014})},\ \Eprint {http://arxiv.org/abs/1402.5916}
  {arXiv:1402.5916 [astro-ph.CO]} \BibitemShut {NoStop}%
\bibitem [{\citenamefont {Angulo}\ \emph
  {et~al.}(2015{\natexlab{b}})\citenamefont {Angulo}, \citenamefont {Fasiello},
  \citenamefont {Senatore},\ and\ \citenamefont {Vlah}}]{Angulo:2015eqa}%
  \BibitemOpen
  \bibfield  {author} {\bibinfo {author} {\bibfnamefont {R.}~\bibnamefont
  {Angulo}}, \bibinfo {author} {\bibfnamefont {M.}~\bibnamefont {Fasiello}},
  \bibinfo {author} {\bibfnamefont {L.}~\bibnamefont {Senatore}}, \ and\
  \bibinfo {author} {\bibfnamefont {Z.}~\bibnamefont {Vlah}},\ }\href {\doibase
  10.1088/1475-7516/2015/09/029, 10.1088/1475-7516/2015/9/029} {\bibfield
  {journal} {\bibinfo  {journal} {JCAP}\ }\textbf {\bibinfo {volume} {1509}},\
  \bibinfo {pages} {029} (\bibinfo {year} {2015}{\natexlab{b}})},\ \Eprint
  {http://arxiv.org/abs/1503.08826} {arXiv:1503.08826 [astro-ph.CO]}
  \BibitemShut {NoStop}%
\bibitem [{\citenamefont {Abolhasani}\ \emph {et~al.}(2016)\citenamefont
  {Abolhasani}, \citenamefont {Mirbabayi},\ and\ \citenamefont
  {Pajer}}]{Abolhasani:2015mra}%
  \BibitemOpen
  \bibfield  {author} {\bibinfo {author} {\bibfnamefont {A.~A.}\ \bibnamefont
  {Abolhasani}}, \bibinfo {author} {\bibfnamefont {M.}~\bibnamefont
  {Mirbabayi}}, \ and\ \bibinfo {author} {\bibfnamefont {E.}~\bibnamefont
  {Pajer}},\ }\href {\doibase 10.1088/1475-7516/2016/05/063} {\bibfield
  {journal} {\bibinfo  {journal} {JCAP}\ }\textbf {\bibinfo {volume} {05}},\
  \bibinfo {pages} {063} (\bibinfo {year} {2016})},\ \Eprint
  {http://arxiv.org/abs/1509.07886} {arXiv:1509.07886 [hep-th]} \BibitemShut
  {NoStop}%
\bibitem [{\citenamefont {Assassi}\ \emph
  {et~al.}(2015{\natexlab{a}})\citenamefont {Assassi}, \citenamefont {Baumann},
  \citenamefont {Pajer}, \citenamefont {Welling},\ and\ \citenamefont {van~der
  Woude}}]{Assassi:2015jqa}%
  \BibitemOpen
  \bibfield  {author} {\bibinfo {author} {\bibfnamefont {V.}~\bibnamefont
  {Assassi}}, \bibinfo {author} {\bibfnamefont {D.}~\bibnamefont {Baumann}},
  \bibinfo {author} {\bibfnamefont {E.}~\bibnamefont {Pajer}}, \bibinfo
  {author} {\bibfnamefont {Y.}~\bibnamefont {Welling}}, \ and\ \bibinfo
  {author} {\bibfnamefont {D.}~\bibnamefont {van~der Woude}},\ }\href {\doibase
  10.1088/1475-7516/2015/11/024} {\bibfield  {journal} {\bibinfo  {journal}
  {JCAP}\ }\textbf {\bibinfo {volume} {11}},\ \bibinfo {pages} {024} (\bibinfo
  {year} {2015}{\natexlab{a}})},\ \Eprint {http://arxiv.org/abs/1505.06668}
  {arXiv:1505.06668 [astro-ph.CO]} \BibitemShut {NoStop}%
\bibitem [{\citenamefont {Foreman}\ and\ \citenamefont
  {Senatore}(2016)}]{Foreman:2015uva}%
  \BibitemOpen
  \bibfield  {author} {\bibinfo {author} {\bibfnamefont {S.}~\bibnamefont
  {Foreman}}\ and\ \bibinfo {author} {\bibfnamefont {L.}~\bibnamefont
  {Senatore}},\ }\href {\doibase 10.1088/1475-7516/2016/04/033} {\bibfield
  {journal} {\bibinfo  {journal} {JCAP}\ }\textbf {\bibinfo {volume} {04}},\
  \bibinfo {pages} {033} (\bibinfo {year} {2016})},\ \Eprint
  {http://arxiv.org/abs/1503.01775} {arXiv:1503.01775 [astro-ph.CO]}
  \BibitemShut {NoStop}%
\bibitem [{\citenamefont {Assassi}\ \emph
  {et~al.}(2015{\natexlab{b}})\citenamefont {Assassi}, \citenamefont
  {Baumann},\ and\ \citenamefont {Schmidt}}]{Assassi:2015fma}%
  \BibitemOpen
  \bibfield  {author} {\bibinfo {author} {\bibfnamefont {V.}~\bibnamefont
  {Assassi}}, \bibinfo {author} {\bibfnamefont {D.}~\bibnamefont {Baumann}}, \
  and\ \bibinfo {author} {\bibfnamefont {F.}~\bibnamefont {Schmidt}},\ }\href
  {\doibase 10.1088/1475-7516/2015/12/043} {\bibfield  {journal} {\bibinfo
  {journal} {JCAP}\ }\textbf {\bibinfo {volume} {12}},\ \bibinfo {pages} {043}
  (\bibinfo {year} {2015}{\natexlab{b}})},\ \Eprint
  {http://arxiv.org/abs/1510.03723} {arXiv:1510.03723 [astro-ph.CO]}
  \BibitemShut {NoStop}%
\bibitem [{\citenamefont {Blas}\ \emph
  {et~al.}(2016{\natexlab{a}})\citenamefont {Blas}, \citenamefont {Garny},
  \citenamefont {Ivanov},\ and\ \citenamefont {Sibiryakov}}]{Blas:2015qsi}%
  \BibitemOpen
  \bibfield  {author} {\bibinfo {author} {\bibfnamefont {D.}~\bibnamefont
  {Blas}}, \bibinfo {author} {\bibfnamefont {M.}~\bibnamefont {Garny}},
  \bibinfo {author} {\bibfnamefont {M.~M.}\ \bibnamefont {Ivanov}}, \ and\
  \bibinfo {author} {\bibfnamefont {S.}~\bibnamefont {Sibiryakov}},\ }\href
  {\doibase 10.1088/1475-7516/2016/07/052} {\bibfield  {journal} {\bibinfo
  {journal} {JCAP}\ }\textbf {\bibinfo {volume} {1607}},\ \bibinfo {pages}
  {052} (\bibinfo {year} {2016}{\natexlab{a}})},\ \Eprint
  {http://arxiv.org/abs/1512.05807} {arXiv:1512.05807 [astro-ph.CO]}
  \BibitemShut {NoStop}%
\bibitem [{\citenamefont {Baldauf}\ \emph
  {et~al.}(2015{\natexlab{b}})\citenamefont {Baldauf}, \citenamefont
  {Mirbabayi}, \citenamefont {Simonovi{\'c}},\ and\ \citenamefont
  {Zaldarriaga}}]{Baldauf:2015xfa}%
  \BibitemOpen
  \bibfield  {author} {\bibinfo {author} {\bibfnamefont {T.}~\bibnamefont
  {Baldauf}}, \bibinfo {author} {\bibfnamefont {M.}~\bibnamefont {Mirbabayi}},
  \bibinfo {author} {\bibfnamefont {M.}~\bibnamefont {Simonovi{\'c}}}, \ and\
  \bibinfo {author} {\bibfnamefont {M.}~\bibnamefont {Zaldarriaga}},\ }\href
  {\doibase 10.1103/PhysRevD.92.043514} {\bibfield  {journal} {\bibinfo
  {journal} {Phys. Rev.}\ }\textbf {\bibinfo {volume} {D92}},\ \bibinfo {pages}
  {043514} (\bibinfo {year} {2015}{\natexlab{b}})},\ \Eprint
  {http://arxiv.org/abs/1504.04366} {arXiv:1504.04366 [astro-ph.CO]}
  \BibitemShut {NoStop}%
\bibitem [{\citenamefont {Lewandowski}\ \emph {et~al.}(2018)\citenamefont
  {Lewandowski}, \citenamefont {Senatore}, \citenamefont {Prada}, \citenamefont
  {Zhao},\ and\ \citenamefont {Chuang}}]{Lewandowski:2015ziq}%
  \BibitemOpen
  \bibfield  {author} {\bibinfo {author} {\bibfnamefont {M.}~\bibnamefont
  {Lewandowski}}, \bibinfo {author} {\bibfnamefont {L.}~\bibnamefont
  {Senatore}}, \bibinfo {author} {\bibfnamefont {F.}~\bibnamefont {Prada}},
  \bibinfo {author} {\bibfnamefont {C.}~\bibnamefont {Zhao}}, \ and\ \bibinfo
  {author} {\bibfnamefont {C.-H.}\ \bibnamefont {Chuang}},\ }\href {\doibase
  10.1103/PhysRevD.97.063526} {\bibfield  {journal} {\bibinfo  {journal} {Phys.
  Rev. D}\ }\textbf {\bibinfo {volume} {97}},\ \bibinfo {pages} {063526}
  (\bibinfo {year} {2018})},\ \Eprint {http://arxiv.org/abs/1512.06831}
  {arXiv:1512.06831 [astro-ph.CO]} \BibitemShut {NoStop}%
\bibitem [{\citenamefont {Foreman}\ \emph {et~al.}(2016)\citenamefont
  {Foreman}, \citenamefont {Perrier},\ and\ \citenamefont
  {Senatore}}]{Foreman:2015lca}%
  \BibitemOpen
  \bibfield  {author} {\bibinfo {author} {\bibfnamefont {S.}~\bibnamefont
  {Foreman}}, \bibinfo {author} {\bibfnamefont {H.}~\bibnamefont {Perrier}}, \
  and\ \bibinfo {author} {\bibfnamefont {L.}~\bibnamefont {Senatore}},\ }\href
  {\doibase 10.1088/1475-7516/2016/05/027} {\bibfield  {journal} {\bibinfo
  {journal} {JCAP}\ }\textbf {\bibinfo {volume} {05}},\ \bibinfo {pages} {027}
  (\bibinfo {year} {2016})},\ \Eprint {http://arxiv.org/abs/1507.05326}
  {arXiv:1507.05326 [astro-ph.CO]} \BibitemShut {NoStop}%
\bibitem [{\citenamefont {Vlah}\ \emph
  {et~al.}(2016{\natexlab{a}})\citenamefont {Vlah}, \citenamefont {Seljak},
  \citenamefont {Chu},\ and\ \citenamefont {Feng}}]{Vlah:2015zda}%
  \BibitemOpen
  \bibfield  {author} {\bibinfo {author} {\bibfnamefont {Z.}~\bibnamefont
  {Vlah}}, \bibinfo {author} {\bibfnamefont {U.}~\bibnamefont {Seljak}},
  \bibinfo {author} {\bibfnamefont {M.~Y.}\ \bibnamefont {Chu}}, \ and\
  \bibinfo {author} {\bibfnamefont {Y.}~\bibnamefont {Feng}},\ }\href {\doibase
  10.1088/1475-7516/2016/03/057} {\bibfield  {journal} {\bibinfo  {journal}
  {JCAP}\ }\textbf {\bibinfo {volume} {1603}},\ \bibinfo {pages} {057}
  (\bibinfo {year} {2016}{\natexlab{a}})},\ \Eprint
  {http://arxiv.org/abs/1509.02120} {arXiv:1509.02120 [astro-ph.CO]}
  \BibitemShut {NoStop}%
\bibitem [{\citenamefont {Vlah}\ \emph {et~al.}(2015)\citenamefont {Vlah},
  \citenamefont {White},\ and\ \citenamefont {Aviles}}]{Vlah:2015sea}%
  \BibitemOpen
  \bibfield  {author} {\bibinfo {author} {\bibfnamefont {Z.}~\bibnamefont
  {Vlah}}, \bibinfo {author} {\bibfnamefont {M.}~\bibnamefont {White}}, \ and\
  \bibinfo {author} {\bibfnamefont {A.}~\bibnamefont {Aviles}},\ }\href
  {\doibase 10.1088/1475-7516/2015/09/014} {\bibfield  {journal} {\bibinfo
  {journal} {JCAP}\ }\textbf {\bibinfo {volume} {09}},\ \bibinfo {pages} {014}
  (\bibinfo {year} {2015})},\ \Eprint {http://arxiv.org/abs/1506.05264}
  {arXiv:1506.05264 [astro-ph.CO]} \BibitemShut {NoStop}%
\bibitem [{\citenamefont {Baldauf}\ \emph
  {et~al.}(2015{\natexlab{c}})\citenamefont {Baldauf}, \citenamefont
  {Mercolli},\ and\ \citenamefont {Zaldarriaga}}]{Baldauf:2015aha}%
  \BibitemOpen
  \bibfield  {author} {\bibinfo {author} {\bibfnamefont {T.}~\bibnamefont
  {Baldauf}}, \bibinfo {author} {\bibfnamefont {L.}~\bibnamefont {Mercolli}}, \
  and\ \bibinfo {author} {\bibfnamefont {M.}~\bibnamefont {Zaldarriaga}},\
  }\href {\doibase 10.1103/PhysRevD.92.123007} {\bibfield  {journal} {\bibinfo
  {journal} {Phys. Rev.}\ }\textbf {\bibinfo {volume} {D92}},\ \bibinfo {pages}
  {123007} (\bibinfo {year} {2015}{\natexlab{c}})},\ \Eprint
  {http://arxiv.org/abs/1507.02256} {arXiv:1507.02256 [astro-ph.CO]}
  \BibitemShut {NoStop}%
\bibitem [{\citenamefont {Bertolini}\ \emph
  {et~al.}(2016{\natexlab{a}})\citenamefont {Bertolini}, \citenamefont
  {Schutz}, \citenamefont {Solon}, \citenamefont {Walsh},\ and\ \citenamefont
  {Zurek}}]{Bertolini:2015fya}%
  \BibitemOpen
  \bibfield  {author} {\bibinfo {author} {\bibfnamefont {D.}~\bibnamefont
  {Bertolini}}, \bibinfo {author} {\bibfnamefont {K.}~\bibnamefont {Schutz}},
  \bibinfo {author} {\bibfnamefont {M.~P.}\ \bibnamefont {Solon}}, \bibinfo
  {author} {\bibfnamefont {J.~R.}\ \bibnamefont {Walsh}}, \ and\ \bibinfo
  {author} {\bibfnamefont {K.~M.}\ \bibnamefont {Zurek}},\ }\href {\doibase
  10.1103/PhysRevD.93.123505} {\bibfield  {journal} {\bibinfo  {journal} {Phys.
  Rev.}\ }\textbf {\bibinfo {volume} {D93}},\ \bibinfo {pages} {123505}
  (\bibinfo {year} {2016}{\natexlab{a}})},\ \Eprint
  {http://arxiv.org/abs/1512.07630} {arXiv:1512.07630 [astro-ph.CO]}
  \BibitemShut {NoStop}%
\bibitem [{\citenamefont {Lazeyras}\ \emph {et~al.}(2016)\citenamefont
  {Lazeyras}, \citenamefont {Wagner}, \citenamefont {Baldauf},\ and\
  \citenamefont {Schmidt}}]{Lazeyras:2015lgp}%
  \BibitemOpen
  \bibfield  {author} {\bibinfo {author} {\bibfnamefont {T.}~\bibnamefont
  {Lazeyras}}, \bibinfo {author} {\bibfnamefont {C.}~\bibnamefont {Wagner}},
  \bibinfo {author} {\bibfnamefont {T.}~\bibnamefont {Baldauf}}, \ and\
  \bibinfo {author} {\bibfnamefont {F.}~\bibnamefont {Schmidt}},\ }\href
  {\doibase 10.1088/1475-7516/2016/02/018} {\bibfield  {journal} {\bibinfo
  {journal} {JCAP}\ }\textbf {\bibinfo {volume} {1602}},\ \bibinfo {pages}
  {018} (\bibinfo {year} {2016})},\ \Eprint {http://arxiv.org/abs/1511.01096}
  {arXiv:1511.01096 [astro-ph.CO]} \BibitemShut {NoStop}%
\bibitem [{\citenamefont {Baldauf}\ \emph
  {et~al.}(2016{\natexlab{a}})\citenamefont {Baldauf}, \citenamefont {Schaan},\
  and\ \citenamefont {Zaldarriaga}}]{Baldauf:2015tla}%
  \BibitemOpen
  \bibfield  {author} {\bibinfo {author} {\bibfnamefont {T.}~\bibnamefont
  {Baldauf}}, \bibinfo {author} {\bibfnamefont {E.}~\bibnamefont {Schaan}}, \
  and\ \bibinfo {author} {\bibfnamefont {M.}~\bibnamefont {Zaldarriaga}},\
  }\href {\doibase 10.1088/1475-7516/2016/03/017} {\bibfield  {journal}
  {\bibinfo  {journal} {JCAP}\ }\textbf {\bibinfo {volume} {03}},\ \bibinfo
  {pages} {017} (\bibinfo {year} {2016}{\natexlab{a}})},\ \Eprint
  {http://arxiv.org/abs/1505.07098} {arXiv:1505.07098 [astro-ph.CO]}
  \BibitemShut {NoStop}%
\bibitem [{\citenamefont {Baldauf}\ \emph
  {et~al.}(2016{\natexlab{b}})\citenamefont {Baldauf}, \citenamefont {Schaan},\
  and\ \citenamefont {Zaldarriaga}}]{Baldauf:2015zga}%
  \BibitemOpen
  \bibfield  {author} {\bibinfo {author} {\bibfnamefont {T.}~\bibnamefont
  {Baldauf}}, \bibinfo {author} {\bibfnamefont {E.}~\bibnamefont {Schaan}}, \
  and\ \bibinfo {author} {\bibfnamefont {M.}~\bibnamefont {Zaldarriaga}},\
  }\href {\doibase 10.1088/1475-7516/2016/03/007} {\bibfield  {journal}
  {\bibinfo  {journal} {JCAP}\ }\textbf {\bibinfo {volume} {03}},\ \bibinfo
  {pages} {007} (\bibinfo {year} {2016}{\natexlab{b}})},\ \Eprint
  {http://arxiv.org/abs/1507.02255} {arXiv:1507.02255 [astro-ph.CO]}
  \BibitemShut {NoStop}%
\bibitem [{\citenamefont {Desjacques}\ \emph
  {et~al.}(2018{\natexlab{a}})\citenamefont {Desjacques}, \citenamefont
  {Jeong},\ and\ \citenamefont {Schmidt}}]{Desjacques:2016bnm}%
  \BibitemOpen
  \bibfield  {author} {\bibinfo {author} {\bibfnamefont {V.}~\bibnamefont
  {Desjacques}}, \bibinfo {author} {\bibfnamefont {D.}~\bibnamefont {Jeong}}, \
  and\ \bibinfo {author} {\bibfnamefont {F.}~\bibnamefont {Schmidt}},\ }\href
  {\doibase 10.1016/j.physrep.2017.12.002} {\bibfield  {journal} {\bibinfo
  {journal} {Phys. Rept.}\ }\textbf {\bibinfo {volume} {733}},\ \bibinfo
  {pages} {1} (\bibinfo {year} {2018}{\natexlab{a}})},\ \Eprint
  {http://arxiv.org/abs/1611.09787} {arXiv:1611.09787 [astro-ph.CO]}
  \BibitemShut {NoStop}%
\bibitem [{\citenamefont {Fasiello}\ and\ \citenamefont
  {Vlah}(2016)}]{Fasiello:2016qpn}%
  \BibitemOpen
  \bibfield  {author} {\bibinfo {author} {\bibfnamefont {M.}~\bibnamefont
  {Fasiello}}\ and\ \bibinfo {author} {\bibfnamefont {Z.}~\bibnamefont
  {Vlah}},\ }\href {\doibase 10.1103/PhysRevD.94.063516} {\bibfield  {journal}
  {\bibinfo  {journal} {Phys. Rev.}\ }\textbf {\bibinfo {volume} {D94}},\
  \bibinfo {pages} {063516} (\bibinfo {year} {2016})},\ \Eprint
  {http://arxiv.org/abs/1604.04612} {arXiv:1604.04612 [astro-ph.CO]}
  \BibitemShut {NoStop}%
\bibitem [{\citenamefont {Blas}\ \emph
  {et~al.}(2016{\natexlab{b}})\citenamefont {Blas}, \citenamefont {Garny},
  \citenamefont {Ivanov},\ and\ \citenamefont {Sibiryakov}}]{Blas:2016sfa}%
  \BibitemOpen
  \bibfield  {author} {\bibinfo {author} {\bibfnamefont {D.}~\bibnamefont
  {Blas}}, \bibinfo {author} {\bibfnamefont {M.}~\bibnamefont {Garny}},
  \bibinfo {author} {\bibfnamefont {M.~M.}\ \bibnamefont {Ivanov}}, \ and\
  \bibinfo {author} {\bibfnamefont {S.}~\bibnamefont {Sibiryakov}},\ }\href
  {\doibase 10.1088/1475-7516/2016/07/028} {\bibfield  {journal} {\bibinfo
  {journal} {JCAP}\ }\textbf {\bibinfo {volume} {1607}},\ \bibinfo {pages}
  {028} (\bibinfo {year} {2016}{\natexlab{b}})},\ \Eprint
  {http://arxiv.org/abs/1605.02149} {arXiv:1605.02149 [astro-ph.CO]}
  \BibitemShut {NoStop}%
\bibitem [{\citenamefont {Bertolini}\ \emph
  {et~al.}(2016{\natexlab{b}})\citenamefont {Bertolini}, \citenamefont
  {Schutz}, \citenamefont {Solon},\ and\ \citenamefont
  {Zurek}}]{Bertolini:2016bmt}%
  \BibitemOpen
  \bibfield  {author} {\bibinfo {author} {\bibfnamefont {D.}~\bibnamefont
  {Bertolini}}, \bibinfo {author} {\bibfnamefont {K.}~\bibnamefont {Schutz}},
  \bibinfo {author} {\bibfnamefont {M.~P.}\ \bibnamefont {Solon}}, \ and\
  \bibinfo {author} {\bibfnamefont {K.~M.}\ \bibnamefont {Zurek}},\ }\href
  {\doibase 10.1088/1475-7516/2016/06/052} {\bibfield  {journal} {\bibinfo
  {journal} {JCAP}\ }\textbf {\bibinfo {volume} {06}},\ \bibinfo {pages} {052}
  (\bibinfo {year} {2016}{\natexlab{b}})},\ \Eprint
  {http://arxiv.org/abs/1604.01770} {arXiv:1604.01770 [astro-ph.CO]}
  \BibitemShut {NoStop}%
\bibitem [{\citenamefont {Vlah}\ \emph
  {et~al.}(2016{\natexlab{b}})\citenamefont {Vlah}, \citenamefont {Castorina},\
  and\ \citenamefont {White}}]{Vlah:2016bcl}%
  \BibitemOpen
  \bibfield  {author} {\bibinfo {author} {\bibfnamefont {Z.}~\bibnamefont
  {Vlah}}, \bibinfo {author} {\bibfnamefont {E.}~\bibnamefont {Castorina}}, \
  and\ \bibinfo {author} {\bibfnamefont {M.}~\bibnamefont {White}},\ }\href
  {\doibase 10.1088/1475-7516/2016/12/007} {\bibfield  {journal} {\bibinfo
  {journal} {JCAP}\ }\textbf {\bibinfo {volume} {12}},\ \bibinfo {pages} {007}
  (\bibinfo {year} {2016}{\natexlab{b}})},\ \Eprint
  {http://arxiv.org/abs/1609.02908} {arXiv:1609.02908 [astro-ph.CO]}
  \BibitemShut {NoStop}%
\bibitem [{\citenamefont {Nadler}\ \emph {et~al.}(2018)\citenamefont {Nadler},
  \citenamefont {Perko},\ and\ \citenamefont {Senatore}}]{Nadler:2017qto}%
  \BibitemOpen
  \bibfield  {author} {\bibinfo {author} {\bibfnamefont {E.~O.}\ \bibnamefont
  {Nadler}}, \bibinfo {author} {\bibfnamefont {A.}~\bibnamefont {Perko}}, \
  and\ \bibinfo {author} {\bibfnamefont {L.}~\bibnamefont {Senatore}},\ }\href
  {\doibase 10.1088/1475-7516/2018/02/058} {\bibfield  {journal} {\bibinfo
  {journal} {JCAP}\ }\textbf {\bibinfo {volume} {02}},\ \bibinfo {pages} {058}
  (\bibinfo {year} {2018})},\ \Eprint {http://arxiv.org/abs/1710.10308}
  {arXiv:1710.10308 [astro-ph.CO]} \BibitemShut {NoStop}%
\bibitem [{\citenamefont {Lewandowski}\ and\ \citenamefont
  {Senatore}(2017)}]{Lewandowski:2017kes}%
  \BibitemOpen
  \bibfield  {author} {\bibinfo {author} {\bibfnamefont {M.}~\bibnamefont
  {Lewandowski}}\ and\ \bibinfo {author} {\bibfnamefont {L.}~\bibnamefont
  {Senatore}},\ }\href {\doibase 10.1088/1475-7516/2017/08/037} {\bibfield
  {journal} {\bibinfo  {journal} {JCAP}\ }\textbf {\bibinfo {volume} {08}},\
  \bibinfo {pages} {037} (\bibinfo {year} {2017})},\ \Eprint
  {http://arxiv.org/abs/1701.07012} {arXiv:1701.07012 [astro-ph.CO]}
  \BibitemShut {NoStop}%
\bibitem [{\citenamefont {Senatore}\ and\ \citenamefont
  {Trevisan}(2018)}]{Senatore:2017pbn}%
  \BibitemOpen
  \bibfield  {author} {\bibinfo {author} {\bibfnamefont {L.}~\bibnamefont
  {Senatore}}\ and\ \bibinfo {author} {\bibfnamefont {G.}~\bibnamefont
  {Trevisan}},\ }\href {\doibase 10.1088/1475-7516/2018/05/019} {\bibfield
  {journal} {\bibinfo  {journal} {JCAP}\ }\textbf {\bibinfo {volume} {1805}},\
  \bibinfo {pages} {019} (\bibinfo {year} {2018})},\ \Eprint
  {http://arxiv.org/abs/1710.02178} {arXiv:1710.02178 [astro-ph.CO]}
  \BibitemShut {NoStop}%
\bibitem [{\citenamefont {Lazeyras}\ and\ \citenamefont
  {Schmidt}(2018)}]{Lazeyras:2017hxw}%
  \BibitemOpen
  \bibfield  {author} {\bibinfo {author} {\bibfnamefont {T.}~\bibnamefont
  {Lazeyras}}\ and\ \bibinfo {author} {\bibfnamefont {F.}~\bibnamefont
  {Schmidt}},\ }\href {\doibase 10.1088/1475-7516/2018/09/008} {\bibfield
  {journal} {\bibinfo  {journal} {JCAP}\ }\textbf {\bibinfo {volume} {1809}},\
  \bibinfo {pages} {008} (\bibinfo {year} {2018})},\ \Eprint
  {http://arxiv.org/abs/1712.07531} {arXiv:1712.07531 [astro-ph.CO]}
  \BibitemShut {NoStop}%
\bibitem [{\citenamefont {Senatore}\ and\ \citenamefont
  {Zaldarriaga}(2017)}]{Senatore:2017hyk}%
  \BibitemOpen
  \bibfield  {author} {\bibinfo {author} {\bibfnamefont {L.}~\bibnamefont
  {Senatore}}\ and\ \bibinfo {author} {\bibfnamefont {M.}~\bibnamefont
  {Zaldarriaga}},\ }\href@noop {} {\  (\bibinfo {year} {2017})},\ \Eprint
  {http://arxiv.org/abs/1707.04698} {arXiv:1707.04698 [astro-ph.CO]}
  \BibitemShut {NoStop}%
\bibitem [{\citenamefont {Ivanov}\ and\ \citenamefont
  {Sibiryakov}(2018)}]{Ivanov:2018gjr}%
  \BibitemOpen
  \bibfield  {author} {\bibinfo {author} {\bibfnamefont {M.~M.}\ \bibnamefont
  {Ivanov}}\ and\ \bibinfo {author} {\bibfnamefont {S.}~\bibnamefont
  {Sibiryakov}},\ }\href {\doibase 10.1088/1475-7516/2018/07/053} {\bibfield
  {journal} {\bibinfo  {journal} {JCAP}\ }\textbf {\bibinfo {volume} {1807}},\
  \bibinfo {pages} {053} (\bibinfo {year} {2018})},\ \Eprint
  {http://arxiv.org/abs/1804.05080} {arXiv:1804.05080 [astro-ph.CO]}
  \BibitemShut {NoStop}%
\bibitem [{\citenamefont {Lewandowski}\ and\ \citenamefont
  {Senatore}(2020)}]{Lewandowski:2018ywf}%
  \BibitemOpen
  \bibfield  {author} {\bibinfo {author} {\bibfnamefont {M.}~\bibnamefont
  {Lewandowski}}\ and\ \bibinfo {author} {\bibfnamefont {L.}~\bibnamefont
  {Senatore}},\ }\href {\doibase 10.1088/1475-7516/2020/03/018} {\bibfield
  {journal} {\bibinfo  {journal} {JCAP}\ }\textbf {\bibinfo {volume} {03}},\
  \bibinfo {pages} {018} (\bibinfo {year} {2020})},\ \Eprint
  {http://arxiv.org/abs/1810.11855} {arXiv:1810.11855 [astro-ph.CO]}
  \BibitemShut {NoStop}%
\bibitem [{\citenamefont {Vlah}\ and\ \citenamefont
  {White}(2019)}]{Vlah:2018ygt}%
  \BibitemOpen
  \bibfield  {author} {\bibinfo {author} {\bibfnamefont {Z.}~\bibnamefont
  {Vlah}}\ and\ \bibinfo {author} {\bibfnamefont {M.}~\bibnamefont {White}},\
  }\href {\doibase 10.1088/1475-7516/2019/03/007} {\bibfield  {journal}
  {\bibinfo  {journal} {JCAP}\ }\textbf {\bibinfo {volume} {1903}},\ \bibinfo
  {pages} {007} (\bibinfo {year} {2019})},\ \Eprint
  {http://arxiv.org/abs/1812.02775} {arXiv:1812.02775 [astro-ph.CO]}
  \BibitemShut {NoStop}%
\bibitem [{\citenamefont {Eggemeier}\ \emph {et~al.}(2018)\citenamefont
  {Eggemeier}, \citenamefont {Scoccimarro},\ and\ \citenamefont
  {Smith}}]{Eggemeier:2018qae}%
  \BibitemOpen
  \bibfield  {author} {\bibinfo {author} {\bibfnamefont {A.}~\bibnamefont
  {Eggemeier}}, \bibinfo {author} {\bibfnamefont {R.}~\bibnamefont
  {Scoccimarro}}, \ and\ \bibinfo {author} {\bibfnamefont {R.~E.}\ \bibnamefont
  {Smith}},\ }\href@noop {} {\  (\bibinfo {year} {2018})},\ \Eprint
  {http://arxiv.org/abs/1812.03208} {arXiv:1812.03208 [astro-ph.CO]}
  \BibitemShut {NoStop}%
\bibitem [{\citenamefont {Konstandin}\ \emph {et~al.}(2019)\citenamefont
  {Konstandin}, \citenamefont {Porto},\ and\ \citenamefont
  {Rubira}}]{Konstandin:2019bay}%
  \BibitemOpen
  \bibfield  {author} {\bibinfo {author} {\bibfnamefont {T.}~\bibnamefont
  {Konstandin}}, \bibinfo {author} {\bibfnamefont {R.~A.}\ \bibnamefont
  {Porto}}, \ and\ \bibinfo {author} {\bibfnamefont {H.}~\bibnamefont
  {Rubira}},\ }\href {\doibase 10.1088/1475-7516/2019/11/027} {\bibfield
  {journal} {\bibinfo  {journal} {JCAP}\ }\textbf {\bibinfo {volume} {11}},\
  \bibinfo {pages} {027} (\bibinfo {year} {2019})},\ \Eprint
  {http://arxiv.org/abs/1906.00997} {arXiv:1906.00997 [astro-ph.CO]}
  \BibitemShut {NoStop}%
\bibitem [{\citenamefont {Lazeyras}\ and\ \citenamefont
  {Schmidt}(2019)}]{Lazeyras:2019dcx}%
  \BibitemOpen
  \bibfield  {author} {\bibinfo {author} {\bibfnamefont {T.}~\bibnamefont
  {Lazeyras}}\ and\ \bibinfo {author} {\bibfnamefont {F.}~\bibnamefont
  {Schmidt}},\ }\href@noop {} {\  (\bibinfo {year} {2019})},\ \Eprint
  {http://arxiv.org/abs/1904.11294} {arXiv:1904.11294 [astro-ph.CO]}
  \BibitemShut {NoStop}%
\bibitem [{\citenamefont {Abidi}\ and\ \citenamefont
  {Baldauf}(2018)}]{Abidi:2018eyd}%
  \BibitemOpen
  \bibfield  {author} {\bibinfo {author} {\bibfnamefont {M.~M.}\ \bibnamefont
  {Abidi}}\ and\ \bibinfo {author} {\bibfnamefont {T.}~\bibnamefont
  {Baldauf}},\ }\href {\doibase 10.1088/1475-7516/2018/07/029} {\bibfield
  {journal} {\bibinfo  {journal} {JCAP}\ }\textbf {\bibinfo {volume} {1807}},\
  \bibinfo {pages} {029} (\bibinfo {year} {2018})},\ \Eprint
  {http://arxiv.org/abs/1802.07622} {arXiv:1802.07622 [astro-ph.CO]}
  \BibitemShut {NoStop}%
\bibitem [{\citenamefont {Cabass}\ and\ \citenamefont
  {Schmidt}(2019)}]{Cabass:2018hum}%
  \BibitemOpen
  \bibfield  {author} {\bibinfo {author} {\bibfnamefont {G.}~\bibnamefont
  {Cabass}}\ and\ \bibinfo {author} {\bibfnamefont {F.}~\bibnamefont
  {Schmidt}},\ }\href {\doibase 10.1088/1475-7516/2019/05/031} {\bibfield
  {journal} {\bibinfo  {journal} {JCAP}\ }\textbf {\bibinfo {volume} {05}},\
  \bibinfo {pages} {031} (\bibinfo {year} {2019})},\ \Eprint
  {http://arxiv.org/abs/1812.02731} {arXiv:1812.02731 [astro-ph.CO]}
  \BibitemShut {NoStop}%
\bibitem [{\citenamefont {Schmittfull}\ \emph {et~al.}(2019)\citenamefont
  {Schmittfull}, \citenamefont {Simonovi\'c}, \citenamefont {Assassi},\ and\
  \citenamefont {Zaldarriaga}}]{Schmittfull:2018yuk}%
  \BibitemOpen
  \bibfield  {author} {\bibinfo {author} {\bibfnamefont {M.}~\bibnamefont
  {Schmittfull}}, \bibinfo {author} {\bibfnamefont {M.}~\bibnamefont
  {Simonovi\'c}}, \bibinfo {author} {\bibfnamefont {V.}~\bibnamefont
  {Assassi}}, \ and\ \bibinfo {author} {\bibfnamefont {M.}~\bibnamefont
  {Zaldarriaga}},\ }\href {\doibase 10.1103/PhysRevD.100.043514} {\bibfield
  {journal} {\bibinfo  {journal} {Phys. Rev. D}\ }\textbf {\bibinfo {volume}
  {100}},\ \bibinfo {pages} {043514} (\bibinfo {year} {2019})},\ \Eprint
  {http://arxiv.org/abs/1811.10640} {arXiv:1811.10640 [astro-ph.CO]}
  \BibitemShut {NoStop}%
\bibitem [{\citenamefont {Schmidt}\ \emph {et~al.}(2019)\citenamefont
  {Schmidt}, \citenamefont {Elsner}, \citenamefont {Jasche}, \citenamefont
  {Nguyen},\ and\ \citenamefont {Lavaux}}]{Schmidt:2018bkr}%
  \BibitemOpen
  \bibfield  {author} {\bibinfo {author} {\bibfnamefont {F.}~\bibnamefont
  {Schmidt}}, \bibinfo {author} {\bibfnamefont {F.}~\bibnamefont {Elsner}},
  \bibinfo {author} {\bibfnamefont {J.}~\bibnamefont {Jasche}}, \bibinfo
  {author} {\bibfnamefont {N.~M.}\ \bibnamefont {Nguyen}}, \ and\ \bibinfo
  {author} {\bibfnamefont {G.}~\bibnamefont {Lavaux}},\ }\href {\doibase
  10.1088/1475-7516/2019/01/042} {\bibfield  {journal} {\bibinfo  {journal}
  {JCAP}\ }\textbf {\bibinfo {volume} {01}},\ \bibinfo {pages} {042} (\bibinfo
  {year} {2019})},\ \Eprint {http://arxiv.org/abs/1808.02002} {arXiv:1808.02002
  [astro-ph.CO]} \BibitemShut {NoStop}%
\bibitem [{\citenamefont {de~Belsunce}\ and\ \citenamefont
  {Senatore}(2019)}]{deBelsunce:2018xtd}%
  \BibitemOpen
  \bibfield  {author} {\bibinfo {author} {\bibfnamefont {R.}~\bibnamefont
  {de~Belsunce}}\ and\ \bibinfo {author} {\bibfnamefont {L.}~\bibnamefont
  {Senatore}},\ }\href {\doibase 10.1088/1475-7516/2019/02/038} {\bibfield
  {journal} {\bibinfo  {journal} {JCAP}\ }\textbf {\bibinfo {volume} {1902}},\
  \bibinfo {pages} {038} (\bibinfo {year} {2019})},\ \Eprint
  {http://arxiv.org/abs/1804.06849} {arXiv:1804.06849 [astro-ph.CO]}
  \BibitemShut {NoStop}%
\bibitem [{\citenamefont {Desjacques}\ \emph
  {et~al.}(2018{\natexlab{b}})\citenamefont {Desjacques}, \citenamefont
  {Jeong},\ and\ \citenamefont {Schmidt}}]{Desjacques:2018pfv}%
  \BibitemOpen
  \bibfield  {author} {\bibinfo {author} {\bibfnamefont {V.}~\bibnamefont
  {Desjacques}}, \bibinfo {author} {\bibfnamefont {D.}~\bibnamefont {Jeong}}, \
  and\ \bibinfo {author} {\bibfnamefont {F.}~\bibnamefont {Schmidt}},\ }\href
  {\doibase 10.1088/1475-7516/2018/12/035} {\bibfield  {journal} {\bibinfo
  {journal} {JCAP}\ }\textbf {\bibinfo {volume} {1812}},\ \bibinfo {pages}
  {035} (\bibinfo {year} {2018}{\natexlab{b}})},\ \Eprint
  {http://arxiv.org/abs/1806.04015} {arXiv:1806.04015 [astro-ph.CO]}
  \BibitemShut {NoStop}%
\bibitem [{\citenamefont {Vlah}\ \emph {et~al.}(2020)\citenamefont {Vlah},
  \citenamefont {Chisari},\ and\ \citenamefont {Schmidt}}]{Vlah:2019byq}%
  \BibitemOpen
  \bibfield  {author} {\bibinfo {author} {\bibfnamefont {Z.}~\bibnamefont
  {Vlah}}, \bibinfo {author} {\bibfnamefont {N.~E.}\ \bibnamefont {Chisari}}, \
  and\ \bibinfo {author} {\bibfnamefont {F.}~\bibnamefont {Schmidt}},\ }\href
  {\doibase 10.1088/1475-7516/2020/01/025} {\bibfield  {journal} {\bibinfo
  {journal} {JCAP}\ }\textbf {\bibinfo {volume} {01}},\ \bibinfo {pages} {025}
  (\bibinfo {year} {2020})},\ \Eprint {http://arxiv.org/abs/1910.08085}
  {arXiv:1910.08085 [astro-ph.CO]} \BibitemShut {NoStop}%
\bibitem [{\citenamefont {Elsner}\ \emph {et~al.}(2020)\citenamefont {Elsner},
  \citenamefont {Schmidt}, \citenamefont {Jasche}, \citenamefont {Lavaux},\
  and\ \citenamefont {Nguyen}}]{Elsner:2019rql}%
  \BibitemOpen
  \bibfield  {author} {\bibinfo {author} {\bibfnamefont {F.}~\bibnamefont
  {Elsner}}, \bibinfo {author} {\bibfnamefont {F.}~\bibnamefont {Schmidt}},
  \bibinfo {author} {\bibfnamefont {J.}~\bibnamefont {Jasche}}, \bibinfo
  {author} {\bibfnamefont {G.}~\bibnamefont {Lavaux}}, \ and\ \bibinfo {author}
  {\bibfnamefont {N.-M.}\ \bibnamefont {Nguyen}},\ }\href {\doibase
  10.1088/1475-7516/2020/01/029} {\bibfield  {journal} {\bibinfo  {journal}
  {JCAP}\ }\textbf {\bibinfo {volume} {01}},\ \bibinfo {pages} {029} (\bibinfo
  {year} {2020})},\ \Eprint {http://arxiv.org/abs/1906.07143} {arXiv:1906.07143
  [astro-ph.CO]} \BibitemShut {NoStop}%
\bibitem [{\citenamefont {Beutler}\ \emph {et~al.}(2019)\citenamefont
  {Beutler}, \citenamefont {Biagetti}, \citenamefont {Green}, \citenamefont
  {Slosar},\ and\ \citenamefont {Wallisch}}]{Beutler:2019ojk}%
  \BibitemOpen
  \bibfield  {author} {\bibinfo {author} {\bibfnamefont {F.}~\bibnamefont
  {Beutler}}, \bibinfo {author} {\bibfnamefont {M.}~\bibnamefont {Biagetti}},
  \bibinfo {author} {\bibfnamefont {D.}~\bibnamefont {Green}}, \bibinfo
  {author} {\bibfnamefont {A.}~\bibnamefont {Slosar}}, \ and\ \bibinfo {author}
  {\bibfnamefont {B.}~\bibnamefont {Wallisch}},\ }\href@noop {} {\  (\bibinfo
  {year} {2019})},\ \Eprint {http://arxiv.org/abs/1906.08758} {arXiv:1906.08758
  [astro-ph.CO]} \BibitemShut {NoStop}%
\bibitem [{\citenamefont {Ivanov}\ \emph
  {et~al.}(2020{\natexlab{a}})\citenamefont {Ivanov}, \citenamefont
  {Simonovi\'c},\ and\ \citenamefont {Zaldarriaga}}]{Ivanov:2019pdj}%
  \BibitemOpen
  \bibfield  {author} {\bibinfo {author} {\bibfnamefont {M.~M.}\ \bibnamefont
  {Ivanov}}, \bibinfo {author} {\bibfnamefont {M.}~\bibnamefont {Simonovi\'c}},
  \ and\ \bibinfo {author} {\bibfnamefont {M.}~\bibnamefont {Zaldarriaga}},\
  }\href {\doibase 10.1088/1475-7516/2020/05/042} {\bibfield  {journal}
  {\bibinfo  {journal} {JCAP}\ }\textbf {\bibinfo {volume} {05}},\ \bibinfo
  {pages} {042} (\bibinfo {year} {2020}{\natexlab{a}})},\ \Eprint
  {http://arxiv.org/abs/1909.05277} {arXiv:1909.05277 [astro-ph.CO]}
  \BibitemShut {NoStop}%
\bibitem [{\citenamefont {D'Amico}\ \emph {et~al.}(2019)\citenamefont
  {D'Amico}, \citenamefont {Gleyzes}, \citenamefont {Kokron}, \citenamefont
  {Markovic}, \citenamefont {Senatore}, \citenamefont {Zhang}, \citenamefont
  {Beutler},\ and\ \citenamefont {Gil-Marín}}]{DAmico:2019fhj}%
  \BibitemOpen
  \bibfield  {author} {\bibinfo {author} {\bibfnamefont {G.}~\bibnamefont
  {D'Amico}}, \bibinfo {author} {\bibfnamefont {J.}~\bibnamefont {Gleyzes}},
  \bibinfo {author} {\bibfnamefont {N.}~\bibnamefont {Kokron}}, \bibinfo
  {author} {\bibfnamefont {D.}~\bibnamefont {Markovic}}, \bibinfo {author}
  {\bibfnamefont {L.}~\bibnamefont {Senatore}}, \bibinfo {author}
  {\bibfnamefont {P.}~\bibnamefont {Zhang}}, \bibinfo {author} {\bibfnamefont
  {F.}~\bibnamefont {Beutler}}, \ and\ \bibinfo {author} {\bibfnamefont
  {H.}~\bibnamefont {Gil-Marín}},\ }\href@noop {} {\  (\bibinfo {year}
  {2019})},\ \Eprint {http://arxiv.org/abs/1909.05271} {arXiv:1909.05271
  [astro-ph.CO]} \BibitemShut {NoStop}%
\bibitem [{\citenamefont {Cabass}\ and\ \citenamefont
  {Schmidt}(2020{\natexlab{a}})}]{Cabass:2019lqx}%
  \BibitemOpen
  \bibfield  {author} {\bibinfo {author} {\bibfnamefont {G.}~\bibnamefont
  {Cabass}}\ and\ \bibinfo {author} {\bibfnamefont {F.}~\bibnamefont
  {Schmidt}},\ }\href {\doibase 10.1088/1475-7516/2020/04/042} {\bibfield
  {journal} {\bibinfo  {journal} {JCAP}\ }\textbf {\bibinfo {volume} {04}},\
  \bibinfo {pages} {042} (\bibinfo {year} {2020}{\natexlab{a}})},\ \Eprint
  {http://arxiv.org/abs/1909.04022} {arXiv:1909.04022 [astro-ph.CO]}
  \BibitemShut {NoStop}%
\bibitem [{\citenamefont {Vasudevan}\ \emph {et~al.}(2019)\citenamefont
  {Vasudevan}, \citenamefont {Ivanov}, \citenamefont {Sibiryakov},\ and\
  \citenamefont {Lesgourgues}}]{Vasudevan:2019ewf}%
  \BibitemOpen
  \bibfield  {author} {\bibinfo {author} {\bibfnamefont {A.}~\bibnamefont
  {Vasudevan}}, \bibinfo {author} {\bibfnamefont {M.~M.}\ \bibnamefont
  {Ivanov}}, \bibinfo {author} {\bibfnamefont {S.}~\bibnamefont {Sibiryakov}},
  \ and\ \bibinfo {author} {\bibfnamefont {J.}~\bibnamefont {Lesgourgues}},\
  }\href {\doibase 10.1088/1475-7516/2019/09/037} {\bibfield  {journal}
  {\bibinfo  {journal} {JCAP}\ }\textbf {\bibinfo {volume} {09}},\ \bibinfo
  {pages} {037} (\bibinfo {year} {2019})},\ \Eprint
  {http://arxiv.org/abs/1906.08697} {arXiv:1906.08697 [astro-ph.CO]}
  \BibitemShut {NoStop}%
\bibitem [{\citenamefont {Schmidt}(2021{\natexlab{a}})}]{Schmidt:2020ovm}%
  \BibitemOpen
  \bibfield  {author} {\bibinfo {author} {\bibfnamefont {F.}~\bibnamefont
  {Schmidt}},\ }\href {\doibase 10.1088/1475-7516/2021/04/033} {\bibfield
  {journal} {\bibinfo  {journal} {JCAP}\ }\textbf {\bibinfo {volume} {04}},\
  \bibinfo {pages} {033} (\bibinfo {year} {2021}{\natexlab{a}})},\ \Eprint
  {http://arxiv.org/abs/2012.09837} {arXiv:2012.09837 [astro-ph.CO]}
  \BibitemShut {NoStop}%
\bibitem [{\citenamefont {Cabass}\ and\ \citenamefont
  {Schmidt}(2020{\natexlab{b}})}]{Cabass:2020nwf}%
  \BibitemOpen
  \bibfield  {author} {\bibinfo {author} {\bibfnamefont {G.}~\bibnamefont
  {Cabass}}\ and\ \bibinfo {author} {\bibfnamefont {F.}~\bibnamefont
  {Schmidt}},\ }\href {\doibase 10.1088/1475-7516/2020/07/051} {\bibfield
  {journal} {\bibinfo  {journal} {JCAP}\ }\textbf {\bibinfo {volume} {07}},\
  \bibinfo {pages} {051} (\bibinfo {year} {2020}{\natexlab{b}})},\ \Eprint
  {http://arxiv.org/abs/2004.00617} {arXiv:2004.00617 [astro-ph.CO]}
  \BibitemShut {NoStop}%
\bibitem [{\citenamefont {Schmidt}\ \emph {et~al.}(2020)\citenamefont
  {Schmidt}, \citenamefont {Cabass}, \citenamefont {Jasche},\ and\
  \citenamefont {Lavaux}}]{Schmidt:2020viy}%
  \BibitemOpen
  \bibfield  {author} {\bibinfo {author} {\bibfnamefont {F.}~\bibnamefont
  {Schmidt}}, \bibinfo {author} {\bibfnamefont {G.}~\bibnamefont {Cabass}},
  \bibinfo {author} {\bibfnamefont {J.}~\bibnamefont {Jasche}}, \ and\ \bibinfo
  {author} {\bibfnamefont {G.}~\bibnamefont {Lavaux}},\ }\href {\doibase
  10.1088/1475-7516/2020/11/008} {\bibfield  {journal} {\bibinfo  {journal}
  {JCAP}\ }\textbf {\bibinfo {volume} {11}},\ \bibinfo {pages} {008} (\bibinfo
  {year} {2020})},\ \Eprint {http://arxiv.org/abs/2004.06707} {arXiv:2004.06707
  [astro-ph.CO]} \BibitemShut {NoStop}%
\bibitem [{\citenamefont {Schmidt}(2021{\natexlab{b}})}]{Schmidt:2020tao}%
  \BibitemOpen
  \bibfield  {author} {\bibinfo {author} {\bibfnamefont {F.}~\bibnamefont
  {Schmidt}},\ }\href {\doibase 10.1088/1475-7516/2021/04/032} {\bibfield
  {journal} {\bibinfo  {journal} {JCAP}\ }\textbf {\bibinfo {volume} {04}},\
  \bibinfo {pages} {032} (\bibinfo {year} {2021}{\natexlab{b}})},\ \Eprint
  {http://arxiv.org/abs/2009.14176} {arXiv:2009.14176 [astro-ph.CO]}
  \BibitemShut {NoStop}%
\bibitem [{\citenamefont {Nguyen}\ \emph {et~al.}(2021)\citenamefont {Nguyen},
  \citenamefont {Schmidt}, \citenamefont {Lavaux},\ and\ \citenamefont
  {Jasche}}]{Nguyen:2020hxe}%
  \BibitemOpen
  \bibfield  {author} {\bibinfo {author} {\bibfnamefont {N.-M.}\ \bibnamefont
  {Nguyen}}, \bibinfo {author} {\bibfnamefont {F.}~\bibnamefont {Schmidt}},
  \bibinfo {author} {\bibfnamefont {G.}~\bibnamefont {Lavaux}}, \ and\ \bibinfo
  {author} {\bibfnamefont {J.}~\bibnamefont {Jasche}},\ }\href {\doibase
  10.1088/1475-7516/2021/03/058} {\bibfield  {journal} {\bibinfo  {journal}
  {JCAP}\ }\textbf {\bibinfo {volume} {03}},\ \bibinfo {pages} {058} (\bibinfo
  {year} {2021})},\ \Eprint {http://arxiv.org/abs/2011.06587} {arXiv:2011.06587
  [astro-ph.CO]} \BibitemShut {NoStop}%
\bibitem [{\citenamefont {Cabass}(2021)}]{Cabass:2020jqo}%
  \BibitemOpen
  \bibfield  {author} {\bibinfo {author} {\bibfnamefont {G.}~\bibnamefont
  {Cabass}},\ }\href {\doibase 10.1088/1475-7516/2021/01/067} {\bibfield
  {journal} {\bibinfo  {journal} {JCAP}\ }\textbf {\bibinfo {volume} {01}},\
  \bibinfo {pages} {067} (\bibinfo {year} {2021})},\ \Eprint
  {http://arxiv.org/abs/2007.14988} {arXiv:2007.14988 [astro-ph.CO]}
  \BibitemShut {NoStop}%
\bibitem [{\citenamefont {Fujita}\ and\ \citenamefont
  {Vlah}(2020)}]{Fujita:2020xtd}%
  \BibitemOpen
  \bibfield  {author} {\bibinfo {author} {\bibfnamefont {T.}~\bibnamefont
  {Fujita}}\ and\ \bibinfo {author} {\bibfnamefont {Z.}~\bibnamefont {Vlah}},\
  }\href {\doibase 10.1088/1475-7516/2020/10/059} {\bibfield  {journal}
  {\bibinfo  {journal} {JCAP}\ }\textbf {\bibinfo {volume} {10}},\ \bibinfo
  {pages} {059} (\bibinfo {year} {2020})},\ \Eprint
  {http://arxiv.org/abs/2003.10114} {arXiv:2003.10114 [astro-ph.CO]}
  \BibitemShut {NoStop}%
\bibitem [{\citenamefont {Nishimichi}\ \emph {et~al.}(2020)\citenamefont
  {Nishimichi}, \citenamefont {D'Amico}, \citenamefont {Ivanov}, \citenamefont
  {Senatore}, \citenamefont {Simonovi\'c}, \citenamefont {Takada},
  \citenamefont {Zaldarriaga},\ and\ \citenamefont
  {Zhang}}]{Nishimichi:2020tvu}%
  \BibitemOpen
  \bibfield  {author} {\bibinfo {author} {\bibfnamefont {T.}~\bibnamefont
  {Nishimichi}}, \bibinfo {author} {\bibfnamefont {G.}~\bibnamefont {D'Amico}},
  \bibinfo {author} {\bibfnamefont {M.~M.}\ \bibnamefont {Ivanov}}, \bibinfo
  {author} {\bibfnamefont {L.}~\bibnamefont {Senatore}}, \bibinfo {author}
  {\bibfnamefont {M.}~\bibnamefont {Simonovi\'c}}, \bibinfo {author}
  {\bibfnamefont {M.}~\bibnamefont {Takada}}, \bibinfo {author} {\bibfnamefont
  {M.}~\bibnamefont {Zaldarriaga}}, \ and\ \bibinfo {author} {\bibfnamefont
  {P.}~\bibnamefont {Zhang}},\ }\href {\doibase 10.1103/PhysRevD.102.123541}
  {\bibfield  {journal} {\bibinfo  {journal} {Phys. Rev. D}\ }\textbf {\bibinfo
  {volume} {102}},\ \bibinfo {pages} {123541} (\bibinfo {year} {2020})},\
  \Eprint {http://arxiv.org/abs/2003.08277} {arXiv:2003.08277 [astro-ph.CO]}
  \BibitemShut {NoStop}%
\bibitem [{\citenamefont {Chudaykin}\ \emph {et~al.}(2021)\citenamefont
  {Chudaykin}, \citenamefont {Ivanov},\ and\ \citenamefont
  {Simonovi\'c}}]{Chudaykin:2020hbf}%
  \BibitemOpen
  \bibfield  {author} {\bibinfo {author} {\bibfnamefont {A.}~\bibnamefont
  {Chudaykin}}, \bibinfo {author} {\bibfnamefont {M.~M.}\ \bibnamefont
  {Ivanov}}, \ and\ \bibinfo {author} {\bibfnamefont {M.}~\bibnamefont
  {Simonovi\'c}},\ }\href {\doibase 10.1103/PhysRevD.103.043525} {\bibfield
  {journal} {\bibinfo  {journal} {Phys. Rev. D}\ }\textbf {\bibinfo {volume}
  {103}},\ \bibinfo {pages} {043525} (\bibinfo {year} {2021})},\ \Eprint
  {http://arxiv.org/abs/2009.10724} {arXiv:2009.10724 [astro-ph.CO]}
  \BibitemShut {NoStop}%
\bibitem [{\citenamefont {Chen}\ \emph
  {et~al.}(2021{\natexlab{a}})\citenamefont {Chen}, \citenamefont {Vlah},
  \citenamefont {Castorina},\ and\ \citenamefont {White}}]{Chen:2020zjt}%
  \BibitemOpen
  \bibfield  {author} {\bibinfo {author} {\bibfnamefont {S.-F.}\ \bibnamefont
  {Chen}}, \bibinfo {author} {\bibfnamefont {Z.}~\bibnamefont {Vlah}}, \bibinfo
  {author} {\bibfnamefont {E.}~\bibnamefont {Castorina}}, \ and\ \bibinfo
  {author} {\bibfnamefont {M.}~\bibnamefont {White}},\ }\href {\doibase
  10.1088/1475-7516/2021/03/100} {\bibfield  {journal} {\bibinfo  {journal}
  {JCAP}\ }\textbf {\bibinfo {volume} {03}},\ \bibinfo {pages} {100} (\bibinfo
  {year} {2021}{\natexlab{a}})},\ \Eprint {http://arxiv.org/abs/2012.04636}
  {arXiv:2012.04636 [astro-ph.CO]} \BibitemShut {NoStop}%
\bibitem [{\citenamefont {Chen}\ \emph
  {et~al.}(2020{\natexlab{a}})\citenamefont {Chen}, \citenamefont {Vlah},\ and\
  \citenamefont {White}}]{Chen:2020fxs}%
  \BibitemOpen
  \bibfield  {author} {\bibinfo {author} {\bibfnamefont {S.-F.}\ \bibnamefont
  {Chen}}, \bibinfo {author} {\bibfnamefont {Z.}~\bibnamefont {Vlah}}, \ and\
  \bibinfo {author} {\bibfnamefont {M.}~\bibnamefont {White}},\ }\href
  {\doibase 10.1088/1475-7516/2020/07/062} {\bibfield  {journal} {\bibinfo
  {journal} {JCAP}\ }\textbf {\bibinfo {volume} {07}},\ \bibinfo {pages} {062}
  (\bibinfo {year} {2020}{\natexlab{a}})},\ \Eprint
  {http://arxiv.org/abs/2005.00523} {arXiv:2005.00523 [astro-ph.CO]}
  \BibitemShut {NoStop}%
\bibitem [{\citenamefont {Donath}\ and\ \citenamefont
  {Senatore}(2020)}]{Donath:2020abv}%
  \BibitemOpen
  \bibfield  {author} {\bibinfo {author} {\bibfnamefont {Y.}~\bibnamefont
  {Donath}}\ and\ \bibinfo {author} {\bibfnamefont {L.}~\bibnamefont
  {Senatore}},\ }\href {\doibase 10.1088/1475-7516/2020/10/039} {\bibfield
  {journal} {\bibinfo  {journal} {JCAP}\ }\textbf {\bibinfo {volume} {10}},\
  \bibinfo {pages} {039} (\bibinfo {year} {2020})},\ \Eprint
  {http://arxiv.org/abs/2005.04805} {arXiv:2005.04805 [astro-ph.CO]}
  \BibitemShut {NoStop}%
\bibitem [{\citenamefont {Bragan\c{c}a}\ \emph {et~al.}(2021)\citenamefont
  {Bragan\c{c}a}, \citenamefont {Lewandowski}, \citenamefont {Sekera},
  \citenamefont {Senatore},\ and\ \citenamefont {Sgier}}]{Braganca:2020nhv}%
  \BibitemOpen
  \bibfield  {author} {\bibinfo {author} {\bibfnamefont {D.~P.~L.}\
  \bibnamefont {Bragan\c{c}a}}, \bibinfo {author} {\bibfnamefont
  {M.}~\bibnamefont {Lewandowski}}, \bibinfo {author} {\bibfnamefont
  {D.}~\bibnamefont {Sekera}}, \bibinfo {author} {\bibfnamefont
  {L.}~\bibnamefont {Senatore}}, \ and\ \bibinfo {author} {\bibfnamefont
  {R.}~\bibnamefont {Sgier}},\ }\href {\doibase 10.1088/1475-7516/2021/10/074}
  {\bibfield  {journal} {\bibinfo  {journal} {JCAP}\ }\textbf {\bibinfo
  {volume} {10}},\ \bibinfo {pages} {074} (\bibinfo {year} {2021})},\ \Eprint
  {http://arxiv.org/abs/2010.02929} {arXiv:2010.02929 [astro-ph.CO]}
  \BibitemShut {NoStop}%
\bibitem [{\citenamefont {Steele}\ and\ \citenamefont
  {Baldauf}(2021{\natexlab{a}})}]{Steele:2020tak}%
  \BibitemOpen
  \bibfield  {author} {\bibinfo {author} {\bibfnamefont {T.}~\bibnamefont
  {Steele}}\ and\ \bibinfo {author} {\bibfnamefont {T.}~\bibnamefont
  {Baldauf}},\ }\href {\doibase 10.1103/PhysRevD.103.023520} {\bibfield
  {journal} {\bibinfo  {journal} {Phys. Rev. D}\ }\textbf {\bibinfo {volume}
  {103}},\ \bibinfo {pages} {023520} (\bibinfo {year} {2021}{\natexlab{a}})},\
  \Eprint {http://arxiv.org/abs/2009.01200} {arXiv:2009.01200 [astro-ph.CO]}
  \BibitemShut {NoStop}%
\bibitem [{\citenamefont {Schmittfull}\ \emph {et~al.}(2021)\citenamefont
  {Schmittfull}, \citenamefont {Simonovi\'c}, \citenamefont {Ivanov},
  \citenamefont {Philcox},\ and\ \citenamefont
  {Zaldarriaga}}]{Schmittfull:2020trd}%
  \BibitemOpen
  \bibfield  {author} {\bibinfo {author} {\bibfnamefont {M.}~\bibnamefont
  {Schmittfull}}, \bibinfo {author} {\bibfnamefont {M.}~\bibnamefont
  {Simonovi\'c}}, \bibinfo {author} {\bibfnamefont {M.~M.}\ \bibnamefont
  {Ivanov}}, \bibinfo {author} {\bibfnamefont {O.~H.~E.}\ \bibnamefont
  {Philcox}}, \ and\ \bibinfo {author} {\bibfnamefont {M.}~\bibnamefont
  {Zaldarriaga}},\ }\href {\doibase 10.1088/1475-7516/2021/05/059} {\bibfield
  {journal} {\bibinfo  {journal} {JCAP}\ }\textbf {\bibinfo {volume} {05}},\
  \bibinfo {pages} {059} (\bibinfo {year} {2021})},\ \Eprint
  {http://arxiv.org/abs/2012.03334} {arXiv:2012.03334 [astro-ph.CO]}
  \BibitemShut {NoStop}%
\bibitem [{\citenamefont {Chen}\ \emph
  {et~al.}(2020{\natexlab{b}})\citenamefont {Chen}, \citenamefont {Vlah},\ and\
  \citenamefont {White}}]{Chen:2020ckc}%
  \BibitemOpen
  \bibfield  {author} {\bibinfo {author} {\bibfnamefont {S.-F.}\ \bibnamefont
  {Chen}}, \bibinfo {author} {\bibfnamefont {Z.}~\bibnamefont {Vlah}}, \ and\
  \bibinfo {author} {\bibfnamefont {M.}~\bibnamefont {White}},\ }\href
  {\doibase 10.1088/1475-7516/2020/11/035} {\bibfield  {journal} {\bibinfo
  {journal} {JCAP}\ }\textbf {\bibinfo {volume} {11}},\ \bibinfo {pages} {035}
  (\bibinfo {year} {2020}{\natexlab{b}})},\ \Eprint
  {http://arxiv.org/abs/2007.00704} {arXiv:2007.00704 [astro-ph.CO]}
  \BibitemShut {NoStop}%
\bibitem [{\citenamefont {D'Amico}\ \emph {et~al.}(2021)\citenamefont
  {D'Amico}, \citenamefont {Marinucci}, \citenamefont {Pietroni},\ and\
  \citenamefont {Vernizzi}}]{DAmico:2021rdb}%
  \BibitemOpen
  \bibfield  {author} {\bibinfo {author} {\bibfnamefont {G.}~\bibnamefont
  {D'Amico}}, \bibinfo {author} {\bibfnamefont {M.}~\bibnamefont {Marinucci}},
  \bibinfo {author} {\bibfnamefont {M.}~\bibnamefont {Pietroni}}, \ and\
  \bibinfo {author} {\bibfnamefont {F.}~\bibnamefont {Vernizzi}},\ }\href
  {\doibase 10.1088/1475-7516/2021/10/069} {\bibfield  {journal} {\bibinfo
  {journal} {JCAP}\ }\textbf {\bibinfo {volume} {10}},\ \bibinfo {pages} {069}
  (\bibinfo {year} {2021})},\ \Eprint {http://arxiv.org/abs/2109.09573}
  {arXiv:2109.09573 [astro-ph.CO]} \BibitemShut {NoStop}%
\bibitem [{\citenamefont {Steele}\ and\ \citenamefont
  {Baldauf}(2021{\natexlab{b}})}]{Steele:2021lnz}%
  \BibitemOpen
  \bibfield  {author} {\bibinfo {author} {\bibfnamefont {T.}~\bibnamefont
  {Steele}}\ and\ \bibinfo {author} {\bibfnamefont {T.}~\bibnamefont
  {Baldauf}},\ }\href {\doibase 10.1103/PhysRevD.103.103518} {\bibfield
  {journal} {\bibinfo  {journal} {Phys. Rev. D}\ }\textbf {\bibinfo {volume}
  {103}},\ \bibinfo {pages} {103518} (\bibinfo {year} {2021}{\natexlab{b}})},\
  \Eprint {http://arxiv.org/abs/2101.10289} {arXiv:2101.10289 [astro-ph.CO]}
  \BibitemShut {NoStop}%
\bibitem [{\citenamefont {Ivanov}\ \emph {et~al.}(2022)\citenamefont {Ivanov},
  \citenamefont {Philcox}, \citenamefont {Nishimichi}, \citenamefont
  {Simonovi\'c}, \citenamefont {Takada},\ and\ \citenamefont
  {Zaldarriaga}}]{Ivanov:2021kcd}%
  \BibitemOpen
  \bibfield  {author} {\bibinfo {author} {\bibfnamefont {M.~M.}\ \bibnamefont
  {Ivanov}}, \bibinfo {author} {\bibfnamefont {O.~H.~E.}\ \bibnamefont
  {Philcox}}, \bibinfo {author} {\bibfnamefont {T.}~\bibnamefont {Nishimichi}},
  \bibinfo {author} {\bibfnamefont {M.}~\bibnamefont {Simonovi\'c}}, \bibinfo
  {author} {\bibfnamefont {M.}~\bibnamefont {Takada}}, \ and\ \bibinfo {author}
  {\bibfnamefont {M.}~\bibnamefont {Zaldarriaga}},\ }\href {\doibase
  10.1103/PhysRevD.105.063512} {\bibfield  {journal} {\bibinfo  {journal}
  {Phys. Rev. D}\ }\textbf {\bibinfo {volume} {105}},\ \bibinfo {pages}
  {063512} (\bibinfo {year} {2022})},\ \Eprint
  {http://arxiv.org/abs/2110.10161} {arXiv:2110.10161 [astro-ph.CO]}
  \BibitemShut {NoStop}%
\bibitem [{\citenamefont {Baldauf}\ \emph {et~al.}(2021)\citenamefont
  {Baldauf}, \citenamefont {Garny}, \citenamefont {Taule},\ and\ \citenamefont
  {Steele}}]{Baldauf:2021zlt}%
  \BibitemOpen
  \bibfield  {author} {\bibinfo {author} {\bibfnamefont {T.}~\bibnamefont
  {Baldauf}}, \bibinfo {author} {\bibfnamefont {M.}~\bibnamefont {Garny}},
  \bibinfo {author} {\bibfnamefont {P.}~\bibnamefont {Taule}}, \ and\ \bibinfo
  {author} {\bibfnamefont {T.}~\bibnamefont {Steele}},\ }\href {\doibase
  10.1103/PhysRevD.104.123551} {\bibfield  {journal} {\bibinfo  {journal}
  {Phys. Rev. D}\ }\textbf {\bibinfo {volume} {104}},\ \bibinfo {pages}
  {123551} (\bibinfo {year} {2021})},\ \Eprint
  {http://arxiv.org/abs/2110.13930} {arXiv:2110.13930 [astro-ph.CO]}
  \BibitemShut {NoStop}%
\bibitem [{\citenamefont {Ivanov}(2023)}]{Ivanov:2022mrd}%
  \BibitemOpen
  \bibfield  {author} {\bibinfo {author} {\bibfnamefont {M.~M.}\ \bibnamefont
  {Ivanov}},\ }\enquote {\bibinfo {title} {{Effective Field Theory for
  Large-Scale Structure}},}\ \ (\bibinfo {year} {2023})\ \Eprint
  {http://arxiv.org/abs/2212.08488} {arXiv:2212.08488 [astro-ph.CO]}
  \BibitemShut {NoStop}%
\bibitem [{\citenamefont {Senatore}\ \emph {et~al.}(2010)\citenamefont
  {Senatore}, \citenamefont {Smith},\ and\ \citenamefont
  {Zaldarriaga}}]{Senatore:2009gt}%
  \BibitemOpen
  \bibfield  {author} {\bibinfo {author} {\bibfnamefont {L.}~\bibnamefont
  {Senatore}}, \bibinfo {author} {\bibfnamefont {K.~M.}\ \bibnamefont {Smith}},
  \ and\ \bibinfo {author} {\bibfnamefont {M.}~\bibnamefont {Zaldarriaga}},\
  }\href {\doibase 10.1088/1475-7516/2010/01/028} {\bibfield  {journal}
  {\bibinfo  {journal} {JCAP}\ }\textbf {\bibinfo {volume} {01}},\ \bibinfo
  {pages} {028} (\bibinfo {year} {2010})},\ \Eprint
  {http://arxiv.org/abs/0905.3746} {arXiv:0905.3746 [astro-ph.CO]} \BibitemShut
  {NoStop}%
\bibitem [{\citenamefont {Cahn}\ \emph {et~al.}(2023)\citenamefont {Cahn},
  \citenamefont {Slepian},\ and\ \citenamefont {Hou}}]{Cahn:2021ltp}%
  \BibitemOpen
  \bibfield  {author} {\bibinfo {author} {\bibfnamefont {R.~N.}\ \bibnamefont
  {Cahn}}, \bibinfo {author} {\bibfnamefont {Z.}~\bibnamefont {Slepian}}, \
  and\ \bibinfo {author} {\bibfnamefont {J.}~\bibnamefont {Hou}},\ }\href
  {\doibase 10.1103/PhysRevLett.130.201002} {\bibfield  {journal} {\bibinfo
  {journal} {Phys. Rev. Lett.}\ }\textbf {\bibinfo {volume} {130}},\ \bibinfo
  {pages} {201002} (\bibinfo {year} {2023})},\ \Eprint
  {http://arxiv.org/abs/2110.12004} {arXiv:2110.12004 [astro-ph.CO]}
  \BibitemShut {NoStop}%
\bibitem [{\citenamefont {Hou}\ \emph {et~al.}(2023)\citenamefont {Hou},
  \citenamefont {Slepian},\ and\ \citenamefont {Cahn}}]{Hou:2022wfj}%
  \BibitemOpen
  \bibfield  {author} {\bibinfo {author} {\bibfnamefont {J.}~\bibnamefont
  {Hou}}, \bibinfo {author} {\bibfnamefont {Z.}~\bibnamefont {Slepian}}, \ and\
  \bibinfo {author} {\bibfnamefont {R.~N.}\ \bibnamefont {Cahn}},\ }\href
  {\doibase 10.1093/mnras/stad1062} {\bibfield  {journal} {\bibinfo  {journal}
  {Mon. Not. Roy. Astron. Soc.}\ }\textbf {\bibinfo {volume} {522}},\ \bibinfo
  {pages} {5701} (\bibinfo {year} {2023})},\ \Eprint
  {http://arxiv.org/abs/2206.03625} {arXiv:2206.03625 [astro-ph.CO]}
  \BibitemShut {NoStop}%
\bibitem [{\citenamefont {Philcox}(2022)}]{Philcox:2022hkh}%
  \BibitemOpen
  \bibfield  {author} {\bibinfo {author} {\bibfnamefont {O.~H.~E.}\
  \bibnamefont {Philcox}},\ }\href {\doibase 10.1103/PhysRevD.106.063501}
  {\bibfield  {journal} {\bibinfo  {journal} {Phys. Rev. D}\ }\textbf {\bibinfo
  {volume} {106}},\ \bibinfo {pages} {063501} (\bibinfo {year} {2022})},\
  \Eprint {http://arxiv.org/abs/2206.04227} {arXiv:2206.04227 [astro-ph.CO]}
  \BibitemShut {NoStop}%
\bibitem [{\citenamefont {Cabass}\ \emph
  {et~al.}(2023{\natexlab{c}})\citenamefont {Cabass}, \citenamefont {Ivanov},\
  and\ \citenamefont {Philcox}}]{Cabass:2022oap}%
  \BibitemOpen
  \bibfield  {author} {\bibinfo {author} {\bibfnamefont {G.}~\bibnamefont
  {Cabass}}, \bibinfo {author} {\bibfnamefont {M.~M.}\ \bibnamefont {Ivanov}},
  \ and\ \bibinfo {author} {\bibfnamefont {O.~H.~E.}\ \bibnamefont {Philcox}},\
  }\href {\doibase 10.1103/PhysRevD.107.023523} {\bibfield  {journal} {\bibinfo
   {journal} {Phys. Rev. D}\ }\textbf {\bibinfo {volume} {107}},\ \bibinfo
  {pages} {023523} (\bibinfo {year} {2023}{\natexlab{c}})},\ \Eprint
  {http://arxiv.org/abs/2210.16320} {arXiv:2210.16320 [astro-ph.CO]}
  \BibitemShut {NoStop}%
\bibitem [{\citenamefont {Philcox}(2023{\natexlab{a}})}]{Philcox:2023ffy}%
  \BibitemOpen
  \bibfield  {author} {\bibinfo {author} {\bibfnamefont {O.~H.~E.}\
  \bibnamefont {Philcox}},\ }\href {\doibase 10.1103/PhysRevLett.131.181001}
  {\bibfield  {journal} {\bibinfo  {journal} {Phys. Rev. Lett.}\ }\textbf
  {\bibinfo {volume} {131}},\ \bibinfo {pages} {181001} (\bibinfo {year}
  {2023}{\natexlab{a}})},\ \Eprint {http://arxiv.org/abs/2303.12106}
  {arXiv:2303.12106 [astro-ph.CO]} \BibitemShut {NoStop}%
\bibitem [{\citenamefont {Cabass}\ \emph
  {et~al.}(2022{\natexlab{b}})\citenamefont {Cabass}, \citenamefont {Ivanov},
  \citenamefont {Philcox}, \citenamefont {Simonovi\'c},\ and\ \citenamefont
  {Zaldarriaga}}]{Cabass:2022wjy}%
  \BibitemOpen
  \bibfield  {author} {\bibinfo {author} {\bibfnamefont {G.}~\bibnamefont
  {Cabass}}, \bibinfo {author} {\bibfnamefont {M.~M.}\ \bibnamefont {Ivanov}},
  \bibinfo {author} {\bibfnamefont {O.~H.~E.}\ \bibnamefont {Philcox}},
  \bibinfo {author} {\bibfnamefont {M.}~\bibnamefont {Simonovi\'c}}, \ and\
  \bibinfo {author} {\bibfnamefont {M.}~\bibnamefont {Zaldarriaga}},\ }\href
  {\doibase 10.1103/PhysRevLett.129.021301} {\bibfield  {journal} {\bibinfo
  {journal} {Phys. Rev. Lett.}\ }\textbf {\bibinfo {volume} {129}},\ \bibinfo
  {pages} {021301} (\bibinfo {year} {2022}{\natexlab{b}})},\ \Eprint
  {http://arxiv.org/abs/2201.07238} {arXiv:2201.07238 [astro-ph.CO]}
  \BibitemShut {NoStop}%
\bibitem [{\citenamefont {D'Amico}\ \emph
  {et~al.}(2022{\natexlab{a}})\citenamefont {D'Amico}, \citenamefont
  {Lewandowski}, \citenamefont {Senatore},\ and\ \citenamefont
  {Zhang}}]{DAmico:2022gki}%
  \BibitemOpen
  \bibfield  {author} {\bibinfo {author} {\bibfnamefont {G.}~\bibnamefont
  {D'Amico}}, \bibinfo {author} {\bibfnamefont {M.}~\bibnamefont
  {Lewandowski}}, \bibinfo {author} {\bibfnamefont {L.}~\bibnamefont
  {Senatore}}, \ and\ \bibinfo {author} {\bibfnamefont {P.}~\bibnamefont
  {Zhang}},\ }\href@noop {} {\  (\bibinfo {year} {2022}{\natexlab{a}})},\
  \Eprint {http://arxiv.org/abs/2201.11518} {arXiv:2201.11518 [astro-ph.CO]}
  \BibitemShut {NoStop}%
\bibitem [{\citenamefont {Cabass}\ \emph
  {et~al.}(2022{\natexlab{c}})\citenamefont {Cabass}, \citenamefont {Ivanov},
  \citenamefont {Philcox}, \citenamefont {Simonovi\'c},\ and\ \citenamefont
  {Zaldarriaga}}]{Cabass:2022ymb}%
  \BibitemOpen
  \bibfield  {author} {\bibinfo {author} {\bibfnamefont {G.}~\bibnamefont
  {Cabass}}, \bibinfo {author} {\bibfnamefont {M.~M.}\ \bibnamefont {Ivanov}},
  \bibinfo {author} {\bibfnamefont {O.~H.~E.}\ \bibnamefont {Philcox}},
  \bibinfo {author} {\bibfnamefont {M.}~\bibnamefont {Simonovi\'c}}, \ and\
  \bibinfo {author} {\bibfnamefont {M.}~\bibnamefont {Zaldarriaga}},\ }\href
  {\doibase 10.1103/PhysRevD.106.043506} {\bibfield  {journal} {\bibinfo
  {journal} {Phys. Rev. D}\ }\textbf {\bibinfo {volume} {106}},\ \bibinfo
  {pages} {043506} (\bibinfo {year} {2022}{\natexlab{c}})},\ \Eprint
  {http://arxiv.org/abs/2204.01781} {arXiv:2204.01781 [astro-ph.CO]}
  \BibitemShut {NoStop}%
\bibitem [{\citenamefont {Xianyu}\ and\ \citenamefont
  {Zang}(2024)}]{Xianyu:2023ytd}%
  \BibitemOpen
  \bibfield  {author} {\bibinfo {author} {\bibfnamefont {Z.-Z.}\ \bibnamefont
  {Xianyu}}\ and\ \bibinfo {author} {\bibfnamefont {J.}~\bibnamefont {Zang}},\
  }\href {\doibase 10.1007/JHEP03(2024)070} {\bibfield  {journal} {\bibinfo
  {journal} {JHEP}\ }\textbf {\bibinfo {volume} {03}},\ \bibinfo {pages} {070}
  (\bibinfo {year} {2024})},\ \Eprint {http://arxiv.org/abs/2309.10849}
  {arXiv:2309.10849 [hep-th]} \BibitemShut {NoStop}%
\bibitem [{\citenamefont {Qin}\ and\ \citenamefont
  {Xianyu}(2023)}]{Qin:2023ejc}%
  \BibitemOpen
  \bibfield  {author} {\bibinfo {author} {\bibfnamefont {Z.}~\bibnamefont
  {Qin}}\ and\ \bibinfo {author} {\bibfnamefont {Z.-Z.}\ \bibnamefont
  {Xianyu}},\ }\href {\doibase 10.1007/JHEP07(2023)001} {\bibfield  {journal}
  {\bibinfo  {journal} {JHEP}\ }\textbf {\bibinfo {volume} {07}},\ \bibinfo
  {pages} {001} (\bibinfo {year} {2023})},\ \Eprint
  {http://arxiv.org/abs/2301.07047} {arXiv:2301.07047 [hep-th]} \BibitemShut
  {NoStop}%
\bibitem [{\citenamefont {Pinol}\ \emph
  {et~al.}(2023{\natexlab{b}})\citenamefont {Pinol}, \citenamefont
  {Renaux-Petel},\ and\ \citenamefont {Werth}}]{Pinol:2023oux}%
  \BibitemOpen
  \bibfield  {author} {\bibinfo {author} {\bibfnamefont {L.}~\bibnamefont
  {Pinol}}, \bibinfo {author} {\bibfnamefont {S.}~\bibnamefont {Renaux-Petel}},
  \ and\ \bibinfo {author} {\bibfnamefont {D.}~\bibnamefont {Werth}},\
  }\href@noop {} {\  (\bibinfo {year} {2023}{\natexlab{b}})},\ \Eprint
  {http://arxiv.org/abs/2312.06559} {arXiv:2312.06559 [astro-ph.CO]}
  \BibitemShut {NoStop}%
\bibitem [{\citenamefont {Werth}\ \emph
  {et~al.}(2024{\natexlab{a}})\citenamefont {Werth}, \citenamefont {Pinol},\
  and\ \citenamefont {Renaux-Petel}}]{Werth:2023pfl}%
  \BibitemOpen
  \bibfield  {author} {\bibinfo {author} {\bibfnamefont {D.}~\bibnamefont
  {Werth}}, \bibinfo {author} {\bibfnamefont {L.}~\bibnamefont {Pinol}}, \ and\
  \bibinfo {author} {\bibfnamefont {S.}~\bibnamefont {Renaux-Petel}},\ }\href
  {\doibase 10.1103/PhysRevLett.133.141002} {\bibfield  {journal} {\bibinfo
  {journal} {Phys. Rev. Lett.}\ }\textbf {\bibinfo {volume} {133}},\ \bibinfo
  {pages} {141002} (\bibinfo {year} {2024}{\natexlab{a}})},\ \Eprint
  {http://arxiv.org/abs/2302.00655} {arXiv:2302.00655 [hep-th]} \BibitemShut
  {NoStop}%
\bibitem [{\citenamefont {Jazayeri}\ \emph {et~al.}(2023)\citenamefont
  {Jazayeri}, \citenamefont {Renaux-Petel},\ and\ \citenamefont
  {Werth}}]{Jazayeri:2023xcj}%
  \BibitemOpen
  \bibfield  {author} {\bibinfo {author} {\bibfnamefont {S.}~\bibnamefont
  {Jazayeri}}, \bibinfo {author} {\bibfnamefont {S.}~\bibnamefont
  {Renaux-Petel}}, \ and\ \bibinfo {author} {\bibfnamefont {D.}~\bibnamefont
  {Werth}},\ }\href {\doibase 10.1088/1475-7516/2023/12/035} {\bibfield
  {journal} {\bibinfo  {journal} {JCAP}\ }\textbf {\bibinfo {volume} {12}},\
  \bibinfo {pages} {035} (\bibinfo {year} {2023})},\ \Eprint
  {http://arxiv.org/abs/2307.01751} {arXiv:2307.01751 [hep-th]} \BibitemShut
  {NoStop}%
\bibitem [{\citenamefont {Chakraborty}\ and\ \citenamefont
  {Stout}(2024)}]{Chakraborty:2023eoq}%
  \BibitemOpen
  \bibfield  {author} {\bibinfo {author} {\bibfnamefont {P.}~\bibnamefont
  {Chakraborty}}\ and\ \bibinfo {author} {\bibfnamefont {J.}~\bibnamefont
  {Stout}},\ }\href {\doibase 10.1007/JHEP03(2024)149} {\bibfield  {journal}
  {\bibinfo  {journal} {JHEP}\ }\textbf {\bibinfo {volume} {03}},\ \bibinfo
  {pages} {149} (\bibinfo {year} {2024})},\ \Eprint
  {http://arxiv.org/abs/2311.09219} {arXiv:2311.09219 [hep-th]} \BibitemShut
  {NoStop}%
\bibitem [{\citenamefont {Werth}\ \emph
  {et~al.}(2024{\natexlab{b}})\citenamefont {Werth}, \citenamefont {Pinol},\
  and\ \citenamefont {Renaux-Petel}}]{Werth:2024aui}%
  \BibitemOpen
  \bibfield  {author} {\bibinfo {author} {\bibfnamefont {D.}~\bibnamefont
  {Werth}}, \bibinfo {author} {\bibfnamefont {L.}~\bibnamefont {Pinol}}, \ and\
  \bibinfo {author} {\bibfnamefont {S.}~\bibnamefont {Renaux-Petel}},\ }\href
  {\doibase 10.1088/1361-6382/ad6740} {\bibfield  {journal} {\bibinfo
  {journal} {Class. Quant. Grav.}\ }\textbf {\bibinfo {volume} {41}},\ \bibinfo
  {pages} {175015} (\bibinfo {year} {2024}{\natexlab{b}})},\ \Eprint
  {http://arxiv.org/abs/2402.03693} {arXiv:2402.03693 [astro-ph.CO]}
  \BibitemShut {NoStop}%
\bibitem [{\citenamefont {Philcox}\ and\ \citenamefont
  {Ivanov}(2021)}]{Philcox:2021kcw}%
  \BibitemOpen
  \bibfield  {author} {\bibinfo {author} {\bibfnamefont {O.~H.~E.}\
  \bibnamefont {Philcox}}\ and\ \bibinfo {author} {\bibfnamefont {M.~M.}\
  \bibnamefont {Ivanov}},\ }\href@noop {} {\  (\bibinfo {year} {2021})},\
  \Eprint {http://arxiv.org/abs/2112.04515} {arXiv:2112.04515 [astro-ph.CO]}
  \BibitemShut {NoStop}%
\bibitem [{\citenamefont {Ivanov}\ \emph {et~al.}(2023)\citenamefont {Ivanov},
  \citenamefont {Philcox}, \citenamefont {Cabass}, \citenamefont {Nishimichi},
  \citenamefont {Simonovi\'c},\ and\ \citenamefont
  {Zaldarriaga}}]{Ivanov:2023qzb}%
  \BibitemOpen
  \bibfield  {author} {\bibinfo {author} {\bibfnamefont {M.~M.}\ \bibnamefont
  {Ivanov}}, \bibinfo {author} {\bibfnamefont {O.~H.~E.}\ \bibnamefont
  {Philcox}}, \bibinfo {author} {\bibfnamefont {G.}~\bibnamefont {Cabass}},
  \bibinfo {author} {\bibfnamefont {T.}~\bibnamefont {Nishimichi}}, \bibinfo
  {author} {\bibfnamefont {M.}~\bibnamefont {Simonovi\'c}}, \ and\ \bibinfo
  {author} {\bibfnamefont {M.}~\bibnamefont {Zaldarriaga}},\ }\href {\doibase
  10.1103/PhysRevD.107.083515} {\bibfield  {journal} {\bibinfo  {journal}
  {Phys. Rev. D}\ }\textbf {\bibinfo {volume} {107}},\ \bibinfo {pages}
  {083515} (\bibinfo {year} {2023})},\ \Eprint
  {http://arxiv.org/abs/2302.04414} {arXiv:2302.04414 [astro-ph.CO]}
  \BibitemShut {NoStop}%
\bibitem [{\citenamefont {Scoccimarro}(2000)}]{Scoccimarro:2000sn}%
  \BibitemOpen
  \bibfield  {author} {\bibinfo {author} {\bibfnamefont {R.}~\bibnamefont
  {Scoccimarro}},\ }\href {\doibase 10.1086/317248} {\bibfield  {journal}
  {\bibinfo  {journal} {Astrophys. J.}\ }\textbf {\bibinfo {volume} {544}},\
  \bibinfo {pages} {597} (\bibinfo {year} {2000})},\ \Eprint
  {http://arxiv.org/abs/astro-ph/0004086} {arXiv:astro-ph/0004086} \BibitemShut
  {NoStop}%
\bibitem [{\citenamefont {Sefusatti}\ \emph {et~al.}(2006)\citenamefont
  {Sefusatti}, \citenamefont {Crocce}, \citenamefont {Pueblas},\ and\
  \citenamefont {Scoccimarro}}]{Sefusatti:2006pa}%
  \BibitemOpen
  \bibfield  {author} {\bibinfo {author} {\bibfnamefont {E.}~\bibnamefont
  {Sefusatti}}, \bibinfo {author} {\bibfnamefont {M.}~\bibnamefont {Crocce}},
  \bibinfo {author} {\bibfnamefont {S.}~\bibnamefont {Pueblas}}, \ and\
  \bibinfo {author} {\bibfnamefont {R.}~\bibnamefont {Scoccimarro}},\ }\href
  {\doibase 10.1103/PhysRevD.74.023522} {\bibfield  {journal} {\bibinfo
  {journal} {Phys. Rev.}\ }\textbf {\bibinfo {volume} {D74}},\ \bibinfo {pages}
  {023522} (\bibinfo {year} {2006})},\ \Eprint
  {http://arxiv.org/abs/astro-ph/0604505} {arXiv:astro-ph/0604505 [astro-ph]}
  \BibitemShut {NoStop}%
\bibitem [{\citenamefont {Oddo}\ \emph {et~al.}(2020)\citenamefont {Oddo},
  \citenamefont {Sefusatti}, \citenamefont {Porciani}, \citenamefont {Monaco},\
  and\ \citenamefont {S\'anchez}}]{Oddo:2019run}%
  \BibitemOpen
  \bibfield  {author} {\bibinfo {author} {\bibfnamefont {A.}~\bibnamefont
  {Oddo}}, \bibinfo {author} {\bibfnamefont {E.}~\bibnamefont {Sefusatti}},
  \bibinfo {author} {\bibfnamefont {C.}~\bibnamefont {Porciani}}, \bibinfo
  {author} {\bibfnamefont {P.}~\bibnamefont {Monaco}}, \ and\ \bibinfo {author}
  {\bibfnamefont {A.~G.}\ \bibnamefont {S\'anchez}},\ }\href {\doibase
  10.1088/1475-7516/2020/03/056} {\bibfield  {journal} {\bibinfo  {journal}
  {JCAP}\ }\textbf {\bibinfo {volume} {03}},\ \bibinfo {pages} {056} (\bibinfo
  {year} {2020})},\ \Eprint {http://arxiv.org/abs/1908.01774} {arXiv:1908.01774
  [astro-ph.CO]} \BibitemShut {NoStop}%
\bibitem [{\citenamefont {Eggemeier}\ \emph {et~al.}(2021)\citenamefont
  {Eggemeier}, \citenamefont {Scoccimarro}, \citenamefont {Smith},
  \citenamefont {Crocce}, \citenamefont {Pezzotta},\ and\ \citenamefont
  {S\'anchez}}]{Eggemeier:2021cam}%
  \BibitemOpen
  \bibfield  {author} {\bibinfo {author} {\bibfnamefont {A.}~\bibnamefont
  {Eggemeier}}, \bibinfo {author} {\bibfnamefont {R.}~\bibnamefont
  {Scoccimarro}}, \bibinfo {author} {\bibfnamefont {R.~E.}\ \bibnamefont
  {Smith}}, \bibinfo {author} {\bibfnamefont {M.}~\bibnamefont {Crocce}},
  \bibinfo {author} {\bibfnamefont {A.}~\bibnamefont {Pezzotta}}, \ and\
  \bibinfo {author} {\bibfnamefont {A.~G.}\ \bibnamefont {S\'anchez}},\
  }\href@noop {} {\  (\bibinfo {year} {2021})},\ \Eprint
  {http://arxiv.org/abs/2102.06902} {arXiv:2102.06902 [astro-ph.CO]}
  \BibitemShut {NoStop}%
\bibitem [{\citenamefont {Alkhanishvili}\ \emph {et~al.}(2021)\citenamefont
  {Alkhanishvili}, \citenamefont {Porciani}, \citenamefont {Sefusatti},
  \citenamefont {Biagetti}, \citenamefont {Lazanu}, \citenamefont {Oddo},\ and\
  \citenamefont {Yankelevich}}]{Alkhanishvili:2021pvy}%
  \BibitemOpen
  \bibfield  {author} {\bibinfo {author} {\bibfnamefont {D.}~\bibnamefont
  {Alkhanishvili}}, \bibinfo {author} {\bibfnamefont {C.}~\bibnamefont
  {Porciani}}, \bibinfo {author} {\bibfnamefont {E.}~\bibnamefont {Sefusatti}},
  \bibinfo {author} {\bibfnamefont {M.}~\bibnamefont {Biagetti}}, \bibinfo
  {author} {\bibfnamefont {A.}~\bibnamefont {Lazanu}}, \bibinfo {author}
  {\bibfnamefont {A.}~\bibnamefont {Oddo}}, \ and\ \bibinfo {author}
  {\bibfnamefont {V.}~\bibnamefont {Yankelevich}},\ }\href@noop {} {\
  (\bibinfo {year} {2021})},\ \Eprint {http://arxiv.org/abs/2107.08054}
  {arXiv:2107.08054 [astro-ph.CO]} \BibitemShut {NoStop}%
\bibitem [{\citenamefont {Oddo}\ \emph {et~al.}(2021)\citenamefont {Oddo},
  \citenamefont {Rizzo}, \citenamefont {Sefusatti}, \citenamefont {Porciani},\
  and\ \citenamefont {Monaco}}]{Oddo:2021iwq}%
  \BibitemOpen
  \bibfield  {author} {\bibinfo {author} {\bibfnamefont {A.}~\bibnamefont
  {Oddo}}, \bibinfo {author} {\bibfnamefont {F.}~\bibnamefont {Rizzo}},
  \bibinfo {author} {\bibfnamefont {E.}~\bibnamefont {Sefusatti}}, \bibinfo
  {author} {\bibfnamefont {C.}~\bibnamefont {Porciani}}, \ and\ \bibinfo
  {author} {\bibfnamefont {P.}~\bibnamefont {Monaco}},\ }\href {\doibase
  10.1088/1475-7516/2021/11/038} {\bibfield  {journal} {\bibinfo  {journal}
  {JCAP}\ }\textbf {\bibinfo {volume} {11}},\ \bibinfo {pages} {038} (\bibinfo
  {year} {2021})},\ \Eprint {http://arxiv.org/abs/2108.03204} {arXiv:2108.03204
  [astro-ph.CO]} \BibitemShut {NoStop}%
\bibitem [{\citenamefont {Chen}\ \emph
  {et~al.}(2021{\natexlab{b}})\citenamefont {Chen}, \citenamefont {Vlah},\ and\
  \citenamefont {White}}]{Chen:2021wdi}%
  \BibitemOpen
  \bibfield  {author} {\bibinfo {author} {\bibfnamefont {S.-F.}\ \bibnamefont
  {Chen}}, \bibinfo {author} {\bibfnamefont {Z.}~\bibnamefont {Vlah}}, \ and\
  \bibinfo {author} {\bibfnamefont {M.}~\bibnamefont {White}},\ }\href@noop {}
  {\  (\bibinfo {year} {2021}{\natexlab{b}})},\ \Eprint
  {http://arxiv.org/abs/2110.05530} {arXiv:2110.05530 [astro-ph.CO]}
  \BibitemShut {NoStop}%
\bibitem [{\citenamefont {Xu}\ \emph {et~al.}(2022)\citenamefont {Xu},
  \citenamefont {Mu\~noz},\ and\ \citenamefont {Dvorkin}}]{Xu:2021rwg}%
  \BibitemOpen
  \bibfield  {author} {\bibinfo {author} {\bibfnamefont {W.~L.}\ \bibnamefont
  {Xu}}, \bibinfo {author} {\bibfnamefont {J.~B.}\ \bibnamefont {Mu\~noz}}, \
  and\ \bibinfo {author} {\bibfnamefont {C.}~\bibnamefont {Dvorkin}},\ }\href
  {\doibase 10.1103/PhysRevD.105.095029} {\bibfield  {journal} {\bibinfo
  {journal} {Phys. Rev. D}\ }\textbf {\bibinfo {volume} {105}},\ \bibinfo
  {pages} {095029} (\bibinfo {year} {2022})},\ \Eprint
  {http://arxiv.org/abs/2107.09664} {arXiv:2107.09664 [astro-ph.CO]}
  \BibitemShut {NoStop}%
\bibitem [{\citenamefont {Philcox}(2021{\natexlab{a}})}]{Philcox:2020vbm}%
  \BibitemOpen
  \bibfield  {author} {\bibinfo {author} {\bibfnamefont {O.~H.~E.}\
  \bibnamefont {Philcox}},\ }\href {\doibase 10.1103/PhysRevD.103.103504}
  {\bibfield  {journal} {\bibinfo  {journal} {Phys. Rev. D}\ }\textbf {\bibinfo
  {volume} {103}},\ \bibinfo {pages} {103504} (\bibinfo {year}
  {2021}{\natexlab{a}})},\ \Eprint {http://arxiv.org/abs/2012.09389}
  {arXiv:2012.09389 [astro-ph.CO]} \BibitemShut {NoStop}%
\bibitem [{\citenamefont {Philcox}(2021{\natexlab{b}})}]{Philcox:2021ukg}%
  \BibitemOpen
  \bibfield  {author} {\bibinfo {author} {\bibfnamefont {O.~H.~E.}\
  \bibnamefont {Philcox}},\ }\href {\doibase 10.1103/PhysRevD.104.123529}
  {\bibfield  {journal} {\bibinfo  {journal} {Phys. Rev. D}\ }\textbf {\bibinfo
  {volume} {104}},\ \bibinfo {pages} {123529} (\bibinfo {year}
  {2021}{\natexlab{b}})},\ \Eprint {http://arxiv.org/abs/2107.06287}
  {arXiv:2107.06287 [astro-ph.CO]} \BibitemShut {NoStop}%
\bibitem [{\citenamefont {Akitsu}(2024)}]{future_kaz}%
  \BibitemOpen
  \bibfield  {author} {\bibinfo {author} {\bibfnamefont {K.}~\bibnamefont
  {Akitsu}},\ }\href@noop {} {\  (\bibinfo {year} {2024})},\ \Eprint
  {http://arxiv.org/abs/2410.08998} {arXiv:2410.08998 [astro-ph.CO]}
  \BibitemShut {NoStop}%
\bibitem [{\citenamefont {Ivanov}\ \emph
  {et~al.}(2024{\natexlab{a}})\citenamefont {Ivanov}, \citenamefont
  {Cuesta-Lazaro}, \citenamefont {Mishra-Sharma}, \citenamefont {Obuljen},\
  and\ \citenamefont {Toomey}}]{Ivanov:2024hgq}%
  \BibitemOpen
  \bibfield  {author} {\bibinfo {author} {\bibfnamefont {M.~M.}\ \bibnamefont
  {Ivanov}}, \bibinfo {author} {\bibfnamefont {C.}~\bibnamefont
  {Cuesta-Lazaro}}, \bibinfo {author} {\bibfnamefont {S.}~\bibnamefont
  {Mishra-Sharma}}, \bibinfo {author} {\bibfnamefont {A.}~\bibnamefont
  {Obuljen}}, \ and\ \bibinfo {author} {\bibfnamefont {M.~W.}\ \bibnamefont
  {Toomey}},\ }\href@noop {} {\  (\bibinfo {year} {2024}{\natexlab{a}})},\
  \Eprint {http://arxiv.org/abs/2402.13310} {arXiv:2402.13310 [astro-ph.CO]}
  \BibitemShut {NoStop}%
\bibitem [{\citenamefont {Ivanov}\ \emph
  {et~al.}(2024{\natexlab{b}})\citenamefont {Ivanov}, \citenamefont {Obuljen},
  \citenamefont {Cuesta-Lazaro},\ and\ \citenamefont
  {Toomey}}]{Ivanov:2024xgb}%
  \BibitemOpen
  \bibfield  {author} {\bibinfo {author} {\bibfnamefont {M.~M.}\ \bibnamefont
  {Ivanov}}, \bibinfo {author} {\bibfnamefont {A.}~\bibnamefont {Obuljen}},
  \bibinfo {author} {\bibfnamefont {C.}~\bibnamefont {Cuesta-Lazaro}}, \ and\
  \bibinfo {author} {\bibfnamefont {M.~W.}\ \bibnamefont {Toomey}},\
  }\href@noop {} {\  (\bibinfo {year} {2024}{\natexlab{b}})},\ \Eprint
  {http://arxiv.org/abs/2409.10609} {arXiv:2409.10609 [astro-ph.CO]}
  \BibitemShut {NoStop}%
\bibitem [{\citenamefont {Philcox}\ \emph {et~al.}(2022)\citenamefont
  {Philcox}, \citenamefont {Ivanov}, \citenamefont {Cabass}, \citenamefont
  {Simonovi\'c}, \citenamefont {Zaldarriaga},\ and\ \citenamefont
  {Nishimichi}}]{Philcox:2022frc}%
  \BibitemOpen
  \bibfield  {author} {\bibinfo {author} {\bibfnamefont {O.~H.~E.}\
  \bibnamefont {Philcox}}, \bibinfo {author} {\bibfnamefont {M.~M.}\
  \bibnamefont {Ivanov}}, \bibinfo {author} {\bibfnamefont {G.}~\bibnamefont
  {Cabass}}, \bibinfo {author} {\bibfnamefont {M.}~\bibnamefont {Simonovi\'c}},
  \bibinfo {author} {\bibfnamefont {M.}~\bibnamefont {Zaldarriaga}}, \ and\
  \bibinfo {author} {\bibfnamefont {T.}~\bibnamefont {Nishimichi}},\ }\href
  {\doibase 10.1103/PhysRevD.106.043530} {\bibfield  {journal} {\bibinfo
  {journal} {Phys. Rev. D}\ }\textbf {\bibinfo {volume} {106}},\ \bibinfo
  {pages} {043530} (\bibinfo {year} {2022})},\ \Eprint
  {http://arxiv.org/abs/2206.02800} {arXiv:2206.02800 [astro-ph.CO]}
  \BibitemShut {NoStop}%
\bibitem [{\citenamefont {Bonifacio}\ \emph {et~al.}(2020)\citenamefont
  {Bonifacio}, \citenamefont {Hinterbichler}, \citenamefont {Johnson},
  \citenamefont {Joyce},\ and\ \citenamefont {Rosen}}]{Bonifacio:2019rpv}%
  \BibitemOpen
  \bibfield  {author} {\bibinfo {author} {\bibfnamefont {J.}~\bibnamefont
  {Bonifacio}}, \bibinfo {author} {\bibfnamefont {K.}~\bibnamefont
  {Hinterbichler}}, \bibinfo {author} {\bibfnamefont {L.~A.}\ \bibnamefont
  {Johnson}}, \bibinfo {author} {\bibfnamefont {A.}~\bibnamefont {Joyce}}, \
  and\ \bibinfo {author} {\bibfnamefont {R.~A.}\ \bibnamefont {Rosen}},\ }\href
  {\doibase 10.1007/JHEP07(2020)056} {\bibfield  {journal} {\bibinfo  {journal}
  {JHEP}\ }\textbf {\bibinfo {volume} {07}},\ \bibinfo {pages} {056} (\bibinfo
  {year} {2020})},\ \Eprint {http://arxiv.org/abs/1911.04490} {arXiv:1911.04490
  [hep-th]} \BibitemShut {NoStop}%
\bibitem [{\citenamefont {Philcox}\ \emph
  {et~al.}(2021{\natexlab{a}})\citenamefont {Philcox}, \citenamefont {Ivanov},
  \citenamefont {Zaldarriaga}, \citenamefont {Simonovic},\ and\ \citenamefont
  {Schmittfull}}]{Philcox:2020zyp}%
  \BibitemOpen
  \bibfield  {author} {\bibinfo {author} {\bibfnamefont {O.~H.~E.}\
  \bibnamefont {Philcox}}, \bibinfo {author} {\bibfnamefont {M.~M.}\
  \bibnamefont {Ivanov}}, \bibinfo {author} {\bibfnamefont {M.}~\bibnamefont
  {Zaldarriaga}}, \bibinfo {author} {\bibfnamefont {M.}~\bibnamefont
  {Simonovic}}, \ and\ \bibinfo {author} {\bibfnamefont {M.}~\bibnamefont
  {Schmittfull}},\ }\href {\doibase 10.1103/PhysRevD.103.043508} {\bibfield
  {journal} {\bibinfo  {journal} {Phys. Rev. D}\ }\textbf {\bibinfo {volume}
  {103}},\ \bibinfo {pages} {043508} (\bibinfo {year} {2021}{\natexlab{a}})},\
  \Eprint {http://arxiv.org/abs/2009.03311} {arXiv:2009.03311 [astro-ph.CO]}
  \BibitemShut {NoStop}%
\bibitem [{\citenamefont {Sohn}\ \emph {et~al.}(2024)\citenamefont {Sohn},
  \citenamefont {Wang}, \citenamefont {Fergusson},\ and\ \citenamefont
  {Shellard}}]{Sohn:2024xzd}%
  \BibitemOpen
  \bibfield  {author} {\bibinfo {author} {\bibfnamefont {W.}~\bibnamefont
  {Sohn}}, \bibinfo {author} {\bibfnamefont {D.-G.}\ \bibnamefont {Wang}},
  \bibinfo {author} {\bibfnamefont {J.~R.}\ \bibnamefont {Fergusson}}, \ and\
  \bibinfo {author} {\bibfnamefont {E.~P.~S.}\ \bibnamefont {Shellard}},\
  }\href {\doibase 10.1088/1475-7516/2024/09/016} {\bibfield  {journal}
  {\bibinfo  {journal} {JCAP}\ }\textbf {\bibinfo {volume} {09}},\ \bibinfo
  {pages} {016} (\bibinfo {year} {2024})},\ \Eprint
  {http://arxiv.org/abs/2404.07203} {arXiv:2404.07203 [astro-ph.CO]}
  \BibitemShut {NoStop}%
\bibitem [{\citenamefont {Creminelli}\ \emph {et~al.}(2014)\citenamefont
  {Creminelli}, \citenamefont {Gleyzes}, \citenamefont {Nore{\~n}a},\ and\
  \citenamefont {Vernizzi}}]{Creminelli:2014wna}%
  \BibitemOpen
  \bibfield  {author} {\bibinfo {author} {\bibfnamefont {P.}~\bibnamefont
  {Creminelli}}, \bibinfo {author} {\bibfnamefont {J.}~\bibnamefont {Gleyzes}},
  \bibinfo {author} {\bibfnamefont {J.}~\bibnamefont {Nore{\~n}a}}, \ and\
  \bibinfo {author} {\bibfnamefont {F.}~\bibnamefont {Vernizzi}},\ }\href
  {\doibase 10.1103/PhysRevLett.113.231301} {\bibfield  {journal} {\bibinfo
  {journal} {Phys. Rev. Lett.}\ }\textbf {\bibinfo {volume} {113}},\ \bibinfo
  {pages} {231301} (\bibinfo {year} {2014})},\ \Eprint
  {http://arxiv.org/abs/1407.8439} {arXiv:1407.8439 [astro-ph.CO]} \BibitemShut
  {NoStop}%
\bibitem [{\citenamefont {Bordin}\ \emph {et~al.}(2017)\citenamefont {Bordin},
  \citenamefont {Cabass}, \citenamefont {Creminelli},\ and\ \citenamefont
  {Vernizzi}}]{Bordin:2017hal}%
  \BibitemOpen
  \bibfield  {author} {\bibinfo {author} {\bibfnamefont {L.}~\bibnamefont
  {Bordin}}, \bibinfo {author} {\bibfnamefont {G.}~\bibnamefont {Cabass}},
  \bibinfo {author} {\bibfnamefont {P.}~\bibnamefont {Creminelli}}, \ and\
  \bibinfo {author} {\bibfnamefont {F.}~\bibnamefont {Vernizzi}},\ }\href
  {\doibase 10.1088/1475-7516/2017/09/043} {\bibfield  {journal} {\bibinfo
  {journal} {JCAP}\ }\textbf {\bibinfo {volume} {1709}},\ \bibinfo {pages}
  {043} (\bibinfo {year} {2017})},\ \Eprint {http://arxiv.org/abs/1706.03758}
  {arXiv:1706.03758 [astro-ph.CO]} \BibitemShut {NoStop}%
\bibitem [{\citenamefont {Bordin}\ and\ \citenamefont
  {Cabass}(2020)}]{Bordin:2020eui}%
  \BibitemOpen
  \bibfield  {author} {\bibinfo {author} {\bibfnamefont {L.}~\bibnamefont
  {Bordin}}\ and\ \bibinfo {author} {\bibfnamefont {G.}~\bibnamefont
  {Cabass}},\ }\href {\doibase 10.1088/1475-7516/2020/07/014} {\bibfield
  {journal} {\bibinfo  {journal} {JCAP}\ }\textbf {\bibinfo {volume} {07}},\
  \bibinfo {pages} {014} (\bibinfo {year} {2020})},\ \Eprint
  {http://arxiv.org/abs/2004.00619} {arXiv:2004.00619 [astro-ph.CO]}
  \BibitemShut {NoStop}%
\bibitem [{\citenamefont {Aghanim}\ \emph {et~al.}(2018)\citenamefont {Aghanim}
  \emph {et~al.}}]{Aghanim:2018eyx}%
  \BibitemOpen
  \bibfield  {author} {\bibinfo {author} {\bibfnamefont {N.}~\bibnamefont
  {Aghanim}} \emph {et~al.} (\bibinfo {collaboration} {Planck}),\ }\href@noop
  {} {\  (\bibinfo {year} {2018})},\ \Eprint {http://arxiv.org/abs/1807.06209}
  {arXiv:1807.06209 [astro-ph.CO]} \BibitemShut {NoStop}%
\bibitem [{\citenamefont {Babich}\ \emph {et~al.}(2004)\citenamefont {Babich},
  \citenamefont {Creminelli},\ and\ \citenamefont
  {Zaldarriaga}}]{Babich:2004gb}%
  \BibitemOpen
  \bibfield  {author} {\bibinfo {author} {\bibfnamefont {D.}~\bibnamefont
  {Babich}}, \bibinfo {author} {\bibfnamefont {P.}~\bibnamefont {Creminelli}},
  \ and\ \bibinfo {author} {\bibfnamefont {M.}~\bibnamefont {Zaldarriaga}},\
  }\href {\doibase 10.1088/1475-7516/2004/08/009} {\bibfield  {journal}
  {\bibinfo  {journal} {JCAP}\ }\textbf {\bibinfo {volume} {08}},\ \bibinfo
  {pages} {009} (\bibinfo {year} {2004})},\ \Eprint
  {http://arxiv.org/abs/astro-ph/0405356} {arXiv:astro-ph/0405356} \BibitemShut
  {NoStop}%
\bibitem [{\citenamefont {Gwyn}\ \emph {et~al.}(2013)\citenamefont {Gwyn},
  \citenamefont {Palma}, \citenamefont {Sakellariadou},\ and\ \citenamefont
  {Sypsas}}]{Gwyn:2012mw}%
  \BibitemOpen
  \bibfield  {author} {\bibinfo {author} {\bibfnamefont {R.}~\bibnamefont
  {Gwyn}}, \bibinfo {author} {\bibfnamefont {G.~A.}\ \bibnamefont {Palma}},
  \bibinfo {author} {\bibfnamefont {M.}~\bibnamefont {Sakellariadou}}, \ and\
  \bibinfo {author} {\bibfnamefont {S.}~\bibnamefont {Sypsas}},\ }\href
  {\doibase 10.1088/1475-7516/2013/04/004} {\bibfield  {journal} {\bibinfo
  {journal} {JCAP}\ }\textbf {\bibinfo {volume} {04}},\ \bibinfo {pages} {004}
  (\bibinfo {year} {2013})},\ \Eprint {http://arxiv.org/abs/1210.3020}
  {arXiv:1210.3020 [hep-th]} \BibitemShut {NoStop}%
\bibitem [{\citenamefont {Alishahiha}\ \emph {et~al.}(2004)\citenamefont
  {Alishahiha}, \citenamefont {Silverstein},\ and\ \citenamefont
  {Tong}}]{Alishahiha:2004eh}%
  \BibitemOpen
  \bibfield  {author} {\bibinfo {author} {\bibfnamefont {M.}~\bibnamefont
  {Alishahiha}}, \bibinfo {author} {\bibfnamefont {E.}~\bibnamefont
  {Silverstein}}, \ and\ \bibinfo {author} {\bibfnamefont {D.}~\bibnamefont
  {Tong}},\ }\href {\doibase 10.1103/PhysRevD.70.123505} {\bibfield  {journal}
  {\bibinfo  {journal} {Phys. Rev. D}\ }\textbf {\bibinfo {volume} {70}},\
  \bibinfo {pages} {123505} (\bibinfo {year} {2004})},\ \Eprint
  {http://arxiv.org/abs/hep-th/0404084} {arXiv:hep-th/0404084} \BibitemShut
  {NoStop}%
\bibitem [{\citenamefont {Cabass}\ \emph
  {et~al.}(2023{\natexlab{d}})\citenamefont {Cabass}, \citenamefont {Ivanov},
  \citenamefont {Philcox}, \citenamefont {Simonovic},\ and\ \citenamefont
  {Zaldarriaga}}]{Cabass:2022epm}%
  \BibitemOpen
  \bibfield  {author} {\bibinfo {author} {\bibfnamefont {G.}~\bibnamefont
  {Cabass}}, \bibinfo {author} {\bibfnamefont {M.~M.}\ \bibnamefont {Ivanov}},
  \bibinfo {author} {\bibfnamefont {O.~H.~E.}\ \bibnamefont {Philcox}},
  \bibinfo {author} {\bibfnamefont {M.}~\bibnamefont {Simonovic}}, \ and\
  \bibinfo {author} {\bibfnamefont {M.}~\bibnamefont {Zaldarriaga}},\ }\href
  {\doibase 10.1016/j.physletb.2023.137912} {\bibfield  {journal} {\bibinfo
  {journal} {Phys. Lett. B}\ }\textbf {\bibinfo {volume} {841}},\ \bibinfo
  {pages} {137912} (\bibinfo {year} {2023}{\natexlab{d}})},\ \Eprint
  {http://arxiv.org/abs/2211.14899} {arXiv:2211.14899 [astro-ph.CO]}
  \BibitemShut {NoStop}%
\bibitem [{\citenamefont {Philcox}\ \emph {et~al.}(2020)\citenamefont
  {Philcox}, \citenamefont {Ivanov}, \citenamefont {Simonovi\'c},\ and\
  \citenamefont {Zaldarriaga}}]{Philcox:2020vvt}%
  \BibitemOpen
  \bibfield  {author} {\bibinfo {author} {\bibfnamefont {O.~H.~E.}\
  \bibnamefont {Philcox}}, \bibinfo {author} {\bibfnamefont {M.~M.}\
  \bibnamefont {Ivanov}}, \bibinfo {author} {\bibfnamefont {M.}~\bibnamefont
  {Simonovi\'c}}, \ and\ \bibinfo {author} {\bibfnamefont {M.}~\bibnamefont
  {Zaldarriaga}},\ }\href {\doibase 10.1088/1475-7516/2020/05/032} {\bibfield
  {journal} {\bibinfo  {journal} {JCAP}\ }\textbf {\bibinfo {volume} {05}},\
  \bibinfo {pages} {032} (\bibinfo {year} {2020})},\ \Eprint
  {http://arxiv.org/abs/2002.04035} {arXiv:2002.04035 [astro-ph.CO]}
  \BibitemShut {NoStop}%
\bibitem [{\citenamefont {Ivanov}\ \emph {et~al.}(2021)\citenamefont {Ivanov},
  \citenamefont {Philcox}, \citenamefont {Simonovi\'c}, \citenamefont
  {Zaldarriaga}, \citenamefont {Nishimichi},\ and\ \citenamefont
  {Takada}}]{Ivanov:2021haa}%
  \BibitemOpen
  \bibfield  {author} {\bibinfo {author} {\bibfnamefont {M.~M.}\ \bibnamefont
  {Ivanov}}, \bibinfo {author} {\bibfnamefont {O.~H.~E.}\ \bibnamefont
  {Philcox}}, \bibinfo {author} {\bibfnamefont {M.}~\bibnamefont
  {Simonovi\'c}}, \bibinfo {author} {\bibfnamefont {M.}~\bibnamefont
  {Zaldarriaga}}, \bibinfo {author} {\bibfnamefont {T.}~\bibnamefont
  {Nishimichi}}, \ and\ \bibinfo {author} {\bibfnamefont {M.}~\bibnamefont
  {Takada}},\ }\href@noop {} {\  (\bibinfo {year} {2021})},\ \Eprint
  {http://arxiv.org/abs/2110.00006} {arXiv:2110.00006 [astro-ph.CO]}
  \BibitemShut {NoStop}%
\bibitem [{\citenamefont {Chudaykin}\ \emph {et~al.}(2020)\citenamefont
  {Chudaykin}, \citenamefont {Ivanov}, \citenamefont {Philcox},\ and\
  \citenamefont {Simonovi\'c}}]{Chudaykin:2020aoj}%
  \BibitemOpen
  \bibfield  {author} {\bibinfo {author} {\bibfnamefont {A.}~\bibnamefont
  {Chudaykin}}, \bibinfo {author} {\bibfnamefont {M.~M.}\ \bibnamefont
  {Ivanov}}, \bibinfo {author} {\bibfnamefont {O.~H.~E.}\ \bibnamefont
  {Philcox}}, \ and\ \bibinfo {author} {\bibfnamefont {M.}~\bibnamefont
  {Simonovi\'c}},\ }\href {\doibase 10.1103/PhysRevD.102.063533} {\bibfield
  {journal} {\bibinfo  {journal} {Phys. Rev. D}\ }\textbf {\bibinfo {volume}
  {102}},\ \bibinfo {pages} {063533} (\bibinfo {year} {2020})},\ \Eprint
  {http://arxiv.org/abs/2004.10607} {arXiv:2004.10607 [astro-ph.CO]}
  \BibitemShut {NoStop}%
\bibitem [{\citenamefont {Simonovi{\'c}}\ \emph {et~al.}(2018)\citenamefont
  {Simonovi{\'c}}, \citenamefont {Baldauf}, \citenamefont {Zaldarriaga},
  \citenamefont {Carrasco},\ and\ \citenamefont
  {Kollmeier}}]{Simonovic:2017mhp}%
  \BibitemOpen
  \bibfield  {author} {\bibinfo {author} {\bibfnamefont {M.}~\bibnamefont
  {Simonovi{\'c}}}, \bibinfo {author} {\bibfnamefont {T.}~\bibnamefont
  {Baldauf}}, \bibinfo {author} {\bibfnamefont {M.}~\bibnamefont
  {Zaldarriaga}}, \bibinfo {author} {\bibfnamefont {J.~J.}\ \bibnamefont
  {Carrasco}}, \ and\ \bibinfo {author} {\bibfnamefont {J.~A.}\ \bibnamefont
  {Kollmeier}},\ }\href {\doibase 10.1088/1475-7516/2018/04/030} {\bibfield
  {journal} {\bibinfo  {journal} {JCAP}\ }\textbf {\bibinfo {volume} {1804}},\
  \bibinfo {pages} {030} (\bibinfo {year} {2018})},\ \Eprint
  {http://arxiv.org/abs/1708.08130} {arXiv:1708.08130 [astro-ph.CO]}
  \BibitemShut {NoStop}%
\bibitem [{\citenamefont {Baumann}\ and\ \citenamefont
  {Green}(2022)}]{Baumann:2021ykm}%
  \BibitemOpen
  \bibfield  {author} {\bibinfo {author} {\bibfnamefont {D.}~\bibnamefont
  {Baumann}}\ and\ \bibinfo {author} {\bibfnamefont {D.}~\bibnamefont
  {Green}},\ }\href {\doibase 10.1088/1475-7516/2022/08/061} {\bibfield
  {journal} {\bibinfo  {journal} {JCAP}\ }\textbf {\bibinfo {volume} {08}},\
  \bibinfo {pages} {061} (\bibinfo {year} {2022})},\ \Eprint
  {http://arxiv.org/abs/2112.14645} {arXiv:2112.14645 [astro-ph.CO]}
  \BibitemShut {NoStop}%
\bibitem [{\citenamefont {Barreira}\ \emph {et~al.}(2021)\citenamefont
  {Barreira}, \citenamefont {Lazeyras},\ and\ \citenamefont
  {Schmidt}}]{Barreira:2021ukk}%
  \BibitemOpen
  \bibfield  {author} {\bibinfo {author} {\bibfnamefont {A.}~\bibnamefont
  {Barreira}}, \bibinfo {author} {\bibfnamefont {T.}~\bibnamefont {Lazeyras}},
  \ and\ \bibinfo {author} {\bibfnamefont {F.}~\bibnamefont {Schmidt}},\
  }\href@noop {} {\  (\bibinfo {year} {2021})},\ \Eprint
  {http://arxiv.org/abs/2105.02876} {arXiv:2105.02876 [astro-ph.CO]}
  \BibitemShut {NoStop}%
\bibitem [{\citenamefont {Sullivan}\ \emph {et~al.}(2021)\citenamefont
  {Sullivan}, \citenamefont {Seljak},\ and\ \citenamefont
  {Singh}}]{Sullivan:2021sof}%
  \BibitemOpen
  \bibfield  {author} {\bibinfo {author} {\bibfnamefont {J.~M.}\ \bibnamefont
  {Sullivan}}, \bibinfo {author} {\bibfnamefont {U.}~\bibnamefont {Seljak}}, \
  and\ \bibinfo {author} {\bibfnamefont {S.}~\bibnamefont {Singh}},\ }\href
  {\doibase 10.1088/1475-7516/2021/11/026} {\bibfield  {journal} {\bibinfo
  {journal} {JCAP}\ }\textbf {\bibinfo {volume} {11}},\ \bibinfo {pages} {026}
  (\bibinfo {year} {2021})},\ \Eprint {http://arxiv.org/abs/2104.10676}
  {arXiv:2104.10676 [astro-ph.CO]} \BibitemShut {NoStop}%
\bibitem [{\citenamefont {Lazeyras}\ \emph {et~al.}(2021)\citenamefont
  {Lazeyras}, \citenamefont {Barreira},\ and\ \citenamefont
  {Schmidt}}]{Lazeyras:2021dar}%
  \BibitemOpen
  \bibfield  {author} {\bibinfo {author} {\bibfnamefont {T.}~\bibnamefont
  {Lazeyras}}, \bibinfo {author} {\bibfnamefont {A.}~\bibnamefont {Barreira}},
  \ and\ \bibinfo {author} {\bibfnamefont {F.}~\bibnamefont {Schmidt}},\ }\href
  {\doibase 10.1088/1475-7516/2021/10/063} {\bibfield  {journal} {\bibinfo
  {journal} {JCAP}\ }\textbf {\bibinfo {volume} {10}},\ \bibinfo {pages} {063}
  (\bibinfo {year} {2021})},\ \Eprint {http://arxiv.org/abs/2106.14713}
  {arXiv:2106.14713 [astro-ph.CO]} \BibitemShut {NoStop}%
\bibitem [{\citenamefont {Alam}\ \emph {et~al.}(2017)\citenamefont {Alam} \emph
  {et~al.}}]{Alam:2016hwk}%
  \BibitemOpen
  \bibfield  {author} {\bibinfo {author} {\bibfnamefont {S.}~\bibnamefont
  {Alam}} \emph {et~al.} (\bibinfo {collaboration} {BOSS}),\ }\href {\doibase
  10.1093/mnras/stx721} {\bibfield  {journal} {\bibinfo  {journal} {Mon. Not.
  Roy. Astron. Soc.}\ }\textbf {\bibinfo {volume} {470}},\ \bibinfo {pages}
  {2617} (\bibinfo {year} {2017})},\ \Eprint {http://arxiv.org/abs/1607.03155}
  {arXiv:1607.03155 [astro-ph.CO]} \BibitemShut {NoStop}%
\bibitem [{\citenamefont {Eisenstein}\ \emph {et~al.}(2011)\citenamefont
  {Eisenstein} \emph {et~al.}}]{SDSS:2011jap}%
  \BibitemOpen
  \bibfield  {author} {\bibinfo {author} {\bibfnamefont {D.~J.}\ \bibnamefont
  {Eisenstein}} \emph {et~al.} (\bibinfo {collaboration} {SDSS}),\ }\href
  {\doibase 10.1088/0004-6256/142/3/72} {\bibfield  {journal} {\bibinfo
  {journal} {Astron. J.}\ }\textbf {\bibinfo {volume} {142}},\ \bibinfo {pages}
  {72} (\bibinfo {year} {2011})},\ \Eprint {http://arxiv.org/abs/1101.1529}
  {arXiv:1101.1529 [astro-ph.IM]} \BibitemShut {NoStop}%
\bibitem [{\citenamefont {Kitaura}\ \emph {et~al.}(2016)\citenamefont {Kitaura}
  \emph {et~al.}}]{Kitaura:2015uqa}%
  \BibitemOpen
  \bibfield  {author} {\bibinfo {author} {\bibfnamefont {F.-S.}\ \bibnamefont
  {Kitaura}} \emph {et~al.},\ }\href {\doibase 10.1093/mnras/stv2826}
  {\bibfield  {journal} {\bibinfo  {journal} {Mon. Not. Roy. Astron. Soc.}\
  }\textbf {\bibinfo {volume} {456}},\ \bibinfo {pages} {4156} (\bibinfo {year}
  {2016})},\ \Eprint {http://arxiv.org/abs/1509.06400} {arXiv:1509.06400
  [astro-ph.CO]} \BibitemShut {NoStop}%
\bibitem [{\citenamefont {Bernardeau}\ \emph {et~al.}(2002)\citenamefont
  {Bernardeau}, \citenamefont {Colombi}, \citenamefont {Gaztanaga},\ and\
  \citenamefont {Scoccimarro}}]{Bernardeau:2001qr}%
  \BibitemOpen
  \bibfield  {author} {\bibinfo {author} {\bibfnamefont {F.}~\bibnamefont
  {Bernardeau}}, \bibinfo {author} {\bibfnamefont {S.}~\bibnamefont {Colombi}},
  \bibinfo {author} {\bibfnamefont {E.}~\bibnamefont {Gaztanaga}}, \ and\
  \bibinfo {author} {\bibfnamefont {R.}~\bibnamefont {Scoccimarro}},\ }\href
  {\doibase 10.1016/S0370-1573(02)00135-7} {\bibfield  {journal} {\bibinfo
  {journal} {Phys. Rept.}\ }\textbf {\bibinfo {volume} {367}},\ \bibinfo
  {pages} {1} (\bibinfo {year} {2002})},\ \Eprint
  {http://arxiv.org/abs/astro-ph/0112551} {arXiv:astro-ph/0112551 [astro-ph]}
  \BibitemShut {NoStop}%
\bibitem [{\citenamefont {Sefusatti}\ and\ \citenamefont
  {Komatsu}(2007)}]{Sefusatti:2007ih}%
  \BibitemOpen
  \bibfield  {author} {\bibinfo {author} {\bibfnamefont {E.}~\bibnamefont
  {Sefusatti}}\ and\ \bibinfo {author} {\bibfnamefont {E.}~\bibnamefont
  {Komatsu}},\ }\href {\doibase 10.1103/PhysRevD.76.083004} {\bibfield
  {journal} {\bibinfo  {journal} {Phys. Rev. D}\ }\textbf {\bibinfo {volume}
  {76}},\ \bibinfo {pages} {083004} (\bibinfo {year} {2007})},\ \Eprint
  {http://arxiv.org/abs/0705.0343} {arXiv:0705.0343 [astro-ph]} \BibitemShut
  {NoStop}%
\bibitem [{\citenamefont {Sefusatti}(2009)}]{Sefusatti:2009qh}%
  \BibitemOpen
  \bibfield  {author} {\bibinfo {author} {\bibfnamefont {E.}~\bibnamefont
  {Sefusatti}},\ }\href {\doibase 10.1103/PhysRevD.80.123002} {\bibfield
  {journal} {\bibinfo  {journal} {Phys. Rev. D}\ }\textbf {\bibinfo {volume}
  {80}},\ \bibinfo {pages} {123002} (\bibinfo {year} {2009})},\ \Eprint
  {http://arxiv.org/abs/0905.0717} {arXiv:0905.0717 [astro-ph.CO]} \BibitemShut
  {NoStop}%
\bibitem [{\citenamefont {Taruya}\ \emph {et~al.}(2008)\citenamefont {Taruya},
  \citenamefont {Koyama},\ and\ \citenamefont {Matsubara}}]{Taruya:2008pg}%
  \BibitemOpen
  \bibfield  {author} {\bibinfo {author} {\bibfnamefont {A.}~\bibnamefont
  {Taruya}}, \bibinfo {author} {\bibfnamefont {K.}~\bibnamefont {Koyama}}, \
  and\ \bibinfo {author} {\bibfnamefont {T.}~\bibnamefont {Matsubara}},\ }\href
  {\doibase 10.1103/PhysRevD.78.123534} {\bibfield  {journal} {\bibinfo
  {journal} {Phys. Rev. D}\ }\textbf {\bibinfo {volume} {78}},\ \bibinfo
  {pages} {123534} (\bibinfo {year} {2008})},\ \Eprint
  {http://arxiv.org/abs/0808.4085} {arXiv:0808.4085 [astro-ph]} \BibitemShut
  {NoStop}%
\bibitem [{\citenamefont {Schmidt}(2013)}]{Schmidt:2013nsa}%
  \BibitemOpen
  \bibfield  {author} {\bibinfo {author} {\bibfnamefont {F.}~\bibnamefont
  {Schmidt}},\ }\href {\doibase 10.1103/PhysRevD.87.123518} {\bibfield
  {journal} {\bibinfo  {journal} {Phys. Rev. D}\ }\textbf {\bibinfo {volume}
  {87}},\ \bibinfo {pages} {123518} (\bibinfo {year} {2013})},\ \Eprint
  {http://arxiv.org/abs/1304.1817} {arXiv:1304.1817 [astro-ph.CO]} \BibitemShut
  {NoStop}%
\bibitem [{\citenamefont {Cabass}\ \emph {et~al.}(2018)\citenamefont {Cabass},
  \citenamefont {Pajer},\ and\ \citenamefont {Schmidt}}]{Cabass:2018roz}%
  \BibitemOpen
  \bibfield  {author} {\bibinfo {author} {\bibfnamefont {G.}~\bibnamefont
  {Cabass}}, \bibinfo {author} {\bibfnamefont {E.}~\bibnamefont {Pajer}}, \
  and\ \bibinfo {author} {\bibfnamefont {F.}~\bibnamefont {Schmidt}},\ }\href
  {\doibase 10.1088/1475-7516/2018/09/003} {\bibfield  {journal} {\bibinfo
  {journal} {JCAP}\ }\textbf {\bibinfo {volume} {09}},\ \bibinfo {pages} {003}
  (\bibinfo {year} {2018})},\ \Eprint {http://arxiv.org/abs/1804.07295}
  {arXiv:1804.07295 [astro-ph.CO]} \BibitemShut {NoStop}%
\bibitem [{\citenamefont {Schmidt}\ and\ \citenamefont
  {Kamionkowski}(2010)}]{Schmidt:2010gw}%
  \BibitemOpen
  \bibfield  {author} {\bibinfo {author} {\bibfnamefont {F.}~\bibnamefont
  {Schmidt}}\ and\ \bibinfo {author} {\bibfnamefont {M.}~\bibnamefont
  {Kamionkowski}},\ }\href {\doibase 10.1103/PhysRevD.82.103002} {\bibfield
  {journal} {\bibinfo  {journal} {Phys. Rev. D}\ }\textbf {\bibinfo {volume}
  {82}},\ \bibinfo {pages} {103002} (\bibinfo {year} {2010})},\ \Eprint
  {http://arxiv.org/abs/1008.0638} {arXiv:1008.0638 [astro-ph.CO]} \BibitemShut
  {NoStop}%
\bibitem [{\citenamefont {Ivanov}\ \emph
  {et~al.}(2020{\natexlab{b}})\citenamefont {Ivanov}, \citenamefont
  {Simonovi\'c},\ and\ \citenamefont {Zaldarriaga}}]{Ivanov:2019hqk}%
  \BibitemOpen
  \bibfield  {author} {\bibinfo {author} {\bibfnamefont {M.~M.}\ \bibnamefont
  {Ivanov}}, \bibinfo {author} {\bibfnamefont {M.}~\bibnamefont {Simonovi\'c}},
  \ and\ \bibinfo {author} {\bibfnamefont {M.}~\bibnamefont {Zaldarriaga}},\
  }\href {\doibase 10.1103/PhysRevD.101.083504} {\bibfield  {journal} {\bibinfo
   {journal} {Phys. Rev. D}\ }\textbf {\bibinfo {volume} {101}},\ \bibinfo
  {pages} {083504} (\bibinfo {year} {2020}{\natexlab{b}})},\ \Eprint
  {http://arxiv.org/abs/1912.08208} {arXiv:1912.08208 [astro-ph.CO]}
  \BibitemShut {NoStop}%
\bibitem [{\citenamefont {Alcock}\ and\ \citenamefont
  {Paczynski}(1979)}]{Alcock:1979mp}%
  \BibitemOpen
  \bibfield  {author} {\bibinfo {author} {\bibfnamefont {C.}~\bibnamefont
  {Alcock}}\ and\ \bibinfo {author} {\bibfnamefont {B.}~\bibnamefont
  {Paczynski}},\ }\href {\doibase 10.1038/281358a0} {\bibfield  {journal}
  {\bibinfo  {journal} {Nature}\ }\textbf {\bibinfo {volume} {281}},\ \bibinfo
  {pages} {358} (\bibinfo {year} {1979})}\BibitemShut {NoStop}%
\bibitem [{\citenamefont {Laureijs}\ \emph {et~al.}(2011)\citenamefont
  {Laureijs} \emph {et~al.}}]{Laureijs:2011gra}%
  \BibitemOpen
  \bibfield  {author} {\bibinfo {author} {\bibfnamefont {R.}~\bibnamefont
  {Laureijs}} \emph {et~al.} (\bibinfo {collaboration} {EUCLID}),\ }\href@noop
  {} {\  (\bibinfo {year} {2011})},\ \Eprint {http://arxiv.org/abs/1110.3193}
  {arXiv:1110.3193 [astro-ph.CO]} \BibitemShut {NoStop}%
\bibitem [{\citenamefont {Aghamousa}\ \emph {et~al.}(2016)\citenamefont
  {Aghamousa} \emph {et~al.}}]{Aghamousa:2016zmz}%
  \BibitemOpen
  \bibfield  {author} {\bibinfo {author} {\bibfnamefont {A.}~\bibnamefont
  {Aghamousa}} \emph {et~al.} (\bibinfo {collaboration} {DESI}),\ }\href@noop
  {} {\  (\bibinfo {year} {2016})},\ \Eprint {http://arxiv.org/abs/1611.00036}
  {arXiv:1611.00036 [astro-ph.IM]} \BibitemShut {NoStop}%
\bibitem [{\citenamefont {Schlegel}\ \emph {et~al.}(2019)\citenamefont
  {Schlegel} \emph {et~al.}}]{Schlegel:2019eqc}%
  \BibitemOpen
  \bibfield  {author} {\bibinfo {author} {\bibfnamefont {D.~J.}\ \bibnamefont
  {Schlegel}} \emph {et~al.},\ }\href@noop {} {\  (\bibinfo {year} {2019})},\
  \Eprint {http://arxiv.org/abs/1907.11171} {arXiv:1907.11171 [astro-ph.IM]}
  \BibitemShut {NoStop}%
\bibitem [{\citenamefont {Karkare}\ \emph {et~al.}(2022)\citenamefont
  {Karkare}, \citenamefont {Dizgah}, \citenamefont {Keating}, \citenamefont
  {Breysse},\ and\ \citenamefont {Chung}}]{Karkare:2022bai}%
  \BibitemOpen
  \bibfield  {author} {\bibinfo {author} {\bibfnamefont {K.~S.}\ \bibnamefont
  {Karkare}}, \bibinfo {author} {\bibfnamefont {A.~M.}\ \bibnamefont {Dizgah}},
  \bibinfo {author} {\bibfnamefont {G.~K.}\ \bibnamefont {Keating}}, \bibinfo
  {author} {\bibfnamefont {P.}~\bibnamefont {Breysse}}, \ and\ \bibinfo
  {author} {\bibfnamefont {D.~T.}\ \bibnamefont {Chung}} (\bibinfo
  {collaboration} {Snowmass Cosmic Frontier 5 Topical Group}),\ }in\ \href@noop
  {} {\emph {\bibinfo {booktitle} {{Snowmass 2021}}}}\ (\bibinfo {year}
  {2022})\ \Eprint {http://arxiv.org/abs/2203.07258} {arXiv:2203.07258
  [astro-ph.CO]} \BibitemShut {NoStop}%
\bibitem [{\citenamefont {Fl\"oss}\ \emph {et~al.}(2023)\citenamefont
  {Fl\"oss}, \citenamefont {Biagetti},\ and\ \citenamefont
  {Meerburg}}]{Floss:2022wkq}%
  \BibitemOpen
  \bibfield  {author} {\bibinfo {author} {\bibfnamefont {T.}~\bibnamefont
  {Fl\"oss}}, \bibinfo {author} {\bibfnamefont {M.}~\bibnamefont {Biagetti}}, \
  and\ \bibinfo {author} {\bibfnamefont {P.~D.}\ \bibnamefont {Meerburg}},\
  }\href {\doibase 10.1103/PhysRevD.107.023528} {\bibfield  {journal} {\bibinfo
   {journal} {Phys. Rev. D}\ }\textbf {\bibinfo {volume} {107}},\ \bibinfo
  {pages} {023528} (\bibinfo {year} {2023})},\ \Eprint
  {http://arxiv.org/abs/2206.10458} {arXiv:2206.10458 [astro-ph.CO]}
  \BibitemShut {NoStop}%
\bibitem [{\citenamefont {D'Amico}\ \emph
  {et~al.}(2022{\natexlab{b}})\citenamefont {D'Amico}, \citenamefont {Donath},
  \citenamefont {Lewandowski}, \citenamefont {Senatore},\ and\ \citenamefont
  {Zhang}}]{DAmico:2022osl}%
  \BibitemOpen
  \bibfield  {author} {\bibinfo {author} {\bibfnamefont {G.}~\bibnamefont
  {D'Amico}}, \bibinfo {author} {\bibfnamefont {Y.}~\bibnamefont {Donath}},
  \bibinfo {author} {\bibfnamefont {M.}~\bibnamefont {Lewandowski}}, \bibinfo
  {author} {\bibfnamefont {L.}~\bibnamefont {Senatore}}, \ and\ \bibinfo
  {author} {\bibfnamefont {P.}~\bibnamefont {Zhang}},\ }\href@noop {} {\
  (\bibinfo {year} {2022}{\natexlab{b}})},\ \Eprint
  {http://arxiv.org/abs/2206.08327} {arXiv:2206.08327 [astro-ph.CO]}
  \BibitemShut {NoStop}%
\bibitem [{\citenamefont {D'Amico}\ \emph
  {et~al.}(2022{\natexlab{c}})\citenamefont {D'Amico}, \citenamefont {Donath},
  \citenamefont {Lewandowski}, \citenamefont {Senatore},\ and\ \citenamefont
  {Zhang}}]{DAmico:2022ukl}%
  \BibitemOpen
  \bibfield  {author} {\bibinfo {author} {\bibfnamefont {G.}~\bibnamefont
  {D'Amico}}, \bibinfo {author} {\bibfnamefont {Y.}~\bibnamefont {Donath}},
  \bibinfo {author} {\bibfnamefont {M.}~\bibnamefont {Lewandowski}}, \bibinfo
  {author} {\bibfnamefont {L.}~\bibnamefont {Senatore}}, \ and\ \bibinfo
  {author} {\bibfnamefont {P.}~\bibnamefont {Zhang}},\ }\href@noop {} {\
  (\bibinfo {year} {2022}{\natexlab{c}})},\ \Eprint
  {http://arxiv.org/abs/2211.17130} {arXiv:2211.17130 [astro-ph.CO]}
  \BibitemShut {NoStop}%
\bibitem [{\citenamefont {Philcox}\ \emph
  {et~al.}(2021{\natexlab{b}})\citenamefont {Philcox}, \citenamefont {Hou},\
  and\ \citenamefont {Slepian}}]{Philcox:2021hbm}%
  \BibitemOpen
  \bibfield  {author} {\bibinfo {author} {\bibfnamefont {O.~H.~E.}\
  \bibnamefont {Philcox}}, \bibinfo {author} {\bibfnamefont {J.}~\bibnamefont
  {Hou}}, \ and\ \bibinfo {author} {\bibfnamefont {Z.}~\bibnamefont
  {Slepian}},\ }\href@noop {} {\  (\bibinfo {year} {2021}{\natexlab{b}})},\
  \Eprint {http://arxiv.org/abs/2108.01670} {arXiv:2108.01670 [astro-ph.CO]}
  \BibitemShut {NoStop}%
\bibitem [{\citenamefont {Sohn}\ \emph {et~al.}(2023)\citenamefont {Sohn},
  \citenamefont {Fergusson},\ and\ \citenamefont {Shellard}}]{Sohn:2023fte}%
  \BibitemOpen
  \bibfield  {author} {\bibinfo {author} {\bibfnamefont {W.}~\bibnamefont
  {Sohn}}, \bibinfo {author} {\bibfnamefont {J.~R.}\ \bibnamefont {Fergusson}},
  \ and\ \bibinfo {author} {\bibfnamefont {E.~P.~S.}\ \bibnamefont
  {Shellard}},\ }\href@noop {} {\  (\bibinfo {year} {2023})},\ \Eprint
  {http://arxiv.org/abs/2305.14646} {arXiv:2305.14646 [astro-ph.CO]}
  \BibitemShut {NoStop}%
\bibitem [{\citenamefont {Philcox}(2023{\natexlab{b}})}]{Philcox:2023uwe}%
  \BibitemOpen
  \bibfield  {author} {\bibinfo {author} {\bibfnamefont {O.~H.~E.}\
  \bibnamefont {Philcox}},\ }\href {\doibase 10.1103/PhysRevD.107.123516}
  {\bibfield  {journal} {\bibinfo  {journal} {Phys. Rev. D}\ }\textbf {\bibinfo
  {volume} {107}},\ \bibinfo {pages} {123516} (\bibinfo {year}
  {2023}{\natexlab{b}})},\ \Eprint {http://arxiv.org/abs/2303.08828}
  {arXiv:2303.08828 [astro-ph.CO]} \BibitemShut {NoStop}%
\bibitem [{\citenamefont {Philcox}(2023{\natexlab{c}})}]{Philcox:2023psd}%
  \BibitemOpen
  \bibfield  {author} {\bibinfo {author} {\bibfnamefont {O.~H.~E.}\
  \bibnamefont {Philcox}},\ }\href {\doibase 10.1103/PhysRevD.108.063506}
  {\bibfield  {journal} {\bibinfo  {journal} {Phys. Rev. D}\ }\textbf {\bibinfo
  {volume} {108}},\ \bibinfo {pages} {063506} (\bibinfo {year}
  {2023}{\natexlab{c}})},\ \Eprint {http://arxiv.org/abs/2306.03915}
  {arXiv:2306.03915 [astro-ph.CO]} \BibitemShut {NoStop}%
\bibitem [{\citenamefont {Blas}\ \emph {et~al.}(2011)\citenamefont {Blas},
  \citenamefont {Lesgourgues},\ and\ \citenamefont {Tram}}]{Blas:2011rf}%
  \BibitemOpen
  \bibfield  {author} {\bibinfo {author} {\bibfnamefont {D.}~\bibnamefont
  {Blas}}, \bibinfo {author} {\bibfnamefont {J.}~\bibnamefont {Lesgourgues}}, \
  and\ \bibinfo {author} {\bibfnamefont {T.}~\bibnamefont {Tram}},\ }\href
  {\doibase 10.1088/1475-7516/2011/07/034} {\bibfield  {journal} {\bibinfo
  {journal} {JCAP}\ }\textbf {\bibinfo {volume} {1107}},\ \bibinfo {pages}
  {034} (\bibinfo {year} {2011})},\ \Eprint {http://arxiv.org/abs/1104.2933}
  {arXiv:1104.2933 [astro-ph.CO]} \BibitemShut {NoStop}%
\bibitem [{\citenamefont {Audren}\ \emph {et~al.}(2013)\citenamefont {Audren},
  \citenamefont {Lesgourgues}, \citenamefont {Benabed},\ and\ \citenamefont
  {Prunet}}]{Audren:2012wb}%
  \BibitemOpen
  \bibfield  {author} {\bibinfo {author} {\bibfnamefont {B.}~\bibnamefont
  {Audren}}, \bibinfo {author} {\bibfnamefont {J.}~\bibnamefont {Lesgourgues}},
  \bibinfo {author} {\bibfnamefont {K.}~\bibnamefont {Benabed}}, \ and\
  \bibinfo {author} {\bibfnamefont {S.}~\bibnamefont {Prunet}},\ }\href
  {\doibase 10.1088/1475-7516/2013/02/001} {\bibfield  {journal} {\bibinfo
  {journal} {JCAP}\ }\textbf {\bibinfo {volume} {1302}},\ \bibinfo {pages}
  {001} (\bibinfo {year} {2013})},\ \Eprint {http://arxiv.org/abs/1210.7183}
  {arXiv:1210.7183 [astro-ph.CO]} \BibitemShut {NoStop}%
\bibitem [{\citenamefont {Brinckmann}\ and\ \citenamefont
  {Lesgourgues}(2019)}]{Brinckmann:2018cvx}%
  \BibitemOpen
  \bibfield  {author} {\bibinfo {author} {\bibfnamefont {T.}~\bibnamefont
  {Brinckmann}}\ and\ \bibinfo {author} {\bibfnamefont {J.}~\bibnamefont
  {Lesgourgues}},\ }\href {\doibase 10.1016/j.dark.2018.100260} {\bibfield
  {journal} {\bibinfo  {journal} {Phys. Dark Univ.}\ }\textbf {\bibinfo
  {volume} {24}},\ \bibinfo {pages} {100260} (\bibinfo {year} {2019})},\
  \Eprint {http://arxiv.org/abs/1804.07261} {arXiv:1804.07261 [astro-ph.CO]}
  \BibitemShut {NoStop}%
\bibitem [{\citenamefont {Lewis}(2019)}]{Lewis:2019xzd}%
  \BibitemOpen
  \bibfield  {author} {\bibinfo {author} {\bibfnamefont {A.}~\bibnamefont
  {Lewis}},\ }\href@noop {} {\  (\bibinfo {year} {2019})},\ \Eprint
  {http://arxiv.org/abs/1910.13970} {arXiv:1910.13970 [astro-ph.IM]}
  \BibitemShut {NoStop}%
\bibitem [{\citenamefont {Akitsu}\ \emph {et~al.}(2023)\citenamefont {Akitsu},
  \citenamefont {Li},\ and\ \citenamefont {Okumura}}]{Akitsu:2023eqa}%
  \BibitemOpen
  \bibfield  {author} {\bibinfo {author} {\bibfnamefont {K.}~\bibnamefont
  {Akitsu}}, \bibinfo {author} {\bibfnamefont {Y.}~\bibnamefont {Li}}, \ and\
  \bibinfo {author} {\bibfnamefont {T.}~\bibnamefont {Okumura}},\ }\href
  {\doibase 10.1088/1475-7516/2023/08/068} {\bibfield  {journal} {\bibinfo
  {journal} {JCAP}\ }\textbf {\bibinfo {volume} {08}},\ \bibinfo {pages} {068}
  (\bibinfo {year} {2023})},\ \Eprint {http://arxiv.org/abs/2306.00969}
  {arXiv:2306.00969 [astro-ph.CO]} \BibitemShut {NoStop}%
\bibitem [{\citenamefont {Schmittfull}\ \emph {et~al.}(2015)\citenamefont
  {Schmittfull}, \citenamefont {Baldauf},\ and\ \citenamefont
  {Seljak}}]{Schmittfull:2014tca}%
  \BibitemOpen
  \bibfield  {author} {\bibinfo {author} {\bibfnamefont {M.}~\bibnamefont
  {Schmittfull}}, \bibinfo {author} {\bibfnamefont {T.}~\bibnamefont
  {Baldauf}}, \ and\ \bibinfo {author} {\bibfnamefont {U.}~\bibnamefont
  {Seljak}},\ }\href {\doibase 10.1103/PhysRevD.91.043530} {\bibfield
  {journal} {\bibinfo  {journal} {Phys. Rev.}\ }\textbf {\bibinfo {volume}
  {D91}},\ \bibinfo {pages} {043530} (\bibinfo {year} {2015})},\ \Eprint
  {http://arxiv.org/abs/1411.6595} {arXiv:1411.6595 [astro-ph.CO]} \BibitemShut
  {NoStop}%
\end{thebibliography}%

\end{document}